\documentclass[11pt, a4paper]{book}

\usepackage[UKenglish]{babel}
\usepackage{amsmath, amsfonts, graphicx, hyperref}
\usepackage[hmargin=2.2cm, bottom=3.4cm, top=3.9cm]{geometry}

\allowdisplaybreaks

\hypersetup{
    colorlinks=true,
    linkcolor=black,
    citecolor=black,
    filecolor=black,
    urlcolor=black
}

\def\ssl#1{\rlap{\hbox{$\mskip 3 mu /$}}#1}

\DeclareMathOperator{\arsinh}{arsinh}
\DeclareMathOperator{\Arsinh}{Arsinh}
\DeclareMathOperator{\arcosh}{arcosh}

\DeclareMathOperator{\tr}{tr}
\DeclareMathOperator{\im}{Im}
\DeclareMathOperator{\re}{Re}
\DeclareMathOperator{\res}{res}
\DeclareMathOperator{\const}{const}
\DeclareMathOperator{\diag}{diag}
\DeclareMathOperator{\sign}{sign}

\begin{document}

\pagenumbering{Roman}

\begin{titlepage}

\null\vfill

\begin{center}

\vspace{1cm}

	{\huge \sc wall{\Huge-}crossing in supersymmetric \\ gauge theories \par}

\vspace{1.8cm}

	\includegraphics[width=17mm]{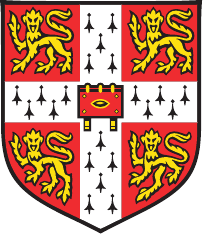}

\vspace{1.5cm}

	{\Large \sc kirill petunin}

\vspace{0.25cm}

	{\sc trinity college}

\vspace{1.5cm}

	{\it Dissertation submitted for the degree of}

\vspace{0.5cm}

	{\Large \sc doctor of philosophy}

\vspace{0.5cm}

	{\it at the}

\vspace{0.5cm}

	{\Large \sc university of cambridge}

\vspace{5cm}

	{\sc November 2011}

\end{center}

\null\vfill

\end{titlepage}

\newpage

\thispagestyle{empty}

\frontmatter

\paragraph{}

This dissertation is the result of my own work and includes nothing which is the outcome of work done in collaboration except where specifically indicated in the text.
It is not substantially the same as any that I have submitted or is being concurrently submitted for a degree, diploma, or other qualification.
The research described in this dissertation was carried out at the Department of Applied Mathematics and Theoretical Physics of the University of Cambridge between October 2008 and November 2011.
Except where reference is made to the work of others, all the results are original, mainly based on the following works:

\paragraph{}

$\bullet$ \quad
  H.~Y.~Chen, N.~Dorey and K.~Petunin,
  ``Wall crossing and instantons in compactified gauge theory.''

\paragraph{}

$\bullet$ \quad
  H.~Y.~Chen and K.~Petunin,
  ``Notes on wall crossing and instanton in compactified gauge theory with matter.''

\paragraph{}

$\bullet$ \quad
  H.~Y.~Chen, N.~Dorey and K.~Petunin,
  ``Moduli space and wall-crossing formulae in higher-rank gauge theories.''

\paragraph{}

$\bullet$ \quad
  N.~Dorey and K.~Petunin,
  ``On the BPS spectrum at the root of the Higgs branch.''

\paragraph{}

Sections 1.1--2.4 are a review, sections 2.5--6.2 contain original research.

\paragraph{}
\paragraph{}

\quad Kirill Petunin

\paragraph{}

\quad November 2011

\thispagestyle{empty}

\newpage

\thispagestyle{empty}

\mainmatter

\chapter*{Abstract}

\paragraph{}

We study $\mathcal{N}=2$ supersymmetric Yang--Mills theory in four dimensions and then compactify it on $\mathbb{R}^{3}\times S^{1}$.
The gauge symmetry of the theory is broken by a vacuum expectation value of the scalar field, which parametrises the moduli space.
The spectrum of BPS states, carrying electric and magnetic charges, is piece-wise constant, changing only when the vacuum expectation value crosses the so-called walls of marginal stability.
For gauge group $SU(2)$, there is only one wall, separating the strong- and weak-coupling regions of the moduli space; for gauge groups $SU(n)$ with $n>2$, there are walls extending into the weak-coupling region, allowing us to study wall-crossing semiclassically.

\paragraph{}

Kontsevich and Soibelman proposed an algebraic construction relating BPS spectra on both sides of a wall of marginal stability.
Given the BPS spectrum on one side of the wall, in principle, one can determine the BPS spectrum on the other side of the wall using the wall-crossing formula.
These formulae are known to correctly relate the strong- and weak-coupling spectra in theories with gauge group $SU(2)$ with and without fundamental flavours; we generalise this result to gauge group $SU(n)$ without flavours in the weak-coupling regime.
In addition, we find the walls of marginal stability in the $SU(n)$ theory at the root of the Higgs branch and, employing the wall-crossing formula, determine the BPS spectrum in all regions of the moduli space.

\paragraph{}

Gaiotto, Moore, and Neitzke (GMN) proposed an ansatz expressing the moduli space metric of $\mathcal{N}=2$ theory on $\mathbb{R}^{3}\times S^{1}$ in terms of a set of integral equations.
It was shown that although the BPS spectrum jumps across the walls, the predicted moduli space metric remains continuous because the BPS spectra in different regions are related by the wall-crossing formulae.
Using the GMN ansatz, we find perturbative and instanton corrections in $\mathbb{R}^{3}\times S^{1}$ for gauge group $SU(2)$ with or without flavours and for gauge group $SU(n)$ without flavours.
For gauge group $SU(n)$, we also demonstrate that the predicted two-instanton metric is continuous across the walls.
Then, we calculate instanton corrections from first principles.
We find that the overall factor of fluctuations of fermionic and bosonic fields in $\mathbb{R}^{3}\times S^{1}$ is a non-trivial function, reproducing the factor coming from purely electrically charged states in the GMN ansatz.
We find perfect agreement between the GMN prediction and the first-principles result.
We also take the limit of small radius of the compactified dimension finding one- and two-instanton corrections in three dimensions, recovering some of the previously known semiclassical results.

\tableofcontents

\chapter{N=2 supersymmetric Yang--Mills theory} 
\label{ch: supersymmetry}

\paragraph{}

In this chapter, we review some facts about $\mathcal{N}=2$ supersymmetric Yang--Mills theories and introduce our conventions.
We start by constructing the microscopic action for $\mathcal{N}=2$ theory using the $\mathcal{N}=1$ formalism.
Then, we show how the gauge symmetry of the theory is broken by the Higgs mechanism and discuss the resulting moduli space.
We also discuss the electric-magnetic duality and central charges in the $\mathcal{N}=2$ supersymmetry algebra.

\paragraph{}

In particular, we focus on the easiest non-trivial case of the theory with gauge group $SU(2)$, considered by Seiberg and Witten \cite{SW, SW2}.
At low energy, the gauge symmetry is broken down to $U(1)$ by a vacuum expectation value of the scalar field $\phi$.
The moduli space can be parametrised by the gauge-invariant Casimir operator $u=\langle\tr\phi^{2}\rangle$.
Different vacuum expectation values lead to physically different theories.
The problem of finding the spectrum was solved by considering monodromies of the moduli space, which consists of two regions separated by a wall of marginal stability~\footnote{
Direct semiclassical tests of the Seiberg--Witten solution itself were conducted in \cite{FP, DKM, Nekrasov}.
}.
In the $u$ plane, this wall has the form of a curve topologically equivalent to a circle: outside the curve, in the weak-coupling region, an infinite set of BPS states is present; as the VEV crosses the curve, they decay into a finite set of states in the strong-coupling region: these states are responsible for the strong-coupling singularities where they become massless.

\paragraph{}

In the more general case of a theory with gauge group $SU(n)$, the gauge symmetry breaks down to $U(1)^{n-1}$, and some elements of the analysis performed by Seiberg and Witten are still applicable.
Although this theory has not been completely solved, its weak-coupling spectrum \cite{Fraser Hollowood} as well as strong-coupling spectrum for the $SU(3)$ theory \cite{Taylor, Taylor 2} have been found: solving the monodromies problem, the charges of states can be conveniently expressed in terms of roots of the gauge group.
When $n>2$, the walls of marginal stability exist at weak and at strong coupling, and their structure becomes more complicated.

\paragraph{}

Finally, we compactify the four-dimensional $\mathcal{N}=2$ theory on $\mathbb{R}^{3}\times S^{1}$ \cite{SW3} and show how the resulting moduli space metric can be described in terms of K\"ahler potentials and symplectic forms \cite{GMN}.
We define Darboux coordinates serving as auxiliary functions allowing one to find the metric.
This construction will later be used extensively in computing the perturbative and instanton corrections, corresponding to the BPS states, in theories compactified on $\mathbb{R}^{3}\times S^{1}$.
When the corrections are not taken into account, we find the exact expression for Darboux coordinates.
The BPS corrections will be considered in the following chapters using the Kontsevich--Soibelman wall-crossing formula.

%%%%%%%%%%%%%%%%%%%%%%%%%%%%%%%%%%%%%%%%%%%%%%%%%%%%%%%%%%%%%

\section{N=1 and N=2 supersymmetric actions}

\paragraph{}

First of all, we introduce our conventions and construct the supersymmetric actions.
$\mathcal{N}=2$ super Yang--Mills action for the theory without flavours combines a scalar field $\phi$, a pair of two-component spinor fields $\psi,\lambda$, and a vector field $A_{\mu}$ in Minkowski space in a single supersymmetric multiplet.
One can construct the $\mathcal{N}=1$ action and then promote it to the more constrained $\mathcal{N}=2$ action.

\paragraph{}

In the $\mathcal{N}=2$ theory, it is necessary to consider the fields $\phi$, $\psi$, $\lambda$, and $A_{\mu}$ in the adjoint representation.
We will introduce a basis $T^{a}$ for the Lie algebra of the gauge group in the space of $n\times n$ matrices, so that every field $X$ can be written as a sum of its components counted by index $a$ as $X^{a}T^{a}$ (implying tensor summation for every pair of repeating indices).
We normalise the basis so that $\tr(T^{a}T^{b})=\delta^{ab}$; the commutation relation is $[T^{a},T^{b}]=i f^{abc}T^{c}$, where $f^{abc}$ is antisymmetric.
E.g., in the case of gauge group $SU(2)$, $T^{a}$ are Pauli matrices divided by $\sqrt{2}$, $\{a,b,c\}\subset\{1,2,3\}$, and $f^{123}=\sqrt{2}$.

\paragraph{}

Introduce the set of mutually anticommuting variables $\theta^{\alpha}$ and $\bar\theta_{\dot\alpha}$ ($\alpha\in\{1,2\}$, $\theta\theta=\theta^{\alpha}\theta_{\alpha}$, $\bar\theta\bar\theta=\bar\theta_{\dot\alpha}\bar\theta^{\dot\alpha}$).
These indices can be raised and lowered by acting on them with antisymmetric symbols $\epsilon^{\alpha\beta}$, $\epsilon^{\dot\alpha\dot\beta}$ and $\epsilon_{\alpha\beta}$, $\epsilon_{\dot\alpha\dot\beta}$; we define $\epsilon^{12}=\epsilon^{\dot 1\dot 2}=1$, $\epsilon_{12}=\epsilon_{\dot 1\dot 2}=-1$.
The signature of Minkowski space is defined as $(+,-,-,-)$ with respect to its $0,1,2,3$ space-time components~\footnote{
Dirac matrices are defined as $\gamma^{\mu}=\left(\begin{array}{ll}0&\sigma^{\mu}\\\bar\sigma^{\mu}&0\end{array}\right)$ where $\sigma^{0}=\bar\sigma^{0}=\left(\begin{array}{rr}1&0\\0&1\end{array}\right)$, $\sigma^{1}=-\bar\sigma^{1}=\left(\begin{array}{rr}0&1\\1&0\end{array}\right)$, $\sigma^{2}=-\bar\sigma^{2}=\left(\begin{array}{rr}0&-i\\i&0\end{array}\right)$, $\sigma^{3}=-\bar\sigma^{3}=\left(\begin{array}{rr}1&0\\0&-1\end{array}\right)$.
The indices of $\sigma^{\mu}$ and $\bar\sigma^{\mu}$ are defined as $\sigma_{\alpha\dot\alpha}^{\mu}$ and $\bar\sigma^{\mu\,\dot\alpha\alpha}=\epsilon^{\alpha\beta}\epsilon^{\dot\alpha\dot\beta}\sigma_{\beta\dot\beta}^{\mu}$.
}.
The superspace derivatives $D_{\alpha}$, $\bar D_{\dot\alpha}$ and the supercharges $Q_{\alpha}$, $\bar Q_{\dot\alpha}$ are then given by \cite{Wess Bagger}
\begin{align}
D_{\alpha} = \frac{\partial}{\partial\theta^{\alpha}}+i\sigma_{\alpha\dot\alpha}^{\mu}\bar\theta^{\dot\alpha}\partial_{\mu}
\,
\quad
\bar D_{\dot\alpha} = -\frac{\partial}{\partial\bar\theta^{\dot\alpha}}-i\theta^{\alpha}\sigma_{\alpha\dot\alpha}^{\mu}\partial_{\mu}
\,;
\\
Q_{\alpha} = \frac{\partial}{\partial\theta^{\alpha}}-i\sigma_{\alpha\dot\alpha}^{\mu}\bar\theta^{\dot\alpha}\partial_{\mu}
\,
\quad
\bar Q_{\dot\alpha} = -\frac{\partial}{\partial\bar\theta^{\dot\alpha}}+i\theta^{\alpha}\sigma_{\alpha\dot\alpha}^{\mu}\partial_{\mu}
\,.
\end{align}

\paragraph{}

We start by constructing the $\mathcal{N}=1$ action.
In order to do this, it is convenient to gather the fields $\lambda$ and $A_{\mu}$ in a vector superfield $V=V^{\dagger}$, which can be expressed in Wess--Zumino gauge as
\begin{equation}
V = (\theta\sigma^{\mu}\bar\theta)A_{\mu}+i(\theta\theta)(\bar\theta\bar\lambda)-i(\bar\theta\bar\theta)(\theta\lambda)+\frac{1}{2}D(\theta\theta)(\bar\theta\bar\theta)
\,.
\end{equation}
Using this superfield, one can construct
\begin{equation}
\label{superfield W}
W_{\alpha} = -\frac{1}{8g}(\bar D_{\dot\alpha}\bar D^{\dot\alpha}) \left( e^{-2gV}D_{\alpha}e^{2gV} \right)
\,,
\end{equation}
which is a chiral superfield (i.e., $\bar D_{\dot\beta}W_{\alpha}=0$); here we introduce the gauge coupling constant $g$.
The fields $\phi$ and $\psi$ are grouped together in another chiral superfield, $\Phi$, and we will use the adjoint representation for these fields:
\begin{equation}
\label{superfield Phi}
\Phi = \phi+\sqrt{2}\theta\psi+(\theta\theta)f+i(\theta\sigma^{\mu}\bar\theta)\partial_{\mu}\phi-\frac{i}{\sqrt{2}}(\theta\theta)(\partial_{\mu}\psi\sigma^{\mu}\bar\theta)-\frac{1}{4}(\theta\theta)(\bar\theta\bar\theta)\partial_{\mu}\partial^{\mu}\phi
\,.
\end{equation}
Actions invariant under supersymmetry transformations can be constructed by projecting a chiral superfield on its $F$-term and projecting a general field on its $D$-term: we define the $\mathcal{N}=1$ Lagrangian densities for field and for matter as
\begin{equation}
\label{Lagrangian densities}
\mathcal{L}_{\rm F} = \frac{1}{8\pi} \im \int d^{2}\theta \left(
\tau\tr\left( W^{\alpha}W_{\alpha} \right)
\right)
\,,
\quad
\mathcal{L}_{\rm M} = \frac{1}{g^{2}} \int d^{2}\theta \, d^{2}\bar\theta \tr\left(
\Phi^{\dagger}e^{2gV}\Phi
\right)
\end{equation}
where the integration for anticommuting variables is defined as $\int d^{2}\theta(\theta\theta)=1$, $\int d^{2}\bar\theta(\bar\theta\bar\theta)=1$, and the complex constant is
\begin{equation}
\label{tau}
\tau=\frac{\Theta}{2\pi}+\frac{4\pi i}{g^{2}}
\,.
\end{equation}
Since covariant derivatives of $W_{\alpha}$ with respect to $\bar\theta^{\dot\beta}$ yield zero, $\mathcal{L}_{\rm F}$ does not contain integration over $\bar\theta^{\dot\beta}$.
In the adjoint representation, covariant partial derivatives are given as $\nabla_{\mu}=\partial_{\mu}-ig[A_{\mu},\ ]$, the field strength is $F_{\mu\nu}=\partial_{\mu}A_{\nu}-\partial_{\nu}A_{\mu}-ig[A_{\mu},A_{\nu}]$~\footnote{
Its gauge components are $F_{\mu\nu}^{a}=\partial_{\mu}A_{\nu}^{a}-\partial_{\nu}A_{\mu}^{a}+gf^{abc}A_{\mu}^{b}A_{\nu}^{c}$.
}, and the dual field strength is $\tilde{F}^{\mu\nu}=\frac{1}{2}\epsilon^{\mu\nu\rho\sigma}F_{\rho\sigma}$ where $\epsilon^{\mu\nu\rho\sigma}$ is antisymmetric with $\epsilon^{0123}=1$.
After expanding (\ref{Lagrangian densities}), the two parts of the $\mathcal{N}=1$ action become~\footnote{
In these conventions,
\begin{equation}
W_{\alpha} =
-\frac{1}{4}(\bar D_{\dot\alpha}\bar D^{\dot\alpha})(D_{\alpha}V)+\frac{g}{4}(\bar D_{\dot\alpha}\bar D^{\dot\alpha})[V,D_{\alpha}V] =
-i\lambda_{\alpha}+\theta_{\alpha}D+(\theta\theta)(\sigma_{\alpha\dot\alpha}^{\mu}\nabla_{\mu}\bar\lambda^{\dot\alpha})+(\sigma^{\mu\nu})_{\alpha}{}^{\beta}\theta_{\beta}F_{\mu\nu}
\end{equation}
where $(\sigma^{\mu\nu})_{\alpha}{}^{\beta}=\frac{i}{4}(\sigma_{\alpha\dot\gamma}^{\mu}\bar\sigma^{\nu\,\dot\gamma\beta}-\sigma_{\alpha\dot\gamma}^{\nu}\bar\sigma^{\mu\,\dot\gamma\beta})$ and $(\bar\sigma^{\mu\nu})^{\dot\alpha}{}_{\dot\beta}=\frac{i}{4}(\bar\sigma^{\mu\,\dot\alpha\gamma}\sigma_{\gamma\dot\beta}^{\nu}-\bar\sigma^{\nu\,\dot\alpha\gamma}\sigma_{\gamma\dot\beta}^{\mu})$ are conjugate $2\times 2$ matrices, so that the generators of the Lorentz group are $M^{\mu\nu}=\frac{i}{4}[\gamma^{\mu},\gamma^{\nu}]=\diag(\sigma^{\mu\nu},\bar\sigma^{\mu\nu})$.
}
\begin{align}
\label{action field}
S_{\rm F} & = \frac{1}{g^{2}} \int d^{4}x \, \tr\left(
-\frac{1}{4}F_{\mu\nu}F^{\mu\nu}+\frac{g^{2}\Theta}{32\pi^{2}}F_{\mu\nu}\tilde{F}^{\mu\nu}-i\lambda\sigma^{\mu}\nabla_{\mu}\bar\lambda+\frac{1}{2}D^{2}
\right)
\,,
\\
\label{action matter}
S_{\rm M} & = \frac{1}{g^{2}} \int d^{4}x \, \tr\left(
f^{\dagger}f+g\phi^{\dagger}D\phi+i\sqrt{2}g\phi^{\dagger}(\lambda\psi)-i\sqrt{2}g(\bar\psi\bar\lambda)\phi-i\psi\sigma^{\mu}\nabla_{\mu}\bar\psi+(\nabla_{\mu}\phi)^{\dagger}\nabla^{\mu}\phi
\right)
\,.
\end{align}

\paragraph{}

Let us now turn to the $\mathcal{N}=2$ supersymmetry, which imposes additional requirements on the Lagrangian.
The new Lagrangian must be symmetric with respect to the $SU(2)_{R}$ rotation of the spinor fields $\psi$ and $\lambda$.
From now on, all fields are in the adjoint representation, so the terms involving $D$ in (\ref{action field}) and (\ref{action matter}) become $\frac{1}{g^{2}}\int d^{4}x\tr\left(\frac{1}{2}D^{2}+gD[\phi^{\dagger},\phi]\right)$: this expression has no derivatives of $D$, allowing us to eliminate $D$ using its equation of motion in favour of $\phi$.
The auxiliary field $f$ gives no contribution and can be eliminated.

\paragraph{}

After these refinements, the action of $\mathcal{N}=2$ supersymmetric Yang--Mills theory in four dimensions can be expressed as
\begin{equation}
\label{action N2}
\begin{aligned}
S = \frac{1}{g^{2}} \int d^{4}x \, \tr & \left(
-\frac{1}{4}F_{\mu\nu}F^{\mu\nu}+\frac{g^{2}\Theta}{32\pi^{2}}F_{\mu\nu}\tilde{F}^{\mu\nu}+(\nabla_{\mu} \phi)^{\dagger}\nabla^{\mu}\phi-i\lambda\sigma^{\mu}\nabla_{\mu}\bar\lambda-i\psi\sigma^{\mu}\nabla_{\mu}\bar\psi
\right.
\\
& \left.
+i\sqrt{2}g\phi^{\dagger}\{\lambda,\psi\}-i\sqrt{2}g\{\bar\psi,\bar\lambda\}\phi-\frac{1}{2}g^{2}[\phi^{\dagger},\phi]^{2}
\right)
\,.
\end{aligned}
\end{equation}

\paragraph{}

In addition to the $\mathcal{N}=2$ vector multiplet of the pure theory, there can also be $\mathcal{N}=2$ matter hypermultiplets.
In terms of $\mathcal{N}=1$ notations, each hypermultiplet contains a chiral superfield $Q_{i}$ and an antichiral superfield $\tilde Q_{i}^{\dagger}$, both transforming under the same representation of the gauge group, where $i$ labels different hypermultiplets.
The Lagrangian density for hypermultiplets in $\mathcal{N}=2$ supersymmetric theory with $N_{f}$ flavours including mixing terms with the vector multiplet has the following form:
\begin{equation}
\begin{aligned}
\mathcal{L}_{\rm H} = \frac{1}{g^{2}} \sum_{i=1}^{N_{f}} & \left(
\int d^{2}\theta \, d^{2}\bar\theta \tr \left(
Q_{i}^{\dagger}e^{2gV}Q_{i}+\tilde Q_{i}^{\dagger}e^{2gV}\tilde Q_{i}
\right) +
\right.
\\
& \left.
\int d^{2}\theta \tr \left(
\tilde Q_{i}(\Phi+m_{i})Q_{i}
\right) +
\int d^{2}\bar\theta \tr \left(
Q_{i}^{\dagger}(\Phi^{\dagger}+\bar m_{i})\tilde Q_{i}^{\dagger}
\right)
\right)
\end{aligned}
\end{equation}
where the flavour masses $m_{i}$ are complex parameters, which we will denote collectively as $\vec m=(m_{1},\dots,m_{N_{f}})$.

%%%%%%%%%%%%%%%%%%%%%%%%%%%%%%%%%%%%%%%%%%%%%%%%%%%%%%%%%%%%%

\section{Gauge symmetry breaking in N=2 theory}

\paragraph{}

The $\mathcal{N}=2$ action (\ref{action N2}) contains the bosonic term
\begin{equation}
\label{bosonic potential}
U(\phi) = \frac{1}{2} \int d^{4}x \, \tr \, [\phi^{\dagger},\phi]^{2}
\,.
\end{equation}
It is always non-negative and, therefore, breaks supersymmetry except the case when it is always zero.
However, it does not mean that one must require $\phi=0$: it is necessary and sufficient that $\phi$ and $\phi^{\dagger}$ commute, in other words, that the vacuum expectation value (VEV), $\langle\phi\rangle$, belongs to the Cartan subalgebra of the gauge group, in effect, breaking the gauge symmetry by the Higgs mechanism to $U(1)^{r}$ where $r$ is the rank of the Cartan subalgebra (for $SU(n)$, we have $r=n-1$).

\paragraph{}

The moduli space can be parametrised by $r$ complex variables.
Introduce a basis $\hat H_{I}$ of diagonal matrices generating the Cartan subalgebra ($1\le I\le r$; from now on, hats above letters denote matrices of the gauge group).
$\langle\phi\rangle$ can be chosen to be a linear combination of $\hat H_{I}$ since the VEV $\langle\phi\rangle$ is given up to a gauge transformation.
The commutator of the elements of the Cartan subalgebra with an element $\hat A$ corresponding to a root $\vec\alpha_{\hat A}$ of the algebra is
\begin{equation}
[\vec{\hat H},\hat A] = \vec\alpha_{\hat A}\hat A
\,.
\end{equation}
Let $\vec a$ be the VEV in the above-mentioned basis.
One obvious consequence of non-vanishing $\langle\phi\rangle$ is generating masses by the kinetic term, $(\nabla_{\mu}\phi)^{\dagger}\nabla^{\mu}\phi$.
Explicitly, for every root $\vec\alpha_{\hat A}$, the mass of the corresponding $W$ boson charged under $\vec\alpha_{\hat A}$ is equal to $|\vec a\vec\alpha_{\hat A}|$~\footnote{
$\vec a\,\vec b$ denotes the scalar product of two vectors, $\vec a$ and $\vec b$, here and everywhere else in the text (we omit the dot in $\vec a\cdot\vec b$).
}.

\paragraph{}

The special case of the $\mathcal{N}=2$ theory with gauge group $SU(2)$ was solved by Seiberg and Witten \cite{SW, SW2}.
In this case, the condition $[\phi^{\dagger},\phi]=0$ implies that by performing gauge transformations, we may choose the scalar field to be proportional to the third Pauli matrix.
We define the vacuum expectation value as $\langle\phi\rangle=a\sigma^{3}/2$ with a complex parameter $a$ (also referred to as VEV).
For $a\ne 0$, the gauge symmetry is broken to $U(1)$.
It is convenient to describe gauge inequivalent vacua using a gauge invariant complex expectation value, $u=\langle\tr\phi^{2}\rangle$, whose classical value is simply $a^{2}/2$.
Then, the perturbative spectrum consists of $W^{\pm}$ bosons with electric charge $\pm 1$ and quarks with electric charge $\pm 1/2$ and one non-zero flavour charge $1$ or $-1$.

\paragraph{}

In Lagrangian formalism, the $\mathcal{N}=2$ supersymmetry can be made manifest by using an extra set of anticommuting variables, $\tilde\theta^{\alpha}$ and $\bar{\tilde\theta}_{\dot\alpha}$ (with the same conventions as above).
We define the $\mathcal{N}=2$ prepotential $\mathcal{F}(\Psi)$ as some function of a chiral superfield $\Psi$ and construct the general action constrained by the $\mathcal{N}=2$ supersymmetry \cite{Grimm Sohnius Wess}, following the approach of \cite{Seiberg}:
\begin{equation}
\label{action N2 general}
S = \frac{1}{4\pi}\im\int d^{4}x \, d^{2}\theta \, d^{2}\tilde\theta \, \mathcal{F}(\Psi)
\,.
\end{equation}
The chiral $\mathcal{N}=2$ superfield $\Psi$ can be expanded as
\begin{equation}
\label{N2 chiral}
\Psi\left(x,\theta,\bar\theta,\tilde\theta,\bar{\tilde\theta}\right) = \Phi(\tilde y,\theta)+\sqrt{2}\tilde\theta^{\alpha}W_{\alpha}(\tilde y,\theta)+\tilde\theta^{\alpha}\tilde\theta_{\alpha}G(\tilde y,\theta)
\end{equation}
where the constituent superfields are functions of $\tilde y^{\mu}=x^{\mu} +i\theta\sigma^{\mu}\bar\theta+i\tilde\theta\sigma^{\mu}\bar{\tilde\theta}$, ensuring that they are chiral.
After expanding the prepotential in powers of $\tilde\theta$ and integrating over $\tilde\theta$, the low-energy effective action becomes
\begin{equation}
\label{action N2 effective}
S = \frac{1}{8\pi} \im \int d^{4}x \left(
\int d^{2}\theta \, \mathcal{F}_{ab}(\Phi)W^{a\alpha}W_{\alpha}^{b}+
2\int d^{2}\theta \, G^{a}\mathcal{F}_{a}(\Phi)
\right)
\,,
\end{equation}
\begin{equation}
\mathcal{F}_{a}(\Phi) = \frac{\partial\mathcal{F}(\Phi)}{\partial\Phi^{a}}
\,,
\quad
\mathcal{F}_{ab}(\Phi) = \frac{\partial^{2}\mathcal{F}(\Phi)}{\partial\Phi^{a}\partial\Phi^{b}}
\,.
\end{equation}
From (\ref{Lagrangian densities}), one can see that to obtain the classical $\mathcal{N}=2$ supersymmetric Yang--Mills action (\ref{action N2}) from this expression, the prepotential should be defined as
\begin{equation}
\mathcal{F}(\Psi) = \frac{1}{2}\tr\left( \tau\Psi^{2} \right)
\end{equation}
where the complex parameter $\tau$ is given by its classical value (\ref{tau}), the superfields $\Phi$ and $W_{\alpha}$ in (\ref{N2 chiral}) should be identified with the $\mathcal{N}=1$ expressions (\ref{superfield Phi}) and (\ref{superfield W}), and $G$ should be defined as
\begin{equation}
G(\tilde y,\theta) = \int d^{2}\bar\theta\left( \Phi(\tilde y-i\theta\sigma\bar\theta,\theta,\bar\theta) \right)^{\dagger} \exp\left( 2gV(\tilde y-i\theta\sigma\bar\theta,\theta,\bar\theta) \right)
\,.
\end{equation}

%%%%%%%%%%%%%%%%%%%%%%%%%%%%%%%%%%%%%%%%%%%%%%%%%%%%%%%%%%%%%

\section{Moduli space and duality}

\paragraph{}

Consider the theory with gauge group $U(1)$ after gauge symmetry breaking.
The $\mathcal{N}=2$ action (\ref{action N2 effective}) is
\begin{equation}
\label{action N2 effective abelian}
S = \frac{1}{8\pi} \im \int d^{4}x \left(
\int d^{2}\theta \, \mathcal{F}''(\Phi)W^{\alpha}W_{\alpha}+
2\int d^{2}\theta \, d^{2}\bar\theta \, \Phi^{\dagger}\mathcal{F}'(\Phi)
\right)
\,.
\end{equation}
The action of the scalar field $\phi$ is encoded in the second term of (\ref{action N2 effective abelian}).
To extract it, we write out $\Phi$ and Taylor-expand $\mathcal{F}'(\Phi)$ at $\phi$ using (\ref{superfield Phi}); then, integrating over the anticommuting variables, we obtain
\begin{equation}
\label{flat metric}
\frac{1}{4\pi} \im \int d^{4}x \, \mathcal{F}''(\phi) \, \partial_{\mu}\bar\phi \, \partial^{\mu}\phi
\,.
\end{equation}
This is a sigma model whose metric is given by
\begin{equation}
g = \im \, \mathcal{F}''(a) \, da \, d\bar a = \im \, \tau(a) \, da \, d\bar a
\end{equation}
where $a$ is the VEV, $\bar a$ is its complex conjugate, and $\tau(a)=\mathcal{F}''(a)$ is the complex coupling.

\paragraph{}

Let us now show how to dualise \cite{SW} the low-energy action (\ref{action N2 effective abelian}), following \cite{Bilal}.
Define a superfield $\Phi_{D}$ dual to $\Phi$ by setting $\Phi_{D}=\mathcal{F}'(\Phi)$ and a prepotential $\mathcal{F}_{D}'(\Phi_{D})$ dual to $\mathcal{F}(\Phi)$ by setting $\mathcal{F}_{D}'(\Phi_{D}) = -\Phi$.
This transformation obeys
\begin{equation}
\mathcal{F}_{D}(\Phi_{D}) = \mathcal{F}(\Phi)-\Phi\Phi_{D}
\,.
\end{equation}
Making use of these relations, we can rewrite the second term in (\ref{action N2 effective abelian}) in terms of the dual variables:
\begin{equation}
\begin{aligned}
\frac{1}{4\pi} \im \int d^{4}x \, d^{2}\theta \, d^{2}\bar\theta \, \Phi^{\dagger}\mathcal{F}'(\Phi)
& = \frac{1}{4\pi} \im \int d^{4}x \, d^{2}\theta \, d^{2}\bar\theta \left( -\mathcal{F}_{D}'(\Phi_{D}) \right)^{\dagger} \Phi_{D}
\\
& = \frac{1}{4\pi} \im \int d^{4}x \, d^{2}\theta \, d^{2}\bar\theta \, \Phi_{D}^{\dagger} \, \mathcal{F}_{D}'(\Phi_{D})
\,.
\end{aligned}
\end{equation}
Now, consider the first term in (\ref{action N2 effective abelian}).
To dualise the action, we introduce a Lagrange multiplier in the functional integral:
\begin{equation}
\begin{aligned}
& \int \mathcal{D}V \exp\left(
\frac{i}{8\pi} \im \int d^{4}x \, d^{2}\theta \, \mathcal{F}''(\Phi)W^{\alpha}W_{\alpha}
\right) =
\\
& \int \mathcal{D}W \mathcal{D}V_{D} \exp\left(
\frac{i}{8\pi} \im \int d^{4}x \left( \int d^{2}\theta \, \mathcal{F}''(\Phi)W^{\alpha}W_{\alpha} + \frac{1}{2} \int d^{2}\theta \, d^{2}\bar\theta \, V_{D}D_{\alpha}W^{\alpha} \right)
\right)
\,.
\end{aligned}
\end{equation}
Using that $\bar D_{\dot\beta}W^{\alpha}=0$, we modify the second term as
\begin{equation}
\begin{aligned}
& \int d^{2}\theta \, d^{2}\bar\theta \, V_{D}D_{\alpha}W^{\alpha} =
-\int d^{2}\theta \, d^{2}\bar\theta \, (D_{\alpha}V_{D})W^{\alpha} =
\\
& -\int d^{2}\theta \, d^{2}\bar\theta \, (\bar D_{\dot\alpha}\bar D^{\dot\alpha}D_{\alpha}V_{D})W^{\alpha} =
4 \int d^{2}\theta \, (W_{D})_{\alpha}W^{\alpha}
\,.
\end{aligned}
\end{equation}
Integrating over $W^{\alpha}$ in the functional integral, we get
\begin{equation}
\int \mathcal{D}V_{D} \exp\left(
\frac{i}{8\pi} \im \int d^{4}x \, d^{2}\theta \left( -\frac{1}{\mathcal{F}''(\Phi)}W_{D}^{\alpha}W_{D\,\alpha} \right)
\right)
\,.
\end{equation}
We have obtained a dual action where the initial effective coupling $\tau(a)=\mathcal{F}''(a)$ is replaced by $-1/\tau(a)=-1/\mathcal{F}''(a)$.
One can also show that changing $W^{\alpha}\to W_{D}^{\alpha}$ corresponds to changing $F_{\mu\nu}\to \tilde F_{\mu\nu}$; this generalises the electromagnetic Montonen--Olive duality \cite{Montonen Olive}.
Since $\mathcal{F}_{D}''(\Phi_{D})=-d\Phi/d\Phi_{D}=-1/\mathcal{F}''(\Phi)$, we can relate the two complex couplings:
\begin{equation}
-\frac{1}{\tau(a)} = \tau_{D}(a_{D})
\end{equation}
where $a_{D}$, the VEV of the dual scalar field, is the magnetic dual of $a$.
Finally, we see that the full action (\ref{action N2 effective abelian}) is invariant under substituting all its constituent fields by their duals.
Using both $a$ and $a_{D}=\mathcal{F}'(a)$, we can rewrite the metric in a symmetric way:
\begin{equation}
g = \im (da_{D} \, d\bar a)
\,,
\end{equation}
where $a_{D}$ and $a$ are non-trivial functions of $u$.

\paragraph{}

The theory is symmetric under $F_{\mu\nu}\leftrightarrow\tilde{F}_{\mu\nu}$ (for the electric field $E$ and the magnetic field $B$, this means $E\to B$, $B\to -B$).
To preserve this symmetry, Dirac introduced magnetic monopoles with charge $q_{m}$ in addition to the electric charge $q_{e}$ (so that the duality exchanges $q_{m}$ and $q_{e}$).
It can be shown that this construction can be consistent only if all charges obey $q_{m}q_{e}\in 2\pi\mathbb{Z}$, hence, all electric and magnetic charges quantise, and their minimal values are related as $q_{m}q_{e}=2\pi$.

\paragraph{}

In our notations, each particle in the theory with gauge group $U(1)$ has an electric charge $\gamma_{e}$ and a magnetic charge $\gamma_{m}$ \cite{Hooft, Polyakov, Julia Zee, Prasad Sommerfield}, denoted together as $\gamma=(\gamma_{e},\gamma_{m})$.
In general, for a theory with gauge group of rank $r$ after symmetry breaking, BPS states have $r$ electric charges $\gamma_{e\,I}$ and $r$ magnetic charges $\gamma_{m}^{I}$ under the residual $U(1)^{r}$ gauge symmetry whose components are labelled by $I$; in addition, when flavours are present, BPS states have flavour charges $s_{i}$ for each hypermultiplet labelled by $i$:
\begin{equation}
\label{charge}
\gamma = ({\vec\gamma}_{e},{\vec\gamma}_{m},\vec s\,) = \left( (\gamma_{e\,1},\dots,\gamma_{e\,r}),(\gamma_{m}^{1},\dots,\gamma_{m}^{r}),(s_{1},\dots,s_{N_{f}}) \right)
\,.
\end{equation}

\paragraph{}

In the $\mathcal{N}=2$ supersymmetry algebra, the anticommutators of the supercharges depend on the central charge $Z$ \cite{Haag Lopuszanski Sohnius}:
\begin{equation}
\{Q_{\alpha},\tilde Q_{\beta}\} = 2\epsilon_{\alpha\beta}Z
\,,
\quad
\{\bar Q_{\dot\alpha},\bar{\tilde Q}_{\dot\beta}\} = 2\epsilon_{\dot\alpha\dot\beta}Z
\,.
\end{equation}
The mass of any state obeys $M\ge |Z|$, with equality holding only for BPS states, i.e., states belonging to short representations of the superalgebra.
We will focus on the BPS case.
In the simplest case of Seiberg--Witten theory with gauge group $SU(2)$ without flavours, the central charge \cite{Alvarez-Gaume Hassan} of a particle with charge $\gamma$ is given by
\begin{equation}
\label{central charge 1}
Z_\gamma = a\gamma_{e} + a_{D}\gamma_{m}
\end{equation}
where $Z(u)=(a(u),a_{D}(u))$.
Thus, the mass $M_{(\gamma_{e},\gamma_{m})}$ of a particle with charge $(\gamma_{e},\gamma_{m})$ is given by the length of $a(u)\gamma_{e}+a_{D}(u)\gamma_{m}$ in the complex plane \cite{Witten Olive}.
Analogously, for an $\mathcal{N}=2$ theory with gauge group of rank $r$ and with $N_{f}$ flavours, the central charge takes the form
\begin{equation}
\label{central charge}
Z_{\gamma} = \vec a\vec\gamma_{e}+\vec a_{D}\vec\gamma_{m}+\vec m\vec s = \sum_{I=1}^{r} \left( a^{I}\gamma_{e\,I}+a_{D\,I}\gamma_{m}^{I}\right)+\sum_{i=1}^{N_{f}}m_{i}s_{i}
\end{equation}
where $\vec a$ is the VEV of the scalars in the Cartan subalgebra, $\vec a_{D}$ is its dual, and $\vec m$ is the vector of complex masses of hypermultiplets.

\paragraph{}

Let us find the conditions required for a decay process $\gamma_{0}\to\sum_{i=1}^{p}\gamma_{i}$ to take place (conversely, we can consider creation of a state).
First, all total electric, magnetic, and flavour charges must be conserved; this also implies conservation of the total central charge: $Z_{\gamma_{0}}=\sum_ {i=1}^{p}Z_{\gamma_{i}}$.
Second, since the total mass must remain the same, there is another constraint on the central charges: $|Z_{\gamma_{0}}|=\sum_{i=1}^{p}|Z_{\gamma_{i}}|$.
Applying the extended triangle inequality to the central charges, we see that these two conditions are simultaneously satisfied when
\begin{equation}
\label{decay condition}
\arg Z_{\gamma_{i}} = \arg Z_{\gamma_{0}}
\,,
\quad
\forall \ i
\,.
\end{equation}
The regions in the moduli space where such processes are possible are called walls of marginal stability.
They form hyper-surfaces.
In particular, they must contain all singular points where one of the states $\gamma$ becomes massless ($Z_{\gamma}=0$, its complex argument is not well-defined, and rotating around the singularity, $\arg Z_{\gamma}$ jumps by a non-zero multiple of $2\pi$, therefore (\ref{decay condition}) is satisfied somewhere near the singularity).

\paragraph{}

\begin{figure}[ht]
\centering
\includegraphics[width=65mm]{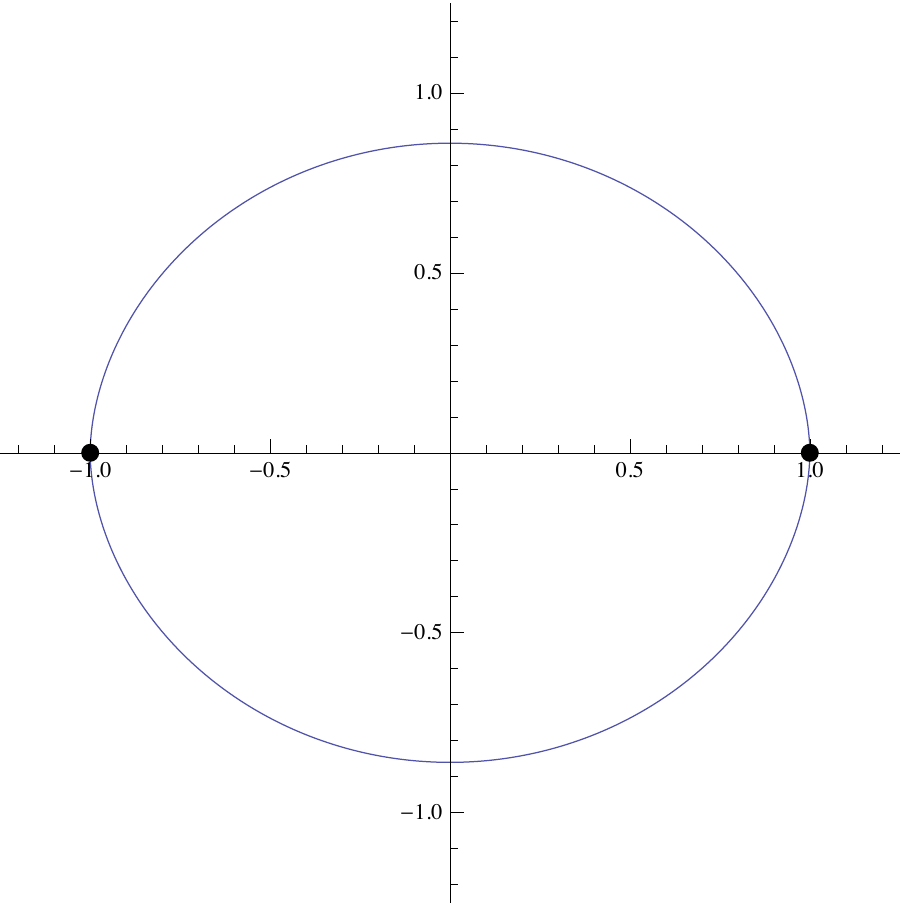}
\caption{
The wall of marginal stability in the pure $\mathcal{N}=2$ $SU(2)$ theory in the $u$ plane \cite{Bilal Ferrari}.
}
\label{fig: WMS pure}
\end{figure}

In the pure $SU(2)$ theory, the wall determined by (\ref{decay condition}) has a particularly simple form: it is a curve in the $u$ plane where the VEV and its dual are aligned in the complex plane:
\begin{equation}
\label{decay condition 1}
r(u) = \frac{a_{D}(u)}{a(u)} \in \mathbb{R}
\,.
\end{equation}
The general exact solution for $a(u)$ and $a_{D}(u)$ was found in \cite{Bilal} by solving a differential equation with periodic boundary conditions corresponding to the monodromies around singularities (which will be discussed below).
The solution is given in terms of hypergeometric functions:
\begin{equation}
\label{SW charge parameters}
\begin{aligned}
a(u) & = \sqrt{2(u+1)} \, F\left( -\frac{1}{2},\frac{1}{2},1;\frac{2}{u+1} \right)
\,,
\\
a_{D}(u) & = \frac{i}{2} \, (u-1) \,F\left( \frac{1}{2},\frac{1}{2},2;\frac{1-u}{2} \right)
\end{aligned}
\end{equation}
where the dynamically generated scale has been set as $\Lambda=1$.
This result allows one to recover the wall of marginal stability: the curve in the complex $u$ plane is approximately (up to $10^{-2}$), although not exactly, an ellipse \cite{Bilal Ferrari} with axes $[-1,1]$ and $\approx[-0.86,0.86]$ (figure~\ref{fig: WMS pure}).
Let us consider how $r$ (\ref{decay condition 1}) changes along the curve.
As we move from $u=-1$ clockwise, $r$ grows monotonically from $r=-2$ (when $u=-1+i\varepsilon$) to $r=2$ (when $u=-1-i\varepsilon$), with $r=0$ at $u=1$.
The wall divides the complex plane into two disconnected regions: the weak-coupling region corresponds to large $|u|$, the strong-coupling region lies inside the curve, when $|u|$ is small.
In theories with flavours, the solution and the walls of marginal stability, which are topologically equivalent to a circle, were found in \cite{Bilal Ferrari 2, Bilal Ferrari 3} using a similar approach.

\paragraph{}

This solution (\ref{SW charge parameters}) has singularities at $u=\pm 1$ and at infinity.
Near these points, the VEV and its dual can be approximated as \cite{Bilal Ferrari}
\begin{equation}
\left.
\begin{aligned}
a(u) & \simeq \sqrt{2u}
\\
a_{D}(u) & \simeq \frac{i}{\pi}\sqrt{2u} \, (\log u+3\log 2-2)
\end{aligned}
\right\}
\quad
u \to\infty
\,,
\end{equation}
\begin{equation}
\left.
\begin{aligned}
a(u) & \simeq \frac{4}{\pi}-\frac{u-1}{2\pi}\log\frac{u-1}{2}+\frac{u-1}{2\pi}(-1+4\log 2)
\\
a_{D}(u) & \simeq \frac{i(u-1)}{2}
\end{aligned}
\right\}
\quad
u \to 1
\,,
\end{equation}
\begin{equation}
\left.
\begin{aligned}
a(u) & \simeq \frac{i}{\pi}\left( \epsilon\frac{u+1}{2}\log\frac{u+1}{2}+\frac{u+1}{2} \, (-i\pi-\epsilon(1+4\log 2))+4\epsilon \right)
\\
a_{D}(u) & \simeq \frac{i}{\pi}\left( -\frac{u+1}{2}\log\frac{u+1}{2}+\frac{u+1}{2} \, (1+4\log 2)-4 \right)
\end{aligned}
\right\}
\quad
u \to -1
\end{equation}
where $\epsilon=\sign\im u$.
The singularity at $u=1$ arises when $\pm(0,1)$ becomes massless; $u=-1$ corresponds to the massless particle described as $\pm(-1,1)$ for $\im u<0$ and $\pm(1,1)$ for $\im u>0$.

%%%%%%%%%%%%%%%%%%%%%%%%%%%%%%%%%%%%%%%%%%%%%%%%%%%%%%%%%%%%%

\section{Spectrum and monodromies}

\paragraph{}

We start by computing the BPS spectrum of the $SU(2)$ theory \cite{SW} by requiring that it is consistent with the monodromies corresponding to the singularities in the moduli space.
Then, following \cite{Fraser Hollowood}, we compute the weak-coupling BPS spectrum of the $SU(n)$ theory from the semiclassical monodromies of the associated Seiberg--Witten curve \cite{Klemm Lerche Yankielowicz Theisen, Argyres Faraggi} (this analysis can also be extended to all other gauge groups \cite{Danielsson Sundborg, Brandhuber Landsteiner, Argyres Shapere, Danielsson Sundborg 2, Abolhasani Alishahiha Ghezelbash}).

%%%%%%%%%%%%%%%%%%%%%%%%%%%%%%

\subsection{SU(2) theory}

\paragraph{}

Consider what happens as we turn clockwise around the singular points: for $u=\pm 1$, $\log(u\mp 1)\to\log(u\mp 1)+2\pi i$; for $u\to\infty$, $\log u\to\log u+2\pi i$ (and $\sqrt{u}\to -\sqrt{u}$).
Thus, the values $a$ and $a_{D}$ jump: their changes are characterised by the monodromy matrices \cite{SW}
\begin{equation}
M_{1} =
\left(
 \begin{array}{rr}
 1 & -2 \\
 0 & 1
 \end{array}
\right)
\,,
\quad
M_{-1} =
\left(
 \begin{array}{rr}
 3 & -2 \\
 2 & -1
 \end{array}
\right)
\,,
\quad
M_{\infty} =
\left(
 \begin{array}{rr}
 -1 & 0 \\
 2 & -1
 \end{array}
\right)
\end{equation}
acting on the vector $(a,a_{D})$ from the left where $M_{1}$, $M_{-1}$, and $M_{\infty}$ denote clockwise rotations around $u=1$, $u=-1$, and $u\to\infty$.
Since the central charge is given by $(\gamma_{e},\gamma_{m})\cdot(a,a_{D})^{T}$, these monodromies can be thought of as transformations acting on the electromagnetic charge $(\gamma_{e},\gamma_{m})$ from the right.

\paragraph{}

Each of the two strong-coupling singularities, $u=1$ and $u=-1$, corresponds to a state with vanishing mass: its electromagnetic charge must be invariant under the corresponding monodromy. 
We assume that at strong coupling, there are only two states satisfying this condition: $(0,1)$ is a left eigenvector of $M_{1}$, and $(-1,1)$ is a left eigenvectors of $M_{-1}$.
We also observe that the contour around $u\to\infty$ can be split into a contour around $u=-1$ and a contour around $u=1$; as there are no other singularities, this means that the monodromies are related by
\begin{equation}
M_{\infty}=M_{1}M_{-1}
\,,
\end{equation}
and one can see that the condition is satisfied.
The monodromy at infinity transforms electromagnetic charges as $(\gamma_{e},\gamma_{m})\to(-\gamma_{e}+2\gamma_{m},\gamma_{m})$.
As a result, the weak-coupling spectrum is larger than the strong-coupling spectrum: semiclassically, there is an infinite tower of dyons $\pm(m,1)$, $m\in\mathbb{Z}$, and $W^{\pm}$ bosons with charge $\pm(1,0)$.
In other words, the monodromy at infinity and the two states existing in the strong-coupling region give rise to the tower of semiclassical Julia--Zee dyons \cite{Julia Zee}.

%%%%%%%%%%%%%%%%%%%%%%%%%%%%%%

\subsection{SU(n) theory}

\paragraph{}

Now we are prepared to proceed to the $\mathcal{N}=2$ theory with gauge group $SU(n)$.
We will restrict ourselves to the weak-coupling regime, constructing the $SU(n)$ spectrum by acting with monodromies on the $SU(2)$ spectrum.
Let $r$ be the dimension of the Cartan subalgebra of the gauge group: for the $SU(n)$ group, $r=n-1$.
The gauge symmetry is maximally broken as $SU(n)\to U(1)^{n-1}$ by the vacuum expectation value of the scalar field $\phi$.
$\langle\phi\rangle=\vec a \vec H$, where $\vec a$ is called the electric coordinate, $\vec H$ is the vector of matrices generating the Cartan subalgebra (all vectors are $(n-1)$-dimensional).
As a result, the spectrum of the theory has $n(n-1)/2$ massive pairs of $W^{\pm}$ bosons.

\paragraph{}

Denote the set of all roots of the gauge group $SU(n)$ as $\Phi$, the set of the $r=n-1$ simple roots as $\Phi_{0}$,
and the set of the $n(n-1)/2$ positive roots as $\Phi_{+}$.
Each $W^{+}$ boson corresponds to a positive root $\vec\alpha_{A}\in\Phi_{+}$ (and vice versa), so that it has charge $W_{A}=(\vec\alpha_{A},\vec 0)$ and mass $M_{W_{A}}=|\vec\alpha_{A}\vec a|$ (analogously, anti-bosons are paired with negative roots).
In the $SU(n)$ case, we normalise every root $\vec\alpha$ as $\|\vec\alpha\|=1$.
By default, all roots will be denoted by Greek letters ($\vec\alpha$, \dots), positive roots will have capital Latin indices ($A$, \dots), and simple roots will have small Latin indices ($i$, \dots).
In terms of an orthonormal basis $\vec e_{i}$~\footnote{
This basis has $n=r+1$ dimensions, whereas all other vectors being considered are restricted to lie in $n-1=r$ dimensions.
}, simple roots for the $SU(n)$ group can be set as
\begin{equation}
\vec\alpha_{i} = \frac{1}{\sqrt{2}} \left( \vec e_{i}-\vec e_{i+1} \right)
\,,
\quad
1\le i\le n
\,.
\end{equation}

\paragraph{}

E.g., for the $SU(3)$ gauge group, there exists a set of 3 positive roots, which may be chosen as $\Phi_{+}=\{\vec\alpha_{1}=(1,0),\ \vec\alpha_{2}=(-1/2,\sqrt{3}/2),\ \vec\alpha_{3}=(1/2,\sqrt{3}/2)\}$, where $\Phi_{0}=\{(1,0),(-1/2,\sqrt{3}/2)\}$ is the set of simple roots for the two-dimensional Cartan subalgebra.
For the $SU(2)$ gauge group, there is only one positive root, $1$, which is also simple.

\paragraph{}

We will be dealing with the weak-coupling region, setting
\begin{equation}
\left| \frac{\vec\alpha_{A}\vec a}{\Lambda} \right| \gg 1
\,,
\quad
\forall \ \vec\alpha_{A} \in \Phi_{+}
\end{equation}
where $\Lambda$ is the dynamical scale.
The global gauge transformations are not completely fixed: one can still perform discrete transformations in the Weyl group.
This discrete degree of freedom can be eliminated by requiring that $\re\vec a$ lies in the fundamental Weyl chamber corresponding to some choice of positive roots:
\begin{equation}
\label{Weyl chamber}
\re\left(\vec\alpha_{i}\vec a\right) \ge 0
\,,
\quad
\forall \ \vec\alpha_{i} \in \Phi_{0}
\,.
\end{equation}

\paragraph{}

The spectrum of dyons whose magnetic charge-vectors are given by simple roots (``simple dyons'') is analogous to the $SU(2)$ case:
\begin{equation}
\label{simple dyon}
\left( p\vec\alpha_{i},\vec\alpha_{i} \right)
\,,
\quad
\vec\alpha_{i} \in \Phi_{0}
\,, \
p \in \mathbb{Z}
\,.
\end{equation}

\paragraph{}

The tree-level and one-loop corrections were analysed in \cite{Seiberg}.
Higher order perturbative corrections are absent.
Instanton corrections produce new terms, which are suppressed in this limit, and we will ignore them.
In this limit, the prepotential is \cite{Seiberg 2, Seiberg 3}
\begin{equation}
\label{prepotential}
\mathcal{F}(\vec a) \simeq \frac{i}{2\pi} \sum_{\vec\alpha_{A}\in\Phi_{+}} (\vec\alpha_{A}\vec a)^{2}
\log\left( \frac{\vec\alpha_{A}\vec a}{\Lambda} \right)^{2}
\,.
\end{equation}
The coefficient of the logarithm follows from the one-loop beta function and preserves the anomalous $U(1)_{\mathcal{R}}$ symmetry.
To the leading order, the magnetic coordinate $\vec a_{D}=\nabla_{\vec a}\mathcal{F}(\vec a)$ is given as
\footnote{
In \cite{Fraser Tong}, the convention is $\|\vec\alpha_{A}\|=2$, and therefore, the resulting coefficient is divided by 2 with respect to our conventions.
}
\begin{equation}
\label{coordinate magnetic}
\vec a_{D} \simeq \frac{i}{\pi}\sum_{\vec\alpha_{A}\in\Phi_{+}} \vec\alpha_{A}(\vec\alpha_{A}\vec a) \,
\log\left( \frac{\vec\alpha_{A}\vec a}{\Lambda} \right)^{2} =
\hat\tau_{\rm eff} \vec a
\end{equation}
where $\hat\tau_{\rm eff}$ is the effective complex coupling.
The important feature of this expression is that it has singularities when one of the bosons becomes massless.
For each singularity $\vec\alpha_{i}\vec a=0$, there should be a Weyl reflection \cite{Fraser Hollowood} acting on the VEV when $\re\vec\alpha_{i}\vec a=0$ to ensure that it stays within the fundamental Weyl chamber (\ref{Weyl chamber}).
This transformation reflects the projection of $\vec a$ onto $\vec\alpha_{i}$:
\begin{equation}
\vec a(t) = \vec a-\vec\alpha_{i} \left( \vec\alpha_{i}\vec a \right) \left( 1-e^{it} \right)
\,,
\quad
0 \le t \le \pi
\end{equation}
where $t=0$ and $t=\pi$ correspond to the initial and the final position (as $t$ increases, $\vec a(t)$ moves counterclockwise).
The associated monodromy matrix $\hat{M}_{i}$ acting on the vector $\left(\vec a,\vec a_{D}\right)$ from the left, and its inverse are given as
\begin{equation}
\hat{M}_{i} =
\left( 
\begin{array}{cc} 
\hat{1}-2\vec\alpha_{i}\otimes\vec\alpha_{i} & \hat{0} \\ 
-2\vec\alpha_{i}\otimes\vec\alpha_{i} & \hat{1}-2\vec\alpha_{i}\otimes\vec\alpha_{i}
\end{array}
\right)
\,,
\quad
\hat{M}_{i}^{-1} =
\left( 
\begin{array}{cc} 
\hat{1}-2\vec\alpha_{i}\otimes\vec\alpha_{i} & \hat{0} \\ 
2\vec\alpha_{i}\otimes\vec\alpha_{i} & \hat{1}-2\vec\alpha_{i}\otimes\vec\alpha_{i}
\end{array}
\right)
\,.
\end{equation}

\paragraph{}

We shall follow the approach in \cite{Fraser Hollowood} to obtain the full spectrum of dyons.
All dyons whose magnetic charges are not simple roots (``composite dyons''), up to their overall sign, are generated by acting on simple dyons with these monodromies (from the right).
Some possibilities are
\begin{equation}
\label{composite dyon}
\begin{aligned}
\left( p\vec\alpha_{i},\vec\alpha_{i} \right)
\hat{M}_{i+1}^{\epsilon_{i+1}}\hat{M}_{i+2}^{\epsilon_{i+2}}\dots\hat{M}_{j-1}^{\epsilon_{j-1}} & =
\left(
p\sum_{m=i}^{j-1}\vec\alpha_{m} +
\sum_{l=i+1}^{j-1}\epsilon_{l} \sum_{m=l}^{j-1}\vec\alpha_{m}
\,,\,
\sum_{m=i}^{j-1}\vec\alpha_{m}
\right)
\\
& = \frac{1}{\sqrt{2}} \left(
p \left( \vec e_{i}-\vec e_{j} \right) +
\sum_{l=i+1}^{j-1}\epsilon_{l} \left( \vec e_{l}-\vec e_{j} \right)
\,,\,
\vec e_{i}-\vec e_{j}
\right)
\end{aligned}
\end{equation}
where $\epsilon_{l}=\pm 1$ (as for $|\epsilon_{l}|>1$, the VEV would cross a wall of marginal stability), $1\le j\le n$.
The most general product of monodromies acting on a state whose magnetic charge is a simple root is given as
\begin{equation}
\begin{aligned}
& \left( \tilde{p}\vec\alpha_{k},\vec\alpha_{k} \right)
\hat{M}_{k+1}^{\tilde{\epsilon}_{k+1}}\hat{M}_{k+2}^{\tilde{\epsilon}_{k+2}}\dots\hat{M}_{j-1}^{\tilde{\epsilon}_{j-1}}
\cdot
\hat{M}_{k-1}^{\tilde{\epsilon}_{k-1}}\hat{M}_{k-2}^{\tilde{\epsilon}_{i-2}}\dots\hat{M}_{i}^{\tilde{\epsilon}_{i}}
\\
& = \left(
\tilde{p}\sum_{m=i}^{j-1}\vec\alpha_{m} +
\sum_{l=k+1}^{j-1}\tilde{\epsilon}_{l} \sum_{m=l}^{j-1}\vec\alpha_{m} +
\sum_{l=i}^{k-1}\tilde{\epsilon}_{l} \sum_{m=i}^{l}\vec\alpha_{m}
\,,\,
\sum_{m=i}^{j-1}\vec\alpha_{m}
\right)
\,.
\end{aligned}
\end{equation}
This is, in fact, equal to the previous result, (\ref{composite dyon}), if we set $\tilde{\epsilon}_{l}=-\epsilon_{l+1}$ for $i\le l\le k-1$, $\tilde{\epsilon}_{l}=\epsilon_{l}$ for $k\le l\le j-1$, and $\tilde{p}=p+\sum_{l=i+1}^{k}\epsilon_{l}$.

\paragraph{}

In the case of gauge group $SU(3)$ (considered explicitly in \cite{GKPY}), (\ref{composite dyon}) has only one monodromy matrix, $\hat M_{2}$, and the composite dyons are $(p\vec\alpha_{3}\pm\vec\alpha_{2},\vec\alpha_{3})$, where $\vec\alpha_{3}=\vec\alpha_{1}+\vec\alpha_{2}$, depending on whether one acts with $\hat M_{2}$ or $\hat M_{2}^{-1}$ on $(p\vec\alpha_{1},\vec\alpha_{1})$.
This demonstrates that the moduli space of the theory at weak coupling consists of two separate regions.

\paragraph{}

Summing up, the spectrum is given by the sets of simple dyons (\ref{simple dyon}), composite dyons (\ref{composite dyon}), $W$ bosons whose charges are $(\vec\alpha_{A},\vec 0)$, and their antiparticles.

%%%%%%%%%%%%%%%%%%%%%%%%%%%%%%%%%%%%%%%%%%%%%%%%%%%%%%%%%%%%%

\section{Dimensional reduction}

\paragraph{}

Following \cite{SW3}, we can compactify the Euclidean $\mathbb{R}^{4}$ theory on $\mathbb{R}^{3}\times S^{1}$ along $x^{0}$ imposing periodic conditions on bosonic and fermionic fields, which preserve supersymmetry.
First of all, let us restrict our attention to the bosonic part of the $\mathbb{R}^{4}$ action:
\begin{equation}
\label{action bosonic 4D}
S_{\rm B} = \int d^{4} x \left(
\frac{1}{4 g^{2}}F_{\mu\nu}F^{\mu\nu} +
\frac{i\Theta}{32\pi^{2}}F_{\mu\nu}\tilde F^{\mu\nu} +
\frac{1}{2g^{2}}\partial_{\mu}\bar\phi \, \partial^{\mu}\phi
\right)
\,,
\end{equation}
where the complex parameter that we are using is (\ref{tau}).
Define the Wilson line:
\begin{equation}
\label{Wilson line}
\theta_{e} = \oint_{S_{R}^{1}} dx^{0} A_{0}
\,.
\end{equation}
The theory is invariant under the following gauge transformation:
\begin{equation}
\label{field gauge invariance}
A_{\mu} \to A_{\mu}+\partial_{\mu}w
\,.
\end{equation}
In $\mathbb{R}^{3}\times S^{1}$, one must require $e^{iw(x_{0}=0)}=e^{iw(x_{0}=2\pi R)}$ to preserve periodicity along the compactified dimension.
For the Wilson line, this means that it is defined up to shifting $\theta_{e}\to\theta_{e}+2\pi n,\ n\in\mathbb{Z}$.
The first two terms of (\ref{action bosonic 4D}) compactified on $\mathbb{R}^{3}\times S^{1}$ at low energy (where we are retaining only zero modes, ignoring Kaluza--Klein massive terms) are given by
\begin{equation}
\label{compactified 1}
S_{\rm B} = \int d^{3} x \left(
\frac{\pi R}{2 g^{2}}F_{\mu\nu}F^{\mu\nu}+
\frac{1}{4\pi R g^{2}}\partial_{\mu}\theta_{e} \, \partial^{\mu}\theta_{e}+
\frac{i\Theta}{16\pi^{2}}\epsilon^{\mu\nu\rho}F_{\mu\nu} \, \partial_{\rho}\theta_{e}
\right)
\end{equation}
where the spatial indices run over $1,2,3$.

\paragraph{}

We can now dualise the three-dimensional photon: to achieve this, we modify the Lagrangian by defining a new field strength $B_{\mu\nu}=\partial_{\mu}C_{\nu}-\partial_{\nu}C_{\mu}$, so that in addition to the original gauge invariance (\ref{field gauge invariance}), the action is invariant under
\begin{equation}
\begin{aligned}
A_{\mu} & \to A_{\mu}+C_{\mu}
\,,
\\
B_{\mu\nu} & \to B_{\mu\nu}+\partial_{\mu}C_{\nu}-\partial_{\nu}C_{\mu}
\,.
\end{aligned}
\end{equation}
Introduce a periodic scalar field $\theta_{m}\in[0,2\pi]$ which will serve as a Lagrange multiplier.
The initial action (\ref{compactified 1}) can then be rewritten as
\begin{equation}
\label{compactified 2}
\begin{aligned}
S_{\rm B} = \int d^{3} x & \left(
\frac{\pi R}{2 g^{2}}(F_{\mu\nu}-B_{\mu\nu})(F^{\mu\nu}-B^{\mu\nu})+
\frac{1}{4\pi R g^{2}}\partial_{\mu}\theta_{e}\,\partial^{\mu}\theta_{e}
+\right. \\ & + \left.
\frac{i\Theta}{16\pi^{2}}\epsilon^{\mu\nu\rho}(F_{\mu\nu}-B_{\mu\nu})\partial_ {\rho}\theta_{e}+
\frac{i}{8\pi}\epsilon^{\mu\nu\rho}B_{\mu\nu}\,\partial_{\rho}\theta_{m}
\right).
\end{aligned}
\end{equation}
Dirac quantisation for magnetic charges in three dimensions is
\begin{equation}
\label{magnetic charge quantisation}
\frac{1}{8\pi} \int d^{3} x \, \epsilon^{\mu\nu\rho}\partial_{\mu} F_{\nu\rho} \in \mathbb{Z}
\,,
\end{equation}
hence, the dual photon $\theta_{m}$ is invariant under shifting $\theta_{m}\to\theta_{m}+2\pi n$, $n\in\mathbb{Z}$.
In the path integral, integrating over $\theta_{m}$ allows one to set $B_{\mu\nu}=0$ and recover the original action.
On the other hand, one can use the extended gauge invariance to set $F_{\mu\nu}=0$ and integrate over $B_{\mu\nu}$.
Then, the compactified bosonic action becomes
\begin{equation}
\label{action bosonic compactified abelian}
S_{\rm B} = \frac{1}{4} \int d^{3}x \left(
\frac{g^{2}}{16\pi^{3} R} \partial_{\mu}\bar z \, \partial^{\mu}z+
\frac{4\pi R}{g^{2}} \partial_{\mu}\bar\phi \, \partial^{\mu}\phi
\right)
\end{equation}
where $z = \theta_{m}-\tau\theta_{e}$.
The fermionic part of the low-energy action after compactifying one dimension is
\begin{equation}
\label{action fermionic compactified abelian}
S_{\rm F} = \frac{2\pi R}{g^{2}} \int d^{3}x \left(
i\bar\psi\bar\sigma^{\mu}\partial_{\mu}\psi+
i\bar\lambda\bar\sigma^{\mu}\partial_{\mu}\lambda
\right)
\,.
\end{equation}

\paragraph{}

For generic gauge group, we can define $r$ Wilson lines $\theta_{e}^{I}$ and $r$ dual photons $\theta_{m\,I}$ (where $I$ counts gauge components); it is also convenient to introduce a new field mixing Wilson lines and dual photons:
\begin{equation}
z_{I} = \theta_{m\,I}-\sum_{J=1}^{r}\tau_{IJ}\theta_{e}^{J}
\,.
\end{equation}
For the $r$-vectors of Wilson lines and dual photons, $\vec\theta_{e}=(\theta_{e}^{1},\dots,\theta_{e}^{r})$ and $\vec\theta_{m}=(\theta_{m\,1},\dots,\theta_{m\,r})$, we define
\begin{equation}
\label{surface charge}
\theta_{\gamma} = \vec\theta_{e}\vec\gamma_{e}+\vec\theta_{m}\vec\gamma_{m} = \sum_{I=1}^{r}\left( \theta_{e}^{I}\gamma_{e\,I}+\theta_{m\,I}\gamma_{m}^{I} \right)
\,.
\end{equation}
In general, (\ref{action bosonic compactified abelian}) and (\ref{action fermionic compactified abelian}) contain summations over all gauge components where one should treat the effective complex parameter $\im\tau_{\rm eff}=4\pi/g_{\rm eff}^{2}$ (\ref{tau}) as a matrix $(\im\tau_{\rm eff})_{IJ}$ where $(\tau_{\rm eff})_{IJ}=\partial^{2}\mathcal{F}(\vec a)/\partial a^{I}\partial a^{J}$ (in our case, the prepotential is given in (\ref{prepotential})):
\begin{equation}
\label{action bosonic compactified}
S_{\rm B} = \frac{1}{4} \int d^{3}x \sum_{I=1}^{r}\sum_{J=1}^{r} \left(
\frac{1}{4\pi^{2} R} \left( (\im\tau_{\rm eff})^{-1} \right)^{IJ} \partial_{\mu}\bar z_{I} \, \partial^{\mu}z_{J}+
R (\im\tau_{\rm eff})_{IJ} \partial_{\mu}\bar\phi^{I} \, \partial^{\mu}\phi^{J}
\right)
\,,
\end{equation}
\begin{equation}
\label{action fermionic compactified}
S_{\rm F} = \frac{R}{2} \int d^{3}x \sum_{I=1}^{r}\sum_{J=1}^{r} (\im\tau_{\rm eff})_{IJ} \left(
i\bar\psi^{I}\bar\sigma^{\mu}\partial_{\mu}\psi^{J}+
i\bar\lambda^{I}\bar\sigma^{\mu}\partial_{\mu}\lambda^{J}
\right)
\,.
\end{equation}
The real dimension of the Coulomb branch is $4r$ (parametrised by complex $\phi^{I}$ and $z_{I}$ for all $I$).

%%%%%%%%%%%%%%%%%%%%%%%%%%%%%%%%%%%%%%%%%%%%%%%%%%%%%%%%%%%%%

\section{Hyper-K\"ahler description}

\paragraph{}

As has been shown above, the compactified low-energy effective theory on the Coulomb branch is a three-dimensional sigma model.
It is known that the metric of the target space of the theory is hyper-K\"ahler \cite{Alvarez-Gaume Freedman}.
This is ensured by the presence of 8 real supercharges.
In this section, using this property, we will describe the metric via a symplectic product depending on Darboux coordinates.

\paragraph{}

Let us recall some definitions.
A K\"ahler manifold has a potential $K$ defining the corresponding symplectic form
\begin{equation}
\label{K form}
\omega_{3} = i \, \frac{\partial^{2}K}{\partial z^{a}\partial z^{\bar b}} \, dz^{a}\wedge dz^{\bar b}
\,,
\end{equation}
which is related to the metric
\begin{equation}
\label{K metric}
g = 2 \, \frac{\partial^{2}K}{\partial z^{a}\partial z^{\bar b}} \, dz^{a}dz^{\bar b} = 2 \, g_{a\bar b} \, dz^{a}dz^{\bar b}
\,.
\end{equation}
After dimensional reduction (\ref{action bosonic compactified}), the leading behaviour of the metric of the moduli space is given by its semiflat component:
\begin{equation}
\label{metric semiflat}
g^{\rm sf} = R(\im\tau)|da|^{2} + \frac{1}{4\pi^{2} R}(\im\tau)^{-1}|dz|^{2}
\end{equation}
where gauge indices are suppressed (the term ``semiflat'' refers to the two-torus spanned by $\theta_{e}^{I}$ and $\theta_{m\,I}$ being flat).
A hyper-K\"ahler manifold \cite{HKLR, Ivanov Rocek} is defined as a K\"ahler manifold with respect to a triplet of complex structures $\vec{J}$ obeying the quaternion relations
\begin{equation}
J_{1} J_{2}=J_{3},\quad J_{2} J_{3}=J_{1},\quad J_{3} J_{1}=J_{2},\quad J_a^{2}=-1
\end{equation}
for $a=1,2,3$.
Each complex structure has a corresponding K\"ahler form $\omega_{a}$.
More generally, a hyper-K\"ahler manifold is K\"ahler for any complex structure $v^{a}J_{a}$ parametrised by a vector $\vec v$ such that $||\vec v||=1$. Its corresponding K\"ahler form is $v^{a}\omega_{a}$.

\paragraph{}

The three K\"ahler forms can be combined in a single form depending on an auxiliary complex parameter $\zeta$:
\begin{equation}
\label{symplectic form decomposition}
\omega(\zeta) = -\frac{i}{2\zeta}\omega_{+}+\omega_{3}-\frac{i\zeta}{2}\omega_{-}
\end{equation}
where we have introduced $\omega_{\pm}=\omega_{1}\pm i\omega_{2}$.
After adding infinity, $\zeta$ parametrises ${\mathbb{CP}}^{1}$.

\paragraph{}

For an abelian gauge theory of rank $r$, this form can be expressed in terms of Darboux coordinates ($\mathcal{X}_{e}^{I}$ and $\mathcal{X}_{m\,I}$ are called electric and magnetic components, $I$ is the index counting gauge group components):
\begin{equation}
\label{symplectic form}
\omega(\zeta) = -\frac{1}{4\pi^{2}R} \, \sum_{I=1}^{r} \,
\frac{d\mathcal{X}_{e}^{I}(\zeta)}{\mathcal{X}_{e}^{I}(\zeta)}
\wedge
\frac{d\mathcal{X}_{m\,I}(\zeta)}{\mathcal{X}_{m\,I}(\zeta)}
\,.
\end{equation}
We can expand this definition of Darboux coordinates to any charge $\gamma$:
\begin{equation}
\label{Darboux coordinate}
\mathcal{X}_{\gamma}(\zeta) = \prod_{I=1}^{r} \left(\mathcal{X}_{e}^{I}(\zeta)\right)^{\gamma_{e\,I}} \left(\mathcal{X}_{m\,I}(\zeta)\right)^{\gamma_{m}^{I}}
\end{equation}
where $\mathcal{X}_{e}^{I}$ and $\mathcal{X}_{m\,J}$ are Darboux coordinates corresponding to the charges with only one non-zero component, $\gamma_{e\,I}=1$ and $\gamma_{m}^{J}=1$, respectively.
Thus, we demand that the logarithm of Darboux coordinates is linear with respect to all components of $\gamma$; as we will see later, this is always the case.
We also introduce the symplectic product of two charges, $(\vec\alpha_{e},\vec\alpha_{m})$ and $(\vec\beta_{e},\vec\beta_{m})$:
\begin{equation}
\label{symplectic product}
\langle(\vec\alpha_{e},\vec\alpha_{m}),(\vec\beta_{e},\vec\beta_{m})\rangle =
-\vec\alpha_{e}\vec\beta_{m}+\vec\alpha_{m}\vec\beta_{e}
\,.
\end{equation}

\paragraph{}

To make use of this formalism, we need to match the symplectic form with the semiflat metric (\ref{metric semiflat}).
In \cite{GMN}, it was shown that this condition is satisfied by the following choice of coordinates in (\ref{symplectic form}):
\begin{equation}
\label{coordinate semiflat}
\mathcal{X}_{(\vec\gamma_{e},\vec\gamma_{m},\vec s\,)}^{\rm sf}(\zeta) = \mathcal{X}_{(\vec\gamma_{e},\vec\gamma_{m},\vec 0)}^{\rm sf}(\zeta) = \exp\left( \pi R \zeta^{-1}Z_{(\vec\gamma_{e},\vec\gamma_{m})}+i\theta_{(\vec\gamma_{e},\vec\gamma_{m})}+\pi R\zeta\bar Z_{(\vec\gamma_{e},\vec\gamma_{m})} \right)
\end{equation}
(in this section, we can ignore all flavour charges).
Indeed, reinterpreting the magnetic central charge as
\begin{equation}
Z_{(\vec 0,\vec\gamma_{m})} = \vec a_{D}\vec\gamma_{m} = \sum_{I=1}^{r}\gamma_{m}^{I} \, \frac{\partial\mathcal{F}}{\partial a^{I}}
\,,
\end{equation}
we re-express the symplectic form as
\begin{equation}
\begin{aligned}
\omega^{\rm sf}(\zeta)
= & \frac{1}{4\pi^{2} R} \sum_{I=1}^{r}
d\left( \pi R\zeta^{-1}\frac{\partial\mathcal{F}}{\partial a^{I}}+i\theta_{m\,I}+\pi R\zeta\overline{\frac{\partial\mathcal{F}}{\partial a^{I}}} \right)
\wedge
d\left( \pi R\zeta^{-1}a^{I}+i\theta_{e}^{I}+\pi R\zeta\bar a^{I} \right)
\\
= & \frac{1}{4\pi^{2} R} \sum_{I=1}^{r} \sum_{J=1}^{r}
\left( \pi R\left( \zeta^{-1}\frac{\partial^{2}\mathcal{F}}{\partial a^{I}\partial a^{J}}da^{J}+\zeta\overline{\frac{\partial^{2}\mathcal{F}}{\partial a^{I}\partial a^{J}}}d\bar a^{J} \right)+id\theta_{m\,I} \right)
\\
& \wedge
\left( \pi R(\zeta^{-1}da^{I}+\zeta d\bar a^{I})+id\theta_{e}^{I} \right)
\,.
\end{aligned}
\end{equation}
We use the fact that $\tau_{IJ}=\frac{\partial^{2}\mathcal{F}}{\partial a^{I}\partial a^{J}}$ and introduce the inverse matrix for $\im\tau_{IJ}$ as $(\im\tau^{-1})^{IJ}=(\im\tau^{-1})^{JI}$, so that $d\theta_{m\,I}\wedge d\theta_{e}^{I}=\frac{1}{2i} \, dz_{I}\wedge d\bar z_{J}\left( (\im\tau)^{-1} \right)^{IJ}$~\footnote{
To prove this, we can expand the wedge-products as
\begin{equation}
\begin{aligned}
dz_{I} \wedge d\bar z_{J}
\left( (\im\tau)^{-1} \right)^{IJ}
& =
\left( d\theta_{m\,I}-(\re\tau_{IK}+i\im\tau_{IK})d\theta_{e}^{K} \right)
\wedge
\left( d\theta_{m\,J}-(\re\tau_{JL}-i\im\tau_{JL})d\theta_{e}^{L} \right)
\left( (\im\tau)^{-1} \right)^{IJ}
\\
& = 2i \, d\theta_{m\,I} \wedge d\theta_{e}^{L}
\im\tau_{JL} \left( (\im\tau)^{-1} \right)^{IJ}
= 2i \, d\theta_{m\,I} \wedge d\theta_{e}^{I}
\,,
\end{aligned}
\end{equation}
where the real part of $\tau$ disappears due to antisymmetricity of the wedge-product.
}.
The three components of the semiflat symplectic form are
\begin{align}
\omega_{3}^{\rm sf} & = \frac{i}{2} R(\im\tau)_{IJ}da^{I}\wedge d\bar a^{J}+\frac{1}{4\pi^{2} R}\left( (\im\tau)^{-1} \right)^{IJ} dz_{I}\wedge d\bar z_{J}
\,,
\\
\omega_{+}^{\rm sf} & = \frac{1}{2\pi}da^{I}\wedge dz_{I}
\,,
\\
\omega_{-}^{\rm sf} & = \frac{1}{2\pi}d\bar a^{I}\wedge dz_{I}
\,.
\end{align}
Thus, we have matched the semiflat metric (\ref{metric semiflat}) and the symplectic form $\omega_{3}^{\rm sf}$ corresponding to $J_{3}^{\rm sf}$.
Hence, the manifold is hyper-K\"ahler with respect to the triplet $\vec{J}$ defined above.

\paragraph{}

The semiflat symplectic form can be written in short as
\begin{equation}
\begin{aligned}
\omega^{\rm sf}(\zeta) & = \frac{1}{8\pi} \left( \frac{i}{\zeta}\langle dZ',d\theta\rangle+\left( 2\pi R\langle dZ',d\bar Z'\rangle-\frac{1}{\pi R}\langle d\theta',d\theta'\rangle \right)+i\zeta\langle d\bar{Z}',d\theta'\rangle \right)
\,,
\\
Z' & = (\vec a, \vec a_{D})
\,,
\quad
\theta' = (\vec\theta_{e},\vec\theta_{m})
\,.
\end{aligned}
\end{equation}
We also note that the symplectic form has no terms of order $\zeta^{\pm 2}$ as $\langle dZ',dZ'\rangle=0$, ensuring self-consistency of this construction.

\paragraph{}

If we ignored all instanton-like corrections from BPS states of the compactified theory, $g^{\rm sf}$ would be the final answer for the moduli space metric.
However, the metric in $\mathbb{R}^{3}\times S^{1}$ receives corrections from the BPS states whose worldlines wrap around the compactified dimension.
In order to take these corrections into account, it was suggested in \cite{GMN} that the Darboux coordinates $\mathcal{X}_{m\,I}(\zeta)$ and $\mathcal{X}_{e}^{I}(\zeta)$ in (\ref{symplectic form}) should have discontinuities corresponding to the BPS states belonging to the spectrum.
We will discuss how to construct these general Darboux coordinates in chapter \ref{ch: walls}.

\chapter{Wall-crossing formulae}
\label{ch: walls}

\paragraph{}

As has been discussed in chapter \ref{ch: supersymmetry}, the BPS spectrum of $\mathcal{N}=2$ supersymmetric theory is not constant in the moduli space, rather, it jumps across the walls of marginal stability.
This wall-crossing phenomenon also appears in other contexts: it was first discovered in two-dimensional theories with $\mathcal{N}=(2,2)$ supersymmetry \cite{CFIV} with a class of explicit formulae considered in \cite{CV}, the decays $\gamma\to\gamma_{1}+\gamma_{2}$ in supergravity were considered in \cite{Denef, Denef 2, Denef 3}.
Kontsevich and Soibelman conjectured an exact formula relating the BPS spectra on both sides of any wall of marginal stability \cite{KS}.
Knowing the spectrum on one side of a given wall and using the wall-crossing formula, technically, one is able to predict the spectrum on the other side of the wall.

\paragraph{}

It was suggested by Gaiotto, Moore, and Neitzke that the group elements in the wall-crossing formula should lead to discontinuities of the Darboux coordinates on $\mathbb{R}^{3}\times S^{1}$ with respect to the $\mathbb{CP}^{1}$ parameter $\zeta$ introduced above \cite{GMN}.
These discontinuities and the asymptotic behaviour of Darboux coordinates for large radius of the compactified dimension defines a Riemann--Hilbert problem whose solution is a set of non-linear integral equations of rank $2r$ where $r$ is the rank of the unbroken gauge group.
These equations express Darboux coordinates as functions of their semiflat values and convolutions depending on the set of BPS charges and their multiplicities.
The discontinuities of the Darboux coordinates constructed according to this method depend on the BPS spectrum, which has jumps across the walls of marginal stability, however, the predicted moduli space metric is manifestly continuous: this is ensured by the Kontsevich--Soibelman wall-crossing formula relating the BPS spectra in every region of the moduli space.
The integral equations for Darboux coordinates appeared in another form in string theory: the problem in $\mathcal{N}=2$ supersymmetric Yang--Mills theory with one electric charge turns out to be mathematically the same as the problem of finding the surface with minimal area ending on a polygon at the boundary of the $AdS$ space, studied in \cite{Alday Maldacena}.
This area defines the amplitude of a gluon scattering at strong coupling.

\paragraph{}

We start by discussing the Kontsevich--Soibelman algebra and constructing the group elements which will be used in the wall-crossing formula.
Then, we provide the Gaiotto--Moore--Neitzke solution for Darboux coordinates and show that it correctly reproduces the discontinuities generated by Kontsevich--Soibelman operators.
After reviewing the general method, we construct and prove several formulae for actual physical theories.
First, we consider wall-crossing formulae in $\mathcal{N}=2$ theories with gauge group $SU(2)$ and up to three fundamental flavours \cite{KS}: in this case, there is only one wall, and these formulae are known to correctly relate the weak- and strong-coupling spectra.
Then, we generalise the result to the pure theory with gauge group $SU(n)$ by considering decay processes at the walls of marginal stability, which extend into the weak-coupling region for $n>2$, and verify that wall-crossing formulae relate the spectra on both sides of each wall; combining these individual formulae, we find general equalities relating the BPS spectra in different regions of the moduli space.

%%%%%%%%%%%%%%%%%%%%%%%%%%%%%%%%%%%%%%%%%%%%%%%%%%%%%%%%%%%%%

\section{Kontsevich--Soibelman algebra}

\paragraph{}

The wall-crossing formula, which was first considered by Kontsevich and Soibelman in \cite{KS, KS2}, can be described in terms of a Lie algebra with generators $e_{\gamma}$, which will be paired with transformations preserving the symplectic form, whose arguments are $\gamma=(\gamma_{e},\gamma_{m})$, $\{\gamma_{e},\gamma_{m}\}\subset\mathbb{Z}$.
We will then show that $\gamma_{e}$ and $\gamma_{m}$ should be identified with electric and magnetic charges~\footnote{
In this section, we require all electric charges to be integers for conciseness.
In the rest of the paper, the default convention is to allow half-integer charges for theories with flavours, so that $W$ bosons have electric charge one.
Although here we deal with abelian group of rank one, generalising the results of this section to gauge groups of higher ranks is straightforward and amounts to redefining Darboux coordinates as in (\ref{Darboux coordinate}) and symplectic products as in (\ref{symplectic product}).
}.
For our purposes, it will be sufficient to construct the group elements which serve as operators in the wall-crossing formula.
The commutation relation for two elements of the algebra is given by
\begin{equation}
\label{KS algebra}
[e_{\gamma_{1}},e_{\gamma_{2}}] =
(-1)^{\langle\gamma_{1},\gamma_{2}\rangle} \langle\gamma_{1},\gamma_{2}\rangle \, e_{\gamma_{1}+\gamma_{2}}
\end{equation}
where the symplectic product of $\gamma_{1}=(\gamma_{e_{1}},\gamma_{m_{1}})$ and $\gamma_{2}=(\gamma_{e_{2}},\gamma_{m_{2}})$ is defined as
\begin{equation}
\langle\gamma_{1},\gamma_{2}\rangle = -\gamma_{e_{1}}\gamma_{m_{2}}+\gamma_{m_{1}}\gamma_{e_{2}}
\,,
\end{equation}
we will prove the Jacobi identity later.
We define an ``electric'' coordinate $\mathcal{X}_{e}=\mathcal{X}_{(1,0)}$ and a ``magnetic'' coordinate $\mathcal{X}_{m}=\mathcal{X}_{(0,1)}$; the product of two elements is defined as $\mathcal{X}_{\gamma_{1}}\mathcal{X}_{\gamma_{2}}=\mathcal{X}_{\gamma_{1}+\gamma_{2}}$ (so that $\mathcal{X}_{(p,q)}=\mathcal{X}_{e}^{p}\mathcal{X}_{m}^{q}$).
For this definition to be consistent, one must require that $\log\mathcal{X}_{(p,q)}$ is linear with respect to $p$ and $q$.

\paragraph{}

We introduce the symplectic form as
\begin{equation}
\tilde\omega = \langle\gamma_{1},\gamma_{2}\rangle^{-1} \, \frac{d\mathcal{X}_{\gamma_{1}}}{\mathcal{X}_{\gamma_{1}}} \wedge \frac{d\mathcal{X}_{\gamma_{2}}}{\mathcal{X}_{\gamma_{2}}}
\,.
\end{equation}
for any pair of charges $\gamma_{1}$ and $\gamma_{2}$ obeying $\langle\gamma_{1},\gamma_{2}\rangle\ne 0$.
Then, we identify $e_{\gamma}$ with the infinitesimal symplectomorphism generated by the Hamiltonian $\mathcal{X}_{\gamma}$ preserving this form, and the Poisson bracket is defined as
\begin{equation}
\{\mathcal{X}_{\gamma_{1}},\mathcal{X}_{\gamma_{2}}\} =
(\tilde\omega)_{\gamma_{1}\gamma_{2}}^{-1} =
\langle\gamma_{1},\gamma_{2}\rangle \, \mathcal{X}_{\gamma_{1}} \mathcal{X}_{\gamma_{2}}
\,.
\end{equation}
The Hamilton equations corresponding to this symplectomorphism must have the following form:
\begin{equation}
\delta_{\gamma} \mathcal{X}_{\gamma'} = \{\mathcal{X}_{\gamma},\mathcal{X}_{\gamma'}\}
\end{equation}
where $\delta_{\gamma}$ means variation with respect to charge $\gamma$ (it acts as an equivalent of the time derivative in classical mechanics).
Using the Jacobi identity for Poisson brackets, we find the commutator of two variations:
\begin{equation}
(\delta_{\gamma_{1}}\delta_{\gamma_{2}} - \delta_{\gamma_{2}}\delta_{\gamma_{1}}) \mathcal{X}_{\gamma} =
\langle\gamma_{1},\gamma_{2}\rangle \,
\delta_{\gamma_{1}+\gamma_{2}} \mathcal{X}_{\gamma}
\,.
\end{equation}
In order to connect this construction with the Kontsevich--Soibelman algebra (\ref{KS algebra}), one needs to find a quadratic refinement $\tilde\sigma(\gamma)$ obeying
\begin{equation}
\tilde\sigma(\gamma_{1}) \, \tilde\sigma(\gamma_{2}) =
(-1)^{\langle\gamma_{1},\gamma_{2}\rangle} \tilde\sigma(\gamma_{1}+\gamma_{2})
\,.
\end{equation}
Although $\tilde\sigma(\gamma)$ is not uniquely defined, it is sufficient to set 
\begin{equation}
\tilde\sigma(\gamma) = (-1)^{\gamma_{e}\gamma_{m}}
\,.
\end{equation}
Then, we can see that the correspondence between the symplectomorphisms and the Lie algebra is
\begin{equation}
e_{\gamma}\ \leftrightarrow\ \tilde\sigma(\gamma) \, \delta_{\gamma}
\,,
\end{equation}
and the commutation relation in (\ref{KS algebra}) can be re-expressed in terms of these variations:
\begin{equation}
[e_{\gamma_{1}},e_{\gamma_{2}}]\ \leftrightarrow\ \tilde\sigma(\gamma_{1}) \, \tilde\sigma(\gamma_{2}) \,(\delta_{\gamma_{1}}\delta_{\gamma_{2}} - \delta_{\gamma_{2}}\delta_{\gamma_{1}})
\,.
\end{equation}
This identification also shows that the Jacobi identity holds for the Lie algebra.

\paragraph{}

For every charge $\gamma$, Kontsevich and Soibelman associate a group element defined as
\begin{equation}
\tilde{\mathcal{K}}_{\gamma} = \exp\left(\sum_{n=1}^{+\infty} \frac{1}{n^{2}} e_{n\gamma}\right)
\,.
\end{equation}
These operators act on Darboux coordinates as
\begin{equation}
\begin{aligned}
\tilde{\mathcal{K}}_{\gamma} \mathcal{X}_{\gamma'} & =
\exp\left( \sum_{n=1}^{+\infty} \frac{1}{n^{2}} \tilde\sigma^{n}(\gamma)\{\mathcal{X}_{n\gamma},\log\mathcal{X}_{\gamma'}\} \right)=
\exp\left( \sum_{n=1}^{+\infty} \frac{1}{n} \langle\gamma,\gamma'\rangle \left( \tilde\sigma(\gamma) \, \mathcal{X}_{\gamma} \right)^{n} \log\mathcal{X}_{\gamma'} \right)
\\
& = (1-\tilde\sigma(\gamma)\,\mathcal{X}_{\gamma})^{\langle\gamma',\gamma\rangle} \mathcal{X}_{\gamma'}
\,,
\end{aligned}
\end{equation}
where in the last equality, we used the fact that the infinite sum is the Taylor series of a logarithm.
Using the definition of symplectic product, one can easily see how powers of these operators act on Darboux coordinates:
\begin{equation}
\tilde{\mathcal{K}}_{(p,q)}^{n} \quad \colon \quad (x,y) \to \left( (1-(-1)^{pq}x^{p}y^{q})^{-nq}, (1-(-1)^{pq}x^{p}y^{q})^{np} \right)
\,,
\quad
n \in \mathbb{Z}
\,.
\end{equation}
Kontsevich--Soibelman operators given in this form and their generalisations will serve as building blocks in constructing the wall-crossing formula.

%%%%%%%%%%%%%%%%%%%%%%%%%%%%%%%%%%%%%%%%%%%%%%%%%%%%%%%%%%%%%

\section{Kontsevich--Soibelman wall-crossing formula}

\paragraph{}

The wall-crossing phenomenon can be described in terms of Kontsevich--Soibelman operators introduced above \cite{KS}, where each group element $\mathcal{K}_{\gamma}$ corresponds to a BPS particle with charge $\gamma$.
In theories without flavours and with gauge group of any rank, we reparametrise the Kontsevich--Soibelman operator as
\begin{equation}
\label{KS operator}
\mathcal{K}_{\gamma} \quad \colon \quad
\mathcal{X}_{\beta} \to \mathcal{X}_{\beta} \left( 1-\sigma(\gamma)\mathcal{X}_{\gamma} \right)^{2\langle\beta,\gamma\rangle}
\end{equation}
where the quadratic refinement is now given by
\begin{equation}
\label{quadratic refinement}
\sigma(\gamma) = (-1)^{2\vec\gamma_{e}\vec\gamma_{m}} = (-1)^{2\sum_{I=1}^{r}\gamma_{e\,I}\gamma_{m}^{I}}
\end{equation}
(for gauge group of rank $r=1$, $\sigma(\gamma)=1$ for all charges).
Our default conventions differ from \cite{KS, GMN} (where $\tilde{\mathcal{K}}_{\gamma}$ and $\tilde\sigma(\gamma)=(-1)^{\vec\gamma_{e}\vec\gamma_{m}}$ from the previous section were used instead of $\mathcal{K}_{\gamma}$ and $\sigma(\gamma)$), because we allow electric charges to be any integers (not necessarily even).
In theories with flavours, however, it will often be convenient to use $\tilde{\mathcal{K}}_{\gamma}$ and $\tilde\sigma(\gamma)$, so that all electric charges are integers.

\paragraph{}

For a BPS particle with charge $\gamma$ belonging to the spectrum $\Gamma(\vec a)$, for later convenience, we associate a BPS ray $l_{\gamma}$ in the $\zeta$ plane (where $\zeta\in\mathbb{C}$ is an auxiliary parameter), determined by the central charge of the particle~\footnote{
$\mathbb{R}_{+}$, $\mathbb{R}_{-}$ denote positive and negative real numbers, in integrals, we will imply integration from $0$ to $+\infty$ and from $0$ to $-\infty$.
}:
\begin{equation}
\label{BPS ray}
l_{\gamma} = \left\{ \zeta \ : \ \frac{Z_{\gamma}(\vec a)}{\zeta} \in \mathbb{R}_{-} \right\}
\,.
\end{equation}

\paragraph{}

The Darboux coordinates $\mathcal{X}_{\gamma}(\zeta)$ (for any $\gamma$) are discontinuous along every ray $l$ which is aligned with one or more BPS rays $l_{\gamma'}$ with $\gamma'\in\Gamma(\vec a)$ and $\langle\gamma,\gamma'\rangle\ne 0$.
Explicitly, the jump is given as
\begin{equation}
\label{discontinuity}
\mathcal{X}_{\gamma}^{\text{cw}(l)}(\zeta) = S_{l} \, \mathcal{X}_{\gamma}^{\text{ccw}(l)}(\zeta)
\,,
\quad
S_{l} = \prod_{\gamma'\in\Gamma(\vec a):\,l_{\gamma'}=l}^{\rm (cw)} \mathcal{K}_{\gamma'}^{\Omega(\gamma',\vec a)}
\end{equation}
where $\mathcal{X}_{\gamma}^{\text{cw}(l)}(\zeta) $ and $\mathcal{X}_{\gamma}^{\text{ccw}(l)}(\zeta)$ denote the limits of $\mathcal{X}_{\gamma}(\zeta)$ as it approaches $l$ clockwise (cw) and counterclockwise (ccw) in the complex $\zeta$ plane, $\Omega(\gamma,\vec a)$ is the degeneracy of the BPS state with charge $\gamma$; all operators in products (i.e., their BPS rays) are ordered clockwise where the counting starts from the right operator (equivalently, their central charges as complex vectors are ordered counterclockwise).
Explicitly, the $\mathcal{N}=2$ index $\Omega(\gamma,\vec a)$ is \cite{CFIV}
\begin{equation}
\label{multiplicity}
\Omega(\gamma,\vec a) = -\frac{1}{2} \tr_{\mathcal{H}_{\gamma,\rm BPS}} (-1)^{2J_{3}} (2J_{3})^{2}
\end{equation}
where $J_{3}$ is a generator of the rotational subgroup of the massive little group.
When $\vec a$ does not lie on a wall of marginal stability, no BPS rays coincide, and these discontinuities reduce to
\begin{equation}
\mathcal{X}_{\gamma}^{\text{cw}(l_{\gamma'})}(\zeta) = \mathcal{K}_{\gamma'}^{\Omega(\gamma',\vec a)} \, \mathcal{X}_{\gamma}^{\text{ccw}(l_{\gamma'})}(\zeta)
\,,
\quad
\gamma'\in\Gamma(\vec a)
\,.
\end{equation}

\paragraph{}

The BPS rays change their position as we vary $\vec a$.
When $\vec a$ is on a wall of marginal stability, (\ref{decay condition}) is satisfied (reducing to (\ref{decay condition 1}) for gauge group of rank 1), and there is a set of charges $\gamma$ for which $l_{\gamma}$ become aligned.
The set of charges with aligned BPS rays can be parametrised as $\gamma=n_{1}\gamma_{1}+n_{2}\gamma_{2}$ with $\{n_{1},n_{2}\}\subset\mathbb{N}$ for some basis $\{\gamma_{1},\gamma_{2}\}$ with $Z_{\gamma_{1}}/Z_{\gamma_{2}}\in\mathbb{R_{+}}$.
Near the wall, we form the product of operators corresponding to every aligned BPS ray at the wall:
\begin{equation}
P_{\gamma_{1},\gamma_{2}} = \prod_{\gamma=n_{1}\gamma_{1}+n_{2}\gamma_{2}\in\Gamma(\vec a):\,\{n_{1},n_{2}\}\in\mathbb{N}}^{\rm (cw)} \mathcal{K}_{\gamma}^{\Omega(\gamma,\vec a)} = \const
\,.
\end{equation}
The statement of the wall-crossing formula \cite{KS} is that as $\vec a$ crosses the wall of marginal stability, some multiplicities $\Omega(\gamma,\vec a)$ jump, the order of operators reverses, but the total product $P_{\gamma_{1},\gamma_{2}}$ remains invariant.
If the number of operators is infinite, the Lie algebra must be truncated by setting $e_{n\gamma_{1}+m\gamma_{2}}=0$ for $n+m>L$, and the infinite product can be understood as taking the limit $L\to\infty$.
In principle, if $\Omega(\gamma,\vec a)$ on one side of the wall are known, one can calculate them on the other side, however, this is difficult to implement in practice.
When dealing with several walls of marginal stability, the statement above is equivalent to requiring that the clockwise-ordered product of operators corresponding to all BPS states is conserved throughout the moduli space:
\begin{equation}
S = \prod_{\gamma\in\Gamma(\vec a)}^{\rm (cw)} \mathcal{K}_{\gamma}^{\Omega(\gamma,\vec a)} = \const
\,.
\end{equation}

\paragraph{}

It is not difficult to extend this construction to include massive hypermultiplets \cite{GMN}: for a vector of flavour charges $\vec s=(s_{1},\dots,s_{N_{f}})$, define its corresponding mass as
\begin{equation}
m_{\vec s} = \sum_{i=1}^{N_{f}} s_{i}m_{i}
\,.
\end{equation}
The resulting central charge (\ref{central charge}) has additional terms depending on the flavour masses $m_{i}$ and charges $s_{i}$:
\begin{equation}
\label{central charge flavours shift}
Z_{(\vec\gamma_{e},\vec\gamma_{m},\vec s)} (\vec a) = Z_{(\vec\gamma_{e},\vec\gamma_{m},\vec 0)}(\vec a)+m_{\vec s}
\,,
\end{equation}
where $Z_{(\vec\gamma_{e},\vec\gamma_{m},\vec 0)}(\vec a)$ is the central charge of the pure theory, and $m_{\vec s}$ is the shift corresponding to the massive flavours.
After compactifying the theory on $\mathbb{R}^{3}\times S^{1}$, in addition to the complex mass $m_{i}=m_{i\,1}+im_{i\,2}$, $\{m_{i\,1},m_{i\,2}\}\in\mathbb{R}$, an extra periodic mass parameter $m_{i\,3}$ appears, and we introduce the flavour Wilson line as $\psi_{i}=2\pi Rm_{i\,3}$ with period $2\pi R$.
Define a new factor $\mu_{\gamma}$ acting as a semiflat Darboux coordinate (\ref{coordinate semiflat}) for flavours, so that $\log(\mu_{\gamma}\mathcal{X}_{\gamma})$ is linear with respect to $Z_{\gamma}$ and $\bar Z_{\gamma}$:
\begin{equation}
\mu_{\gamma}(\zeta) = \mu_{\vec s}(\zeta) = \exp\left( \pi R \zeta^{-1}m_{\vec s}+i\psi_{\vec s}+\pi R\zeta\bar m_{\vec s} \right)
\,.
\end{equation}
From (\ref{central charge flavours shift}), we can deduce that Kontsevich--Soibelman operators carrying flavour charges should be defined as~\footnote{
Note that in our conventions, all flavour information is encoded in $\mu_{\gamma}$, and Darboux coordinates $\mathcal{X}_{\gamma}$ do not explicitly depend on flavour masses.
}
\begin{equation}
\mathcal{K}_{\gamma} \quad \colon \quad
\mathcal{X}_{\beta} \to \mathcal{X}_{\beta} \left( 1-\sigma(\gamma)\mu_{\gamma}\mathcal{X}_{\gamma} \right)^{2\langle\beta,\gamma\rangle}
\,.
\end{equation}

%%%%%%%%%%%%%%%%%%%%%%%%%%%%%%%%%%%%%%%%%%%%%%%%%%%%%%%%%%%%%

\section{Gaiotto--Moore--Neitzke equation}

\paragraph{}

The discontinuities of Darboux coordinates along the BPS rays in the complex plane and their asymptotic behaviour define a Riemann--Hilbert problem having a unique solution.
In the present case, the discontinuities of Darboux coordinates $\mathcal{X}$ along (\ref{BPS ray}) are given by (\ref{discontinuity}).
It was demonstrated \cite{GMN} that that this knowledge is sufficient to recover the values of $\mathcal{X}$ for any argument $\zeta$.
As we will show, their values are determined implicitly by a set of integral equations.

\paragraph{}

To completely define the problem, we need to fix the asymptotic behaviour of $\mathcal{X}$ as $\zeta\to 0$ and $\zeta\to\infty$.
To do this, we introduce
\begin{equation}
\Upsilon = \mathcal{X}(\mathcal{X}^{\rm sf})^{-1}
\end{equation}
and require that the limits
\begin{equation}
\label{RH limits}
\lim_{\zeta\to 0}\Upsilon = \Upsilon_{0}
\,,
\quad
\lim_{\zeta\to\infty}\Upsilon = \Upsilon_{\infty}
\end{equation}
exist and obey the reality condition
\begin{equation}
\label{RH asymptotics}
\Upsilon_{0} = \bar{\Upsilon}_{\infty}
\,.
\end{equation}

\paragraph{}

Because of the relation $\Omega(\gamma,\vec a)=\Omega(-\gamma,\vec a)$ between particles and their antiparticles, the problem possesses a discrete symmetry: for any given solution $\mathcal{X}$ of the problem, we can obtain another solution, $\tilde{\mathcal{X}}$, by defining
\begin{equation}
\tilde{\mathcal{X}}_{\gamma}(\zeta) = \overline{\mathcal{X}_{-\gamma}(- 1/\bar{\zeta})}
\,,
\end{equation}
guarantying that (\ref{RH limits}) is satisfied.
In fact, one can see that the initial solution is invariant under this transformation, i.e., that $\tilde{\mathcal{X}}=\mathcal{X}$.
To prove this, consider $Y=\tilde{\mathcal{X}}\mathcal{X}^{-1}$.
As $\mathcal{X}$ and $\tilde{\mathcal{X}}$ have the same discontinuities in the $\zeta$ plane, $Y$ is analytic in $\zeta$.
Equation (\ref{RH asymptotics}) means that $Y\to 1$ as $\zeta\to 0$ and $\zeta\to\infty$; therefore, by Liouville's theorem, $Y=1$, and $\tilde{\mathcal{X}}=\mathcal{X}$.
As a consequence of this,
\begin{equation}
\overline{\omega(-1/\bar\zeta)} = \omega(\zeta)
\,.
\end{equation}

\paragraph{}

The solution of the Riemann--Hilbert problem can be expressed in terms of integral equations.
Choosing the integration kernel that matches the boundary conditions, one can obtain the Gaiotto--Moore--Neitzke equation \cite{GMN, CV}:
\begin{equation}
\label{RH general}
\mathcal{X}_{\gamma}(\zeta) =
\mathcal{X}_{\gamma}^{\rm sf}(\zeta)
\exp\left(
\frac{1}{4\pi i} \sum_{l\in L(\Gamma)}
\int_{l} \frac{d\zeta'}{\zeta'} \frac{\zeta'+\zeta}{\zeta'-\zeta}
\log\frac{\mathcal{X}_{\gamma}(\zeta')}{(S_{l} \, \mathcal{X}_{\gamma})(\zeta')}
\right)
\end{equation}
where $L(\Gamma)$ is the set of different BPS rays for the BPS spectrum $\Gamma$.
The sum is over all rays along which the jump operator $S_{l}$ is a non-trivial product.
Since the solution is unique, in order to prove (\ref{RH general}), it is sufficient to show that it reproduces the discontinuities of Darboux coordinates and their asymptotic behaviour correctly.

\paragraph{}

We can choose one ray, $l$, and verify that the jump across $l$ is reproduced correctly: when $l$ is crossed counterclockwise, the jump of the right-hand side is given in terms of a residue becauses the integrand has a first-order pole at $\zeta'=\zeta$ along the ray:
\begin{equation}
\frac{\mathcal{X}_{\gamma}^{\text{ccw}(l)}(\zeta)}{\mathcal{X}_{\gamma}^{\text{cw}(l)}(\zeta)} =
\exp\left(
\res_{\zeta'=\zeta} \frac{1}{4\pi i}
\int_{l} \frac{d\zeta'}{\zeta'} \frac{\zeta'+\zeta}{\zeta'-\zeta}
\log\frac{\mathcal{X}_{\gamma}(\zeta')}{(S_{l} \, \mathcal{X}_{\gamma})(\zeta')}
\right)
\,,
\end{equation}
here, the residue is equal to the logarithm, hence, the jump is given correctly.
The asymptotic constraint is satisfied by setting the overall coefficient.

\paragraph{}

With respect to the discontinuities that we are considering (\ref{discontinuity}), the solution of the Riemann--Hilbert problem is encoded in a set of $2r$ integral equations~\footnote{
This formula appears in another form in Thermodynamic Bethe Ansatz \cite{Zamolodchikov, Alexandrov Roche}.
}:
\begin{equation}
\label{RH}
\mathcal{X}_{\gamma}(\zeta) =
\mathcal{X}_{\gamma}^{\rm sf}(\zeta)
\exp\left(
-\frac{1}{2\pi i} \sum_{\gamma'\in\Gamma}
\Omega(\gamma',\vec a) \langle\gamma,\gamma'\rangle
\int_{l_{\gamma'}} \frac{d\zeta'}{\zeta'} \frac{\zeta'+\zeta}{\zeta'-\zeta}
\log\left( 1-\sigma(\gamma')\mu_{\gamma'}(\zeta')\mathcal{X}_{\gamma'}(\zeta') \right)
\right)
\,.
\end{equation}
In theories without flavours, one should set all $\mu_{\gamma'}(\zeta')=1$.
This equation allows us to construct the moduli space metric if the relevant BPS spectrum is known.
In the following chapters, we will consider some specific cases.

%%%%%%%%%%%%%%%%%%%%%%%%%%%%%%%%%%%%%%%%%%%%%%%%%%%%%%%%%%%%%

\section{Wall-crossing in SU(2) theories}

\paragraph{}

The easiest example of the wall-crossing formula is the so-called pentagon identity \cite{KS}:
\begin{equation}
\label{WCF1}
\tilde{\mathcal{K}}_{(1,0)}\tilde{\mathcal{K}}_{(0,1)}
=
\tilde{\mathcal{K}}_{(0,1)}\tilde{\mathcal{K}}_{(1,1)}\tilde{\mathcal{K}}_{(1,0)}
\,,
\end{equation}
The formula predicts that as we cross the wall, only one extra particle is created in this case.
To prove the relation, it is sufficient to show that both sides are equal when they act on a basis of Darboux coordinates, say, $\mathcal{X}_{(1,0)}$ and $\mathcal{X}_{(0,1)}$, which is straightforward.
The pentagon formula is atypical in the sense that on both sides of the wall, the spectrum is finite.

\paragraph{}

Another useful example involves an infinite product \cite{KS}:
\begin{equation}
\label{WCF2 initial}
\tilde{\mathcal{K}}_{(0,1)}^{2} \tilde{\mathcal{K}}_{(1,0)}^{2}
=
\tilde{\mathcal{K}}_{(1,0)}^{2} \tilde{\mathcal{K}}_{(2,1)}^{2} \tilde{\mathcal{K}}_{(3,2)}^{2} \tilde{\mathcal{K}}_{(4,3)}^{2}
\dots \tilde{\mathcal{K}}_{(1,1)}^{4} \tilde{\mathcal{K}}_{(2,2)}^{-2} \dots
\tilde{\mathcal{K}}_{(3,4)}^{2} \tilde{\mathcal{K}}_{(2,3)}^{2} \tilde{\mathcal{K}}_{(1,2)}^{2} \tilde{\mathcal{K}}_{(0,1)}^{2}
\,.
\end{equation}
In \cite{GMN}, this formula was proven by using the following recursion:
\begin{equation}
\label{WCF2 recursion}
x_{n+1} x_{n-1} = (1-x_{n})^{2}
\,.
\end{equation}
Its general solutions is
\begin{equation}
\label{WCF2 recursion solution}
x_{n} = -\frac{\cosh^{2}(an+b)}{\sinh^{2}a}
\,.
\end{equation}
It will be convenient to deal with only one infinite product of operators, so we will use a different form of (\ref{WCF2 initial}):
\begin{equation}
\label{WCF2 recursion product}
\dots
\tilde{\mathcal{K}}_{(3,4)}^{2} \tilde{\mathcal{K}}_{(2,3)}^{2} \tilde{\mathcal{K}}_{(1,2)}^{2} \tilde{\mathcal{K}}_{(0,1)}^{2}
\tilde{\mathcal{K}}_{(1,0)}^{-2} \tilde{\mathcal{K}}_{(0,1)}^{-2}
\tilde{\mathcal{K}}_{(1,0)}^{2} \tilde{\mathcal{K}}_{(2,1)}^{2} \tilde{\mathcal{K}}_{(3,2)}^{2} \tilde{\mathcal{K}}_{(4,3)}^{2}
\dots
=
\tilde{\mathcal{K}}_{(2,2)}^{2} \tilde{\mathcal{K}}_{(1,1)}^{-4}
\,.
\end{equation}
To start with, identify $\mathcal{X}_{(1,0)}^{(0)}=x_{0}$, $\mathcal{X}_{(0,1)}^{(0)}=x_{1}^{-1}$.
The superscript $(0)$ indicates Darboux coordinates before acting on them with $\tilde{\mathcal{K}}_{(0,1)}^{2}$; moving left (right) by one operator corresponds to increasing (decreasing) the superscript by 1 (to move right, one needs to invert all operators starting with $\tilde{\mathcal{K}}_{(1,0)}^{-2}$).
First, we act on the Darboux coordinates with $\tilde{\mathcal{K}}_{(0,1)}^{2}$ and get $\mathcal{X}_{(0,1)}^{(1)}=x_{1}^{-1}$, $\mathcal{X}_{(1,2)}^{(1)}=x_{0}x_{1}^{-2}(1-x_{1}^{-1})^{-2}=x_{2}^{-1}$; then, we act with $\tilde{\mathcal{K}}_{(1,2)}^{2}$ and get $\mathcal{X}_{(1,2)}^{(2)}=x_{2}^{-1}$, $\mathcal{X}_{(2,3)}^{(2)}=x_{1}x_{2}^{-2}(1-x_{2}^{-1})^{-2}=x_{3}^{-1}$.
We keep applying $\tilde{\mathcal{K}}_{(n,n+1)}^{2}$ for $n\ge 0$.
After acting with $\tilde{\mathcal{K}}_{(n,n+1)}^{2}$, the coordinates are $\mathcal{X}_{(n,n+1)}^{(n+1)}=x_{n+1}^{-1}$, $\mathcal{X}_{(n+1,n+2)}^{(n+1)}=x_{n+2}^{-1}$: indeed, if $\tilde{\mathcal{K}}_{(n,n+1)}^{2}$ acts on the coordinates for which $\mathcal{X}_{(n-1,n)}^{(n)}=x_{n}^{-1}$ and $\mathcal{X}_{(n,n+1)}^{(n)}=x_{n+1}^{-1}$, we see that $\mathcal{X}_{(n,n+1)}^{(n)}=\mathcal{X}_{(n,n+1)}^{(n+1)}$, and the transformed values are determined by
\begin{equation}
\begin{aligned}
& \mathcal{X}_{(n+1,n+2)}^{(n+1)} =
\mathcal{X}_{(n+1,n+2)}^{(n)}\left( 1-\mathcal{X}_{(n,n+1)}^{(n)} \right)^{2(-(n+1)^{2}+n(n+2))}
\\
& = \left( \mathcal{X}_{(n-1,n)}^{(n)} \right)^{-1}\left( \mathcal{X}_{(n,n+1)}^{(n)} \right)^{2}\left( 1-\mathcal{X}_{(n,n+1)}^{(n)} \right)^{-2}
= x_{n}x_{n+1}^{-2}(1-x_{n+1}^{-1})^{-2} =
x_{n}(1-x_{n+1})^{-2} =
x_{n+2}^{-1}
\,.
\end{aligned}
\end{equation}
In the limit $n\to+\infty$, according to (\ref{WCF2 recursion solution}), we find
\begin{equation}
\begin{aligned}
\mathcal{X}_{(1,1)}^{(+\infty)} & = \mathcal{X}_{(n+1,n+2)}^{(+\infty)} \left( \mathcal{X}_{(n,n+1)}^{(+\infty)} \right)^{-1} = \lim_{n\to+\infty} x_{n+2}^{-1} x_{n+1} = e^{-2a}
\,,
\\
\mathcal{X}_{(0,1)}^{(+\infty)} & = \mathcal{X}_{(n,n+1)}^{(+\infty)} \left( \mathcal{X}_{(1,1)}^{(+\infty)} \right)^{-n} = \lim_{n\to+\infty} x_{n+1}^{-1} e^{2na} = -e^{-2b}(1-e^{-2a})^{2}
\,.
\end{aligned}
\end{equation}
On the other hand, we can go in the opposite direction in (\ref{WCF2 recursion product}) using inverse operators: $\mathcal{X}_{(1,0)}^{(0)}=x_{0}$, $\mathcal{X}_{(0,1)}^{(0)}=x_{1}^{-1}$, after acting with $\tilde{\mathcal{K}}_{(1,0)}^{2}$, become $\mathcal{X}_{(1,0)}^{(-1)}=x_{0}$, $\mathcal{X}_{(0,1)}^{(-1)}=x_{1}^{-1}(1-x_{0})^{2}=x_{-1}$, then, after acting with $\tilde{\mathcal{K}}_{(0,1)}^{2}$ on the result, they become $\mathcal{X}_{(1,0)}^{(-2)}=x_{0}(1-x_{-1})^{-2}=x_{-2}^{-1}$, $\mathcal{X}_{(0,1)}^{(-2)}=x_{-1}$ (and $\mathcal{X}_{(0,-1)}^{(-2)}=x_{-1}^{-1}$).
To shorten the proof, we notice that the two infinite series, $\tilde{\mathcal{K}}_{(1,0)}^{-2},\tilde{\mathcal{K}}_{(2,1)}^{-2},\tilde{\mathcal{K}}_{(3,2)}^{-2},\dots$ and $\tilde{\mathcal{K}}_{(0,1)}^{2},\tilde{\mathcal{K}}_{(1,2)}^{2},\tilde{\mathcal{K}}_{(2,3)}^{2},\dots$, are symmetric under simultaneous swapping of electric and magnetic charges and inverting the powers of operators, which is a symmetry of Kontsevich--Soibelman operators.
Therefore, after applying $\tilde{\mathcal{K}}_{(n+1,n)}^{-2}$ for $n\ge 0$, we have $\mathcal{X}_{(n+1,n)}^{(-n-3)}=x_{-n-2}^{-1}$, $\mathcal{X}_{(n+2,n+1)}^{(-n-3)}=x_{-n-3}^{-1}$ (with $n=-1$ corresponding to the initial state).
In the limit $n\to-\infty$, using (\ref{WCF2 recursion solution}) again, we find
\begin{equation}
\begin{aligned}
\mathcal{X}_{(1,1)}^{(-\infty)} & = \mathcal{X}_{(n+2,n+1)}^{(-\infty)} \left( \mathcal{X}_{(n+1,n)}^{(-\infty)} \right)^{-1} = \lim_{n\to-\infty} x_{-n-3}^{-1} x_{-n-2} = e^{-2a}
\,,
\\
\mathcal{X}_{(0,1)}^{(-\infty)} & = \left( \mathcal{X}_{(n+1,n)}^{(-\infty)} \right)^{-1} \left( \mathcal{X}_{(1,1)}^{(-\infty)} \right)^{n+1} = \lim_{n\to-\infty} x_{-n-2} e^{-2(n+1)a} = -e^{-2b}(1-e^{-2a})^{-2}
\,.
\end{aligned}
\end{equation}
Thus, the full recursion from $n\to-\infty$ to $n\to+\infty$, corresponding to the left-hand side of (\ref{WCF2 recursion product}), transforms the Darboux coordinates as
\begin{equation}
\begin{aligned}
\mathcal{X}_{(1,1)} & \to \mathcal{X}_{(1,1)} = e^{-2a}
\,,
\\
\mathcal{X}_{(0,1)} & \to \mathcal{X}_{(0,1)}(1-e^{-2a})^{4} = \mathcal{X}_{(0,1)}(1-e^{-4a})^{4}(1+e^{-2a})^{-4}
\,.
\end{aligned}
\end{equation}
This is precisely the transformation generated by $\tilde{\mathcal{K}}_{(2,2)}^{2} \tilde{\mathcal{K}}_{(1,1)}^{-4}$ (this follows directly from (\ref{KS operator}) as these two operators commute).
This completes the proof.

\paragraph{}

Changing the basis in (\ref{WCF2 initial}), one can obtain \cite{GMN2}
\begin{equation}
\tilde{\mathcal{K}}_{\gamma_{1}}^{2} \tilde{\mathcal{K}}_{\gamma_{2}}^{2}
=
\tilde{\mathcal{K}}_{\gamma_{2}}^{2} \tilde{\mathcal{K}}_{\gamma_{1}+2\gamma_{2}}^{2} \tilde{\mathcal{K}}_{2\gamma_{1}+3\gamma_{2}}^{2} \tilde{\mathcal{K}}_{3\gamma_{1}+4\gamma_{2}}^{2}
\dots \tilde{\mathcal{K}}_{\gamma_{1}+\gamma_{2}}^{4} \tilde{\mathcal{K}}_{2\gamma_{1}+2\gamma_{2}}^{-2} \dots
\tilde{\mathcal{K}}_{4\gamma_{1}+3\gamma_{2}}^{2} \tilde{\mathcal{K}}_{3\gamma_{1}+2\gamma_{2}}^{2} \tilde{\mathcal{K}}_{2\gamma_{1}+\gamma_{2}}^{2} \tilde{\mathcal{K}}_{\gamma_{1}}^{2}
\end{equation}
for all $(\gamma_{1},\gamma_{2})$ obeying $\langle\gamma_{1},\gamma_{2}\rangle=\pm 1$.
In particular, we can select the basis as $\gamma_{1}=(1,-1)$, $\gamma_{2}=(0,1)$ to match the charges on the right-hand side of this equation with the weak-coupling spectrum of the $SU(2)$ theory with $N_{f} = 2$ flavours (we need to transform the coordinates as $\mathcal{X}_{m}\to -\mathcal{X}_{m}$ in (\ref{WCF2 initial}) prior to changing the basis), then, the left-hand side predicts the strong-coupling spectrum; analogously, one can find the formulae for $N_{f} = 1$ and $N_{f} = 3$; explicitly, the results are
\begin{align}
\label{WCF2 1f}
& N_{f} = 1 \colon \
\tilde{\mathcal{K}}_{(1,-1)} \tilde{\mathcal{K}}_{(1,0)} \tilde{\mathcal{K}}_{(0,1)}
=
\tilde{\mathcal{K}}_{(0,1)} \tilde{\mathcal{K}}_{(1,1)} \tilde{\mathcal{K}}_{(2,1)}
\dots \tilde{\mathcal{K}}_{(1,0)}^{2} \tilde{\mathcal{K}}_{(2,0)}^{-2} \dots
\tilde{\mathcal{K}}_{(4,-1)} \tilde{\mathcal{K}}_{(3,-1)} \tilde{\mathcal{K}}_{(2,-1)} \tilde{\mathcal{K}}_{(1,-1)}
\,,
\\
\label{WCF2 2f}
& N_{f} = 2 \colon \ 
\tilde{\mathcal{K}}_{(1,-1)}^{2} \tilde{\mathcal{K}}_{(0,1)}^{2}
=
\tilde{\mathcal{K}}_{(0,1)}^{2} \tilde{\mathcal{K}}_{(1,1)}^{2} \tilde{\mathcal{K}}_{(2,1)}^{2} \tilde{\mathcal{K}}_{(3,1)}^{2}
\dots \tilde{\mathcal{K}}_{(1,0)}^{4} \tilde{\mathcal{K}}_{(2,0)}^{-2} \dots
\tilde{\mathcal{K}}_{(4,-1)}^{2} \tilde{\mathcal{K}}_{(3,-1)}^{2} \tilde{\mathcal{K}}_{(2,-1)}^{2} \tilde{\mathcal{K}}_{(1,-1)}^{2}
\,,
\\
\label{WCF2 3f}
& N_{f} = 3 \colon \
\tilde{\mathcal{K}}_{(1,-2)} \tilde{\mathcal{K}}_{(0,1)}^{4}
=
\tilde{\mathcal{K}}_{(0,1)}^{4} \tilde{\mathcal{K}}_{(1,2)} \tilde{\mathcal{K}}_{(1,1)}^{4} \tilde{\mathcal{K}}_{(3,2)}
\dots \tilde{\mathcal{K}}_{(1,0)}^{6} \tilde{\mathcal{K}}_{(2,0)}^{-2} \dots
\tilde{\mathcal{K}}_{(2,-1)}^{4} \tilde{\mathcal{K}}_{(3,-2)} \tilde{\mathcal{K}}_{(1,-1)}^{4} \tilde{\mathcal{K}}_{(1,-2)}
\,.
\end{align}
The charges on the left-hand side are indeed the strong-coupling spectrum of the $SU(2)$ theory \cite{SW2}.
In fact, it can be shown that equations (\ref{WCF2 1f}, \ref{WCF2 3f}) can be derived from (\ref{WCF2 2f}) by repeatedly applying the pentagon formula (\ref{WCF1}) \cite{Dimofte Gukov Soibelman}.

\paragraph{}

Changing the Darboux coordinates in (\ref{WCF2 2f}) as $\mathcal{X}_{(x,y)}\to(-1)^{x}\mathcal{X}_{(\frac{x}{2},y)}$ and using $\mathcal{K}$ instead of $\tilde{\mathcal{K}}$, one can get another important formula:
\begin{equation}
\label{WCF2}
N_{f} = 0 \colon \
\mathcal{K}_{(1,-1)} \mathcal{K}_{(0,1)}
=
\mathcal{K}_{(0,1)} \mathcal{K}_{(1,1)} \mathcal{K}_{(2,1)} \mathcal{K}_{(3,1)}
\dots \mathcal{K}_{(1,0)}^{-2} \dots
\mathcal{K}_{(4,-1)} \mathcal{K}_{(3,-1)} \mathcal{K}_{(2,-1)} \mathcal{K}_{(1,-1)}
\,.
\end{equation}
It describes the same theory, but without flavour charges \cite{SW}.

\paragraph{}

In the case of massive flavours, the operators in (\ref{WCF2 1f}, \ref{WCF2 2f}, \ref{WCF2 3f}) must also contain the flavour charges.
We will focus on modifying the massless formula (\ref{WCF2 initial}) in order to derive an analogue of (\ref{WCF2 2f}) applicable in the massive case.
When $N_{f}=2$, the operators contain flavour charges under the Cartan generators of $SU(2)_{A}$ and $SU(2)_{B}$ of the $SO(4)= SU(2)_{A}\times SU(2)_{B}$ flavour symmetry.
The wall-crossing formula should take this into account by alternating the operators for $SU(2)_{A}$ and $SU(2)_{B}$.
The recursion relation (\ref{WCF2 recursion}) should be generalised to \cite{GMN}
\begin{equation}
\label{WCF2 recursion matter}
x_{n+1} x_{n-1} = (1-e^{u+(-1)^{n}v}x_{n}) (1-e^{-u+(-1)^{n}v}x_{n})
\,.
\end{equation}
Its solution, which reduces to (\ref{WCF2 recursion solution}) for $u=v$, is
\begin{equation}
\label{WCF2 recursion solution matter}
\begin{aligned}
x_{n} = & -\frac{\cosh(2an+2b)\sqrt{(\cosh(2a)+\cosh(2u))(\cosh(2a)+\cosh(2v))}}{\sinh^{2}(2a)}
\\
& -\frac{1}{2}\frac{\cosh u\cosh v}{\sinh^{2}a}+(-1)^{n}\frac{1}{2}\frac{\sinh u\sinh v}{\sinh^{2}a}
\,.
\end{aligned}
\end{equation}
To preserve the relations between the recursion elements and Darboux coordinates that we found in the massless case, we should identify the masses under $SU(2)_{A}$ and $SU(2)_{B}$ as $\mu_{A}=e^{u-v}$ and $\mu_{B}=e^{u+v}$.
We also substitute
\begin{equation}
\begin{aligned}
\tilde{\mathcal{K}}_{(2n+1,2n)}^{2} & \to
\tilde{\mathcal{K}}_{(2n+1,2n),\,(1,0)} \tilde{\mathcal{K}}_{(2n+1,2n),\,(-1,0)}
\,,
\\
\tilde{\mathcal{K}}_{(2n,2n-1)}^{2} & \to
\tilde{\mathcal{K}}_{(2n,2n-1),\,(0,1)} \tilde{\mathcal{K}}_{(2n,2n-1),\,(0,-1)}
\end{aligned}
\end{equation}
where $n\in\mathbb{Z}$.
After these changes, the total infinite product in (\ref{WCF2 recursion product}) generates the following transformations:
\begin{equation}
\begin{aligned}
\mathcal{X}_{(1,1)} & \to \mathcal{X}_{(1,1)}
\,,
\\
\mathcal{X}_{(1,0)} & \to \mathcal{X}_{(1,0)}
\frac{(1-\mathcal{X}_{(1,1)}^{2})^{4}}{(1+\mu_{A}\mu_{B}\mathcal{X}_{(1,1)}) (1+\mu_{A}^{-1}\mu_{B}\mathcal{X}_{(1,1)}) (1+\mu_{A}\mu_{B}^{-1}\mathcal{X}_{(1,1)})(1+\mu_{A}^{-1}\mu_{B}^{-1}\mathcal{X}_{(1,1)})}
\,.
\end{aligned}
\end{equation}
Alternatively, these transformations can be generated by the operators corresponding to four hypermultiplets with flavour charges $(\pm 1,\pm 1)$ and $(\pm 1,\mp1)$ and a vector multiplet without flavour charges.
In other words, the operators for hypermultiplets in the original formula (\ref{WCF2 initial}) should be generalised as
\begin{equation}
\tilde{\mathcal{K}}_{(1,1)}^{4} \to
\tilde{\mathcal{K}}_{(1,1),\,(1,1)} \tilde{\mathcal{K}}_{(1,1),\,(-1,-1)} \tilde{\mathcal{K}}_{(1,1),\,(1,-1)} \tilde{\mathcal{K}}_{(1,1),\,(-1,1)}
\,.
\end{equation}

\paragraph{}

Then, we can modify (\ref{WCF2 2f}) when the matter hypermultiplets become massive.
Denote the flavour charges of the four quarks as $(\pm 1,0)$ and $(0,\pm 1)$.
This means that in order to reuse (\ref{WCF2 initial}), we must rotate all flavour charges as $(1,1)\to(1,0)$, $(1,-1)\to(0,1)$, therefore, the dyons $(2n,1)$ and $(2n+1,1)$ (with $n\in\mathbb{Z}$) have flavour charges $(\frac{1}{2},\frac{1}{2})$ and $(\frac{1}{2},-\frac{1}{2})$, respectively.

\paragraph{}

For any pair $\gamma$ and $-\gamma$ of particles from the spectrum, the wall-crossing formulae considered above contain only one operator, $\mathcal{K}_{\gamma}$ or $\mathcal{K}_{-\gamma}$.
To consider all particles preserving the order of operators, both sides of the formula should be multiplied (from the left or from the right) by the same expression but with opposite charges (including flavour charges).

%%%%%%%%%%%%%%%%%%%%%%%%%%%%%%%%%%%%%%%%%%%%%%%%%%%%%%%%%%%%%

\section{Wall-crossing in higher-rank theories}

\paragraph{}

{\it This section is based on \cite{CDP2}.}

\paragraph{}

As we have already mentioned, in $\mathcal{N}=2$ theories with gauge group $SU(n)$, $n\ge 3$, the weak-coupling spectrum is different in different regions of the moduli space.
These regions are separated by walls of marginal stability: on each wall, one composite dyon becomes unstable and decays (or, conversely, gets created).
Such decays are possible when the total central charge (\ref{central charge}) and the total mass are preserved.
For the decay process $\gamma\to\gamma_{1}+\gamma_{2}$, the condition is simply $Z_{\gamma}=Z_{\gamma_{1}}+Z_{\gamma_{2}}$, $|Z_{\gamma}|=|Z_{\gamma_{1}}|+|Z_{\gamma_{2}}|$; this means that $\arg Z_{\gamma_{1}}=\arg Z_{\gamma_{2}}$.
To the leading order at weak coupling, the values of central charges for dyons depend only on their magnetic charges.
Hence, in this limit, the walls of marginal stability are given by
\begin{equation}
\label{weak wall}
\frac{\vec\alpha_{A}\vec a}{\vec\alpha_{B}\vec a} \in \mathbb{R}_{+}
\end{equation}
for some pair of positive roots, $\vec\alpha_{A}$ and $\vec\alpha_{B}$.
The composite dyon given by (\ref{composite dyon}) decays near the wall of marginal stability which can be reparametrised as
\begin{equation}
\frac{\sum_{m=i}^{k}\vec\alpha_{m}\vec a}{\sum_{m=k+1}^{j-1}\vec\alpha_{m}\vec a} \in \mathbb{R}_{+}
\quad \iff \quad
\frac{\vec e_{i}\vec a}{\vec e_{j}\vec a} \in \mathbb{R}_{+}
\,.
\end{equation}

\paragraph{}

When we take into account the electric charges of dyons, there is, in fact, no single wall of marginal stability, but rather, a collection of walls.
For every composite dyon, there is an individual wall where it can decay (these walls extend into the strong-coupling region \cite{Taylor, Taylor 2}).
Precisely at (\ref{weak wall}), no decay reactions take place (although at this wall, $\arg Z_{(\vec\alpha_{A},\vec 0)}=\arg Z_{(\vec\alpha_{B},\vec 0)}=\arg Z_{(\vec\alpha_{A}+\vec\alpha_{B},\vec 0)}$).
On the other hand, taking the effective coupling constant $g_{\rm eff}$ sufficiently small, all these walls can be set infinitely close to each other; this is the reason why for the VEV far from (\ref{weak wall}), they can be treated as a single wall.

\paragraph{}

Using the fact that each composite dyon can be parametrised as (\ref{composite dyon}), we can write down the decay processes:
\begin{equation}
\label{composite dyon decay}
\begin{aligned}
\pm \left(
p\sum_{m=i}^{j-1}\vec\alpha_{m}+\sum_{l=i+1}^{j-1}\epsilon_{l}\sum_{m=l}^{j-1}\vec\alpha_{m}
\,, \
\sum_{m=i}^{j-1}\vec\alpha_{m}
\right)
\to
\pm \left(
p\sum_{m=i}^{k}\vec\alpha_{m}+\sum_{l=i+1}^{k}\epsilon_{l}\sum_{m=l}^{k}\vec\alpha_{m}
\,, \
\sum_{m=i}^{k}\vec\alpha_{m}
\right)
\\
\pm \left(
\left( p+\sum_{l=i+1}^{k}\epsilon_{l} \right)\sum_{m=k+1}^{j-1}\vec\alpha_{m}+\sum_{l=k+1}^{j-1}\epsilon_{l}\sum_{m=l}^{j-1}\vec\alpha_{m}
\,, \
\sum_{m=k+1}^{j-1}\vec\alpha_{m}
\right)
\,,
\end{aligned}
\end{equation}
or, rewriting it in terms of the orthonormal basis introduced above,
\begin{equation}
\begin{aligned}
\pm \frac{1}{\sqrt{2}} & \left(
p\left( \vec e_{i}-\vec e_{j} \right)+\sum_{l=i+1}^{j-1}\epsilon_{l}\left( \vec e_{l}-\vec e_{j} \right)
\,, \
\vec e_{i}-\vec e_{j}
\right)
\\
\to
\pm \frac{1}{\sqrt{2}} & \left(
p\left( \vec e_{i}-\vec e_{k+1} \right)+\sum_{l=i+1}^{k}\epsilon_{l}\left( \vec e_{l}-\vec e_{k+1} \right)
\,, \
\vec e_{i}-\vec e_{k+1}
\right)
\\
\pm \frac{1}{\sqrt{2}} & \left(
\left( p+\sum_{l=i+1}^{k}\epsilon_{l} \right)\left( \vec e_{k+1}-\vec e_{j} \right)+\sum_{l=k+1}^{j-1}\epsilon_{l}\left( \vec e_{l}-\vec e_{j} \right)
\,, \
\vec e_{k+1}-\vec e_{j}
\right)
\,.
\end{aligned}
\end{equation}
In particular, for gauge group $SU(3)$, with one possible weak-coupling wall (\ref{weak wall}), $\vec\alpha_{1}\vec a/\vec\alpha_{2}\vec a\in\mathbb{R}_{+}$, there are two types of decays corresponding to the VEV approaching the wall from different sides:
\begin{equation}
\label{SU(3) composite dyon decay}
\begin{aligned}
\pm(p(\vec\alpha_{1}+\vec\alpha_{2})+\vec\alpha_{1},\vec\alpha_{1}+\vec\alpha_{2}) & \to
\pm((p+1)\vec\alpha_{1},\vec\alpha_{1})\pm(p\vec\alpha_{2},\vec\alpha_{2})
\,,
\\
\pm(p(\vec\alpha_{1}+\vec\alpha_{2})+\vec\alpha_{2},\vec\alpha_{1}+\vec\alpha_{2}) & \to
\pm((p+1)\vec\alpha_{2},\vec\alpha_{2})\pm(p\vec\alpha_{1},\vec\alpha_{1})
\,.
\end{aligned}
\end{equation}

\paragraph{}

Our goal is to express the spectra and decays discussed above in terms of Kontsevich--Soibelman operators \cite{KS} and show that the wall-crossing formulae are satisfied.

\paragraph{}

First, let us show how to change the basis of charge-vectors in a wall-crossing formula.
Any given formula
\begin{equation}
\prod_{k=1}^{K} \mathcal{K}_{\gamma_{k}}=1
\,,
\quad
K\in\mathbb{N}\cup\{+\infty\}
\,,
\end{equation}
can be re-expressd in different coordinates ($\gamma_{k}\to\beta_{k}$) if the transformation of charge-vectors is linear, and if for any pair of charges in the formula, their symplectic product remains the same, i.e., $\langle\beta_{i},\beta_{j}\rangle=\langle\gamma_{i},\gamma_{j}\rangle$.
The formula in these new coordinates is
\begin{equation}
\prod_{k=1}^{K}\mathcal{K}_{\beta_{k}}=1
\,.
\end{equation}

\paragraph{}

Let us prove this statement. Suppose that we change coordinates as $\gamma_{(i)}\to\beta_{(i)}$ for all possible charges $\gamma_{(i)}$.
Linearity of the transformation ensures that all symplectic products are also linear, i.e.,
\begin{equation}
\langle\beta_{(1)}+\beta_{(2)},\beta_{(3)}\rangle = \langle\beta_{(1)},\beta_{(3)}\rangle+\langle\beta_{(2)},\beta_{(3)}\rangle
\,,
\end{equation}
and that changing the coordinates does not violate the condition
\begin{equation}
\mathcal{K}_{\beta_{(1)}+ \beta_{(2)}}=\mathcal{K}_{\beta_{(1)}}\mathcal{K}_{\beta_{(2)}}
\,.
\end{equation}
The operators $\mathcal{K}_{\beta_{k}}$ act depending only on the symplectic products between $\beta_{k}$ and $\beta_{l}$ where $k<l\le K$, which are conserved.

\paragraph{}

Now, consider the standard pentagon wall-crossing formula:
\begin{equation}
\begin{aligned}
\mathcal{K}_{(\frac{1}{2},0)}\mathcal{K}_{(0,1)} =
\mathcal{K}_{(0,1)}\mathcal{K}_{(\frac{1}{2},1)}\mathcal{K}_{(\frac{1}{2},0)}
\,,
\\
\mathcal{K}_{(0,1)}\mathcal{K}_{(\frac{1}{2},0)} =
\mathcal{K}_{(\frac{1}{2},0)}\mathcal{K}_{(\frac{1}{2},1)}\mathcal{K}_{(0,1)}
\,.
\end{aligned}
\end{equation}
In the case of $r$ electric and $r$ magnetic charges, the equations are
\begin{equation}
\begin{aligned}
\mathcal{K}_{((0,0,\dots,0),(1,0,\dots,0))}\mathcal{K}_{((\frac{1}{2},0,\dots,0),(0,0,\dots,0))} =
\mathcal{K}_{((\frac{1}{2},0,\dots,0),(0,0,\dots,0))}\mathcal{K}_{((\frac{1}{2},0,\dots,0),(1,0,\dots,0))}\mathcal{K}_{((0,0,\dots,0),(1,0,\dots,0))}
\,,
\\
\mathcal{K}_{((\frac{1}{2},0,\dots,0),(0,0,\dots,0))}\mathcal{K}_{((0,0,\dots,0),(1,0,\dots,0))} =
\mathcal{K}_{((0,0,\dots,0),(1,0,\dots,0))}\mathcal{K}_{((\frac{1}{2},0,\dots,0),(1,0,\dots,0))}\mathcal{K}_{((\frac{1}{2},0,\dots,0),(0,0,\dots,0))}
\,.
\end{aligned}
\end{equation}
The proof is straightforward: the relation is known to be valid when the left and the right-hand sides act on $\mathcal{X}_{e}^{1}$ and $\mathcal{X}_{m\,1}$; both sides give identity when acting on $\mathcal{X}_{e}^{I}$ and $\mathcal{X}_{m\,I}$ for $I>1$.

\paragraph{}

Therefore, more generally, changing the basis, we obtain the following form of the pentagon formula for any $r$:
\begin{equation}
\label{pentagon}
\mathcal{K}_{\gamma_{1}}\mathcal{K}_{\gamma_{2}} =
\mathcal{K}_{\gamma_{2}}\mathcal{K}_{\gamma_{1}+\gamma_{2}}\mathcal{K}_{\gamma_{1}}
\,,
\quad
\forall \ \langle\gamma_{1},\gamma_{2}\rangle = \pm\frac{1}{2}
\,.
\end{equation}
This is an extension of the $SU(2)$ formula to the charges with any number of components for any basis.

%%%%%%%%%%%%%%%%%%%%%%%%%%%%%%

\subsection{SU(3) theory}

\paragraph{}

Let us start by considering the theory with gauge group $SU(3)$.
The pentagon formula (\ref{pentagon}) can be applied to the decay reaction of the composite dyons.
Indeed, in (\ref{SU(3) composite dyon decay}), symplectic product of the two simple dyons is
\begin{equation}
\left\langle
\pm (p\vec\alpha_{1},\vec\alpha_{1})
\,,
\pm ((p+1)\vec\alpha_{2},\vec\alpha_{2})
\right\rangle
= -\frac{1}{2}
\,,
\end{equation}
and the decay of the composite dyon in (\ref{SU(3) composite dyon decay}) is described by the following formula:
\begin{equation}
\label{SU(3) pentagon}
\mathcal{K}_{\pm(p\vec\alpha_{1},\vec\alpha_{1})}
\mathcal{K}_{\pm((p+1)\vec\alpha_{2},\vec\alpha_{2})}
=
\mathcal{K}_{\pm((p+1)\vec\alpha_{2},\vec\alpha_{2})}
\mathcal{K}_{\pm(p(\vec\alpha_{2}+\vec\alpha_{1})+\vec\alpha_{2},\vec\alpha_{1}+\vec\alpha_{2})}
\mathcal{K}_{\pm(p\vec\alpha_{1},\vec\alpha_{1})}
\,.
\end{equation}

\paragraph{}

Starting with these formulae, we will construct the wall-crossing formula for the pure $SU(3)$ theory at weak coupling.
It is closely related to the wall-crossing formula (\ref{WCF2}) for the pure $SU(2)$ theory.
In the $SU(n)$ case, we will require electric charge-vectors of $W$ bosons and magnetic charge-vectors of dyons to be positive roots, ignoring their antiparticles.
Let us begin by writing out the wall-crossing formula implied by the known spectra of the $SU(3)$ theory \cite{Fraser Hollowood} on either side of the walls of marginal stability (this equation will be proven shortly):
\begin{equation}
\label{WCF3}
\begin{aligned}
& \dots
\mathcal{K}_{(-2\vec\alpha_{1},\vec\alpha_{1})}
\mathcal{K}_{(-3(\vec\alpha_{1}+\vec\alpha_{2})+\vec\alpha_{1},\vec\alpha_{1}+\vec\alpha_{2})}
\mathcal{K}_{(-3\vec\alpha_{2},\vec\alpha_{2})}
\times
\mathcal{K}_{(-\vec\alpha_{1},\vec\alpha_{1})}
\mathcal{K}_{(-2(\vec\alpha_{1}+\vec\alpha_{2})+\vec\alpha_{1},\vec\alpha_{1}+\vec\alpha_{2})}
\mathcal{K}_{(-2\vec\alpha_{2},\vec\alpha_{2})}
\times \\ &
\mathcal{K}_{(\vec 0,\vec\alpha_{1})}
\mathcal{K}_{(-(\vec\alpha_{1}+\vec\alpha_{2})+\vec\alpha_{1},\vec\alpha_{1}+\vec\alpha_{2})}
\mathcal{K}_{(-\vec\alpha_{2},\vec\alpha_{2})}
\times
\mathcal{K}_{(\vec\alpha_{1},\vec\alpha_{1})}
\mathcal{K}_{(\vec\alpha_{1},\vec\alpha_{1}+\vec\alpha_{2})}
\mathcal{K}_{(\vec 0,\vec\alpha_{2})}
\times \\ &
\mathcal{K}_{(2\vec\alpha_{1},\vec\alpha_{1})}
\mathcal{K}_{((\vec\alpha_{1}+\vec\alpha_{2})+\vec\alpha_{1},\vec\alpha_{1}+\vec\alpha_{2})}
\mathcal{K}_{(\vec\alpha_{2},\vec\alpha_{2})}
\times
\mathcal{K}_{(3\vec\alpha_{1},\vec\alpha_{1})}
\mathcal{K}_{(2(\vec\alpha_{1}+\vec\alpha_{2})+\vec\alpha_{1},\vec\alpha_{1}+\vec\alpha_{2})}
\mathcal{K}_{(2\vec\alpha_{2},\vec\alpha_{2})}
\dots
\\
&
\mathcal{K}_{(\vec\alpha_{1},\vec 0)}^{-2}
\mathcal{K}_{(\vec\alpha_{1}+\vec\alpha_{2},\vec 0)}^{-2}
\mathcal{K}_{(\vec\alpha_{2},\vec 0)}^{-2}
\ =
\\
& \dots
\mathcal{K}_{(-2\vec\alpha_{2},\vec\alpha_{2})}
\mathcal{K}_{(-3(\vec\alpha_{1}+\vec\alpha_{2})+\vec\alpha_{2},\vec\alpha_{1}+\vec\alpha_{2})}
\mathcal{K}_{(-3\vec\alpha_{1},\vec\alpha_{1})}
\times
\mathcal{K}_{(-\vec\alpha_{2},\vec\alpha_{2})}
\mathcal{K}_{(-2(\vec\alpha_{1}+\vec\alpha_{2})+\vec\alpha_{2},\vec\alpha_{1}+\vec\alpha_{2})}
\mathcal{K}_{(-2\vec\alpha_{1},\vec\alpha_{1})}
\times \\ &
\mathcal{K}_{(\vec 0,\vec\alpha_{2})}
\mathcal{K}_{(-(\vec\alpha_{1}+\vec\alpha_{2})+\vec\alpha_{2},\vec\alpha_{1}+\vec\alpha_{2})}
\mathcal{K}_{(-\vec\alpha_{1},\vec\alpha_{1})}
\times
\mathcal{K}_{(\vec\alpha_{2},\vec\alpha_{2})}
\mathcal{K}_{(\vec\alpha_{2},\vec\alpha_{1}+\vec\alpha_{2})}
\mathcal{K}_{(\vec 0,\vec\alpha_{1})}
\times \\ &
\mathcal{K}_{(2\vec\alpha_{2},\vec\alpha_{2})}
\mathcal{K}_{((\vec\alpha_{1}+\vec\alpha_{2})+\vec\alpha_{2},\vec\alpha_{1}+\vec\alpha_{2})}
\mathcal{K}_{(\vec\alpha_{1},\vec\alpha_{1})}
\times
\mathcal{K}_{(3\vec\alpha_{2},\vec\alpha_{2})}
\mathcal{K}_{(2(\vec\alpha_{1}+\vec\alpha_{2})+\vec\alpha_{2},\vec\alpha_{1}+\vec\alpha_{2})}
\mathcal{K}_{(2\vec\alpha_{1},\vec\alpha_{1})}
\dots
\\
&
\mathcal{K}_{(\vec\alpha_{2},\vec 0)}^{-2}
\mathcal{K}_{(\vec\alpha_{1}+\vec\alpha_{2},\vec 0)}^{-2}
\mathcal{K}_{(\vec\alpha_{1},\vec 0)}^{-2}
\,,
\end{aligned}
\end{equation}
where ``$\times$'' are used only to group operators.
The BPS ray at which the ordering starts is chosen differently from the $SU(2)$ case for further convenience.

\paragraph{}

We will now verify (\ref{WCF3}) by evaluating both sides.
We can see how both sides of (\ref{WCF3}) change when the VEV passes thorough the walls.
For each decaying composite dyon, there is a corresponding pentagon identity (\ref{SU(3) pentagon}) modifying a fragment (separated by ``$\times$'') in (\ref{WCF3}).
Close to the wall (\ref{weak wall}), when all composite dyons decay, the wall-crossing formula (\ref{WCF3}) reduces to
\begin{equation}
\label{WCF3.1}
\begin{aligned}
& \dots
\mathcal{K}_{(-3\vec\alpha_{2},\vec\alpha_{2})}
\mathcal{K}_{(-2\vec\alpha_{1},\vec\alpha_{1})}
\times
\mathcal{K}_{(-2\vec\alpha_{2},\vec\alpha_{2})}
\mathcal{K}_{(-\vec\alpha_{1},\vec\alpha_{1})}
\times
\mathcal{K}_{(-\vec\alpha_{2},\vec\alpha_{2})}
\mathcal{K}_{(\vec 0,\vec\alpha_{1})}
\times \\ &
\mathcal{K}_{(\vec 0,\vec\alpha_{2})}
\mathcal{K}_{(\vec\alpha_{1},\vec\alpha_{1})}
\times
\mathcal{K}_{(\vec\alpha_{2},\vec\alpha_{2})}
\mathcal{K}_{(2\vec\alpha_{1},\vec\alpha_{1})}
\times
\mathcal{K}_{(2\vec\alpha_{2},\vec\alpha_{2})}
\mathcal{K}_{(3\vec\alpha_{1},\vec\alpha_{1})}
\dots \\ &
\mathcal{K}_{(\vec\alpha_{1},\vec 0)}^{-2}
\mathcal{K}_{(\vec\alpha_{1}+\vec\alpha_{2},\vec 0)}^{-2}
\mathcal{K}_{(\vec\alpha_{2},\vec 0)}^{-2}
\ =
\\
& \dots
\mathcal{K}_{(-3\vec\alpha_{1},\vec\alpha_{1})}
\mathcal{K}_{(-2\vec\alpha_{2},\vec\alpha_{2})}
\times
\mathcal{K}_{(-2\vec\alpha_{1},\vec\alpha_{1})}
\mathcal{K}_{(-\vec\alpha_{2},\vec\alpha_{2})}
\times
\mathcal{K}_{(-\vec\alpha_{1},\vec\alpha_{1})}
\mathcal{K}_{(\vec 0,\vec\alpha_{2})}
\times \\ &
\mathcal{K}_{(\vec 0,\vec\alpha_{1})}
\mathcal{K}_{(\vec\alpha_{2},\vec\alpha_{2})}
\times
\mathcal{K}_{(\vec\alpha_{1},\vec\alpha_{1})}
\mathcal{K}_{(2\vec\alpha_{2},\vec\alpha_{2})}
\times
\mathcal{K}_{(2\vec\alpha_{1},\vec\alpha_{1})}
\mathcal{K}_{(3\vec\alpha_{2},\vec\alpha_{2})}
\dots \\ &
\mathcal{K}_{(\vec\alpha_{2},\vec 0)}^{-2}
\mathcal{K}_{(\vec\alpha_{1}+\vec\alpha_{2},\vec 0)}^{-2}
\mathcal{K}_{(\vec\alpha_{1},\vec 0)}^{-2}
\,.
\end{aligned}
\end{equation}
We can see that this equation is an identity: $\mathcal{K}_{(p\vec\alpha_{1},\vec\alpha_{1})}$ commutes with $\mathcal{K}_{(p\vec\alpha_{2},\vec\alpha_{2})}$, the three purely electric operators commute with each other (two operators commute when symplectic product of their charges is zero); these commuting operators reverse their order precisely at the wall of marginal stability (\ref{weak wall}).
By proving (\ref{WCF3.1}), we have also shown that (\ref{WCF3}) is correct via substituting the pentagon identity (\ref{pentagon}).

%%%%%%%%%%%%%%%%%%%%%%%%%%%%%%

\subsection{SU(n) theory}

\paragraph{}

Let us generalise our results for gauge group $SU(2)$ to any gauge group $SU(n)$, whose weak-coupling spectrum was found in \cite{Fraser Hollowood}.
The approach is very similar.
Consider symplectic product between the decay products in (\ref{composite dyon decay}): after some algebra, we can simplify it as
\begin{equation}
\begin{aligned}
& \quad \left\langle
\pm \left(
p\sum_{m=i}^{k}\vec\alpha_{m}+\sum_{l=i+1}^{k}\epsilon_{l}\sum_{m=l}^{k}\vec\alpha_{m}
\,, \
\sum_{m=i}^{k}\vec\alpha_{m}
\right)
\,,
\right.
\\
& \quad \quad \left.
\pm \left(
\left( p+\sum_{l=i+1}^{k}\epsilon_{l} \right)\sum_{m=k+1}^{j-1}\vec\alpha_{m}+\sum_{l=k+1}^{j-1}\epsilon_{l}\sum_{m=l}^{j-1}\vec\alpha_{m}
\,, \
\sum_{m=k+1}^{j-1}\vec\alpha_{m}
\right)
\right\rangle
\\
= & \left\langle
\left(
\left( p+\sum_{l=i+1}^{k}\epsilon_{l} \right)\vec\alpha_{k}
\,, \
\vec\alpha_{k}
\right)
\,,
\left(
\left( p+\sum_{l=i+1}^{k}\epsilon_{l} \right)\vec\alpha_{k+1}+\epsilon_{k+1}\vec\alpha_{k+1}
\,, \
\vec\alpha_{k+1}
\right)
\right\rangle
= -\frac{\epsilon_{k+1}}{2} = \pm\frac{1}{2}
\,.
\end{aligned}
\end{equation}
Making use of this, we apply the pentagon identity to the decay processes of the $SU(n)$ composite dyons in (\ref{composite dyon decay}) and obtain
\begin{equation}
\label{SU(n) pentagon}
\begin{aligned}
& \mathcal{K}_{
\pm \left(
p\sum_{m=i}^{k}\vec\alpha_{m}+\sum_{l=i+1}^{k}\epsilon_{l}\sum_{m=l}^{k}\vec\alpha_{m}
\,, \
\sum_{m=i}^{k}\vec\alpha_{m}
\right)
}
\\
& \mathcal{K}_{
\pm \left(
\left( p+\sum_{l=i+1}^{k}\epsilon_{l} \right)\sum_{m=k+1}^{j-1}\vec\alpha_{m}+\sum_{l=k+1}^{j-1}\epsilon_{l}\sum_{m=l}^{j-1}\vec\alpha_{m}
\,, \
\sum_{m=k+1}^{j-1}\vec\alpha_{m}
\right)
}
\\
= \,
& \mathcal{K}_{
\pm \left(
\left( p+\sum_{l=i+1}^{k}\epsilon_{l} \right)\sum_{m=k+1}^{j-1}\vec\alpha_{m}+\sum_{l=k+1}^{j-1}\epsilon_{l}\sum_{m=l}^{j-1}\vec\alpha_{m}
\,, \
\sum_{m=k+1}^{j-1}\vec\alpha_{m}
\right)
}
\\
& \mathcal{K}_{
\pm \left(
p\sum_{m=i}^{j-1}\vec\alpha_{m}+\sum_{l=i+1}^{j-1}\epsilon_{l}\sum_{m=l}^{j-1}\vec\alpha_{m}
\,, \
\sum_{m=i}^{j-1}\vec\alpha_{m}
\right)
}
\\
& \mathcal{K}_{
\pm \left(
p\sum_{m=i}^{k}\vec\alpha_{m}+\sum_{l=i+1}^{k}\epsilon_{l}\sum_{m=l}^{k}\vec\alpha_{m}
\,, \
\sum_{m=i}^{k}\vec\alpha_{m}
\right)
}
\,.
\end{aligned}
\end{equation}

\paragraph{}

Suppose that we have a product $S$ of Kontsevich--Soibelman operators for a given vacuum expectation value.
We want to show that all such products are equal.
In order to do this, let us move the VEV continuously into the region where all composite dyons decay.
For each decay process, the product looses one operator according to (\ref{SU(n) pentagon}), but $S$ remains constant.
When VEV is in the region with no composite dyons, this product simplifies to
\begin{equation}
\prod_{p=-\infty}^{+\infty}
\prod_{i=1}^{r}
\mathcal{K}_{(p\vec\alpha_{i},\vec\alpha_{i})}
\times
\prod_{i=1}^{r}
\mathcal{K}_{(\vec\alpha_{i},\vec 0)}^{-2}
\,.
\end{equation}
We have used that for a given $p$ and any $i$, $\mathcal{K}_{(p\vec\alpha_{i},\vec\alpha_{i})}$ commute with each other; purely electric operators commute.
Therefore, every initial product of operators is equal to this expression.
Putting all pieces together, we recover the wall-crossing formula for any weak-coupling region of the moduli space:
\begin{equation}
\label{WCFn}
\begin{aligned}
\mathcal{O}
\left(
\prod_{p=-\infty}^{+\infty}
\prod_{i=1}^{r}
\mathcal{K}_{(p\vec\alpha_{i},\vec\alpha_{i})}
\times
\prod_{i=1}^{r}
\prod_{j=i+1}^{r}
\mathcal{K}_{\left(
p\sum_{m=i}^{j}\vec\alpha_{m}+\sum_{l=i+1}^{j}\epsilon_{l}\sum_{m=l}^{j}\vec\alpha_{m}
\,, \
\sum_{m=i}^{j}\vec\alpha_{m}
\right)}
\times
\prod_{i=1}^{r}
\mathcal{K}_{(\vec\alpha_{i},\vec 0)}^{-2}
\right)
\\
=
\prod_{p=-\infty}^{+\infty}
\prod_{i=1}^{r}
\mathcal{K}_{(p\vec\alpha_{i},\vec\alpha_{i})}
\times
\prod_{i=1}^{r}
\mathcal{K}_{(\vec\alpha_{i},\vec 0)}^{-2}
=
\mathcal{O}
\left(
\prod_{\gamma\in\Gamma(\vec a)}
\mathcal{K}_{\gamma}
\right)
\end{aligned}
\end{equation}
where $\mathcal{O}$ is the clockwise-ordering operator.
This equation relates the spectra far from every wall of marginal stability, in the region with no composite dyons, and located at an arbitrary point in the weak-coupling region (where $\Gamma(\vec a)$ is the set of all particles), respectively.
Our result confirms that the weak-coupling BPS spectrum for the $SU(n)$ theory found in \cite{Fraser Hollowood} is correct.

\chapter{Instantons in the pure theory}
\label{ch: pure}

\paragraph{}

{\it This chapter is based on \cite{CDP}.}

\paragraph{}

The simplest case one can consider to test the predictions of the Gaiotto--Moore--Neitzke equation (\ref{RH}), which was discussed in chapter \ref{ch: walls}, is the pure $\mathcal{N}=2$ supersymmetric Yang--Mills theory with gauge group $SU(2)$ compactified on $\mathbb{R}^{3}\times S^{1}$.
The moduli space is parametrised by a single complex VEV $a$.
We investigate corrections to the moduli space metric in the weak-coupling region \cite{DHKM}, that is, we set $|a|\gg |\Lambda|$ where $\Lambda$ is the dynamical scale of the theory.
These corrections come from BPS states of the theory winding around the compactified dimension of radius $R$.
We start by calculating the perturbative corrections, produced by the $W^{\pm}$ bosons with electric charge $\pm 1$: they reproduce the shift of the effective coupling constant $g_{\rm eff}$ found in \cite{Ooguri Vafa, Seiberg Shenker}, where $g_{\rm eff}\ll 1$ for $|a|\gg |\Lambda|$.
Then, we consider the non-perturbative corrections, produced by the tower of dyons with magnetic charge $\pm 1$: they give rise to a sum of semiclassical three-dimensional instanton contributions, which are exponentially suppressed in the limit $g_{\rm eff}\to 0$.
These corrections are complicated functions of the dimensionless parameters $|a| R$, which is kept finite, and $g_{\rm eff}$; they are also proportional to the factor corresponding to the $W^{\pm}$ bosons.
As found in previous investigations of instanton effects in compactified gauge theory \cite{Dorey 2000, Dorey Parnachev}, the leading semiclassical contribution can be expanded in two distinct ways depending on whether $|a| R$ is finite or infinitely small (corresponding to $R\to 0$).

\paragraph{}

For finite values of $|a| R$, the result can be expanded as a sum over the contributions of magnetic monopoles and Julia--Zee dyons \cite{Julia Zee} regarded as classical solutions of finite Euclidean action on $\mathbb{R}^{3}\times S^{1}$.
Focusing on one-instantons, we calculate the four-fermion correlation function from the metric predicted by the GMN formalism.
Then, we reproduce the correlator via a direct semiclassical computation based on first principles and find that the two result match exactly.
An important feature is that, unlike similar calculations in four dimensions, the functional determinants corresponding to fluctuations of the bosonic and fermionic fields do not cancel \cite{Kaul, DKMTV}.
We evaluate the ratio of fluctuation determinants from first principles and find that it precisely reproduces the prefactor appearing in the weak-coupling expansion of the GMN results corresponding to the $W^{\pm}$ bosons.
Given the relation between the GMN integral equations and the Kontsevich--Soibelman conjecture, the agreement between these can also be regarded as an indirect test of the latter.

\paragraph{}

When $|a| R$ becomes small, the infinite sum over electric charges of dyons diverges and requires Poisson resummation.
As in \cite{Dorey 2000,Dorey Parnachev}, the resulting series also admits an interpretation in terms of classical configurations of finite action.
The relevant configurations are an infinite tower of twisted monopoles obtained by applying large gauge transformations to the BPS monopole \cite{Lee Yi}.
In the context of $\mathcal{N}=4$ supersymmetric Yang--Mills theory realised on the world volume of two parallel D3 branes in IIB string theory, the relation between the two corresponding expansions can be understood as T-duality to an equivalent configuration of D2 branes in the IIA theory \cite{Dorey 2000}.
In the limit $R\to 0$, the twisted monopoles decouple, and only one monopole remains.

\paragraph{}

Another interesting question concerns the three-dimensional limit of the GMN results.
Taking the limit $|a| R\to 0$ with the effective three-dimensional coupling held fixed, we obtain $\mathcal{N}=4$ supersymmetric Yang--Mills theory in three dimensions with gauge group $SU(2)$.
The exact metric on the moduli space of this theory was conjectured to coincide with the Atiyah--Hitchin metric in \cite{SW3}.
This proposal was then tested against an explicit semiclassical calculation of the one-monopole contribution \cite{DKMTV}.
The leading semiclassical contribution to the GMN metric studied here can easily be continued to three dimensions after the Poisson resummation described above.
We show that it reproduces the one-monopole contribution to the Atiyah--Hitchin metric including the correct numerical prefactor.
As explained in \cite{DKMTV}, this coefficient together with the constraints of supersymmetry and other global symmetries uniquely determines the Atiyah--Hitchin metric.

%%%%%%%%%%%%%%%%%%%%%%%%%%%%%%%%%%%%%%%%%%%%%%%%%%%%%%%%%%%%%

\section{Moduli space and BPS spectrum}

\paragraph{}

We consider the pure $\mathcal{N}=2$ supersymmetric gauge theory in four dimensions with gauge group $SU(2)$ and dynamical scale $\Lambda$.
To begin with, let us review the relevant results in this case and set the subsequent notations.
The massless bosonic fields on the Coulomb branch consist of a $U(1)$ gauge field and a complex scalar $a$ whose VEV we also denote as $a$.

\paragraph{}

Each BPS state carries a central charge $Z_{\gamma}(u)$ which lies on a lattice in the complex plane with periods $a$ and $a_{D}$.
The magnetic period is determined by the prepotential $\mathcal{F}(a)$ via
\begin{equation}
\label{prepotential derivative}
a_{D} = \frac{\partial\mathcal{F}(a)}{\partial a}
\,.
\end{equation}
The prepotential $\mathcal{F}(a)$ also determines the low-energy effective
gauge coupling:
\begin{equation}
\tau_{\rm eff}(a)
=
\frac{4\pi i}{g_{\rm eff}^{2}(a)} + \frac{\Theta_{\rm eff}(a)}{2\pi}
=
\frac{\partial^{2}\mathcal{F}(a)}{\partial a^{2}}
\,.
\end{equation}
In this chapter, we will be interested in the weak coupling regime, where $|a|\gg|\Lambda|$.
The effective coupling constant can be approximated by its one-loop value (\ref{prepotential}, \ref{coordinate magnetic}), in this case,
\begin{equation}
\label{tau effective pure}
\tau_{\rm eff}(a) \simeq \frac{i}{\pi}
\log\left( \frac{a}{\Lambda} \right)^{2}
\,,
\end{equation}
up to corrections proportional to powers of $(\Lambda/a)^{4}$ coming from four-dimensional Yang--Mills instantons.
Ignoring these corrections, we can approximate the magnetic central charge as
\begin{equation}
a_{D}\simeq\tau_{\rm eff}a
\,.
\end{equation}

\paragraph{}

The exact mass formula for BPS states of charge $\gamma$ is $M_{\gamma}=|Z_{\gamma}|$ where the central charge $Z_{\gamma}$ was defined in (\ref{central charge 1}).
Seiberg--Witten solution does not immediately specify the set of values of $\gamma$ which are present in the theory.
Formally, this corresponds to determining the values of the second helicity supertrace (\ref{multiplicity}) at each point on the Coulomb branch.
This yields $\Omega(\gamma,a)=-2$ for the vector multiplet and $\Omega(\gamma,a)=+1$ for the half-hypermultiplet.

\paragraph{}

In the weakly coupled region, the BPS spectrum can be determined by semiclassical analysis.
It consists of the $W$ bosons of charges $\pm (1,0)$ and an infinite tower of Julia--Zee dyons $\pm(n,1)$ with unit magnetic charge and arbitrary integer electric charge~\footnote{
We follow the same normalisation convention for electric charges as in \cite{SW}, which differs from the convention used in \cite{GMN} by a factor of $1/2$.
}.
The degeneracies $\Omega(\gamma,u)$ can change discontinuously only when we cross the wall of marginal stability.
However, in the weak-coupling limit of the theory, the spectrum does not experience jumps.
As has been demonstrated in chapter \ref{ch: walls}, the spectrum of the pure theory with gauge group $SU(2)$ can be expressed in terms of Kontsevich--Soibelman operators by (\ref{WCF2}).

\paragraph{}

Let us now turn to the compactified theory.
The Wilson line, which corresponds to the $A_{0}$ component of the $U(1)$ gauge field along the compactified direction, $x^{0}$, is
\begin{equation}
\label{Wilson line 1}
\theta_{e} = \oint_{S_{R}^{1}} dx^{0} A_{0}
\,.
\end{equation}
Large gauge transformations shift the value of $\theta_{e}$ by integer multiples of $2\pi$, so, we periodically identify $\theta_{e}\sim\theta_{e}+2\pi$.
In three dimensions, we can also dualise the $U(1)$ Abelian gauge field $A_{i}$ with $i=1,2,3$ in favour of another real scalar, $\theta_{m}\in [0,2\pi]$, known as the magnetic Wilson line.
In other words, it is a magnetic analogue of (\ref{Wilson line 1}) that can be expressed in terms the dual field strength:
\begin{equation}
\label{Wilson line magnetic 1}
\theta_{m} = \oint_{S_{R}^{1}} dx^{0} A_{D\,0}
\,.
\end{equation}
This new scalar appears in the classical action in the combination $\gamma_{m}\theta_{m}$.
Dirac quantisation requires $\gamma_{m}$ to take integer values, and as a result, the theory is invariant under the shifts of $\theta_{m}$ by integer multiples of $2\pi$.

\paragraph{}

Taking into account all the scalars, $a,\bar a,\theta_{e},\theta_{m}$, the Coulomb branch of the compactified theory is a manifold of real dimension four.
The low-energy effective field theory on the Coulomb branch is then given by a three-dimensional sigma model with moduli space serving as its target space.

\paragraph{}

To reduce the number of terms in the compactified bosonic action, we use the complex combination $z=\theta_{m}-\tau_{\rm eff}(a)\theta_{e}$ parametrising a torus with complex parameter $\tau_{\rm eff}(a)$.
The moduli space of the compactified theory corresponds to a fibration of this torus over the Coulomb branch of the four-dimensional theory.
After compactification (see chapter \ref{ch: supersymmetry} for more details), the real part of the resulting bosonic action in the low-energy theory is given in terms of $a,\bar a,z,\bar z$ as
\begin{equation}
\label{action bosonic 3D pure}
S_{\rm B} = \frac{1}{4} \int d^{3}x \left(
\frac{4\pi R}{g_{\rm eff}^{2}} \partial_{\mu}a \, \partial^{\mu}\bar a +
\frac{g_{\rm eff}^{2}}{16\pi^{3}R} \partial_{\mu}z \, \partial^{\mu}\bar z
\right)
\,.
\end{equation}
In addition to this, surface terms in (\ref{compactified 2}) give rise to imaginary terms in the action depending on the total electric and magnetic charges:
\begin{equation}
\label{action imaginary 3D pure}
S_{\rm Im} = i\left( \gamma_{e}+\frac{\Theta_{\rm eff}}{2\pi} \, \gamma_{m} \right)\theta_{e}+i\gamma_{m}\theta_{m}
\,.
\end{equation}
The term proportional to $\Theta_{\rm eff}$ arises from dimensional reduction of the $F\wedge F$ term in the low-energy action of the four-dimensional theory after replacing $A_{0}$ by $\theta_{e}/2\pi R$.
The corresponding fermionic terms in the action take the form
\begin{equation}
\label{action fermionic 3D pure}
S_{\rm F} = \frac{2\pi R}{g_{\rm eff}^{2}} \int d^{3}x \tr \left(
i\bar\psi\bar\sigma^{\mu}\partial_{\mu}\psi +
i\bar\lambda\bar\sigma^{\mu}\partial_{\mu}\lambda
\right)
\end{equation}
where $\lambda$ and $\psi$ are the fermions after dimensional reduction along the compactified direction of the four-dimensional fermions.

\paragraph{}

As was discussed in chapter \ref{ch: supersymmetry}, the leading-order behaviour of the metric when $R\to\infty$ is given by its semiflat value,
\begin{equation}
\label{metric semiflat pure}
g^{\rm sf} = \frac{4\pi R}{g_{\rm eff}^{2}}|da|^{2} + \frac{g_{\rm eff}^{3}}{16\pi^{3}R}|dz|^{2}
\,.
\end{equation}
The metric (\ref{metric semiflat pure}) also makes apparent that $g^{\rm sf}$ is K\"ahler.
When we consider finite values of the radius $R$, the semiflat metric gets corrected by instanton contributions which arise from the four-dimensional BPS states wrapping around the compactified dimension $S^{1}$.
To take these corrections into account, we will use the Gaiotto--Moore--Neitzke equation (\ref{RH}): the weak-couplng spectrum and discontinuities of the Darboux coordinates are given by the right-hand side of the wall-crossing formula (\ref{WCF2}).

%%%%%%%%%%%%%%%%%%%%%%%%%%%%%%%%%%%%%%%%%%%%%%%%%%%%%%%%%%%%%

\section{Semiclassical limit via the wall-crossing formula}

\paragraph{}

Taking the logarithm in the Gaiotto--Moore--Neitzke equation (\ref{RH}), we see that the right-hand side contains a source term corresponding to the semiflat expression and an integral convolution produced by the BPS particles in $\mathbb{R}^{3}\times S^{1}$.
Although the integral equation is too complicated to be solved exactly, we can approximate the solution iteratively by choosing an appropriate expansion parameter.
For this approach to be valid, the series of corrections must be much smaller than the semiflat terms.
In \cite{GMN}, a simple expansion of this sort was considered for large radius, more precisely, for $R |a|\gg 1$.
It is valid for any point on the Coulomb branch: this is due to the fact that the contribution of a BPS state with charge $\gamma$ is exponentially suppressed by a factor of $\exp(-2\pi R |Z_{\gamma}|)$ for all $\gamma$ when $R |a|\gg 1$.
Here, we are instead interested in a weak-coupling expansion of the integral equation.
Thus, we will restrict our attention to the semiclassical region of the moduli space, that is, we set $|a|\gg\Lambda$, or, equivalently, $g_{\rm eff}^{2}\ll1$, where the dimensionless parameter $R |a|$ is fixed.
The factor of $\exp(-2\pi R |Z_{\gamma}|)$ in this case is suppressed for all states with non-zero magnetic charge.
This is the case for all states except the massive gauge bosons $W^{\pm}$.
These magnetically charged states give rise to a series of instanton corrections.
On the other hand, the contributions of the $W^{\pm}$ bosons have to be calculated exactly.

\paragraph{}

We begin by decomposing the Darboux coordinate $\mathcal{X}_\gamma(\zeta)$ as
\begin{equation}
\mathcal{X}_{\gamma}(\zeta) =
\left( \mathcal{X}_{e}(\zeta) \right)^{\gamma_{e}}
\left( \mathcal{X}_{m}(\zeta) \right)^{\gamma_{m}}
\,,
\quad
\gamma=(\gamma_{e},\gamma_{m})
\,.
\end{equation}
The GMN solution (\ref{RH}) of the problem in this case is given by a set of two integral equations: the electric and the magnetic Darboux coordinates obey
\begin{align}
\label{Darboux electric pure}
\mathcal{X}_{e}(\zeta) & =
\mathcal{X}_{e}^{\rm sf}(\zeta)
\exp\left( -\frac{1}{2\pi i} \sum_{\gamma'\in\Gamma} c_{e}(\gamma') \, \mathcal{I}_{\gamma'}(\zeta) \right)
\,,
\quad
c_{e}(\gamma') = -\Omega(\gamma',a) \, \gamma_{m}'
\,,
\\
\label{Darboux magnetic pure}
\mathcal{X}_{m}(\zeta) & =
\mathcal{X}_{m}^{\rm sf}(\zeta)
\exp\left( -\frac{1}{2\pi i} \sum_{\gamma'\in\Gamma} c_{m}(\gamma') \, \mathcal{I}_{\gamma'}(\zeta)\right)
\,,
\quad
c_{m}(\gamma') = \Omega(\gamma',u) \, \gamma_{e}'
\,,
\end{align}
where $\mathcal{X}_{e}^{\rm sf}(\zeta)$ and $\mathcal{X}_{m}^{\rm sf}(\zeta)$ are given by (\ref{coordinate semiflat}) with $Z_{\gamma}$ replaced by $Z_{e}$ and $Z_{m}$, respectively, $\Gamma$ is the weak-coupling BPS spectrum, and $\mathcal{I}_{\gamma'}(\zeta)$ is given by
\begin{equation}
\mathcal{I}_{\gamma'}(\zeta) = \int_{l_{\gamma'}} \frac{d\zeta'}{\zeta'} \frac{\zeta'+\zeta}{\zeta'-\zeta}
\log\left( 1-\mathcal{X}_{\gamma'}(\zeta') \right)
\,,
\end{equation}
where, since there are no flavour hypermultiplets, we have set the quadratic refinement $\sigma(\gamma')=1$ for all states.
The symplectic form (\ref{symplectic form}) of the $SU(2)$ theory is simply
\begin{equation}
\label{symplectic form pure}
\omega(\zeta) = -\frac{1}{4\pi^{2}R} \,
\frac{d\mathcal{X}_{e}(\zeta)}{\mathcal{X}_{e}(\zeta)}
\wedge
\frac{d\mathcal{X}_{m}(\zeta)}{\mathcal{X}_{m}(\zeta)} =
\frac{1}{4\pi^{2}R} \, \langle\gamma_{1},\gamma_{2}\rangle^{-1} \,
\frac{d\mathcal{X}_{\gamma_{1}}(\zeta)}{\mathcal{X}_{\gamma_{1}}(\zeta)}
\wedge
\frac{d\mathcal{X}_{\gamma_{2}}(\zeta)}{\mathcal{X}_{\gamma_{2}}(\zeta)}
\,.
\end{equation}
Now, taking the weak coupling limit, to the one-loop order, we find
\begin{equation}
\label{Darboux basis pure}
\log\mathcal{X}_{e}^{\rm sf}(\zeta)=\pi R a\zeta^{-1}+i\theta_{e}+\pi R \bar a\zeta
\,,
\quad
\log\mathcal{X}_{m}^{\rm sf}(\zeta)=\pi R a\tau_{\rm eff}(a)\zeta^{-1}+i\theta_{m}+\pi R\overline{a\tau_{\rm eff}(a)} \zeta
\,.
\end{equation}
We can see that in this limit $\log|\mathcal{X}_{m}^{\rm sf}|\gg\log|\mathcal{X}_{e}^{\rm sf}|$.
This has interesting consquences for deriving an iterative solution to $\mathcal{X}_\gamma(\zeta)$.
Explicitly, let us expand $\log\mathcal{X}_{e}(\zeta)$ and $\log \mathcal{X}_{m}(\zeta)$ for the weak-coupling spectrum of $\mathcal{N}=2$ $SU(2)$ gauge theory using (\ref{Darboux electric pure},
\ref{Darboux magnetic pure}):
\begin{equation}
\label{Darboux electric pure 2}
\log\mathcal{X}_{e}(\zeta) =
\log\mathcal{X}_{e}^{\rm sf}(\zeta)
-\sum_{\gamma_{e}'\in\mathbb{Z}}\sum_{\gamma_{m}'=\pm 1} \frac{c_{e}(\gamma')}{2\pi i} \, \mathcal{I}_{(\gamma_{e}',\gamma_{m}')}(\zeta)
\,,
\end{equation}
\begin{equation}
\label{Darboux magnetic pure 2}
\begin{aligned}
\log\mathcal{X}_{m}(\zeta) =
\log\mathcal{X}_{m}^{\rm sf}(\zeta)
-\sum_{\gamma_{e}'\in \mathbb{Z}}\sum_{\gamma_{m}'=\pm 1} \frac{c_{m}(\gamma')}{2\pi i} \, \mathcal{I}_{(\gamma_{e}',\gamma_{m}')}(\zeta)
\\
-\frac{c_{m}(W^{+})}{2\pi i} \, \mathcal{I}_{(1,0)}(\zeta)
-\frac{c_{m}(W^{-})}{2\pi i} \, \mathcal{I}_{(-1,0)}(\zeta)
\,.
\end{aligned}
\end{equation}
The BPS spectrum $\Gamma$, which in this theory is the same for any VEV at weak coupling, consists of the $W$ bosons of charge $\pm(1,0)$, which we denote as $W^{\pm}$, and only contribute to the middle two terms in (\ref{Darboux magnetic pure 2}), and the remaining summations in (\ref{Darboux electric pure 2}) and (\ref{Darboux magnetic pure 2}) are over the
infinite tower of dyons with charges $\pm (n,1),\ n\in\mathbb{Z}$.

\paragraph{}

The central charge can be approximated as $Z_\gamma(a)=a(\gamma_{e}+\gamma_{m}\tau_{\rm eff}(a))$.
The mass of a BPS particle $\gamma=(\gamma_{e},\gamma_{m})$ in the weak-coupling limit is then given by
\begin{equation}
|Z_{\gamma}(a)| = |a| \sqrt{
\left(\gamma_{m} \, \frac{4\pi}{g_{\rm eff}^{2}}\right)^{2} +
\left(\gamma_{e}+\gamma_{m} \, \frac{\Theta_{\rm eff}}{2\pi}\right)^{2}
}
\end{equation}
where $g_{\rm eff}$ and $\Theta_{\rm eff}$ now denote the effective coupling constant and the effective vacuum angle.

\paragraph{}

\begin{figure}[ht]
\centering
\includegraphics[width=65mm]{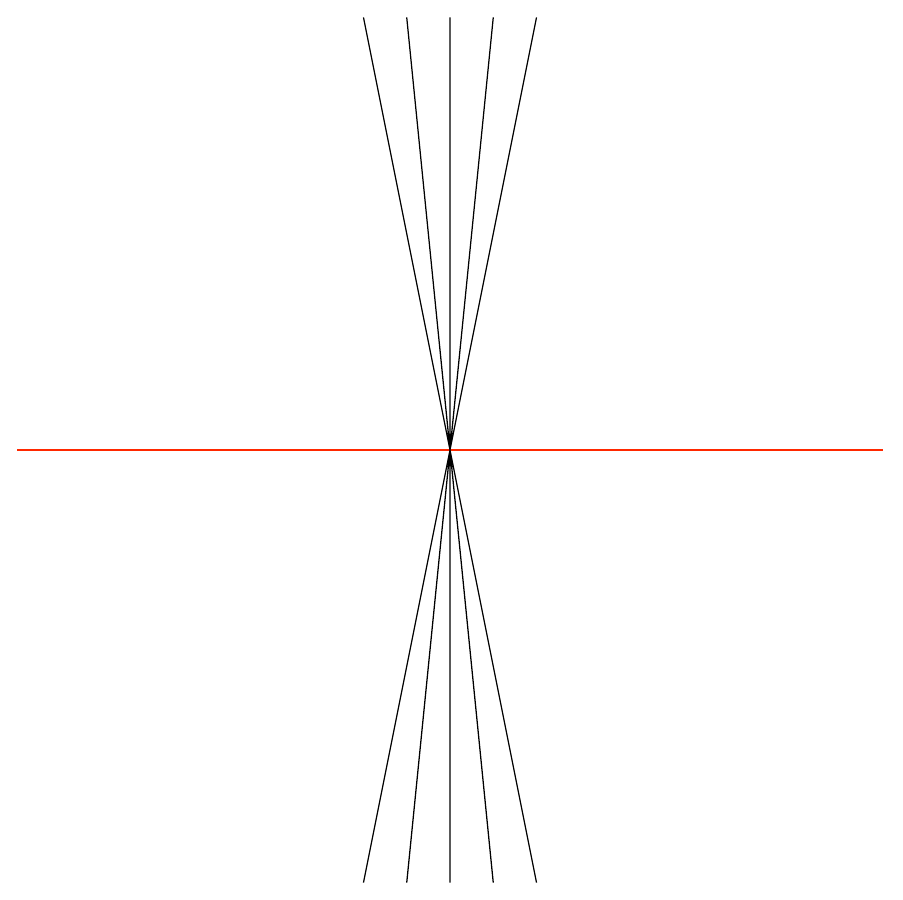}
\caption{
The BPS rays in the $\zeta$ plane for $a\in\mathbb{R}$: the red horizontal line corresponds to the $W$ bosons, all other lines correspond to dyons.
In the weak-coupling limit, the BPS rays for dyons are infinitely close to the imaginary axis.
}
\label{fig: BPS rays pure}
\end{figure}
Let us further describe our iterative approach of finding $\log\mathcal{X}_{\gamma}$ at weak coupling.
At the leading order, we substitute the semiflat coordinates (\ref{Darboux basis pure}) into the right hand side of (\ref{Darboux electric pure 2}) and (\ref{Darboux magnetic pure 2}) ignoring the components negligible for $g_{\rm eff}\to 0$.
In this regime, the magnetic coordinate $\mathcal{X}_{m}$ receives additional order one contribution from the $W$ bosons, while the dyon contributions to both $\mathcal{X}_{e}$ and $\mathcal{X}_{m}$ are exponentially suppressed as $\sim\exp(-c/g_{\rm eff}^{2})$ with some $c>0$.
We denote the resulting coordinates at this order as $\mathcal{X}^{(0)}_{e},\mathcal{X}^{(0)}_{m}$.
Thus, we have
\begin{equation}
\label{Darboux 0 pure}
\log\mathcal{X}_{e}^{(0)}(\zeta) =
\log\mathcal{X}_{e}^{\rm sf}(\zeta)
\,,
\quad
\log\mathcal{X}_{m}^{(0)}(\zeta)=\log\mathcal{X}_{m}^{\rm sf}(\zeta)+\log\mathcal{D}(\zeta)
\,,
\end{equation}
\begin{equation}
\label{D factor pure}
\log\mathcal{D}(\zeta) =
\frac{1}{2\pi i} \left(
\int_{l_{W^{+}}} \frac{d\zeta'}{\zeta'} \frac{\zeta'+\zeta}{\zeta'-\zeta}
\log\left( 1-\mathcal{X}_{e}^{\rm sf}(\zeta') \right)
-\int_{l_{W^{-}}} \frac{d\zeta'}{\zeta'} \frac{\zeta'+\zeta}{\zeta'-\zeta}
\log\left( 1-1/\mathcal{X}_{e}^{\rm sf}(\zeta') \right)
\right)
\,,
\end{equation}
where the BPS rays for the $W^{\pm}$ bosons are given as $l_{W^{\pm}}=\{\zeta':\pm a/\zeta'\in\mathbb{R}_{-}\}$ (thick lines in figure \ref{fig: BPS rays pure}).
We have used the fact that $\Omega(W^{\pm},a)=-2$ as $W$ bosons belong to the vector multiplet, and $c_{m}(W^{\pm})=\mp 2$.
After taking account of the suppressed corrections, $\mathcal{X}_{e}(\zeta)$ and $\mathcal{X}_{m}(\zeta))$ can be rewritten as
\begin{equation}
\begin{aligned}
\log\mathcal{X}_{e}(\zeta) = \log\mathcal{X}_{e}^{(0)}(\zeta)+\delta\log\mathcal{X}_{e}(\zeta)
\,,
\\
\log\mathcal{X}_{m}(\zeta) = \log\mathcal{X}_{m}^{(0)}(\zeta)+\delta\log\mathcal{X}_{m}(\zeta)
\,,
\end{aligned}
\end{equation}
where the second terms will be obtained approximately performing one more iteration.
By inserting $\mathcal{X}_{e}^{(0)}(\zeta)$ and $\mathcal{X}_{m}^{(0)}(\zeta)$ into the right-hand side of (\ref{Darboux electric pure 2}) and (\ref{Darboux magnetic pure 2}), we can get subleading-order corrections to the coordinates:
\begin{align}
\label{Darboux electric suppressed pure}
\delta\log\mathcal{X}_{e}(\zeta) & =
-\frac{1}{2\pi i} \sum_{\gamma_{e}'\in\mathbb{Z}} \sum_{\gamma_{m}'=\pm 1} c_{e}(\gamma') \, \mathcal{I}_{(\gamma_{e}',\gamma_{m}')}^{(0)}(\zeta)
\,,
\\
\label{Darboux magnetic suppressed pure}
\delta \log\mathcal{X}_{m}(\zeta) & =
-\frac{1}{2\pi i} \sum_{\gamma_{e}'\in\mathbb{Z}} \sum_{\gamma_{m}'=\pm 1} c_{m}(\gamma') \, \mathcal{I}_{(\gamma_{e}',\gamma_{m}')}^{(0)}(\zeta)
\,,
\end{align}
where the integration along the dyonic BPS rays (thin lines in figure \ref{fig: BPS rays pure}) takes the form
\begin{equation}
\mathcal{I}_{(\gamma_{e}',\gamma_{m}')}^{(0)}(\zeta) =
\int_{l_{\gamma'}} \frac{d\zeta'}{\zeta'} \frac{\zeta'+\zeta}{\zeta'-\zeta}
\log\left( 1-\left(\mathcal{X}_{e}^{(0)}(\zeta')\right)^{\gamma_{e}'}\left(\mathcal{X}_{m}^{(0)}(\zeta')\right)^{\gamma_{m}'} \right)
\,.
\end{equation}
We will soon see that these terms generate the exponentially suppressed instanton and anti-instanton contributions.

\paragraph{}

To extract the corresponding correction to the metric of the moduli space, we compute the symplectic form (\ref{symplectic form pure}) to the subleading order: we will need to find the corrections (\ref{Darboux electric suppressed pure}) and (\ref{Darboux magnetic suppressed pure}) to (\ref{Darboux 0 pure}).
The result is
\begin{equation}
\label{symplectic form corrections pure}
\begin{aligned}
\omega(\zeta)
& \approx
-\frac{1}{4\pi^{2}R} \,
d\left( \log\mathcal{X}_{e}^{(0)}(\zeta)+\delta\log\mathcal{X}_{e}(\zeta) \right)
\wedge
d\left( \log\mathcal{X}_{m}^{(0)}(\zeta)+\delta\log\mathcal{X}_{m}(\zeta) \right)
\\
& =
\omega^{\rm sf}(\zeta)+\omega^{\rm P}(\zeta)+\omega^{\rm NP}(\zeta)+{\mathcal{O}}(\delta^{2})
\,,
\end{aligned}
\end{equation}
where the semiflat form $\omega^{\rm sf}(\zeta)$ and its perturbative correction $\omega^{\rm P}(\zeta)$ (the one-loop correction due to the $W^{\pm}$ bosons) are the wedge-product of the first terms, the non-perturbative correction $\omega^{\rm NP}(\zeta)$ (the leading-order correction due to the dyons) appears when only one term in the wedge-product has $\delta$, ${\mathcal{O}}(\delta^{2})$ is the wedge-product of the second terms and has greater exponential suppression than the other three terms, allowing us to ignore it in the one-instanton calculation.
Expanding the symplectic forms in (\ref{symplectic form corrections pure}), we get
\begin{align}
\label{symplectic form semiflat pure}
\omega^{\rm sf}(\zeta) & =
-\frac{1}{4\pi^{2}R} \, d\log\mathcal{X}_{e}^{\rm sf}(\zeta) \wedge d\log\mathcal{X}_{m}^{\rm sf}(\zeta)
\,,
\\
\label{symplectic form P pure}
\omega^{\rm P}(\zeta) & =
-\frac{1}{4\pi^{2}R} \, d\log\mathcal{X}_{e}^{\rm sf}(\zeta) \wedge d\log\mathcal{D}(\zeta)
\,,
\\
\label{symplectic form NP pure}
\omega^{\rm NP}(\zeta) & =
-\frac{1}{4\pi^{2}R} \left(
d\delta\log\mathcal{X}_{e}(\zeta) \wedge d\log\mathcal{X}_{m}^{(0)}(\zeta)+
d\log\mathcal{X}_{e}^{(0)}(\zeta) \wedge d\delta\log\mathcal{X}_{m}(\zeta)
\right)
\,.
\end{align}
Let us now calculate the integrals in $\omega^{\rm P}(\zeta)$ and $\omega^{\rm NP}(\zeta)$ when $g_{\rm eff}\to 0$.

%%%%%%%%%%%%%%%%%%%%%%%%%%%%%%

\subsection{Perturbative corrections}

\paragraph{}

We can start by evaluating the perturbative contribution, without having to deal with the full BPS spectrum.
This prediction can then be compared with the previously known results.
$\omega^{\rm P}(\zeta)$ in (\ref{symplectic form P pure}) can be found by adapting the calculation in \cite{GMN} where the contribution of a single electrically charged multiplet was considered.
Differentiating the electric Darboux coordinates under the logarithms in (\ref{D factor pure}), we get
\begin{align}
d\log\left( 1-(\mathcal{X}_{e}^{\rm sf}(\zeta'))^{\pm 1} \right) =
-d\log\left( (\mathcal{X}_{e}^{\rm sf}(\zeta'))^{\pm 1} \right) \frac{(\mathcal{X}_{e}^{\rm sf}(\zeta'))^{\pm 1}}{1-(\mathcal{X}_{e}^{\rm sf}(\zeta'))^{\pm 1}} =
\pm d\log\mathcal{X}_{e}^{\rm sf}(\zeta') \sum_{k=1}^{+\infty}\left( \mathcal{X}_{e}^{\rm sf}(\zeta') \right)^{\mp k}
\,,
\end{align}
where we have Taylor-expanded the fraction (the infinite series is a convergent geometric progression along the corresponding contours of integration, although it does not converge in general).
We will use the following equality:
\begin{equation}
\begin{aligned}
\log\mathcal{X}_{e}^{\rm sf}(\zeta) \wedge \frac{\zeta'+\zeta}{\zeta'-\zeta} \log\mathcal{X}_{e}^{\rm sf}(\zeta') & =
\log\mathcal{X}_{e}^{\rm sf}(\zeta) \wedge \frac{\zeta'+\zeta}{\zeta'-\zeta} \left( \log\mathcal{X}_{e}^{\rm sf}(\zeta') - \log\mathcal{X}_{e}^{\rm sf}(\zeta) \right)
\\
& =
\log\mathcal{X}_{e}^{\rm sf}(\zeta) \wedge \pi R\left( -a\left( \frac{1}{\zeta'}+\frac{1}{\zeta} \right) + \bar a(\zeta'+\zeta) \right)
\,.
\end{aligned}
\end{equation}
Then, evaluating the integrals, we obtain
\begin{align}
\label{symplectic form P pure 2}
\omega^{\rm P}(\zeta) & =
-\frac{i}{4\pi^{2}R} \,
d\log\mathcal{X}_{e}^{\rm sf}(\zeta)
\wedge
\left( 2\pi A^{\rm P}(a,\bar a)+\pi V^{\rm P}(a,\bar a)(\zeta^{-1} da-\zeta d\bar a) \right)
\,,
\\
\label{potential A pure}
A^{\rm P} & =
\frac{R}{\pi} \sum_{k\ne 0} |a| e^{ik\theta_{e}} K_{1}(2\pi R |ka|)
\left( \frac{da}{a}-\frac{d\bar a}{\bar a} \right)
\,,
\\
\label{potential V pure}
V^{\rm P} & =
-\frac{2R}{\pi} \sum_{k\ne 0} e^{ik\theta_{e}} K_{0}(2\pi R |ka|)
\,,
\end{align}
where $K_{\nu}(x)$ are modified Bessel functions of second kind ($K_{\nu}(x)\simeq\sqrt{\frac{\pi}{2x}}e^{-x}$ for $x\gg 1$).
In \cite{GMN}, the limit $R\gg 1/|\Lambda|$ and $2\pi R |a| \gg 1$ was considered: this sets $K_{\nu}(2\pi R |ka|) \sim e^{-2\pi R |ka|}$, and the series in (\ref{potential A pure}) and (\ref{potential V pure}) can be naturally interpreted as series of exponentially suppressed, instanton-like, contributions.
In the weak coupling limit, we only demand $|a/\Lambda| \gg 1$, while keeping $2\pi R |a|$ fixed and arbitrary.

\paragraph{}

If we ignored all non-perturbative corrections, the metric would still be symmetric under $\theta_{m}\to\theta_{m}+\delta$ for any $\delta$, while the symmetry with respect to shifts of $\theta_{e}$ is broken by the presence of $W$ bosons.
The Gibbons--Hawking ansatz for this hyper-K\"ahler metric in this case is \cite{GMN}
\begin{equation}
\begin{aligned}
\label{Gibbons Hawking pure}
& g = V(\vec x)^{-1} \left( \frac{d\theta_{m}}{2\pi}+A(\vec x) \right)^{2} + V(\vec x)d\vec x^{2}
\,,
\\
& a = x_{1}+ix_{2}
\,,
\quad
\theta_{e} = 2\pi Rx_{3}
\,,
\quad
\vec x = (x_{1},x_{2},x_{3})
\,,
\end{aligned}
\end{equation}
where the potentials are given by
\begin{equation}
\begin{aligned}
\label{Gibbons Hawking potentials pure}
V & = V^{\rm sf}+V^{\rm P}
\,,
\quad
V^{\rm sf} = R\im\tau_{\rm eff}
\,,
\\
A & = A^{\rm sf}+A^{\rm P}
\,,
\quad
A^{\rm sf} = -\frac{\re\tau_{\rm eff}}{2\pi}
\,.
\end{aligned}
\end{equation}
For the formulae (\ref{potential A pure}, \ref{potential V pure}) to make sense for small values of $R |a|$, one needs to Poisson-resum the series of Bessel functions \cite{Ooguri Vafa}.
The resulting geometry corresponds to a finite shift of the effective coupling constant (as in \cite{Seiberg Shenker}):
\begin{align}
& \frac{2\pi R}{g_{\rm eff}^{2}} \to
\frac{2\pi R}{g_{\rm eff}^{2}} - \sum_{n\in\mathbb{Z}} \frac{1}{2\pi |M(n)|}
\,,
\\
& |M(n)| = \sqrt{|a|^{2}+\left( \frac{\theta_{e}}{2\pi R}+\frac{n}{R} \right)^{2}}
\,.
\end{align}
In the limit $R |a|\to 0$, this reproduces the three-dimensional shift found in \cite{DKMTV}:
\begin{equation}
\frac{1}{e_{\rm eff}^{2}} \to
\frac{1}{e_{\rm eff}^{2}} - \frac{1}{2\pi M_{W}}
\end{equation}
where $e_{\rm eff}=g_{\rm eff}/\sqrt{2\pi R}$ and $M_{W}=M(0)$ are the gauge coupling and the mass of $W$ boson in three dimensions.

%%%%%%%%%%%%%%%%%%%%%%%%%%%%%%

\subsection{Instanton corrections}

\paragraph{}

Let us now evaluate $\omega^{\rm NP}(\zeta)$ in (\ref{symplectic form NP pure}).
At weak coupling, the integrals contain exponentially suppressed terms, allowing us to use the saddle-point approximation.
First of all, we decompose $\omega^{\rm NP}(\zeta)$ into a series:
\begin{equation}
\omega^{\rm NP}(\zeta) = \sum_{\gamma'=(\gamma_{e}',\pm 1):\,\gamma_{e}'\in\mathbb{Z}} \omega_{\gamma'}^{\rm NP}(\zeta)
\,,
\end{equation}
where for every dyon, we have
\begin{equation}
\label{symplectic form dyon pure}
\begin{aligned}
\omega_{\gamma'}^{\rm NP}(\zeta) & =
-\frac{1}{4\pi^{2}R}
\frac{d\mathcal{X}_{\gamma'}^{(0)}(\zeta)}{\mathcal{X}_{\gamma'}^{(0)}(\zeta)}
\wedge
\left(
\frac{1}{2\pi i}\int_{l_{\gamma'}} \frac{d\zeta'}{\zeta'} \frac{\zeta'+\zeta}{\zeta'-\zeta}
\frac{\mathcal{X}^{(0)}_{\gamma'}(\zeta')}{1-\mathcal{X}^{(0)}_{\gamma'}(\zeta')}
\frac{d\mathcal{X}_{\gamma'}^{(0)}(\zeta')}{\mathcal{X}_{\gamma'}^{(0)}(\zeta')}
\right)
\\
& \approx
-\frac{1}{4\pi^{2}R}
\frac{d\mathcal{X}_{\gamma'}^{\rm sf}(\zeta)}{\mathcal{X}_{\gamma'}^{\rm sf}(\zeta)}
\wedge
\left(
\frac{1}{2\pi i}\int_{l_{\gamma'}} \frac{d\zeta'}{\zeta'} \frac{\zeta'+\zeta}{\zeta'-\zeta}
\frac{\mathcal{X}^{(0)}_{\gamma'}(\zeta')}{1-\mathcal{X}^{(0)}_{\gamma'}(\zeta')}
\frac{d\mathcal{X}_{\gamma'}^{\rm sf}(\zeta')}{\mathcal{X}_{\gamma'}^{\rm sf}(\zeta')}
\right)
\,,
\end{aligned}
\end{equation}
as $\Omega(\gamma',a)=1$ for all the dyon states (a convenient way to obtain (\ref{symplectic form dyon pure}) is to use the last equality in (\ref{symplectic form pure}) setting $\gamma_{1}=\gamma'$ and then apply (\ref{RH}) directly to get $\mathcal{X}_{\gamma_{2}}$).
In the second line, we have further approximated $d\mathcal{X}^{(0)}(\zeta)/\mathcal{X}^{(0)}(\zeta)$ as $d\mathcal{X}_{\gamma}^{\rm sf}(\zeta)/\mathcal{X}_{\gamma}^{\rm sf}(\zeta)$ since they behave as $g_{\rm eff}^{-2}$, and $d\mathcal{D}(\zeta)/\mathcal{D}(\zeta)$ is independent of $g_{\rm eff}$.
Along each integration contour $l_{\gamma'}$, the zeroth order Darboux coordinate $\mathcal{X}^{(0)}_{\gamma'}(\zeta')=\mathcal{X}^{\rm sf}_{\gamma'}(\zeta')\mathcal{D}(\zeta')$ is proportional to $\exp\left(-\pi R |Z_{\gamma'}|(|\zeta'|+1/|\zeta'|)\right)$.
At weak coupling, $|\tau_{\rm eff}| \gg 1$, $|Z_{\gamma'}|\gg 1$, so we can Taylor-expand $\mathcal{X}_{\gamma'}^{(0)}(\zeta')/(1-\mathcal{X}_{\gamma'}^{(0)}(\zeta'))$ in the integrand above into
\begin{equation}
\frac{\mathcal{X}_{\gamma'}^{(0)}(\zeta')}{1-\mathcal{X}_{\gamma'}^{(0)}(\zeta')} =
\sum_{k=1}^{+\infty} \left( \mathcal{X}_{\gamma'}^{(0)}(\zeta') \right)^{k}
\,,
\quad
\zeta'\in l_{\gamma'}
\,,
\end{equation}
and we need to consider only the term $k=1$ as all other terms have higher order of exponential suppression.
To find this term, we perform the saddle-point approximation
\begin{equation}
\label{saddle point}
\int_{a}^{b} e^{f(x)}dx \approx \sqrt\frac{2\pi}{|f''(x_{0})|} \, e^{f(x_{0})}
\end{equation}
for $f(x)$ having sharp peak at $x=x_{0}$, $a<x_{0}<b$.
Along the BPS rays, the saddle-point analysis amounts to finding the extremum of $\exp(-\pi R |Z_{\gamma'}|(|\zeta'|+1/|\zeta'|))$ with respect to $|\zeta'|$.
We can now easily see that the saddle is located at $\zeta'=-Z_{\gamma'}/|Z_{\gamma'}|$ (i.e., $|\zeta'|=1$).
Then, the integrals become Gaussian integrals, and the leading-order expression for the dyonic corrections is given by
\begin{align}
\label{symplectic form dyon pure 2}
\omega_{\gamma'}^{\rm NP}(\zeta) & =
\mathcal{J}_{\gamma'} \,
\frac{d\mathcal{X}_{\gamma'}^{\rm sf}(\zeta)}{\mathcal{X}_{\gamma'}^{\rm sf}(\zeta)}
\wedge
\left(
|Z_{\gamma'}| \left( \frac{dZ_{\gamma'}}{Z_{\gamma'}}-\frac{d\bar{Z}_{\gamma'}}{\bar{Z}_{\gamma'}} \right)
-\left( \frac{dZ_{\gamma'}}{\zeta}-\zeta d\bar{Z}_{\gamma'} \right)
\right)
\,,
\\
\mathcal{J}_{\gamma'} & =
-\frac{1}{8\pi^{2}i} \left( \mathcal{D}(-e^{i\phi_{\gamma'}}) \right)^{\gamma_{m}'} \frac{1}{\sqrt{R |Z_{\gamma'}|}}
\exp\left( -2\pi R |Z_{\gamma'}|+i\theta_{\gamma'} \right)
\,,
\end{align}
\begin{equation}
\label{D factor pure 2}
\begin{aligned}
\gamma_{m}' \log\mathcal{D}(-e^{i\phi_{\gamma'}}) \approx
\pm\log\mathcal{D}(\mp i) =
\frac{1}{\pi i}
\int_{0}^{+\infty} \frac{dy}{y}
\left(
\frac{y+ i}{y-i}
\log\left( 1-e^{-\pi R |a| (y+1/y)+i\theta_{e}} \right)
\right.
\\
\left.
-\frac{y- i}{y+ i}
\log\left(1-e^{-\pi R |a| (y+1/y)-i\theta_{e}} \right)
\right)
\end{aligned}
\end{equation}
where $e^{i\phi_{\gamma'}}=(\gamma_{e}'+\tau_{\rm eff}\gamma_{m}')/|\gamma_{e}'+\tau_{\rm eff}\gamma_{m}'|$, and in the weak-coupling limit, we have further approximated $e^{i\phi_{\gamma'}} \approx i\gamma_{m}'$.
This ensures that $\mathcal{D}$ is real.
By substituting $y=e^{t}$, the equation (\ref{D factor pure 2}) can be re-expressed as
\begin{equation}
\label{D factor pure 3}
\log\mathcal{D}(-i) =
\frac{2}{\pi} \int_{0}^{+\infty} \frac{dt}{\cosh t}
\left(
\log\left( 1-e^{-2\pi R |a|\cosh t+i\theta_{e}} \right) +
\log\left( 1-e^{-2\pi R |a|\cosh t-i\theta_{e}} \right)
\right)
\,.
\end{equation}
In the next section we will show that this expression precisely corresponds to the ratio of one-loop determinants which appear due to small fluctuations around a classical dyon background.

\paragraph{}

We can extract the correction to the metric of the moduli space from the $\zeta$-independent part of $\omega^{\rm NP}(\zeta)$.
Focusing only on instanton (i.e., $\gamma_{m}'= +1$) contributions, we obtain
\begin{equation}
\label{symplectic form 3 instanton pure}
\begin{aligned}
\omega_{3}^{\rm inst} & =
\sum_{\gamma'=(\gamma_{e}',1):\,\gamma_{e}'\in\mathbb{Z}} \mathcal{J}_{\gamma'}
\left(
2\pi R \, dZ_{\gamma'} \wedge d\bar{Z}_{\gamma'}+
i |Z_{\gamma'}| d\theta_{\gamma'} \wedge \left( \frac{dZ_{\gamma'}}{Z_{\gamma'}}-\frac{d\bar{Z}_{\gamma'}}{\bar{Z}_{\gamma'}} \right)
\right)
\\
& =
\sum_{\gamma'=(\gamma_{e}',1):\,\gamma_{e}'\in\mathbb{Z}}
\mathcal{J}_{\gamma'}
\left(
2\pi R \left| \gamma_{e}'+\tau_{\rm eff} \right|^{2}da \wedge d\bar a+
i \left| \gamma_{e}'+\tau_{\rm eff} \right| d\theta_{\gamma'} \wedge |a| \left( \frac{da}{a}-\frac{d\bar a}{\bar a} \right)
\right)
\,,
\end{aligned}
\end{equation}
where $\omega^{\rm NP}_{3}=\omega^{\rm inst}_{3}+\bar\omega^{\rm inst}_{3}$ ($\bar\omega^{\rm inst}_{3}$ corresponds to the $\gamma_{m}'=-1$ contributions).
In the second line of (\ref{symplectic form 3 instanton pure}), we have used the weakly coupled expression for the central charge: $Z_{\gamma'}\simeq a(\gamma_{e}'+\tau_{\rm eff}(a))$.
Explicitly, let us write down the $g_{a\bar a}$ component, which is the dominant term of the weak-coupling metric, using (\ref{symplectic form 3 instanton pure}):
\begin{equation}
\label{metric instanton pure}
g_{a\bar a}^{\rm inst} =
\frac{R^{1/2}}{4\pi} \sum_{\gamma'=(\gamma_{e}',1):\,\gamma_{e}'\in\mathbb{Z}} \frac{\mathcal{D}(-i) | Z_{\gamma'} |^{3/2}}{|a|^{2}}
\exp\left( -2\pi R |Z_{\gamma'}|+ i\theta_{\gamma'} \right)
\,.
\end{equation}
Other metric components, $g_{a\bar z}^{\rm inst}$ and $g_{\bar a z}^{\rm inst}$, which are suppressed by $g_{\rm eff}^{2}$, can also be readily extracted from (\ref{symplectic form 3 instanton pure}).
By including these additional metric components and using the complex coordinates $z$ and $\bar z$ introduced earlier, we can calculate the K\"ahler potential $K^{\rm dyon}$ corresponding to the symplectic form  $\omega_{3}^{\rm dyon}$.
After substituting
\begin{equation}
\theta_{m} = \frac{1}{2}\left( (z+\bar z)+\frac{i\re\tau_{\rm eff}}{\im\tau_{\rm eff}}(z-\bar z) \right)
\,,
\quad
\theta_{e} = \frac{i}{2\im\tau_{\rm eff}}(z-\bar z)
\,,
\end{equation}
we recover the K\"ahler potential (which is a real function):
\begin{equation}
K^{\rm dyon} =
\sum_{\gamma'=(\gamma_{e}',\pm 1):\,\gamma_{e}'\in\mathbb{Z}} \frac{\mathcal{D}(-i)}{4\pi^{3}R^{3/2}|Z_{\gamma'}|^{1/2}}
\exp\left( -2\pi R |Z_{\gamma'}|+i\theta_{\gamma'} \right)
\,.
\end{equation}

\paragraph{}

Finally, we further approximate the metric (\ref{metric instanton pure}) at weak coupling as
\begin{align}
\label{metric instanton pure 2}
g_{a\bar a}^{\rm inst} & \simeq
\frac{2\sqrt{\pi R}}{g_{\rm eff}^{3}}
\sum_{\gamma'=(\gamma_{e}',1):\,\gamma_{e}'\in\mathbb{Z}} \frac{\mathcal{D}(-i)}{\sqrt{|a|}}
\exp\left( -S_{\rm mon}-S_{\varphi}^{(\gamma_{e}')} \right)
\,,
\\
\label{action monopole pure}
S_{\rm mon} & =
\frac{8\pi^{2}R}{g_{\rm eff}^{2}}|a|-i\theta_{m}
\,,
\\
\label{action angular pure}
S_{\varphi}^{(\gamma_{e}')} & =
\frac{g_{\rm eff}^{2}R |a|}{4}\left( \gamma_{e}'+\frac{\Theta_{\rm eff}}{2\pi} \right)^{2}-i\gamma_{e}'\theta_{e}
\,.
\end{align}
The action (\ref{action monopole pure}) in the exponent is the usual Euclidean action of a magnetic monopole can be thought of as a static field configuration on $\mathbb{R}^{3}\times S^{1}$.
The remaining term $S_{\varphi}^{(\gamma_{e}')}$ is the leading correction to the monopole action, coming from the electric charges of dyons belonging to the spectrum.
This is the contribution to the dyon mass appearing due to the slow motion of the monopole in the $S^{1}$ direction \cite{Tomboulis Woo}.
The shift of the electric charge
\begin{equation}
\gamma_{e}' \to \gamma_{e}'+\frac{\Theta_{\rm eff}}{2\pi}
\end{equation}
corresponds to the Witten effect \cite{Witten dyon charge}: we have obtained this for the real part of the action, we will also show that our choice of coordinates implies that the same shift should appear in the imaginary part.

\paragraph{}

This additional shift corresponds to the appropriate choice of global coordinates on the torus fibre of the moduli space \cite{GMN}.
When working near a singularity in the moduli space where a ratio of BPS particle masses vanishes, it is appropriate to change variables to a coordinate which is single-valued in a neighbourhood of the singular point.
In the present case, we are interested in the semiclassical region of the moduli space near infinity and the
singularity corresponds to the logarithm in the one-loop effective coupling (\ref{tau effective pure}).
Adapting eq.\ (4.13) of \cite{GMN} to this case, the corresponding change of variable is
\begin{equation}
\theta_{m} \to \theta_{m}+\frac{\Theta_{\rm eff}}{2\pi} \, \theta_{e}
\,.
\end{equation}
Implementing this replacement, equation (\ref{action angular pure}) takes the form
\begin{equation}
S_{\varphi}^{(\gamma_{e}')} =
\frac{g_{\rm eff}^{2}R |a|}{4}\left( \gamma_{e}'+\frac{\Theta_{\rm eff}}{2\pi} \right)^{2}
-i\left( \gamma_{e}'+\frac{\Theta_{\rm eff}}{2\pi} \right) \theta_{e}
\,,
\end{equation}
and we see that the resulting imaginary part of the total action $S_{\rm mon}+S_{\varphi}^{(\gamma_{e}')}$ agrees with the surface terms of the action (\ref{action imaginary 3D pure}) obtained directly from dimensional reduction of the four-dimensional metric.

\paragraph{}

The hyper-K\"ahler metric on the moduli space completely  determines the low-energy effective action for the massless fields (the action has up to two derivatives and up to four fermions).
This action is a three-dimensional supersymmetric sigma model:
\begin{equation}
\label{action effective 3D}
S_{\rm eff}^{\rm (3D)} =
\frac{1}{4} \int d^{3} x \left(
g_{ij}(X) \left(\partial_{\mu}X^{i}\partial^{\mu} \bar{X}^{j}+i \, \bar{\Omega}^{i}\ssl{D}\Omega^{j} \right)
+\frac{1}{6} R_{ijkl} \left( \bar{\Omega}^{i}\cdot\Omega^{k} \right) \left(\bar{\Omega}^{j}\cdot\Omega^{l} \right)
\right)
\end{equation}
where $X^{i}$ are four bosonic scalar fields, and $\Omega^{i\,\alpha}$ are their Majorana fermionic superpartners.
Specifically, we are interested in finding the instanton correction to the Riemann tensor.

\paragraph{}

The metric of the moduli space is a function of the scalars.
In the semiclassical limit, the metric of the effective action (\ref{action effective 3D}) is given by its semiflat value, and therefore, we should scale the scalar fields here so that the bosonic terms in (\ref{action effective 3D}) reproduce (\ref{action bosonic 3D pure}) in this limit.
One can see that after rescaling the bosonic fields as
\begin{equation}
\label{bosons scaling pure}
X^{1} = \frac{2\sqrt{\pi R}}{g_{\rm eff}} a
\,,
\quad
X^{2} = \frac{g_{\rm eff}}{4\pi\sqrt{\pi R}} z
\,,
\end{equation}
the semiflat metric $g^{\rm sf}$ (\ref{metric semiflat pure}) becomes flat: explicitly, the components of the new metric $\tilde g$ are $\tilde g_{i\bar{j}}=\tilde g_{\bar{j}i}=\delta_{ij}/2$, $\{i,j\}\subset\{1,2\}$.

\paragraph{}

By comparing the fermionic terms in the action with (\ref{action fermionic 3D pure}), we can rewrite the action (\ref{action effective 3D}) using the Weyl spinors $\lambda_\alpha, \bar{\lambda}_{\dot{\alpha}}, \psi_\alpha,\bar{\psi}_{\dot{\alpha}}$.
Following the approach in \cite{DKMTV}, we rewrite the latter in terms of the three-dimensional Majorana fermions $\chi^{\bar a},\bar{\chi}^{a}$ as
\begin{equation}
\lambda_{\alpha} = \chi^{\bar{1}}_{\alpha}
\,
\quad
\epsilon_{\alpha\dot{\beta}}\bar{\lambda}^{\dot{\beta}} = \bar{\chi}^{1}_{\alpha}
\,,
\quad
\psi_{\alpha} = \chi^{\bar{2}}_{\alpha}
\,,
\quad
\epsilon_{\alpha\dot{\beta}}\bar{\psi}^{\dot{\beta}} = \bar{\chi}^{2}_{\alpha}
\,.
\end{equation}
The Majorana fermions $\Omega^{i}$ appearing in (\ref{action effective 3D}) can be then be expressed as
\begin{equation}
\label{fermions transformation pure}
\Omega^{i}_{\alpha} = M^{ic}(X)\bar{\chi}_{c\alpha}
\,,
\quad
\Omega_{\alpha}^{\bar{i}} = M^{\bar{i}\bar{c}}(X)\chi_{\bar{c}\alpha}
\,,
\quad
c \in \{1, 2\}
\,,
\end{equation}
where $M^{ic}(X)$ and $M^{\bar{i}\bar{c}}(X)$ are undetermined
matrices which can depend non-trivially on the bosonic scalars $X^{i}$.
Matching with the fermion kinetic terms in (\ref{action effective 3D}) with (\ref{action fermionic 3D pure}) to the leading order imposes the normalisation condition:
\begin{equation}
\label{fermions constraint pure}
\delta_{i\bar{j}}M^{ia}(X) M^{\bar{j}\bar{b}}(X) =
\left( \frac{8\pi R}{g_{\rm eff}^{2}} \right)\delta^{a\bar{b}}
\,.
\end{equation}
In a vacuum where $\theta_{e}=0$, the relation between the fermions appearing in (\ref{action effective 3D}) and in (\ref{action fermionic 3D pure}) can be made explicit:
\begin{equation}
\label{fermions scaling pure}
\bar{\Omega}^{1}\cdot\Omega^{1} = \left( \frac{8\pi R}{g_{\rm eff}^{2}} \right) \lambda\cdot\bar{\lambda}
\,,
\quad
\bar{\Omega}^{2}\cdot\Omega^{2} = \left(\frac{8\pi R}{g_{\rm eff}^{2}} \right) \psi\cdot\bar{\psi}
\,.
\end{equation}
\paragraph{}
The four-fermion term in the action (\ref{action effective 3D}) involves the Riemann tensor of the hyper-K\"{a}hler moduli space metric.
We shall find the leading order of this term using our result (\ref{metric instanton pure 2}) for the instanton metric.
The non-zero components of the Christoffel symbol on a K\"ahler manifold, have only holomorphic or only antiholomorphic indices:
\begin{equation} 
\Gamma_{bc}^{a} = g^{a\bar d}\partial_{b}g_{c\bar d}
\,,
\quad
\Gamma_{\bar b\bar c}^{\bar a} = g^{\bar ad}\partial_{\bar b}g_{\bar cd}
\,.
\end{equation}
Then, the non-zero components of the Riemann tensor are
\begin{equation}
R_{a\bar b c\bar d} = -R_{\bar b a c\bar d} = -R_{a\bar b\bar d c} = R_{\bar b a \bar d c} =
g_{a\bar f}\partial_{c}\Gamma_{\bar b\bar d}^{\bar f}
\,.
\end{equation}
Since the metric is K\"ahler with respect to $a,\bar a,z,\bar z$, we conclude that at the leading order in $g_{\rm eff}$, the non-vanishing components of the Riemann tensor with four different indices, up to the standard symmetries of a Riemann tensor, are
\begin{equation}
\label{Riemann pure}
\begin{aligned}
R_{a\bar az\bar z} =
R_{a\bar zz\bar a} & =
g_{a\bar p} \partial_{z} \left( g^{\bar pq}\partial_{\bar z}g_{\bar aq} \right) \simeq
g_{a\bar a}^{\rm sf} \partial_{z} \left( g^{{\rm sf}\,\bar aa}\partial_{\bar z}g_{\bar aa}^{\rm inst} \right)
\\
& =
\partial_{z}\partial_{\bar z} g_{a\bar a}^{\rm inst} =
-\frac{1}{4} g_{a\bar a}^{\rm inst}
\,,
\quad
\{p,q\} \subset \{a,z\}
\,,
\end{aligned}
\end{equation}
where we have treated $g^{\rm inst}$ as a perturbation of the dominant $g^{\rm sf}$ (which is independent of the coordinates), the anti-instanton correction has the same form.
Consider all symmetries of the Riemann tensor and the four-fermion product in (\ref{action effective 3D}): in the tensor summation, one has $4!=24$ ways to allocate these four different indices; to obtain a non-zero result, first pair (and, consequently, second pair) of indices should contain one holomorphic and one antiholomorphic index leading to an extra factor of $1/2$; finally, the resulting symmetry factor is 12.
Taking this into account, the action (\ref{action effective 3D}), expressed in terms of $X^{1},\bar X^{1},X^{2},\bar X^{2}$ (\ref{bosons scaling pure}), can be simplified to
\begin{equation}
\label{action effective pure}
S_{\rm eff}^{\rm (3D)} =
\frac{1}{4} \int d^{3} x \left(
\delta_{i\bar j} \left(\partial_{\mu}X^{i}\partial^{\mu} \bar{X}^{j}+i \, \bar{\Omega}^{i}\ssl{D}\Omega^{j} \right)
+ 2 R_{1\bar 2\bar 1 2} \left( \bar\Omega^{1}\cdot\Omega^{1} \right) \left( \bar\Omega^{2}\cdot\Omega^{2} \right)
\right)
\,,
\end{equation}
where the sum is over $i$ and $j$ only.
The Riemann tensor (\ref{Riemann pure}) in these new coordinates is
\begin{equation}
\label{Riemann scaling pure}
R_{1\bar{2}\bar{1}2} = \left| \frac{da}{dX^{1}}\frac{dz}{dX^{2}} \right|^{2}
R_{a\bar z\bar az} = (2\pi)^{2} R_{a\bar z\bar az}
\,.
\end{equation}
Now we can use the above conversion between $\Omega^{1},\Omega^{2}$ and $\lambda,\psi$ (\ref{fermions scaling pure}) to extract the prediction for the four-fermion vertex in the low-energy effective Lagrangian from (\ref{metric instanton pure 2}).
Thus, considering only the leading $k=\gamma_{m}'=1$ sector, we obtain the four-fermion vertex~\footnote{
Strictly speaking, for $\theta_{e}\ne 0$, the matrices appearing in (\ref{fermions transformation pure}) give rise to a rotation which changes the chirality, but preserves the overall normalisation, which is subject to (\ref{fermions constraint pure}).
}:
\begin{equation}
\label{action 4 fermions pure}
S_{\rm 4F} =
\frac{2^{8}\pi^{9/2}R^{5/2}}{|a|^{1/2}g_{\rm eff}^{7}} \, 
\mathcal{D}(-i) \exp\left( -S_{\rm mon} \right)
\sum_{\gamma_{e}'\in\mathbb{Z}} \exp\left( -S_{\varphi}^{(\gamma_{e}')} \right)
\int d^{3}x \left( \psi\cdot\bar\psi \right) \left( \lambda\cdot\bar\lambda \right)
\,.
\end{equation}
We shall verify this term in the effective action via a direct semiclassical calculation.

%%%%%%%%%%%%%%%%%%%%%%%%%%%%%%%%%%%%%%%%%%%%%%%%%%%%%%%%%%%%%

\section{Semiclassical instanton calculation}

\paragraph{}

In this section, we will compute the dyonic contribution to the action in $\mathbb{R}^{3}\times S^{1}$ from first principles.
We focus on the leading instanton contribution coming from magnetic charge $\gamma_{m}=1$ with winding number $k=1$ and arbitrary electric charges $\gamma_{e}\in\mathbb{Z}$; in particular, our goal is to find the appropriate four-fermion correlator.
A similar calculation was performed in \cite{DKMTV} in three dimensions and in \cite{Dorey 2000, Dorey Parnachev} in the corresponding theory with 16 supercharges on $\mathbb{R}^{3}\times S^{1}$.

\paragraph{}

We begin by considering a static BPS monopole of the $\mathcal{N}=2$ theory.
The bosonic moduli of this soliton consist of three coordinates $X^{1},X^{2},X^{3}$ defining the position of its centre in $\mathbb{R}^{3}$ and an additional periodic angle $\varphi\in [0,2\pi)$ describing orientation of the instanton in the unbroken gauge subgroup $U(1)$.
The moduli space is thus $\mathbb{R}^{3}\times S_{\varphi}^{1}$.
There are four fermionic zero modes in the monopole background, which are generated by half of the eight supercharges (Weyl fermions $\psi$ and $\lambda$ each have two independent zero-mode solutions).

\paragraph{}

The left- and right-handed Weyl fermions of the auxiliary theory are denoted $\rho_{\delta}^{\ A}$ and $\bar{\rho}^{\ A}_{\delta}$, respectively, where $A\in\{1,2\}$, $\delta\in\{1,2\}$.
In terms of these fermions, the four zero modes of the instanton are all left-handed, yielding a non-zero contribution to the correlator
\begin{equation}
\label{correlator 4 fermions pure}
\mathcal{G}_{4}(y_{1},y_{2},y_{3},y_{4}) =
\left\langle \prod_{A=1}^{2}\rho^{\ A}_{1}(y_{2A-1})\rho^{\ A}_{2}(y_{2A}) \right\rangle
\,,
\end{equation}
corresponding to a vertex of the form $(\bar{\rho}^{1}\cdot\bar{\rho}^{1})(\bar{\rho}^{2}\cdot\bar{\rho}^{2})$ in the low-energy effective action.
The fermions of the auxiliary theory are related to the original four-dimensional Weyl fermions by an $SO(3)$ $R$-symmetry rotation which mixes left- and right-handed chiralities but preserves the normalisation of the four-fermion vertex in the effective Lagrangian.
In a vacuum where $\theta_{e}=0$, the zero modes of a monopole are chirally symmetric in the original four-dimensional theory, and the explict relation takes the form~\footnote{
As in the previous section, when $\theta_{e}\ne 0$, the rotation leads to chirally asymmetric vertex when written in terms of $\psi$ and $\lambda$.
}
\begin{equation}
\label{fermion transformation pure}
(\bar{\rho}^{1}\cdot\bar{\rho}^{1})(\bar{\rho}^{2}\cdot\bar{\rho}^{2}) =
(\psi\cdot\bar\psi)(\lambda\cdot\bar\lambda)
\,.
\end{equation}

\paragraph{}

In the weak-coupling regime, we can replace the fermions in the correlation function (\ref{correlator 4 fermions pure}) with their zero mode values multiplied by corresponding Grassmann collective coordinates $\xi_{\delta}^{A}$.
The explicit form of the zero modes is given in appendix C of \cite{DKMTV}.
As we are interested in deriving the low-energy effective action, we focus on the long-distance limit of the correlation function and fermionic zero modes.
We can then express the long-distance limit of $\rho_{\alpha}^{\ A}$ in terms of $\xi_{\delta}^{\ A}$ and the three-dimensional Dirac fermion propagator $S_{\rm F}(x)=\gamma^{\mu}x_{\mu}/4\pi |x|^{2}$ as
\begin{equation}
\label{propagator fermion 3D pure}
\rho_{\alpha}^{{\rm (LD)} \, A}(y) = 8\pi \, S_{\rm F}(y-X)_{\alpha}^{\ \beta} \, \xi_{\beta}^{\ A}
\,.
\end{equation}

\paragraph{}

In four dimensions, the semiclassical dynamics of monopoles is described by supersymmetric quantum mechanics on the moduli space \cite{Gauntlett, Blum}.
For a single monopole of mass
\begin{equation}
M = \frac{4\pi}{g^{2}} |a|
\,,
\end{equation}
this corresponds to the dynamics of a free non-relativistic particle moving on $\mathbb{R}^{3}\times S_{\varphi}^{1}$.
These bosonic degrees of freedom have four free fermionic superpartners.
The Lagrangian of collective coordinates takes the form
\begin{align}
\label{Lagrangian monopole pure}
& L_{\rm QM} = L_{X}+L_{\varphi}+L_{\xi}
\,,
\\
& L_{X} = \frac{M}{2}|\dot{\vec X}|^{2}
\,,
\quad
L_{\varphi} = \frac{1}{2}\frac{M}{|a|^{2}}\dot\varphi^{2}
\,,
\quad
L_{\xi} = \frac{M}{2}\dot\xi_{\alpha}^{A}\dot\xi_{A}^{\alpha}
\end{align}
where dot denotes time derivative.
The combination $M/|a|^{2}$ is the moment of inertia of a monopole with respect to global gauge rotation, $L_{\varphi}$ describes a free particle of mass $M/|a|^{2}$ moving along $S_{\varphi}^{1}$ with $\varphi\in [0,2\pi]$.

\paragraph{}

The quantity of interest here is the long-distance behavior of the four-fermion correlation function (\ref{correlator 4 fermions pure}).
To consider the theory on $\mathbb{R}^{3}\times S^{1}$, we Wick-rotate the quantum mechanics for collective coordinates described above so that the Euclidean time is identified with the periodic $x^{0}$ coordinate introduced earlier.
As a result, there are periodic boundary conditions for both bosons and fermions.
To the leading semiclassical order, the fermionic fields in the correlator are replaced by their values in the monopole background.
The resulting long-distance correlation function then takes the following form:
\begin{equation}
\label{correlator 4 fermions pure 2}
\mathcal{G}_{4}(y_{1},y_{2},y_{3},y_{4}) =
\int d\mu \, \prod_{A=1}^{2} \rho^{{\rm (LD)} \, A}_{1}(y_{2A-1})\rho^{{\rm (LD)} \, A}_{2}(y_{2A})
\,,
\end{equation}
\begin{equation}
\label{correlator measure pure}
\int d\mu = \frac{1}{4\pi^{2}} \int d^{3}X(x^{0}) \, d\varphi(x^{0}) \, d^{4}\xi(x^{0}) \, \mathcal{R} \,
\exp\left( -\int_{0}^{2\pi R} dx^{0}L_{\rm QM} \right)
\exp\left( -\frac{8\pi^{2}R |a|}{g^{2}}+i\theta_{m} \right)
\,,
\end{equation}
where we consider the long-distance behaviour of the fermionic zero modes (\ref{propagator fermion 3D pure}).
The actor of $1/4\pi^{2}$ arises from the Jacobian corresponding to the change of variables from bosonic and fermionic measures to the four bosonic  and four fermionic collective coordinates and can be traced to the same factor in \cite{Bernard} given explicitly by eq.\ (114, 125) in \cite{DKMTV}.    
The integration measure $d\mu$ consists of the bosonic $d^{3}X\,d\varphi$ and fermionic $d^{4}\xi$ zero mode measures, and the one-loop factor $\mathcal{R}$ describing the non-zero mode fluctuations.
It is multiplied by the expression corresponding to the monopole action and the collective coordinates Lagrangian (\ref{Lagrangian monopole pure}).
We shall now evaluate various contributions in turns following the approaches in \cite{Dorey Parnachev} and \cite{Kaul}.

\paragraph{}

For the bosonic $d^{3}X(x^{0})$ and fermionic $d^{4}\xi(x^{0})$ zero mode measures, we note that $\vec X(x^{0})$ and $\xi_{\alpha}^{A}(x^{0})$ need to satisfy the periodic boundary conditions $\vec X(x^{0})=\vec X(x^{0}+2\pi R)$ and $\xi_{\alpha}^{A}(x^{0})=\xi_{\alpha}^{A}(x^{0}+2\pi R)$.
This implies that for the free lagragians $L_{X}$ and $L_{\xi}$, the path integrals are dominated by the constant classical paths, which we again denote as $\vec X$ and $\xi_{\alpha}^{A}$.
We can then expand these coordinates around the classical paths:
\begin{align}
\vec X(x^{0}) & = \vec X+\delta\vec X(x^{0})
\,,
\\
\xi_{\alpha}^{A}(x^{0}) & = \xi_{\alpha}^{A}+\delta\xi_{\alpha}^{A}(x^{0})
\,.
\end{align}
Then, we decompose the path integrals as
\begin{align}
\label{instanton measure translation}
\int d^{3}X(x^{0}) \,
\exp\left( -\int_{0}^{2\pi R} dx^{0}\,  L_{X} \right) & =
\int d^{3}X \int d^{3}\delta X(x^{0}) \,
\exp\left( -\int_{0}^{2\pi R} dx^{0} \, \frac{M}{2}(\delta\dot{\vec X}(x^{0}))^{2} \right)
\,,
\\
\label{instanton measure fermions}
\int d^{4}\xi(x^{0}) \,
\exp\left( -\int_{0}^{2\pi R} dx^{0} \, L_{\xi} \right) & =
\int d^{4}\xi \int d^{4}\delta\xi(x^{0}) \,
\exp\left(-\int_{0}^{2\pi R} dx^{0} \, \frac{M}{2} \, \delta\dot{\xi}_{\alpha}^{A}(x^{0})\delta\dot{\xi}_{A}^{\alpha}(x^{0}) \right)
\,.
\end{align}
The Gaussian integrals in (\ref{instanton measure translation}) and (\ref{instanton measure fermions}) over $\delta X(x^{0})$ and $\delta \xi(x^{0})$ can be readily evaluated using standard results, and we obtain
\begin{align}
\label{instanton measure translation 2}
\int d^{3}X(x^{0}) \,
\exp\left( -\int_{0}^{2\pi R} dx^{0} \, L_{X} \right) & =
\int d^{3}X
\left( \frac{1}{2\pi}\sqrt{\frac{M}{R}} \right)^{3}
\,,
\\
\label{instanton measure fermions 2}
\int d^{4}\xi(x^{0}) \,
\exp\left( -\int_{0}^{2\pi R} dx^{0} \, L_{\xi} \right) & =
\int d^{4}\xi
\left( \frac{1}{2\pi}\sqrt{\frac{M}{R}} \right)^{-4}
\,.
\end{align}
Next, consider the path integral for $\varphi$, which encodes the motion of the monopole along $S_{\varphi}^{1}$.
The conjugate momentum $P_{\varphi}=\dot\varphi M/|a|^{2}$ to $\varphi$ is identified with electric charge quantised in integer units.
The corresponding Hamiltonian is, therefore, $H_{\varphi}=P_{\varphi}^{2}|a|^{2}/2M$.
The resulting states in four dimensions carry one unit of magnetic charge and $P_{e}=\gamma_{e}$ units of electric charge; they are naturally identified as the corresponding BPS dyons.
We can then equate the path integral for $d\varphi(x^{0})$ with the quantum-mechanical partition function $\tr\exp(-2\pi RH_{\varphi})$, where the trace sums over the eigenstates corresponding to $H_{\varphi}$:
\begin{equation}
\label{instanton measure rotation}
\int d\varphi(x^{0}) \,
\exp\left( -\int_{0}^{2\pi R} dx^{0} \, L_{\varphi} \right) =
\sum_{\gamma_{e}\in\mathbb{Z}} \exp\left( -\frac{|a|^{2}\pi R}{M} \, \gamma_{e}^{2} \right)
\end{equation}
when $\Theta=0$, $\theta_{e}=0$.
A further phase in the classical action arises from the surface terms coupled to electric and magnetic charges.
Adding the appropriate surface term, a bare vacuum angle $\Theta$, and including the Witten effect, which shifts $\gamma_{e}\to\gamma_{e}+\Theta/2\pi$, the summation in (\ref{instanton measure rotation}) is replaced by
\begin{equation}
\label{instanton measure rotation 2}
\int d\varphi(x^{0}) \,
\exp\left( -\int_{0}^{2\pi R} dx^{0} \, L_{\varphi} \right) =
\sum_{\gamma_{e}\in\mathbb{Z}} \exp\left( -\frac{|a|^{2}\pi R}{M}\left( \gamma_{e}+\frac{\Theta}{2\pi} \right)^{2}+i\left( \gamma_{e}+\frac{\Theta}{2\pi} \right)\theta_{e} \right)
\,.
\end{equation}
We note that this coincides with the corresponding sum appearing in the GMN prediction (\ref{metric instanton pure 2}) up to replacing the bare coupling and vacuum angle by their one-loop renormalised counterparts.

\paragraph{}

To complete the semiclassical integration measure, in additon to the zero modes discussed so far, it is necessary to include the non-zero mode fluctuations, which lead to a non-cancelling factor $\mathcal{R}$ given via a ratio of functional determinants.
Again, we start by reviewing the situation in the four-dimensional theory, where similar fluctuations are taken into account in the calculation \cite{Kaul} of the one-loop corrections to the monopole mass.
In this case, the ratios of determinants corresponding to the flucutations of the scalars, spinors, and ghosts around the static monopole background can be ultimately described in terms of two operators $\Delta_{\pm}$ \cite{Weinberg} given explicitly
as
\begin{align}
\Delta_{+} & = D^{2}_{i}+|a|^{2}
\,,
\\
\Delta_{-} & = D^{2}_{i}+|a|^{2}+2\epsilon_{ijk}\sigma_{i}F_{jk}^{\rm mon}
\,,
\quad
\{i,j\} \subset \{1,2,3\}
\end{align}
where the three-dimensional covariant derivative $D_{i}=\partial_{i}+iA_{i}^{\rm mon}$ is with respect to background static monopole and, as above, $a$ is the VEV of the complex scalar in the $U(1)$ vector multiplet.
The one-loop correction to the monopole mass in four dimensions then involves the ratio $(\det(\Delta_{+})/\det'(\Delta_{-}))^{1/2}$ where the prime indicates removing the zero mode contribution.
In this section, we are interested in the corresponding fluctuations around the monopole, thought of as a static configuration yielding a finite Euclidean action on $\mathbb{R}^{3}\times S^{1}$.
In the absence of Wilson line, the corresponding fluctuation operators in this case are
\begin{equation}
\label{Delta pure}
\mathbb{D}_{\pm} = \Delta_{\pm}+\left( \frac{\partial}{\partial x^{0}} \right)^{2}
\,,
\end{equation}
where the extra derivatives with respect to $x^{0}$ take account of the modes of each fluctuation field on $S^{1}$.
We then identify the corresponding one-loop contribution to the path integral measure as
\begin{equation}
\label{loop factor pure}
\mathcal{R} = \left( \frac{\det\mathbb{D}_{+}}{\det'\mathbb{D}_{-}} \right)^{1/2}
\,.
\end{equation}

\paragraph{}

By translation invariance on $S^{1}$, we can decompose any eigenfunction of $\mathbb{D}_{\pm}$ as $\Phi_{\pm}(\vec x,x^{0})=\phi_{\pm}(\vec x)f_{\pm}(x^{0})$, where $\phi_{\pm}(\vec x)$ satisfies
\begin{equation}
\Delta_{\pm}\phi_{\pm}(\vec x) = \lambda_{\pm}^{2}\phi_{\pm}(\vec x)
\,,
\end{equation}
while $f_{\pm}(x^{0})$ along the compactified circle takes the plane-wave form $f_{\pm}(x^{0})\sim e^{i\omega_{\pm}x^{0}}$.
In a supersymmetric theory, the total number of non-zero eigenvalues for the bosonic and for the fermionic fields is the same; this seems to imply that their contributions cancel completely and that $\mathcal{R}=1$.
However, the spectra of $\mathbb{D}_{\pm}$ contain both normalisable bound states and continuous scattering states, as inherited from $\Delta_{\pm}$; the precise cancellation requires the densities of bosonic and fermionic eigenvalues to be identical.
As discovered by \cite{Kaul}, this is not the case in the monopole background.
The same effect leads to the non-cancelling one-loop factor $\mathcal{R}$ in the three-dimensional instanton calculation of \cite{DKMTV}, and we find a similar effect in the present case of $\mathbb{R}^{3}\times S^{1}$.

\paragraph{}

Splitting the determinants in (\ref{loop factor pure}) as $\det_{x^{0},\vec x}=\det_{\vec x}\det_{x^{0}}$ and making use of the operator identity $\log\det M=\tr\log M$ for $\det_{\vec x}$ while keeping $\det_{x^{0}}$ intact, we can rewrite the one-loop factor $\mathcal{R}$ as the following integral expression:
\begin{equation}
\label{loop factor pure 2}
\begin{aligned}
\mathcal{R} & = (2\pi R )^{-2} \exp\left(
\frac{1}{2}\tr_{\vec x}\log\det{}_{x^{0}}\mathbb{D}_{+} -
\frac{1}{2}\tr_{\vec x}\log\det{}_{x^{0}}\mathbb{D}_{-}
\right)
\\
& = (2\pi R )^{-2} \exp\left(
- \int_{0}^{+\infty} d\lambda^{2}\delta\rho(\lambda)
\log\mathcal{K}(\lambda,2\pi R)
\right)
\end{aligned}
\end{equation}
where the integration kernel for $\theta_{e}=0$ is given by
\begin{equation}
\mathcal{K}(\lambda,2\pi R) = \det{}_{x^{0}}\left( \left( \frac{\partial}{\partial x^{0}} \right)^{2}+\lambda^{2} \right)
\,.
\end{equation}
The overall normalisation constant $(2\pi R)^{-2}$ was introduced so that $\mathcal{R}$ reproduces the corresponding three-dimensional factor which is given as \cite{DKMTV}~\footnote{
To take the three-dimensional limit, we will need to first Poisson-resum the explicit logarithmic expressions arising in (\ref{loop factor pure 2}), cf.\ (\ref{loop factor pure 3}) and (\ref{D factor resummed pure}) in the next section, before setting $R\to 0$.
}
\begin{equation}
\mathcal{R}^{\rm (3D)} =
\left( \frac{\det\Delta_{+}}{\det'\Delta_{-}} \right)^{1/2} = 4M_{W}^{2}
\end{equation}
in the limit $R\to 0$, $g_{\rm eff}^{2}\to 0$ where the gauge coupling in three dimensions $e_{\rm eff}=g_{\rm eff}/\sqrt{2\pi R}$ is held fixed (we will find the three-dimensional value of (\ref{loop factor pure 2}) in the next section).
The quantity $\delta\rho(\lambda)=\rho_{+}(\lambda)-\rho_{-}(\lambda)$ is the difference between the densities of eigenvalues of the operators $\Delta_{+}$ and $\Delta_{-}$.
This quantity was calculated in \cite{Kaul} using the Callias index theorem \cite{Callias}.
In our notations, the result of \cite{Kaul} gives
\begin{equation}
\label{states density difference pure}
d\lambda^{2}\delta\rho(\lambda) =
-\frac{2 |a|}{\pi\lambda^{2}\sqrt{\lambda^{2}-|a|^{2}}} \, \theta(\lambda^{2}-|a|^{2}) \, d\lambda^{2}
\end{equation}
where $\theta(y)$ is a step function such that $\theta(y)=1$ for $y\geq 0$, $\theta(y)=0$ for $y<0$.
For the remaining kernel $\mathcal{K}(\lambda,2\pi R)$, we observe that $\mathcal{K}(\lambda,2\pi R)^{-1}$ is precisely the partition function of harmonic oscillator with frequency $\varpi=\lambda$ at inverse temperature $\beta=2\pi R$.
Summing over all energies, we get
\begin{equation}
\mathcal{K}(\lambda,2\pi R)^{-1} = \frac{\exp(-\pi R\lambda)}{1-\exp(-2\pi R\lambda)}
\,.
\end{equation}
Introducing a non-vanishing Wilson line $\theta_{e}$ corresponds to turning on the compactified component of the gauge field.
This can be incorporated in the operators $\mathbb{D}_{\pm}$ defined in (\ref{Delta pure}) by the minimal coupling prescription
\begin{equation}
\frac{\partial}{\partial x_{0}} \to
\frac{\partial}{\partial x_{0}} + \gamma_{e} \, \frac{\theta_{e}}{2\pi R}
\,,
\end{equation}
which introduces a chemical potential shifting the oscillator frequencies to the complex values $\varpi=\lambda\pm i\theta_{e}/2\pi R$.
Summing over both contributions, we find that the modified kernel in (\ref{loop factor pure 2}) must obey $\mathcal{K}^{2}=\mathcal{K}_{+}\mathcal{K}_{-}$ where
\begin{equation}
\label{kernel pure}
\mathcal{K}_{\pm}(\lambda,\theta_{e},2\pi R)^{-1} = \frac{\exp(-\pi R\lambda\pm i\theta_{e}/2)}{1-\exp(-2\pi R\lambda\pm i\theta_{e})}
\,.
\end{equation}

\paragraph{}

Substituting (\ref{states density difference pure}) and (\ref{kernel pure}) into (\ref{loop factor pure 2}) and changing the variable $\lambda=2|a|\cosh t$, we rewrite the ratio of one-loop determinants $\mathcal{R}$ as
\begin{equation}
\label{loop factor pure 3}
\begin{aligned}
\log\mathcal{R} & =
4R |a|\arcosh\frac{\Lambda_{\rm UV}}{|a|}
-2\log(2\pi R)
\\
& +\frac{2}{\pi} \int_{0}^{+\infty} \frac{dt}{\cosh t}
\log\left( 1-e^{-2\pi R |a|\cosh t+i\theta_{e}} \right)
+\frac{2}{\pi} \int_{0}^{+\infty} \frac{dt}{\cosh t}
\log\left( 1-e^{-2\pi R |a|\cosh t-i\theta_{e}} \right)
\end{aligned}
\end{equation}
where we have evaluated the integral over the eigenvalues with an ultraviolet cutoff $\Lambda_{\rm UV}$.
We can see that after exponentiating (\ref{loop factor pure 3}), the ultraviolet divergence in the first term is precisely that encountered in the four-dimensional calculation of \cite{Kaul}.
The divergence is cancelled by the counter-term responsible for renormalisation of the coupling constant, and the net effect is replacing the classical coupling $g^{2}$ appearing in the monopole mass by the one-loop effective coupling, $g_{\rm eff}^{2}(a)$.
Similarly, the chiral anomaly results in the replacement of the classical vacuum angle $\Theta$ by its effective counterpart $\Theta_{\rm eff}(a)$ defined above.
The remaining finite terms yield a complicated function of the dimensionless parameter $|a|R$.
However, we recognise the integrals in the second line as precisely the same appearing in the expression (\ref{D factor pure 3}) for the perturbative factor $\log\mathcal{D}(-i)$ in the semiclassical expansion of the moduli space metric.

\paragraph{}

Collecting all the pieces and summing over electric charges $\gamma_{e}$, we can extract the four-fermion vertex in the low-energy effective action from examining the large distance behavior of the four-fermion correlation function $\mathcal{G}_{4}(y_{1},y_{2},y_{3},y_{4})$.
Substituting (\ref{propagator fermion 3D pure}), (\ref{instanton measure translation 2}, \ref{instanton measure fermions 2}, \ref{instanton measure rotation 2}), and (\ref{loop factor pure 3}) into
(\ref{correlator 4 fermions pure 2}) and (\ref{correlator measure pure}), we rewrite the four-fermion correlation function as
\begin{equation}
\label{correlator 4 fermions pure 3}
\begin{aligned}
\mathcal{G}_{4}(y_{1},y_{2},y_{3},y_{4}) & =
\frac{2^{13/2}\pi}{R |a|^{1/2}} \,
\mathcal{D}(-i) \left( \frac{2\pi R}{g_{\rm eff}^{2}} \right)^{-1/2} \exp\left( -S_{\rm mon} \right)
\sum_{\gamma_{e}\in\mathbb{Z}} \exp\left( -S_{\varphi}^{(\gamma_{e})} \right)
\\
& \int d^{3}X \epsilon^{\alpha'\beta'}\epsilon^{\gamma'\delta'}
S_{\rm F}(y_{1}-X)_{\alpha\alpha'}S_{\rm F}(y_{2}-X)_{\beta\beta'}S_{\rm F}(y_{3}-X)_{\gamma\gamma'}S_{\rm F}(y_{4}-X)_{\delta\delta'}
\,,
\end{aligned}
\end{equation}
where we have used the relation between $\mathcal{D}(-i)$ and $\mathcal{R}$.
We have also taken into account the one-loop renormalisation effect discussed earlier.
Finally, for consistency, the same renormalisation of the classical coupling $g^{2}$ leading to its replacement by the corresponding effective coupling $g_{\rm eff}^{2}(a)$ in the exponent must also be implemented wherever the coupling appears~\footnote{
Concretely, this renormalisation corresponds to a divergent contribution to the instanton measure arising from loop diagrams of perturbation theory in the monopole background.
}, so that the monopole action $S_{\rm mon}$ and the angular action $S_{\varphi}^{(\gamma_{e})}$, both including surface terms, are given in terms of the effective parameters $g_{\rm eff}$ and $\Theta_{\rm eff}$ as
\begin{align}
S_{\rm mon} & =
\frac{8\pi^{2}R}{g_{\rm eff}^{2}}|a|-i\theta_{m}
\,,
\\
S_{\varphi}^{(\gamma_{e})} & =
\frac{g_{\rm eff}^{2}R |a|}{4}\left( \gamma_{e}+\frac{\Theta_{\rm eff}}{2\pi} \right)^{2}-i\left( \gamma_{e}+\frac{\Theta_{\rm eff}}{2\pi} \right)\theta_{e}
\,.
\end{align}
In the low-energy effective action, the resulting correlator implies the appearance of a four-fermion interaction term which, after expanding the fermion propagators in (\ref{correlator 4 fermions pure 3}), is given by
\begin{equation}
\label{action 4 fermions pure semiclassical}
S_{\rm 4F} =
\frac{2^{9/2}\pi}{R |a|^{1/2}}
\left( \frac{2\pi R}{g_{\rm eff}^{2}} \right)^{7/2}
\mathcal{D}(-i) \exp\left( -S_{\rm mon} \right)
\sum_{\gamma_{e}\in\mathbb{Z}} \exp\left( -S_{\varphi}^{(\gamma_{e})} \right)
\int d^{3}x \left( \psi\cdot\bar\psi \right) \left( \lambda\cdot\bar\lambda \right)
\,.
\end{equation}
We see that this is exactly our prediction (\ref{action 4 fermions pure}) which was obtained from expanding the GMN equation (\ref{RH}).

%%%%%%%%%%%%%%%%%%%%%%%%%%%%%%%%%%%%%%%%%%%%%%%%%%%%%%%%%%%%%

\section{Interpolating to three dimensions}

\paragraph{}

Having matched the predicted instanton action (\ref{action 4 fermions pure}) and the semiclassical result (\ref{action 4 fermions pure semiclassical}), in this section we explain the relation to the semiclassical instanton result in three-dimensional theory, found in \cite{DKMTV}, which confirms that in the limit $R\to 0$, the hyper-K\"ahler metric on the Coulomb branch is given by Atiyah--Hitchin manifold \cite{Atiyah Hitchin}.
To achieve this, we take the semiclassical metric (\ref{metric instanton pure}), which is given by an infinite sum over all electric charges $\gamma_{e}'$ of the dyons with magnetic charge 1, and Poisson-resum it using the standard formula:
\begin{equation}
\label{Poisson resummation}
\sum_{k=-\infty}^{+\infty} f(k) = \sum_{n=-\infty}^{+\infty} \widehat f(n)
\,,
\quad
\widehat f(n) = \int_{-\infty}^{+\infty} f(k) \, e^{-2\pi ink}dk
\,.
\end{equation}
This transformation exchanges the electric charges of dyons with a corresponding set of winding modes \cite{Dorey 2000}.
This resummation is necessary because the sum over electric charges appearing in (\ref{metric instanton pure 2}) and (\ref{action 4 fermions pure semiclassical}) diverges when $R |a| \to 0$.
We can, in fact, directly perform the Poisson resummation on the initial expression (\ref{metric instanton pure}) for the metric component $g_{a\bar a}^{\rm inst}$.
The relevant Fourier transform can be evaluated using eq.\ (6.726-4) in \cite{Gradshteyn Ryzhik}~\footnote{
The required result is obtained by approximating the Bessel function in the integrand of eq.\ (6.726-4) by its asymptotic form for large arguments.
}, allowing us to rewrite the expression as
\begin{equation}
\label{metric instanton resummed pure}
g_{a\bar a}^{\rm inst} =
\frac{4\pi}{g_{\rm eff}^{4}}
\sum_{n\in\mathbb{Z}}
\frac{|a|^{2}\mathcal{D}(-i)}{|M(n)|^{3}}
\exp\left( -\frac{8\pi^{2}R}{g_{\rm eff}^{2}}|M(n)|+i\Psi(n) \right)
\,,
\end{equation}
where we have used the short-hand notation:
\begin{align}
|M(n)| & = \sqrt{|a|^{2}+\left( \frac{\theta_{e}+2\pi n}{2\pi R} \right)^{2}}
\,,
\\
\Psi(n) & = \theta_{m}-\frac{\Theta_{\rm eff}}{2\pi}(\theta_{e}+2\pi n)
\,.
\end{align}
One can also Taylor-expand and Poisson-resum the prefactor $\mathcal{D}(-i)$ (\ref{D factor pure 3}) to demonstrate that it satisfies the following equation:
\begin{equation}
\label{D factor resummed pure}
\frac{d\log\mathcal{D}(-i)}{d(2\pi R |a|)} =
2\left( \sum_{n\in\mathbb{Z}} \frac{1}{2\pi R \, |M(n)|} - \frac{1}{\pi}\arsinh\frac{\Lambda_{\rm UV}}{|a|} \right)
\,.
\end{equation}
The quantity $M(n)$ appearing in the exponent of (\ref{metric instanton resummed pure}) corresponds to the Euclidean action of BPS field configurations in the compactified gauge theory on $\mathbb{R}^{3}\times S^{1}$, which are obtained by applying a large gauge transformation of the form $A_{0}(x)\to A_{0}(x)+\partial\chi(x)$ with $\chi(x_{0}+2\pi R)=\chi(x_{0})+2\pi n$ \cite{Lee Yi, Dorey 1999, Dorey 2000, Dorey Parnachev}.
As a result, the Wilson line (\ref{Wilson line 1}) undergoes periodic shifts
\begin{equation}
\theta_{e} \to \theta_{e}+2\pi n
\,.
\end{equation}
These transformations are topologically non-trivial and are classified by an element of $\pi_{1}(S^{1})=\mathbb{Z}$.
This leads to an infinite tower of field configurations labelled by the winding number $n$; summing over these configurations ensures that the metric retains the correct periodicity with respect to $\theta_{e}$.

\paragraph{}

To compare our prediction with the three-dimensional result, we take the limit $R\to 0$ while keeping the three-dimensional gauge coupling fixed~\footnote{
Note that in \cite{DKMTV}, the three and four-dimensional couplings are related via $1/e^{2}=R/g^{2}$, which differs from our convention by $1/2\pi$.
}: $1/e_{\rm eff}^{2}=2\pi R/g_{\rm eff}^{2}$.
Note that in three dimensions, $|M(n)|\to\infty$ for $n\ne 0$, and thus in (\ref{metric instanton resummed pure}) and (\ref{D factor resummed pure}), only the $n=0$ terms remain.
Since we fix $\theta_{e}/2\pi R$ when $R\to 0$, the surface term containing $\theta_{e}$ vanishes.
In the strict three-dimensional limit, $(\re a,\im a,\theta_{e}/2\pi R)$ transform as a $\bf 3$ under the global $SU(2)_{N}$ symmetry \cite{SW3}, and we can rotate the vector into a vacuum for which $\theta_{e}=0$.
After this rotation, we can easily calculate (\ref{D factor pure 3}) by Taylor-expanding the integrands to the lowest non-zero term to obtain $\mathcal{D}(-i)=(4\pi R |a|)^{2}$; therefore, in generic vacuum, we have
\begin{equation}
\label{D factor 3D pure}
\mathcal{D}(-i) = (4\pi R M_{W})^{2}
\,.
\end{equation}
Using the normalisation factors (\ref{Riemann pure}, \ref{Riemann scaling pure}, \ref{fermions scaling pure}) and the one-loop factor (\ref{D factor 3D pure}), calculated above, we can again evaluate the four-fermion vertex in the low-energy effective action (\ref{action effective pure}):
\begin{equation}
\label{action 4 fermions 3D pure}
S_{\rm 4F} =
\frac{2^{7}\pi^{3}M_{W}}{e_{\rm eff}^{8}}
\exp\left( -\frac{4\pi}{e_{\rm eff}^{2}}M_{W}+i\theta_{m} \right)
\int d^{3}x \left( \psi\cdot\bar{\psi} \right) \left( \lambda\cdot\bar{\lambda} \right)
\end{equation}
where the mass of the $W$ boson in three dimensions is
\begin{equation}
M_{W}=\sqrt{|a|^{2}+\left( \frac{\theta_{e}}{2\pi R} \right)^{2}}
\,.
\end{equation}
Comparing this result with eq.\ (29) and (34) in \cite{DKMTV} obtained from the direct semiclassical analysis of three-dimensional instantons, we see that we have correctly reproduced the four-fermion vertex.
As has been shown in \cite{DKMTV} , the resulting four-fermion action implies that the moduli space of the theory is, in fact, the Atiyah--Hitchin manifold \cite{Atiyah Hitchin}.

\chapter{Instantons in theories with matter}
\label{ch: matter}

\paragraph{}

{\it This chapter is based on \cite{CP}.}

\paragraph{}

In this chapter, we extend the semiclassical analysis conducted in chapter \ref{ch: pure} to theories with $N_{f}\le 4$ fundamental flavours compactified on $\mathbb{R}^{3}\times S^{1}$ focusing on gauge group $SU(2)$ \cite{SW2}.
It is known that theories with $N_{f}<4$ fundamental hypermultiplets are asymptotically free, hence, in the semiclassical regime, we need to set $|a|\gg |\Lambda|$ where $\Lambda$ is the dynamical scale; the theory with $N_{f}=4$ is conformally invariant, and we select small values of the coupling constant.
We expand the Gaiotto--Moore--Neitzke equation (\ref{RH}) on the Coulomb branch at weak coupling and compare the expression with the semiclassical result obtained from first principles.
In theories with flavours, the perturbative corrections to the moduli space metric are produced by $W^{\pm}$ bosons with electric charge $\pm 2$ and by quarks with electric charge $\pm 1$ and one of the flavour charges equal to $\pm 1$ (these charges correspond to $N_{f}$ $U(1)$ flavour symmetries, contributing to the central charge).
They give rise to a shift of the effective coupling constant $g_{\rm eff}$, in particular, reproducing the three-dimensional result derived in \cite{DTV}.
The non-perturbative corrections are produced by dyons: they have integer electric charge and magnetic charge $\pm 1$ for all $N_{f}<4$ and magnetic charge $\pm 2$ for $N_{f}=3$; in the special case of $N_{f}=4$, their charges are $(p,q)$ and $(2p,2q)$ where $p$ and $q$ are relatively prime integers.

\paragraph{}

For finite values of $|a| R$, we expand the non-perturbative corrections to the moduli space metric as a linear combination of terms corresponding to individual dyons.
Analogous expressions have been derived in chapter \ref{ch: pure}, but in theories with flavours, the formulae are more complicated due to the presence of flavour charges.
For one-instantons, we calculate the four-fermion correlation function and verify the result semiclassically.
In the present case, the one-loop factor, which can be found by extending our analysis in chapter \ref{ch: pure}, corresponds to the contributions of $W^{\pm}$ bosons and quarks.

\paragraph{}

For small $|a| R$, we Poisson-resum the moduli space metric.
Then, we demonstrate how the previously known three-dimensional quantities \cite{DTV} can be recovered in the three-dimensional limit $|a| R\to 0$.
From dimensional analysis of the one-loop factor, we show that only one- and two-instanton corrections can exist in three dimensions.
If all flavour hypermultiplets are massive, there are only one-instanton corrections; if exactly one flavour hypermultiplet is massless, there are only two-instanton corrections; if there are more than one massless flavour hypermultiplets, there are no non-perturbative corrections in three dimensions.
Finally, we briefly describe the relevant Hanany--Witten D-brane configuration \cite{Hanany Witten} providing a geometric understanding of some of the field theory results discussed above.

%%%%%%%%%%%%%%%%%%%%%%%%%%%%%%%%%%%%%%%%%%%%%%%%%%%%%%%%%%%%%

\section{Moduli space and BPS spectrum}

\paragraph{}

We start by considering the four-dimensional $\mathcal{N}=2$ supersymmetric gauge theory with gauge group $SU(2)$ and $N_{f}\le 4$ hypermultiplets in the fundamental representation.
In terms of $\mathcal{N}=1$ superfield notations, each $\mathcal{N}=2$ vector multiplet consists of a vector multiplet and an adjoint chiral scalar $\Phi$, while each $\mathcal{N}=2$ hypermultiplet contains two chiral superfields, $Q_{ia}$ and $\tilde Q_{ia}$ ($i=1,\dots, N_{f}$ is the flavour index, and $a=1,2$ is the colour index).
The superpotential preserving $\mathcal{N}=2$ supersymmetry that includes these chiral superfields is given by
\begin{equation}
\mathcal{W} = \sum_{i=1}^{N_{f}} \left( \tilde Q_{i}\Phi Q_{i}+m_{i}\tilde Q_{i}Q_{i} \right)
\end{equation}
where $m_{i}$ are complex masses, and the colour indices are suppressed. 
We consider the Coulomb branch of the theory, where $\phi$, the scalar component of $\Phi$, acquires a VEV $\langle\phi\rangle=a\sigma^{3}/2$ where $\sigma^{3}$ is the third Pauli matrix, and the $SU(2)$ gauge group is spontaneously broken down to $U(1)$.
The VEV parametrises the Coulomb branch as a complex manifold, and the gauge-invariant parameter $u=\langle\tr\phi^{2}\rangle$ provides a globally defined coordinate.

\paragraph{}

The BPS spectrum of the theory on the Coulomb branch contains BPS states of the form $\gamma=(\gamma_{e},\gamma_{m},\vec s\,)$ carrying electric and magnetic charges, $\gamma_{e}$ and $\gamma_{m}$, under the unbroken gauge $U(1)$ and a vector $\vec s$ of flavour charges consisting of $N_{f}$ components $s_{i}$.
Due to additional matter fields, the BPS states also transform under the flavour symmetry: when all $m_{i}=0$, the symmetry is $SO(2N_{f})$, while for distinct $m_{i}\ne 0$, the $SO(2N_{f})$ symmetry is broken down to $U(1)^{N_{f}}$, and the BPS states are distinguished by the charges $s_{i}$ under the $U(1)$ flavour symmetries (the representations are summarised in \cite{DHIZ}).
The central charge (\ref{central charge}) in this case is~\footnote{
Our normalisation of the complex mass $m_{i}$ differs from that in \cite{SW2} by a factor of $1/\sqrt{2}$.
}
\begin{equation}
\label{central charge matter}
Z_{\gamma}(u) = \gamma_{e}a(u)+\gamma_{m}a_{D}(u)+\sum_{i=1}^{N_{f}}m_{i}s_{i}
\,.
\end{equation}
As before, the magnetic coordinate is determined by the prepotential $\mathcal{F}(a)$ \cite{SW} via (\ref{prepotential derivative}).
The prepotential is also related to the low-energy effective complex coupling~\footnote{
In this chapter, we follow the same normalisation convention for electric charges as in \cite{SW2} (i.e., $W^{\pm}$ bosons have electric charges $\pm 2$), so that the complex gauge coupling $\tau$ is multiplied by a factor of 2 with respect to our default convention (\ref{tau}), and $a$ is scaled to $a/2$ to compensate.
}:
\begin{equation}
\label{tau matter}
\tau_{\rm eff}(a)
=
\frac{8\pi i}{g_{\rm eff}^{2}(a)} + \frac{\Theta_{\rm eff}(a)}{\pi}
=
\frac{\partial^{2}\mathcal{F}(a)}{\partial a^{2}}
\end{equation}
where $g_{\rm eff}$ and $\Theta_{\rm eff}$ are the effective coupling constant and the effective vacuum angle.
We will mostly be interested in the weak coupling regime, i.e., we set $|a|\gg |\Lambda|$ for $N_{f}<4$ (so that $g_{\rm eff}\to 0$) and $g\to 0$ for $N_{f}=4$.
The one-loop value of the effective complex coupling for $N_{f}<4$ is
\begin{equation}
\label{tau effective matter}
\tau_{\rm eff}(a) \simeq \frac{i \, (4-N_{f})}{2\pi}
\log\left( \frac{a}{\Lambda} \right)^{2}
\,.
\end{equation}
Semiclassically, $a$, $a_{D}$, and $\tau_{\rm eff}$ are related via
\begin{equation}
a_{D} \simeq \tau_{\rm eff}a
\,,
\end{equation}
up to suppressed corrections coming from four-dimensional Yang--Mills instantons (see \cite{DKM, DKM2} for explicit instanton computations).

\paragraph{}

It is well-known \cite{Henningson, Bilal Ferrari 2, Bilal Ferrari 3} that the $\mathcal{N}=2$ theories with gauge group $SU(2)$ and $N_{f}<4$ fundamental flavours have non-trivial curves of marginal stability.
In the case of massless hypermultiplets for $N_{f}<4$, the curve of marginal stability is given by the locus $\im(a_{D}(u)/a(u))=0$ in the moduli space, and its solution can be obtained numerically using explicit expressions for $(a(u),a_{D}(u))$.
When we further include the masses $m_{i}\ne 0$, the walls of marginal stability become very complicated \cite{Bilal Ferrari 3} because of extra parameters in the problem~\footnote{
Systematic determination of BPS spectrum for general four-dimensional $\mathcal{N}=2$ supersymmetric theories with gauge group $SU(n)$ where $n>2$ is currently lacking.
}.
Such curve divides the Coulomb branch into the weakly and strongly coupled regions and goes through the singular points where BPS particles become massless, and the BPS spectrum is different inside and outside the curve.
The spectrum in theories with $N_{f}<4$ flavours is encoded in the wall-crossing formulae (\ref{WCF2 1f}, \ref{WCF2 2f}, \ref{WCF2 3f}): outside the curve, there is a finite number of $W$ bosons and quarks and an infinite number of dyons, inside the curve, most states decay into a finite number of stable dyons, as explicitly determined in \cite{Bilal Ferrari 2}.

\paragraph{}

In the weak-coupling regime, the perturbative BPS spectrum on the Coulomb branch includes $W$ bosons with electric charge $\gamma_{e}=\pm 2$ and fundamental quarks with electric charge $\gamma_{e}=\pm 1$ and exactly one non-zero flavour charge $s_{i}=\pm 1$ where we consider all possible $i$.
We denote the full set of charges as $(\gamma_{e},\gamma_{m},\vec s\,)$ where $\gamma_{e}$ is electric charge, $\gamma_{m}$ is magnetic charge, $\vec s$ are $N_{f}$ flavour charges, omitting flavour charges when they are all zero.
We will discuss the non-perturbative spectrum for $N_{f}<4$ and $N_{f}=4$ separately.
In the case of $N_{f}<4$, the dyons are $(n,\pm 1,\vec s\,)$ with $n\in\mathbb{Z}$ and $\vec s$ denoting all possible flavour charges for any $N_{f}<4$ and $(2n+1,\pm 2)$ with $n\in\mathbb{Z}$ for $N_{f}=3$.
The rank of the flavour group representation in theories with $N_{f}<4$ for all states with non-zero flavours is given by the multiplicity in the relevant wall-crossing formula (\ref{WCF2 1f}, \ref{WCF2 2f}, \ref{WCF2 3f}) (where the formulae do not include flavour charges).
For $N_{f}=1$, the flavour charge is $+1/2$ for even $n$ and $-1/2$ for odd $n$; for $N_{f}=2$, $\vec s$ is $\pm (1/2,1/2)$ for even $n$ and $\pm (1/2,-1/2)$ for odd $n$; for $N_{f}=3$, $\vec s$ is $\pm (1/2,1/2,1/2)$, $\pm (1/2,1/2,-1/2)$ for even $n$ and $\pm (1/2,-1/2,1/2)$, $\pm (1/2,-1/2,-1/2)$ for odd $n$ \cite{DHIZ}.
In the conformal case of $N_{f}=4$ flavours, the theory is $SL(2,\mathbb{Z})$ invariant \cite{SW2, Gauntlett Harvey, DHIZ}, allowing one to find the full BPS spectrum: for a pair of relatively prime integers $p$ and $q$, the states $(2p,2q)$ (including $W$ bosons) transform as a singlet ${\bf 1}$ under the flavour group $SO(8)$, the states $(2p+1, 2q)$ (including quarks) transform as a vector $\bf 8_{v}$, the dyons $(2p,2q+1)$ are in a spinor representation $\bf 8_{s}$, $(2p+1,2q+1)$ are in the conjugate spinor representation $\bf 8_{c}$.
When $N_{f}=4$, the set of flavour charges of dyons can be determined by considering charges under the four $SU(2)$ subgroups of the full flavour symmetry group $SO(8)$ \cite{Gaiotto} using the fact that the flavour charge-vector is the sum of a self-dual and an anti-self-dual components with sign plus or minus; then, the sets of flavour charges for all three non-trivial representations of the flavour group $SO(8)$ are
\begin{align}
& {\bf 8_{v}}
\,:
\quad
\pm \left( 1,0,0,0 \right)
\,,
\quad
\pm \left( 0,1,0,0 \right)
\,,
\quad
\pm \left( 0,0,1,0 \right)
\,,
\quad
\pm \left( 0,0,0,1 \right)
\,,
\\
& {\bf 8_{s},\ 8_{c}}
\,:
\quad
\pm \left( \frac{1}{2},\frac{1}{2},\frac{1}{2},\frac{1}{2} \right)
\,,
\quad
\pm \left(\frac{1}{2},-\frac{1}{2},\frac{1}{2},-\frac{1}{2} \right)
\,,
\quad
\pm \left(\frac{1}{2},\frac{1}{2},-\frac{1}{2},-\frac{1}{2} \right)
\,,
\quad
\pm \left( \frac{1}{2},-\frac{1}{2},-\frac{1}{2},\frac{1}{2} \right)
\,.
\end{align}
In addition, one needs to specify the degeneracies of all BPS states: for $W$ bosons and all states transforming under $\bf 8_{v}$, $\Omega(\gamma,a)=-2$, for all other BPS states, including fundamental quarks and dyons with a given set of flavour charges, $\Omega(\gamma,a)=+1$.

\paragraph{}

Let us now compactify the $x^{0}$ dimension on $S^{1}$ of radius $R$.
At length scales much larger than $R$, the low-energy effective action on the Coulomb branch becomes three-dimensional.
In addition to the four-dimensional complex scalar $a$, the electric Wilson line, $\theta_{e}\in [0,2\pi]$, and its magnetic dual, $\theta_{m}\in [0,2\pi]$, appear.
As in the pure theory, we periodically identify $\theta_{e}\sim\theta_{e}+2\pi$, $\theta_{m}\sim\theta_{m}+2\pi$; collectively, they describe a torus. Turning to the matter sector, in four dimensions, the mass of a hypermultiplet is defined by a complex parameter, $m_{i}$, but when one dimension is compactified on $S^{1}$, one more real component, $\tilde m_{i}$, appears.
In the brane picture, this additional real mass has a simple interpretation as the separation of the gauge and flavour branes in the dual compactified dimension.

\paragraph{}

The rest of the analysis in this section essentially repeats chapter \ref{ch: pure}, up to modifying numerical coefficients in several formulae.
The leading-order low-energy effective action follows from direct dimensional reduction of the four-dimensional low-energy theory.
To describe the action, we define the complex combination $z=\theta_{m}-\tau_{\rm eff}\theta_{e}$ parametrising a torus with complex structure $\tau_{\rm eff}(a)$.
The real bosonic part of the action is given in terms of $a,\bar a,z,\bar z$ as
\begin{equation}
\label{action bosonic 3D matter}
S_{\rm B} = \frac{1}{4} \int d^{3}x \left(
\frac{8\pi R}{g_{\rm eff}^{2}} \partial_{\mu}a \, \partial^{\mu}\bar a +
\frac{g^{2}_{\rm eff}}{32\pi^{3}R} \partial_{\mu}z \, \partial^{\mu}\bar z
\right)
\,.
\end{equation}
In addition, surface terms lead to the following purely imaginary terms:
\begin{equation}
\label{action imaginary 3D matter}
S_{\rm Im} = i\left( \gamma_{e}+\frac{\Theta_{\rm eff}}{\pi} \, \gamma_{m} \right)\theta_{e}+i \gamma_{m}\theta_{m}
\,.
\end{equation}
The corresponding fermionic terms in the action take the form
\begin{equation}
\label{action fermionic 3D matter}
S_{\text{F}} = \frac{4\pi R}{g_{\rm eff}^{2}} \int d^{3}x \tr \left(
i\bar\psi\bar\sigma^{\mu}\partial_{\mu}\psi +
i\bar \lambda\bar\sigma^{\mu}\partial_{\mu}\lambda
\right).
\end{equation}

\paragraph{}

The bosonic effective Lagrangian (\ref{action bosonic 3D matter}) allows us to extract the leading $R\to\infty$ behaviour of the hyper-K\"ahler metric on the moduli space:
\begin{equation}
\label{metric semiflat matter}
g^{\rm sf} = \frac{8\pi R}{g_{\rm eff}^{2}}|da|^{2} + \frac{g_{\rm eff}^{3}}{32\pi^{3}R}|dz|^{2}
\,.
\end{equation}
The metric (\ref{metric semiflat matter}) is manifestly K\"ahler with respect to the complex structure in which $a,z$ are holomorphic coordinates.
Going to finite radius $R$, this semiflat metric receives corrections from the four-dimensional BPS states whose worldlines wrap around the $S^{1}$ direction: perturbative one-loop corrections have the same form for any number of fundamental flavours, $N_{f}$, however, the form of non-perturbative corrections depends on $N_{f}$ because the set of BPS states with non-zero magnetic charge is a function of $N_{f}$.
We shall discuss these two types of corrections in turns.

%%%%%%%%%%%%%%%%%%%%%%%%%%%%%%%%%%%%%%%%%%%%%%%%%%%%%%%%%%%%%

\section{Semiclassical limit in theories with flavours}

\paragraph{}

We employ the same approach as in chapter \ref{ch: pure} for the pure $SU(2)$ theory, but also introduce some modifications to take into account the hypermultiplets and the extended BPS spectrum.
To approximate the solution of (\ref{RH}), we take the logarithm and generalise the iterative weak-coupling expansion considered in chapter \ref{ch: pure}.
Again, we restrict our attention to the semiclassical region of the moduli space, i.e., we set $|a|\gg\Lambda$ and, equivalently, $g^{2}_{\rm eff}\ll1$, whereas $R|a|$ is kept finite.
As we have explained in chapter \ref{ch: pure}, semiclassically, the quantity $\exp(-2\pi R |Z_{\gamma}|)$ is exponentially suppressed for all states with non-zero magnetic charge.
This is the case for the infinite series of dyons.
The purely electrically charged fundamental quarks and gauge bosons generate the perturbative one-loop corrections to the moduli space metric.

\paragraph{}

We begin by decomposing the Darboux coordinates as $\mathcal{X}_{\gamma}(\zeta)=\left(\mathcal{X}_{e}(\zeta)\right)^{\gamma_{e}}\left(\mathcal{X}_{m}(\zeta)\right)^{\gamma_{m}}$ for every $\gamma=(\gamma_{e},\gamma_{m})$.
In the case of a theory with flavours, it is also necessary to include the mass terms $m_{i}$ and $\tilde m_{i}$ for BPS states having non-zero flavour charges.
The relevant factor, analogous to the semiflat Darboux coordinates (\ref{coordinate semiflat}), which we introduce to account for the flavour contributions to the total central charge, is
\begin{equation}
\mu_{\gamma}(\zeta) = \exp\left( \sum_{i=1}^{N_{f}}s_{i}\left( \frac{\pi R}{\zeta}m_{i}+i\psi_{i}+\pi R{\zeta}\bar{m}_{i} \right) \right)
\,,
\quad
\psi_{i} = 2\pi R\tilde m_{i}
\,.
\end{equation}
The general integral equation (\ref{RH}) for the electric and magnetic Darboux coordinates can be expanded as
\begin{align}
\label{Darboux electric matter}
\mathcal{X}_{e}(\zeta) & =
\mathcal{X}_{e}^{\rm sf}(\zeta)
\exp\left( -\frac{1}{4\pi i} \sum_{\gamma'\in\Gamma} c_{e}(\gamma') \, \mathcal{I}_{\gamma'}(\zeta) \right)
\,,
\quad
c_{e}(\gamma') = -\Omega(\gamma',a) \, \gamma_{m}'
\,,
\\
\label{Darboux magnetic matter}
\mathcal{X}_{m}(\zeta) & =
\mathcal{X}_{m}^{\rm sf}(\zeta)
\exp\left( -\frac{1}{4\pi i} \sum_{\gamma'\in\Gamma} c_{m}(\gamma') \, \mathcal{I}_{\gamma'}(\zeta)\right)
\,,
\quad
c_{m}(\gamma') = \Omega(\gamma',u) \, \gamma_{e}'
\,,
\end{align}
where $\mathcal{X}_{e}^{\rm sf}(\zeta)$ and $\mathcal{X}_{m}^{\rm sf}(\zeta)$ are given by (\ref{coordinate semiflat}) with $(\gamma_{e},\gamma_{m})$ being equal to $(1,0)$ and $(0,1)$, and $\mathcal{I}_{\gamma'}(\zeta)$ is
\begin{equation}
\mathcal{I}_{\gamma'}(\zeta)=\int_{l_{\gamma'}} \frac{d\zeta'}{\zeta'} \frac{\zeta'+\zeta}{\zeta'-\zeta}
\log\left( 1-\sigma(\gamma')\mu_{\gamma'}(\zeta')\mathcal{X}_{\gamma'}(\zeta') \right)
\end{equation}
(cf.\ (\ref{Darboux electric pure}, \ref{Darboux magnetic pure}) in the pure theory).
Here, $\mu_{\gamma'}(\zeta')$ serves as a source term in the integral equations; this factor does not affect the semiflat Darboux coordinates and the semiflat metric, but it is involved in the BPS corrections.
Now, we take the weak-coupling limit: after setting $a_{D}\simeq\tau_{\rm eff}a$ to the one-loop order, we approximate the semiflat Darboux coordinates (\ref{Darboux basis pure}) and notice that $\log|\mathcal{X}^{\rm sf}_{m}|\gg\log|\mathcal{X}_{e}^{\rm sf}|$.
The expansion of $\log\mathcal{X}_{e}(\zeta)$ and $\log\mathcal{X}_{m}(\zeta)$ for the weak-coupling spectrum of $SU(2)$ theories with $N_{f}$ fundamental flavours is
\begin{equation}
\label{Darboux electric matter 2}
\log\mathcal{X}_{e}(\zeta) =
\log\mathcal{X}_{e}^{\rm sf}(\zeta)
-\sum_{\gamma'=(\gamma_{e}',\pm 1,\vec s\,')\in\Gamma} \frac{c_{e}(\gamma')}{4\pi i} \, \mathcal{I}_{(\gamma_{e}',\gamma_{m}',\vec s\,')}(\zeta)
\,,
\end{equation}
\begin{equation}
\label{Darboux magnetic matter 2}
\begin{aligned}
\log & \mathcal{X}_{m}(\zeta) =
\log\mathcal{X}_{m}^{\rm sf}(\zeta)
-\sum_{\gamma'=(\gamma_{e}',\pm 1,\vec s\,')\in\Gamma} \frac{c_{m}(\gamma')}{4\pi i} \, \mathcal{I}_{(\gamma_{e}',\gamma_{m}',\vec s\,')}(\zeta)
\\
&
-\frac{c_{m}(W^{+})}{4\pi i} \, \mathcal{I}_{(2,0)}(\zeta)
-\frac{c_{m}(W^{-})}{4\pi i} \, \mathcal{I}_{(-2,0)}(\zeta)
\\
&
-\sum_{i=1}^{N_{f}} \left(
\frac{c_{m}(q_{i}^{+})}{4\pi i} \, \mathcal{I}_{(1,0,\vec e_{i})}(\zeta) +
\frac{c_{m}(q_{i}^{-})}{4\pi i} \, \mathcal{I}_{(-1,0,\vec e_{i})}(\zeta) +
\frac{c_{m}(\tilde q_{i}^{+})}{4\pi i} \, \mathcal{I}_{(1,0,-\vec e_{i})}(\zeta) +
\frac{c_{m}(\tilde q_{i}^{-})}{4\pi i} \, \mathcal{I}_{(-1,0,-\vec e_{i})}(\zeta)
\right)
\end{aligned}
\end{equation}
where $\vec e_{i}$ denotes unit vector aligned with $i$-th direction.
In the expansion of $\mathcal{X}_{m}(\zeta)$, we have singled out the purely electrically charged BPS states, $W^{\pm},q_{i}^{\pm},\tilde q_{i}^{\pm}$, as they yield greater corrections than the dyons.
Let $\tilde{\Gamma}$ denote the remaining weakly coupled BPS spectrum, in other words, all magnetically charged states.
They act as non-perturbative instanton corrections to the Coulomb branch metric on $\mathbb{R}^{3}\times S^{1}$. 
At weak coupling, the full central charge can be approximated as $Z_{\gamma}(a)\simeq a(\gamma_{e}+\gamma_{m} \tau_{\rm eff}(a))+\sum_{i=1}^{N_{f}}s_{i}m_{i}$.

\paragraph{}

To approximate the corrections to $\log\mathcal{X}_{\gamma}$ at the leading order, we substitute the semiflat coordinates (\ref{Darboux basis pure})
into the right-hand side of (\ref{Darboux electric matter 2}, \ref{Darboux magnetic matter 2}) and ignore the
components that vanish as $g_{\rm eff}\to 0$.
In this limit, BPS contributions from $\tilde{\Gamma}$ to $\mathcal{X}_{e}, \mathcal{X}_{m}$ (i.e., the $\gamma_{m}'=\pm 1$ terms in (\ref{Darboux electric matter 2}) and (\ref{Darboux magnetic matter 2})) are not relevant; $\mathcal{X}_{m}$ does, however, receive order one contributions from purely electrically charged $W^{\pm},q_{i}^{\pm},\tilde q_{i}^{\pm}$, which we can compute directly.
Let us denote the corrected coordinates at this order as $\mathcal{X}^{(0)}_{e},\mathcal{X}^{(0)}_{m}$.
We can reuse our expansion (\ref{Darboux 0 pure}) from the pure $SU(2)$ theory: after including quark contributions, we have
\begin{equation}
\label{D factor matter}
\log\mathcal{D}(\zeta) = \log\mathcal{D}_{W}(\zeta)+\log \mathcal{D}_{q}(\zeta)+\log \mathcal{D}_{\tilde q}(\zeta)
\,.
\end{equation}
Here we have split the electric contributions $\mathcal{D}(\zeta)$ to $\mathcal{X}_{m}(\zeta)$ into three components: $\mathcal{D}_{W}(\zeta)$ comes from the $W$ bosons, $\mathcal{D}_{q}(\zeta)$ and $\mathcal{D}_{\tilde q}(\zeta)$, come from the quarks belonging to fundamental hypermultiplets.
Explicitly, these components are
\begin{equation}
\label{D factor W matter}
\log\mathcal{D}_{W}(\zeta) =
\frac{1}{\pi i} \left(
\int_{l_{W^{+}}} \frac{d\zeta'}{\zeta'} \frac{\zeta'+\zeta}{\zeta'-\zeta}
\log\left( 1-\mathcal{X}_{W^{+}}^{\rm sf}(\zeta') \right)
-\int_{l_{W^{-}}} \frac{d\zeta'}{\zeta'} \frac{\zeta'+\zeta}{\zeta'-\zeta}
\log\left( 1-\mathcal{X}_ {W^{-}}^{\rm sf}(\zeta') \right)
\right)
\,,
\end{equation}
\begin{equation}
\label{D factor q matter}
\begin{aligned}
\log\mathcal{D}_{q}(\zeta) =
-\frac{1}{4\pi i} \sum_{i=1}^{N_{f}} \left(
\int_{l_{q_{i}^{+}}} \frac{d\zeta'}{\zeta'} \frac{\zeta'+\zeta}{\zeta'-\zeta}
\log\left( 1-\mu_{q_{i}^{+}}(\zeta')\mathcal{X}_{q_{i}^{+}}^{\rm sf}(\zeta') \right)
\right.
\\
\left.
-\int_ {l_{q_{i}^{-}}} \frac{d\zeta'}{\zeta'} \frac{\zeta'+\zeta}{\zeta'-\zeta}
\log\left ( 1-\mu_{q_{i}^{-}}(\zeta')\mathcal{X}_{q_{i}^{-}}^{\rm sf}(\zeta') \right)
\right)
\,,
\end{aligned}
\end{equation}
\begin{equation}
\label{D factor qt matter}
\begin{aligned}
\log\mathcal{D}_{\tilde q}(\zeta) =
-\frac{1}{4\pi i} \sum_{i=1}^{N_{f}} \left(
\int_{l_{\tilde q_{i}^{+}}} \frac{d\zeta'}{\zeta'} \frac{\zeta'+\zeta}{\zeta'-\zeta}
\log\left( 1-\mu_{\tilde q_{i}^{+}}(\zeta')\mathcal{X}_{\tilde q_{i}^{+}}^{\rm sf}(\zeta') \right)
\right.
\\
\left.
-\int_ {l_{\tilde q_{i}^{-}}} \frac{d\zeta'}{\zeta'} \frac{\zeta'+\zeta}{\zeta'-\zeta}
\log\left ( 1-\mu_{\tilde q_{i}^{-}}(\zeta')\mathcal{X}_{\tilde q_{i}^{-}}^{\rm sf}(\zeta') \right)
\right)
\,,
\end{aligned}
\end{equation}
where the BPS rays for the $W^{\pm}$ bosons and for the quarks $q_{i}^{\pm}$ and $\tilde q_{i}^{\pm}$ are $l_{W^{\pm}}=\{\zeta':\pm a/\zeta'\in\mathbb{R}_{-}\}$, $l_{q_{i}^{\pm}}=\{\zeta':\pm(a+m_{i})/\zeta'\in\mathbb{R}_{-}\}$, $l_{\tilde q_{i}^{\pm}}=\{\zeta':\pm (a-m_{i})/\zeta'\in\mathbb{R}_{-}\}$.
To obtain (\ref{D factor W matter}, \ref{D factor q matter}, \ref{D factor qt matter}), we have used the fact that the charges are $\pm(2,0)$ for $W^{\pm}$ and $\pm(1,0)$ for $q_{i}^{\pm}$ and $\tilde q_{i}^{\pm}$, and we have set the degeneracies at weak coupling as $\Omega(W^{\pm},a)=-2$, $\Omega(q_{i}^{\pm},a)=\Omega(\tilde q_{i}^{\pm},a)=1$.
The mass parameters for flavours are given by
\begin{align}
\mu_{q^{\pm}_{i}}(\zeta) & = \exp\left( \pm\left( m_{i}\zeta^{-1}+i\psi_{i}+\bar m_{i}\zeta \right) \right)
\,,
\\
\mu_{\tilde q^{\pm}_{i}}(\zeta) & = \exp\left( \mp\left( m_{i}\zeta^{-1}+i\psi_{i}+\bar m_{i}\zeta \right) \right)
\,,
\end{align}
where $\psi_{i}=2\pi R\tilde m_{i}$.
For the fundamental hypermultiplets $q_{i}^{\pm}$ and $\tilde q_{i}^{\pm}$, the flavour charges are $s_{j}=\pm\delta_{ij}$ and $s_{j}=\mp\delta_{ij}$, respectively; for the $W^{\pm}$ bosons, one has $s_{i}=0$ since they are in the vector multiplet.

\paragraph{}

Having discussed the leading-order contributions, we now further expand $\mathcal{X}_{e}(\zeta),\mathcal{X}_{m}(\zeta)$ to extract the non-perturbative corrections:
\begin{equation}
\begin{aligned}
\log\mathcal{X}_{e}(\zeta) = \log\mathcal{X}_{e}^{(0)}(\zeta)+\delta\log\mathcal{X}_{e}(\zeta)
\,,
\\
\log\mathcal{X}_{m}(\zeta) = \log\mathcal{X}_{m}^{(0)}(\zeta)+\delta\log\mathcal{X}_{m}(\zeta)
\,.
\end{aligned}
\end{equation}
We can compute $\delta\mathcal{X}_{e}(\zeta),\delta\mathcal{X}_{m}(\zeta)$ by substituting $\mathcal{X}^{(0)}_{e}(\zeta),\mathcal{X}^{(0)}_{m}(\zeta)$ into (\ref{Darboux electric matter 2}) and (\ref{Darboux magnetic matter 2}):
\begin{align}
\label{Darboux electric suppressed matter}
\delta\log\mathcal{X}_{e}(\zeta) & =
-\frac{1}{4\pi i} \sum_{\gamma'=(\gamma_{e}',\pm 1,\vec s\,')\in\Gamma} c_{e}(\gamma') \, \mathcal{I}_{(\gamma_{e}',\gamma_{m}',\vec s\,')}^{(0)}(\zeta)
\,,
\\
\label{Darboux magnetic suppressed matter}
\delta \log\mathcal{X}_{m}(\zeta) & =
-\frac{1}{4\pi i} \sum_{\gamma'=(\gamma_{e}',\pm 1,\vec s\,')\in\Gamma} c_{m}(\gamma') \, \mathcal{I}_{(\gamma_{e}',\gamma_{m}',\vec s\,')}^{(0)}(\zeta)
\,,
\end{align}
where the integral is given by
\begin{equation}
\mathcal{I}_{(\gamma_{e}',\gamma_{m}',\vec s\,')}^{(0)}(\zeta) =
\int_{l_{\gamma'}} \frac{d\zeta'}{\zeta'} \frac{\zeta'+\zeta}{\zeta'-\zeta}
\log\left( 1-(-1)^{\gamma_{e}'\gamma_{m}'}\mu_{\vec s\,'}(\zeta')\left( \mathcal{X}_{e}^{(0)}(\zeta') \right)^{\gamma_{e}'}\left( \mathcal{X}_{m}^{(0)}(\zeta') \right)^{\gamma_{m}'} \right)
\,,
\end{equation}
the BPS ray $l_{\gamma'}$, which is the integration contour, is defined as in the case without flavours (\ref{BPS ray}) with $Z_{\gamma'}$ given by (\ref{central charge matter}).
We can now follow our approach in chapter \ref{ch: pure}: we insert the corrected Darboux coordinates into the symplectic form (\ref{symplectic form pure}) and then find the corresponding corrections to the metric (\ref{symplectic form corrections pure}, \ref{symplectic form semiflat pure}, \ref{symplectic form P pure}, \ref{symplectic form NP pure}).

%%%%%%%%%%%%%%%%%%%%%%%%%%%%%%

\subsection{Perturbative corrections}

\paragraph{}

The non-perturbative corrections to the metric $\omega^{\rm P}(\zeta)$, coming from the $W^{\pm}$ bosons and the quarks $q_{i}^{\pm}$ and $\tilde q_{i}^{\pm}$, and, using the same approach as in the pure theory, they can be evaluated in terms of modified Bessel functions of second kind, $K_{\nu}(x)$:
\begin{align}
\omega^{\rm P}(\zeta) & =
-\frac{i}{4\pi^{2}R} \,
d\log\mathcal{X}_{e}^{\rm sf}(\zeta)
\wedge
\left( 2\pi A^{\rm P}(a,\bar a)+\pi V^{\rm P}(a,\bar a)(\zeta^{-1} da-\zeta d\bar a) \right)
\,,
\\
\label{potential A matter}
A^{\rm P} & =
-\frac{R}{4\pi} \sum_{k>0} \sum_{\gamma\in\{W^{\pm},q_{i}^{\pm},\tilde q^{\pm}_{i}\}} \gamma_{e} c_{m}(\gamma) |Z_{\gamma}| e^{ik\left( \gamma_{e}\theta_{e}+\sum_{i=1}^{N_{f}}s_{i}\psi_{i} \right)} K_{1}(2\pi R |ka|)
\left( \frac{da}{Z_{\gamma}}-\frac{d\bar a}{\bar Z_{\gamma}} \right)
\,,
\\
\label{potential V matter}
V^{\rm P} & =
\frac{R}{2\pi} \sum_{k>0} \sum_{\gamma\in\{W^{\pm},q_{i}^{\pm},\tilde q^{\pm}_{i}\}} \gamma_{e} c_{m}(\gamma) e^{ik\left( \gamma_{e}\theta_{e}+\sum_{i=1}^{N_{f}}s_{i}\psi_{i} \right)} K_{0}(2\pi R |ka|)
\,,
\end{align}
where the moduli space metric is given as (\ref{Gibbons Hawking pure}, \ref{Gibbons Hawking potentials pure}) (cf.\ (\ref{symplectic form P pure 2}, \ref{potential A pure}, \ref{potential V pure}) in the pure theory).
Setting $\gamma_{e}=\pm 2$ for $W^{\pm}$ and $\gamma_{e}=\pm 1$ for $q_{i}^{\pm}$ and $\tilde q_{i}^{\pm}$, we can see that $c_{m}(W^{\pm})=\mp 4$, $c_{m}(q_{i}^{\pm})=c_{m}(\tilde q_{i}^{\pm})=\pm 1$.
We are taking the limit $g_{\rm eff}\to 0$ while keeping $R |a|$ fixed and arbitrary.
In particular, $K_{\nu}(x)$ diverges when $x\to 0$, and we should Poisson-resum the series of Bessel functions over $k$ in order to obtain a summation over all Kaluza--Klein momentum modes.
After comparing these perturbative contributions with the semiflat components (\ref{metric semiflat matter}), we can extract the shift of the coupling constant from the corrected moduli space metric:
\begin{align}
\label{coupling shift matter}
& \frac{8\pi R}{g_{\rm eff}^{2}} \to
\frac{8\pi R}{g_{\rm eff}^{2}} - \frac{1}{8\pi} \sum_{n\in\mathbb{Z}} \left(
\frac{8}{|M_{W}(n)|} - \sum_{i=1}^{N_{f}} \left( \frac{1}{|M_{q_{i}}(n)|}+\frac{1}{|M_{\tilde q_{i}}(n)|} \right)
\right)
\,,
\\
& |M_{W}(n)| = \sqrt{\left| 2a \right|^{2}+\left( \frac{\theta_{e}}{\pi R}+\frac{n}{R} \right)^{2}}
\,,
\\
& |M_{q_{i}}(n)| = \sqrt{\left| a+{m_{i}} \right|^{2}+\left( \frac{\theta_{e}+\psi_{i}}{2\pi R}+\frac{n}{R} \right)^{2}}
\,,
\quad
|M_{\tilde q_{i}}(n)| = \sqrt{\left| a-{m_{i}} \right|^{2}+\left( \frac{\theta_{e}-\psi_{i}}{2\pi R}+\frac{n}{R} \right)^{2}}
\,,
\end{align}
where in the limit $R\to 0$, all Kaluza--Klein momentum modes with $n\ne 0$ decouple.
On the other hand, if one takes the limit of large mass for one flavour (i.e., $m_{i}\to\infty$ for some $i$), the corresponding matter components decouple, and the number of flavours $N_{f}$ effectively decreases by 1.
Taking the limit $R |a|\to 0$, we obtain the corresponding shift in three dimensions:
\begin{equation}
\frac{1}{e_{\rm eff}^{2}} \to
\frac{1}{e_{\rm eff}^{2}} - \frac{1}{4\pi M_{W}} + \sum_{i=1}^{N_{f}} \left( \frac{1}{16\pi M_{q_{i}}}+\frac{1}{16\pi M_{\tilde q_{i}}} \right)
\end{equation}
where $e_{\rm eff}=g_{\rm eff}/\sqrt{2\pi R}$, $M_{W}=M_{W}(0)$, $M_{q_{i}}=M_{q_{i}}(0)$ and $M_{\tilde q_{i}}=M_{\tilde q_{i}}(0)$ are the gauge coupling and the masses of $W$ boson and quarks in three dimensions.
This matches with the first-principles one-loop computations performed in \cite{DTV} after taking into account the normalisation of the coupling parameter.

%%%%%%%%%%%%%%%%%%%%%%%%%%%%%%

\subsection{Instanton corrections}

\paragraph{}

For the non-perturbative contributions $\omega^{\rm NP}(\zeta)$, as we have already explained in chapter \ref{ch: pure}, the expression contains exponentially suppressed factors, and they can be readily evaluated using the saddle-point approximation:
\begin{equation}
\omega^{\rm NP}(\zeta) = \sum_{\gamma'\in\tilde\Gamma} \Omega(\gamma',a) \, \omega_{\gamma'}^{\rm NP}(\zeta)
\,,
\end{equation}
where all terms depend on the degeneracies of dyon states, $\Omega(\gamma',a)$; assuming that each flavour is massive, one has $\Omega(\gamma',a)=1$ for all dyons (however, this is not so if the $SO(2N_{f})$ flavour symmetry is not completely broken: in such case, we can still formally sum over all flavour charges setting all $\Omega(\gamma',a)=1$ for convenience).
Each dyon produces a correction:
\begin{equation}
\label{symplectic form dyon matter}
\begin{aligned}
\omega_{\gamma'}^{\rm NP}(\zeta) & =
-\frac{1}{4\pi^{2}R}
\frac{d\mathcal{X}_{\gamma'}^{(0)}(\zeta)}{\mathcal{X}_{\gamma'}^{(0)}(\zeta)}
\wedge
\left(
\frac{1}{2\pi i}\int_{l_{\gamma'}} \frac{d\zeta'}{\zeta'} \frac{\zeta'+\zeta}{\zeta'-\zeta}
\sum_{k=1}^{+\infty} \left( \sigma(\gamma')\mu_{\gamma'}(\zeta')\mathcal{X}^{(0)}_{\gamma'}(\zeta') \right)^{k}
\frac{d\mathcal{X}_{\gamma'}^{(0)}(\zeta')}{\mathcal{X}_{\gamma'}^{(0)}(\zeta')}
\right)
\\
& \approx
-\frac{1}{4\pi^{2}R}
\frac{d\mathcal{X}_{\gamma'}^{\rm sf}(\zeta)}{\mathcal{X}_{\gamma'}^{\rm sf}(\zeta)}
\wedge
\left(
\frac{1}{2\pi i}\int_{l_{\gamma'}} \frac{d\zeta'}{\zeta'} \frac{\zeta'+\zeta}{\zeta'-\zeta}
\sum_{k=1}^{+\infty} \left( \sigma(\gamma')\mu_{\gamma'}(\zeta')\mathcal{X}^{(0)}_{\gamma'}(\zeta') \right)^{k}
\frac{d\mathcal{X}_{\gamma'}^{\rm sf}(\zeta')}{\mathcal{X}_{\gamma'}^{\rm sf}(\zeta')}
\right)
\,.
\end{aligned}
\end{equation}
We should also note that along each integration contour $l_{\gamma'}$, the zeroth order Darboux coordinate $\mu_{\gamma}(\zeta')\mathcal{X}^{(0)}_{\gamma}(\zeta')=\mu_{\gamma}(\zeta')\mathcal{X}^{\rm sf}_{\gamma}(\zeta')(\mathcal{D}(\zeta'))^{\gamma_{m}'}$ is proportional to a suppressed exponential factor ensuring convergence of the integral since the infinite sum is a geometric series.
As has been shown in chapter \ref{ch: pure}, at weak coupling, the saddle point of this integral is located at $\zeta'=-Z_{\gamma}/|Z_{\gamma}|\approx-i\sign\gamma_{m}$.
Upon evaluating the Gaussian fluctuation integral around this point, the leading expression for $\omega_{\gamma'}^{\rm NP}(\zeta)$ is given by
\begin{align}
\label{symplectic form dyon matter 2}
\omega_{\gamma'}^{\rm NP}(\zeta) & =
\sum_{k=1}^{+\infty} \omega_{\gamma',k}^{\rm NP}(\zeta)
\,,
\\
\omega_{\gamma',k}^{\rm NP}(\zeta) & =
\mathcal{J}_{\gamma',k} \,
\frac{d\mathcal{X}_{\gamma'}^{\rm sf}(\zeta)}{\mathcal{X}_{\gamma'}^{\rm sf}(\zeta)}
\wedge
\left(
|Z_{\gamma'}| \left( \frac{dZ_{\gamma'}}{Z_{\gamma'}}-\frac{d\bar{Z}_{\gamma'}}{\bar{Z}_{\gamma'}} \right)
-\left( \frac{dZ_{\gamma'}}{\zeta}-\zeta d\bar{Z}_{\gamma'} \right)
\right)
\,,
\\
\mathcal{J}_{\gamma',k} & =
-\frac{1}{16\pi^{2}i} \left( \mathcal{D}(-i) \right)^{k |\gamma_{m}'|} \frac{1}{\sqrt{kR |Z_{\gamma'}|}}
\exp\left( -2\pi kR |Z_{\gamma'}|+ik\left( \theta_{\gamma'}+\sum_{i=1}^{N_{f}}s_{i}'\psi_{i} \right) \right)
\,,
\end{align}
where we have used the identity $\mathcal{D}(\mp i)^{\pm 1}=\mathcal{D}(-i)$ for the one-loop factor (\ref{D factor W matter}, \ref{D factor q matter}, \ref{D factor qt matter}), and we also shift $\theta_{m}\to\theta_{m}+\theta_{e}\Theta_{\rm eff}/\pi$ in $\theta_{\gamma}$ (this is required to correctly define $\theta_{m}$ near the singularity at infinity), so that $\sigma(\gamma')$ is absorbed by the global definition of $\theta_{m}$ (cf.\ eq.\ (4.16b) in \cite{GMN}).
At the saddle point, $\mathcal{D}(-i)=\mathcal{D}_{W}(-i)\mathcal{D}_{q}(-i)\mathcal{D}_{\tilde q}(-i)$ can be evaluated analogously to (\ref{D factor pure 3}) using symmetries of the integrals:
\begin{equation}
\label{D factor W matter 2}
\log\mathcal{D}_{W}(-i) =
\frac{2}{\pi} \int_{0}^{+\infty} \frac{dt}{\cosh t}
\left(
\log\left( 1-e^{-4\pi R |a|\cosh t+2i\theta_{e}} \right) +
\log\left( 1-e^{-4\pi R |a|\cosh t-2i\theta_{e}} \right)
\right)
\,,
\end{equation}
\begin{equation}
\label{D factor q matter 2}
\begin{aligned}
\log\mathcal{D}_{q}(-i) =
-\frac{1}{2\pi} \sum_{i=1}^{N_{f}} \int_{0}^{+\infty} \frac{dt}{\cosh t}
\left(
\log\left( 1-e^{-2\pi R |a+m_{i}|\cosh t+i(\theta_{e}+\psi_{i})} \right) +
\right.
\\
\left.
\log\left( 1-e^{-2\pi R |a+m_{i}|\cosh t-i(\theta_{e}+\psi_{i})} \right)
\right)
\,,
\end{aligned}
\end{equation}
\begin{equation}
\label{D factor qt matter 2}
\begin{aligned}
\log\mathcal{D}_{\tilde q}(-i) =
-\frac{1}{2\pi} \sum_{i=1}^{N_{f}} \int_{0}^{+\infty} \frac{dt}{\cosh t}
\left(
\log\left( 1-e^{-2\pi R |a-m_{i}|\cosh t+i(\theta_{e}-\psi_{i})} \right) +
\right.
\\
\left.
\log\left( 1-e^{-2\pi R |a-m_{i}|\cosh t-i(\theta_{e}-\psi_{i})} \right)
\right)
\,.
\end{aligned}
\end{equation}
In deriving the quark components, we have also used the residual $U(1)^{N_{f}}$ flavour symmetries to set $\im(m_{i}/a)=0$, so that that the quark terms are manifestly real.
In chapter \ref{ch: pure}, it was shown that (\ref{D factor W matter 2}) corresponds to the ratio of one-loop determinants for non-zero mode fluctuations around a monopole in the $SU(2)$ theory.
We will also perform similar semiclassical computations to demonstrate that (\ref{D factor q matter 2}) and (\ref{D factor qt matter 2}) correspond to the fundamental hypermultiplets non-zero mode fluctuations around a monopole.

\paragraph{}

Now, we focus on the leading one-instanton correction to the moduli space, i.e., we set $\gamma_{m}'=1$, $k=1$ in (\ref{symplectic form dyon matter}, \ref{symplectic form dyon matter 2}).
As in the case without flavours, the moduli space metric can be extracted by finding the $\zeta$-independent part of $\omega^{\rm NP}(\zeta)$ for dyons with magnetic charge 1: the corresponding symplectic form is formally given by (\ref{symplectic form 3 instanton pure}) with each term multiplied by the degeneracy of the dyon state, $\Omega(\gamma',a)$.
Using our results above, we can write out the dominant component of the instanton metric:
\begin{equation}
\label{metric instanton matter}
g_{a\bar a}^{\rm inst} =
\frac{R^{1/2}}{8\pi} \sum_{\gamma'=(\gamma_{e}',1,\vec s\,')\in\Gamma} \Omega(\gamma',a) \, \frac{\mathcal{D}(-i) |Z_{\gamma'}|^{3/2}}{|a|^{2}}
\exp\left( -2\pi R |Z_{\gamma'}|+i\left( \theta_{\gamma'}+\sum_{i=1}^{N_{f}}s_{i}'\psi_{i} \right) \right)
\,.
\end{equation}
All other metric components can also be readily extracted from this formula using the fact that the metric is K\"ahler.
However, in order to find the leading order of the four-fermion action, it is sufficient to consider only one component, (\ref{metric instanton matter}), which we can expand at weak coupling as
\begin{align}
\label{metric instanton matter 2}
g_{a\bar a}^{\rm inst} & \simeq
\frac{2\sqrt{2\pi R}}{g_{\rm eff}^{3}}
\sum_{\gamma'=(\gamma_{e}',1,\vec s\,')\in\Gamma} \Omega(\gamma',a) \, \frac{\mathcal{D}(-i)}{\sqrt{|a|}}
\exp\left( -S_{\rm mon}-S_{\varphi}^{(\gamma_{e}',\vec s\,')} \right)
\,,
\\
\label{action monopole matter}
S_{\rm mon} & =
\frac{16\pi^{2}R}{g_{\rm eff}^{2}}|a|-i\theta_{m}
\,,
\\
\label{action angular matter}
S_{\varphi}^{(\gamma_{e}',\vec s\,')} & =
\frac{g_{\rm eff}^{2}R}{8|a|}\left( \left( \gamma_{e}'+\frac{\Theta_{\rm eff}}{\pi} \right)|a|+\sum_{i=1}^{N_{f}}s_{i}'|m_{i}| \right)^{2}-i\left( \left( \gamma_{e}'+\frac{\Theta_{\rm eff}}{\pi} \right)\theta_{e}+\sum_{i=1}^{N_{f}}s_{i}'\psi_{i} \right)
\,.
\end{align}
The term $S_{\rm mon}$ here is the action of a magnetic monopole dimensionally reduced on $\mathbb{R}^{3}\times S^{1}$.
$S_{\varphi}^{(\gamma_{e}',\vec s\,')}$ is the leading contribution from the electric charges.
The $\Theta_{\rm eff}$ piece in the second term of $S_{\varphi}^{(\gamma_{e}',\vec s\,')}$ should be introduced to account for the shift $\theta_{m}\to\theta_{m}+\theta_{e}\Theta_{\rm eff}/\pi$, as discussed above.

\paragraph{}

We can calculate the two-instanton correction analogously.
It comes from two sources: when we set $\gamma_{m}'=2$, $k=1$ and $\gamma_{m}'=1$, $k=2$ in (\ref{symplectic form dyon matter}, \ref{symplectic form dyon matter 2}).
Writing out both types of corrections, we obtain the dominant component of the two-instanton metric:
\begin{equation}
\label{metric two-instanton matter}
\begin{aligned}
g_{a\bar a}^{\rm inst(2)} & =
\frac{R^{1/2}}{8\pi} \sum_{\gamma'=(\gamma_{e}',2,\vec s\,')\in\Gamma} \Omega(\gamma',a) \, \frac{\left( \mathcal{D}(-i) \right)^{2} |Z_{\gamma'}|^{3/2}}{|a|^{2}}
\exp\left( -2\pi R |Z_{\gamma'}|+i\left( \theta_{\gamma'}+\sum_{i=1}^{N_{f}}s_{i}'\psi_{i} \right) \right)
\\
& +
\frac{R^{1/2}}{8\pi} \sum_{\gamma'=(\gamma_{e}',1,\vec s\,')\in\Gamma} \Omega(\gamma',a) \, \frac{\left( \mathcal{D}(-i) \right)^{2} |Z_{\gamma'}|^{3/2}}{|a|^{2} \, 2^{1/2}}
\exp\left( -4\pi R |Z_{\gamma'}|+2i\left( \theta_{\gamma'}+\sum_{i=1}^{N_{f}}s_{i}'\psi_{i} \right) \right)
\,.
\end{aligned}
\end{equation}
Expanding the central charges to the leading order, we get
\begin{equation}
\label{metric two-instanton matter 2}
\begin{aligned}
g_{a\bar a}^{\rm inst(2)} & \approx
\frac{8\sqrt{\pi R}}{g_{\rm eff}^{3}}
\sum_{\gamma'=(\gamma_{e}',2,\vec s\,')\in\Gamma} \Omega(\gamma',a) \, \frac{\left( \mathcal{D}(-i) \right)^{2}}{\sqrt{|a|}}
\exp\left( -2S_{\rm mon}-S_{\varphi}^{(\gamma_{e}',\vec s\,',2)} \right)
\\
& +
\frac{2\sqrt{\pi R}}{g_{\rm eff}^{3}}
\sum_{\gamma'=(\gamma_{e}',1,\vec s\,')\in\Gamma} \Omega(\gamma',a) \, \frac{\left( \mathcal{D}(-i) \right)^{2}}{\sqrt{|a|}}
\exp\left( -2S_{\rm mon}-2S_{\varphi}^{(\gamma_{e}',\vec s\,')} \right)
\,,
\end{aligned}
\end{equation}
\begin{equation}
\label{action angular matter two}
S_{\varphi}^{(\gamma_{e}',\vec s\,',2)} =
\frac{g_{\rm eff}^{2}R}{16|a|}\left( \left( \gamma_{e}'+\frac{\Theta_{\rm eff}}{\pi} \right)|a|+\sum_{i=1}^{N_{f}}s_{i}'|m_{i}| \right)^{2}-i\left( \left( \gamma_{e}'+\frac{\Theta_{\rm eff}}{\pi} \right)\theta_{e}+\sum_{i=1}^{N_{f}}s_{i}'\psi_{i} \right)
\,,
\end{equation}
where the summations are over integer electric charges and flavour charges discussed above.

\paragraph{}

Finally, let us find the leading order of the four-fermion vertex in theories with flavours.
We can follow our approach for the pure theory in chapter \ref{ch: pure}: the low-energy effective action is given by (\ref{action effective 3D}), where the semiflat metric (\ref{metric semiflat matter}) gets corrections from the instanton metric (\ref{metric instanton matter 2}); the leading-order expression for the Riemann tensor components is (\ref{Riemann pure}).
Then, after identifying the bosonic and fermionic fields in three dimensions (\ref{action bosonic 3D matter}, \ref{action fermionic 3D matter}) with those in the effective action (\ref{action effective 3D}) by scaling the fields appropriately, we obtain the four-fermion action:
\begin{equation}
\label{action 4 fermions matter}
S_{\rm 4F} =
\frac{2^{21/2}\pi^{9/2}R^{5/2}}{|a|^{1/2}g_{\rm eff}^{7}} \, 
\mathcal{D}(-i) \exp\left( -S_{\rm mon} \right)
\sum_{\gamma'=(\gamma_{e}',1,\vec s\,')\in\Gamma} \Omega(\gamma',a) \exp\left( -S_{\varphi}^{(\gamma_{e}',\vec s\,')} \right)
\int d^{3}x \left( \psi\cdot\bar\psi \right) \left( \lambda\cdot\bar\lambda \right)
\,.
\end{equation}
We shall verify this result via a semiclassical calculation including flavour hypermultiplets.

%%%%%%%%%%%%%%%%%%%%%%%%%%%%%%%%%%%%%%%%%%%%%%%%%%%%%%%%%%%%%

\section{Semiclassical instanton calculation with matter}

\paragraph{}

In this section, we will perform a first-principles computation for the dyonic contributions to the low-energy effective action in $\mathbb{R}\times S^{1}$ in the theory with $N_{f}$ fundamental hypermultiplets.
We focus on the states with $\gamma_{m}=1$ with winding number $k=1$, arbitrary electric charges $\gamma_{e}\in\mathbb{Z}$ and all permitted flavour charges $\vec s$, which preserve four out of eight supersymmetries; we find their contributions to the four-fermion correlation function, extending our approach in chapter \ref{ch: pure}.
We will reuse many of our previous results, so we will concentrate on highlighting the essential modifications due to the fundamental hypermultiplets.

\paragraph{}

The main object of interest is the four-fermion correlation function of the form (\ref{correlator 4 fermions pure}) where we are using our conventions from chapter \ref{ch: pure}.
We want to evaluate it in the monopole background.
Let us now consider the zero modes in the monopole background.
The Callias index theorem \cite{Callias} tells us there are $4\gamma_{m}$ real bosonic zero modes for monopole configuration of charge $\gamma_{m}$.
In our case, therefore, there are four bosonic zero modes for the monopole: $X^{1}$, $X^2$, $X^{3}$ parametrising its centre in $\mathbb{R}^{3}$ and a global $U(1)$ angle $\varphi$; then, the bosonic moduli space is $\mathbb{R}^{3}\times S^{1}_{\varphi}$. For general $\gamma_{m}$, the remaining $4(\gamma_{m}-1)$ bosonic zero modes parametrise the relative moduli space.
As required by supersymmetry, there are $4\gamma_{m}$ adjoint fermionic zero modes for theories with eight supercharges, in particular, four of them are generated by the action of the four broken supersymmetries.
We denote the corresponding collective fermionic coordinates $\xi^A_{1,2}$; these fermionic zero modes are protected from lifting by the supersymmetry.
The long-distance limit $|y-X|\gg |a|^{-1}$ of $\rho^{A}_{\alpha}$ is given by
\begin{equation}
\label{propagator fermion 3D matter}
\rho_{\alpha}^{{\rm (LD)} \, A}(y) = 16\pi \, S_{\rm F}(y-X)_{\alpha}^{\ \beta} \, \xi_{\beta}^{\ A}
\end{equation}
where $S_F(x)=\gamma^{\mu} x_{\mu}/(4\pi |x|^{2})$ is the three-dimensional Dirac propagator.
The remaining $4(\gamma_{m}-1)$ fermionic zero modes are the supersymmetric partners of the $4(\gamma_{m}-1)$ bosonic coordinates on the relative moduli space.

\paragraph{}

When we include additional $N_{f}$ fundamental hypermultiplets $q_{i},\tilde q_{i}$ with masses $m_{i}$ and $-m_{i}$, they can also contribute additional zero modes in the monopole background, therefore, it is necessary to perform an index computation to count their numbers.
To do so, we follow \cite{Weinberg, Kaul, DTV} to define the following four-dimensional fluctuation operators for the massive fundamental hypermultiplets in the monopole background:
\begin{align}
\label{Delta plus matter}
\Delta_{+,q_{i}}(m_{i}) & = D^{2}_{i}+|a+m_{i}|^{2}
\,,
\\
\label{Delta minus matter}
\Delta_{-,q_{i}}(m_{i}) & = D^{2}_{i}+|a+m_{i}|^{2}+2\epsilon_{ijk}\sigma_{i}F_{jk}^{\rm mon}
\,,
\quad
\{i,j\} \subset \{1,2,3\}
\,.
\end{align}
The three-dimensional covariant derivative $\vec{D}=\vec{\partial}+ i\vec{A}^{\rm Mon.}$ is with respect to background static monopole in $A_{0}=0$ gauge. 
We can define similar operators $\Delta_{\pm}^{\tilde q_{i}}(m_{i})$ for $\tilde q_{i}$ with $|a+m_{i}|\to |a-m_{i}|$.
The number of the (complex) hypermultiplet zero modes coming from $q_{i}$ and $\tilde q_{i}$ then comes from the $\mu^{2}\to 0$ limit of the regularised trace:
\begin{equation}
\label{index matter}
\mathcal{I}_{H}(m_{i}) = \tr_{i} \left( \frac{\mu^{2}}{\Delta_{-,q_{i}}(m_{i})+\mu^{2}}-\frac{\mu^{2}}{\Delta_{+,q_{i}}(m_{i})+\mu^{2}} \right) +
\tr_{i} \left( \frac{\mu^{2}}{\Delta_{-,\tilde q_{i}}(m_{i})+\mu^{2}}-\frac{\mu^{2}}{\Delta_{+,\tilde q_{i}}(m_{i})+\mu^{2}} \right)
\end{equation}
where $ \tr_{i}$ indicates summing over the flavour indices and normalisable states.
The trace in (\ref{index matter}) can be evaluated analogously following the steps in \cite{Weinberg, Weinberg Yi} for monopole of charge $\gamma_{m}$, and we get
\begin{equation}
\label{index matter 2}
\mathcal{I}_{H}(m_{i}) = \sum_{i=1}^{N_{f}} \frac{\gamma_{m}}{2} \left(
\frac{|a|+|m_{i}|}{\left( \left( |a|+|m_{i}| \right)^{2}+\mu^{2} \right)^{1/2}} +
\frac{|a|-|m_{i}|}{\left( \left( |a|-|m_{i}| \right)^{2}+\mu^{2} \right)^{1/2}}
\right)
\,,
\end{equation}
where in writing out $\mathcal{I}_{H}(m_{i})$ we have also used the fact $\im(a/m_{i})=0$.
In the limit $\mu^{2} \to 0$, we have
\begin{equation}
\mathcal{I}_{H}(m_{i}) \to \frac{\gamma_{m}N_{f}}{2} \left( \sign(|a|+|m_{i}|)+\sign(|a|-|m_{i}|) \right)
\,.
\end{equation}
At weak coupling, we set $|a|\gg |m_{i}|$, therefore, there are $2\gamma_{m}N_{f}$ additional real zero modes appearing.
As discussed in \cite{Gauntlett Harvey}, these additional hypermultiplet zero modes facilitate a natural $\mathcal{O}(\gamma_{m})$ bundle over the $\gamma_{m}$ monopole moduli space, and they are required to form bound states with the BPS dyons for them to transform under the flavour symmetry group \cite{SW2}.
In our computation of single monopole $\gamma_{m}=1$, the $\mathcal{O}(1)$ index bundle ${\rm Ind}_{1} =R^{3}\times \text{M\"ob}$, where $\text{M\"ob}$ is the M\"obius bundle over $S^{1}$ of the monopole moduli space.
This bundle is obviously flat with vanishing curvature, however, the non-trivial twisting comes from the fact that the $2\pi$ global rotation about the $S^{1}$ acts as non-trivial element of the centre of $SU(2)$ gauge group \cite{Gauntlett Harvey}.
We shall return to this point shortly in the following discussions.

\paragraph{}

Having discussed the zero modes, the semiclassical dynamics for a single monopole of mass $M=8\pi |a|/g^{2}$ can be described by supersymmetric quantum mechanics on its moduli space \cite{Gauntlett}.
The collective coordinates Lagrangian, including the hypermultiplet zero modes, takes the form \cite{Gauntlett Harvey, Gauntlett Kim Lee Yi}:
\begin{equation}
\label{Lagrangian monopole matter}
L_{\rm QM} = L_{X}+L_{\varphi}+L_{\xi}+L_{\eta}
\,.
\end{equation}
Here, $L_{X}$ and $L_{\varphi}$ are the bosonic Lagrangians introduced in (\ref{Lagrangian monopole pure}), where $\vec{X}$ is the position of the monopole in $\mathbb{R}^{3}$, and $\varphi$ is the angular position; the supersymmetric counterparts of the bosonic degrees of freedom are the adjoint fermionic collective coordinates $\xi_{\alpha}^{A}$, $\{A,\alpha\}\subset\{1,2\}$ with the free Lagrangian introduced in (\ref{Lagrangian monopole pure}).
The $2N_{f}$ real hypermultiplet collective coordinates $\eta^{i}$ are encoded in the Lagrangian
\begin{equation}
L_{\eta} = \frac{1}{2}(\eta_{i} {\mathcal{D}}_{x^{0}}\eta^{i}+m_{i}\eta_{i}\eta^{i})
\,,
\end{equation}
where $\mathcal{D}_{x^{0}}$ is the covariant derivative with respect to the connection on the index bundle, and we have also included the complex mass term.
We can now write down the long-distance form of the four-fermion correlation function: 
\begin{equation}
\label{correlator 4 fermions matter 2}
\mathcal{G}_{4}(y_{1},y_{2},y_{3},y_{4}) =
\int d\mu \, \prod_{A=1}^{2} \rho^{{\rm (LD)} \, A}_{1}(y_{2A-1})\rho^{{\rm (LD)} \, A}_{2}(y_{2A})
\,,
\end{equation}
\begin{equation}
\label{correlator measure matter}
\begin{aligned}
\int d\mu = \frac{1}{4\pi^{2}} \int d^{3}X(x^{0}) \, d\varphi(x^{0}) \, d^{4}\xi(x^{0}) \, d^{2N_{f}}\eta(x^{0}) \, \mathcal{R} \,
\exp\left( -\int_{0}^{2\pi R} dx^{0}L_{\rm QM} \right)
\\
\exp\left( -\frac{16\pi^{2}R |a|}{g^{2}}+i\theta_{m} \right)
\,,
\end{aligned}
\end{equation}
where the long-distance fermionic zero modes $\rho^{{\rm (LD)} \, A}_{1,2}$ are as given in (\ref{propagator fermion 3D matter}); as in the pure theory, the prefactor of $1/(2\pi)^{2}$ is the Jacobian for changing the variables from the initial measures to the four collective coordinates.
The integration measure consists of bosonic collective coordinates $d^{3}Xd\varphi$, fermionic collective coordinates $d^{4}\xi$, and integration over the $2N_{f}$ hypermultiplet collective coordinates, denoted as $d^{2N_{f}}\eta(x^{0})$ . 
The one-loop ratio of determinants $\mathcal{R}$ gets contributions from non-zero mode fluctuations of both vector multiplets and fundamental hypermultiplets, and will be evaluated below. 
The factors involving $L_{\rm QM}$ and $S_{\rm mon}$ in (\ref{Lagrangian monopole matter}) correspond to the monopole action after compactifying along $x^{0}$.

\paragraph{}

Since the fermionic factors $\rho_{\alpha}^{{\rm (LD)} \, A}(y)$ from (\ref{propagator fermion 3D matter}) only depend on the spatial coordinates and the fermionic zero modes, we can split the four-fermion correlation function as
\begin{equation}
\mathcal{G}_{4} = \mathcal{G}_{4}^{\rm COM} \, \mathcal{Z} \, \mathcal{R}
\end{equation}
where the first two factors, for the centre of mass and for all other zero modes, are given by
\begin{equation}
\begin{aligned}
\mathcal{G}_{4}^{\rm COM}(y_{1},y_{2},y_{3},y_{4}) =
\int \frac{d^{3}X(x^{0}) \, d^{4}\xi(x^{0})}{(2\pi)^{3/2}} \, \prod_{A=1}^{2} \rho^{{\rm (LD)} \, A}_{1}(y_{2A-1})\rho^{{\rm (LD)} \, A}_{2}(y_{2A})
\\
\exp\left( -\int_{0}^{2\pi R} dx^{0} \left(L_{X}+L_{\xi} \right)-S_{\rm mon} \right)
\,,
\end{aligned}
\end{equation}
\begin{equation}
\mathcal{Z} = \int \frac{d\varphi(x^{0})}{(2\pi)^{1/2}} \, d^{2N_{f}}\eta(x^{0})
\exp\left( -\int_{0}^{2\pi R} dx^{0} \left( L_{\varphi}+L_{\eta} \right) \right)
\,.
\end{equation}
In calculating $\mathcal{G}^{\rm COM}_{4}$, we can follow the same method as in chapter \ref{ch: pure}: impose usual periodic boundary condition $\vec{X}(x^{0})=\vec{X}(x^{0}+2\pi R)$ and $\xi_{\alpha}^{A}(x^{0})=\xi_{\alpha}^{A}(x^{0}+2\pi R)$, then for the spatial bosonic and fermionic collective coordinate integration measures, integrating over the classical paths, which constitute the dominant contribution, we have (see chapter \ref{ch: pure} for more details)
\begin{equation}
\label{measures}
\int d^{3} X(x^{0}) \, d^{4}\xi(x^{0})
\exp\left( -\int^{2\pi R}_{0} dx^{0} \left( L_{X}+L_{\xi} \right) \right)
=
\int d^{3}X \, d^{4}\xi \left( 2\pi\sqrt{\frac{R}{M}} \right)
\,,
\end{equation}
where we have also used $\vec{X}$ and $\xi_\alpha^{A}$ in the integrations to denote the classical values of the bosonic and fermionic zero modes, respectively.

\paragraph{}

To evaluate $\mathcal{Z}$, it is convenient to recall an alternative interpretation of the four-fermion correlator $\mathcal{G}_{4}(y_{1},y_{2},y_{4},y_{4})$ (\ref{correlator 4 fermions pure}) in the compactfied theory \cite{Dorey 2000, DHK}.
That is, we can instead work in the Hamiltonian formalism and regard it as a generalisation of the Witten index:
\begin{equation}
\left\langle \prod_{A=1}^{2}\rho^{\ A}_{1}(y_{2A-1})\rho^{\ A}_{2}(y_{2A}) \right\rangle =
\tr\left(
\prod_{A=1}^{2}\rho^{\ A}_{1}(y_{2A-1})\rho^{\ A}_{2}(y_{2A})
(-1)^{F} \exp\left( -2\pi R \, H_{\rm QM}-S_{\rm mon} \right)
\right)
\,.
\end{equation}
Here, $H_{\rm QM}$ is the Hamiltonian for the collective coordinates Lagrangian $L_{\rm QM}$, the trace sums over the BPS states, which will be discussed immediately below, each contributes with the exponential suppression factor $\exp(-S_{\rm mon})$.
In particular, the interpretation above allows us to re-express the factor $\mathcal{Z}$ as
\begin{equation}
\label{Z factor matter}
\mathcal{Z} = \tr_{i} \left( (-1)^{F} \mathbb{P}_{\gamma_{e}} \exp\left( -2\pi R\left( H_{\varphi}+H_{\eta} \right) \right) \right)
\end{equation}
where $\mathbb{P}_{\gamma_{e}}$ is a projector which depends on the electric charge of the state, and the subscript $i$ indicates summing over the representations of the flavour group.
In the semiclassical quantisation of monopole quantum mechanics, the wave function of any BPS states in the one-monopole moduli space can be decomposed schematically into a tensor product $|\Psi_{\rm BPS} \rangle =f(\vec{X},\varphi)|\xi\rangle \otimes|\eta_{i}\rangle$ where $f(\vec{X},\varphi)$ is the bosonic part, $|\xi\rangle$ comes from the superpartners, the remaining $|\eta_{i}\rangle$ corresponds to transformations under the flavour group ($SO(2N_{f})$ in the massless case, $U(1)^{N_{f}}$ in the massive case).

\paragraph{}

Recall that without flavours, a $2\pi$ rotation around the $S^{1}$ direction in the moduli space with generator $Q$ leaves the monopole wave function invariant, i.e., $e^{2\pi iQ}=1$.
This gives rise to the whole tower of quantised electric charges $\gamma_{e}\in\mathbb{Z}$ \cite{Tomboulis Woo}, which we can identify with the quantised conjugate momentum: $P_{\varphi}=M\dot\varphi/|a|^{2}=\gamma_{e}$.
The corresponding Hamiltonian is $H_\varphi=|a|^{2}\gamma_{e}^{2}/2M$, and the trace in (\ref{Z factor matter}) means summing over the BPS dyons of the theory.
In the presence of flavours, there is a key modification in the above discussion \cite{SW2, Harvey}: now, the $2\pi$ rotation around the $S^{1}$ direction does not give identity, but a topologically non-trivial gauge transformation, whose eigenvalue is given by $e^{i\pi Q}=e^{i\Theta}(-1)^{H}$~\footnote{
The factor of 2 for the generator $Q$ is to ensure that the fundamental quarks have charge $\pm 1$, and $W$ bosons have charge $\pm 2$.
}, where $(-1)^{H}$ is the centre of the gauge group $SU(2)$, and $\Theta$ is the Witten angle.
If we set the electric charge $Q=\gamma_{e}+\Theta/\pi$ with $\gamma_{e}\in\mathbb{Z}$, this means that the states with odd $\gamma_{e}$ have odd chirality operator $(-1)^{H}$, and the states with even $\gamma_{e}$ have even $(-1)^{H}$.
This also implies that when $m_{i}=0$, $(-1)^{H}$ acts as an analogue of $\gamma^{5}$ in the $SO(2N_{f})$ Clifford Algebra \cite{Harvey}, and we can form the projection operators $\mathbb{P}_{\gamma_{e}}=(1\pm(-1)^{H})/2$ (such as the one inserted into (\ref{Z factor matter})) to decompose the original reducible $2^{N_{f}}$-dimensional spinor representation into two irreducible $2^{N_{f}-1}$-dimensional representations with definite electric charge $\gamma_{e}$.
Similar analysis can also be done for $m_{i}\ne 0$, which decomposes the wave functions into components carrying definite charge under each $U(1)$ of the residual $U(1)^{N_{f}}$ group, whose value depends on $\gamma_{e}$.
The discussion above leads to the following result:
\begin{equation}
\label{instanton measure rotation matter}
\mathcal{Z} = \sum_{\gamma=(\gamma_{e},1,\vec s\,)\in\Gamma} \Omega(\gamma,a)
\exp\left( -\frac{\pi R}{M}\left (\left( \gamma_{e}+\frac{\Theta}{\pi} \right)|a|+\sum_{i=1}^{N_{f}}s_{i}{|m_{i}|} \right)^{2}+i\left( \gamma_{e}+\frac{\Theta}{\pi} \right)\theta_{e}+i\sum_{i=1}^{N_{f}}s_{i}\psi_{i} \right)
\,.
\end{equation}
In above, the degenaracy factor $\Omega((\gamma_{e},1),a)$ comes from tracing over the factor $e^{-2\pi RH_{\eta}}$ over different $|\eta^{i}\rangle$, and it should be identified with $\Omega(\gamma,a)$ appearing in (\ref{metric instanton matter 2}).
We have also included the shift of electric charges due to the mass terms for $m_{i}$ and $\tilde m_{i}$: this can be motivated from our earlier choice $\im(a/m_{i})=0$, and the $S^{1}$ rotation is now a combination of the global $U(1)$ within the gauge group $SU(2)$ and the residual flavour group $U(1)^{N_{f}}$.
A further phase $\gamma_{e}\theta_{e}+\sum_{i=1}^{N_{f}} s_{i} \psi_{i}$ in the classical action appears from the surface terms. 
In summary, we note that this matches the corresponding sum appearing in the GMN prediction (\ref{metric instanton matter 2}) up to replacing the bare coupling and the vacuum angle by their one-loop renormalised counterparts.

\paragraph{}

To obtain the semiclassical integration measure, it is also necessary to evaluate the one-loop ratio of determinants $\mathcal{R}$ accounting for the non-zero mode fluctuations in the monopole background.
In the case of a theory with $N_{f}$ fundamental hypermultiplets on $\mathbb{R}^{3}\times S^{1}$, we can decompose $\mathcal{R}$ into three components:
\begin{equation}
\log\mathcal{R} = \log\mathcal{R}_{W}+\log\mathcal{R}_{q}+\log\mathcal{R}_{\tilde q}
\end{equation}
where $\log\mathcal{R}_{W}$ is the $W$ boson contribution, $\log\mathcal{R}_{q}$ and $\log\mathcal{R}_{\tilde q}$ are the additional contributions corresponding to hypermultiplets.
In chapter \ref{ch: pure}, $\mathcal{R}_{W}$ was explicitly computed using Kaul's earlier result for the density of states of the fluctuations in the monopole background \cite{Kaul}; we will follow similar steps to calculate $\mathcal{R}_{q}$ and $\mathcal{R}_{\tilde q}$, which depend on flavour masses.
Let us begin with $\mathcal{R}_{q}$; the computation for $\mathcal{R}_{\tilde q}$ can be carried out analogously.
The spatial fluctuations of the hypermultiplets can be encoded by the non-zero eigenfunctions of the fluctuation operators $\Delta^{q_{i}}_{\pm}$ (\ref{Delta plus matter}) and (\ref{Delta minus matter}).
To further include the fluctuations along $S^{1}$ ($x^{0}\sim x^{0}+2\pi R$), we define:
\begin{equation}
\label{Delta matter}
{\mathbb{D}}_{\pm,q_{i}} = \Delta_{\pm,q_{i}}+\left( \frac{\partial}{\partial x^{0}} \right)^{2}
\,,
\end{equation}
where the derivative with respect to $x^{0}$ corresponds to the fluctuation modes on $S^{1}$, and we initially set $\theta_{e}=\psi_{i}=0$ to simplify the discussion. 
For $N_{f}$ different flavours, the one-loop factor associated with the $q_{i}$ fluctuations is given by the following ratio:
\begin{equation}
\label{loop factor matter}
\mathcal{R}_{q} = \prod_{i=1}^{N_{f}} \left( \frac{\det\mathbb{D}_{+,q_{i}}}{\det'\mathbb{D}_{-,q_{i}}} \right)^{-1/2}
\,.
\end{equation}

\paragraph{}

To evaluate (\ref{loop factor matter}), we can decompose each eigenfunction of $\mathbb{D}_{\pm,q_{i}}$ as $\Phi_{\pm}(\vec x,x^{0})=\phi_{\pm}(\vec x)f_{\pm}(x^{0})$ by $S^{1}$ translational invariance, where $\phi_{\pm}(\vec x)$ are eigenfunctions of $\Delta_{\pm,q_{i}}$ with eigenvalues $\lambda_{\pm}^{2}$, and $f_{\pm}(x^{0})$ has the form $f_{\pm}(x^{0})\sim e^{i\varpi_{\pm} x^{0}}$.
As we discussed in chapter \ref{ch: pure}, the densities of states for the bosonic and for the fermionic eigenvalues are not equal in the monopole background \cite{Kaul}.
This leads to non-trivial quantum corrections to the monopole mass.
In our case, the operators $\mathbb{D}_{\pm,q_{i}}$ on $\mathbb{R}^{3}\times S^{1}$ also inherit this subtle effect from $\Delta_{\pm,q_{i}}$, giving a non-trivial formula for $\mathcal{R}_{q}$.

\paragraph{}

We can rewrite $\mathcal{R}_{q}$ (\ref{loop factor matter}) as the following integral expression:
\begin{equation}
\label{loop factor matter 2}
\begin{aligned}
\mathcal{R}_{q} & = (2\pi R )^{N_{f}/2} \, \prod_{i=1}^{N_{f}} \, \exp\left(
\frac{1}{2}\tr_{\vec x}\log\det{}_{x^{0}}\mathbb{D}_{-,q_{i}} -
\frac{1}{2}\tr_{\vec x}\log\det{}_{x^{0}}\mathbb{D}_{+,q_{i}}
\right)
\\
& = (2\pi R )^{N_{f}/2} \exp\left(
\sum_{i=1}^{N_{f}} \int_{0}^{+\infty} d\lambda^{2}\delta\rho_{q_{i}}(\lambda)
\log\mathcal{K}_{q_{i}}(\lambda,2\pi R)
\right)
\end{aligned}
\end{equation}
where the integration kernel for $\theta_{e}=0$ and $\theta_{i}=0$ is given by
\begin{equation}
\mathcal{K}_{q_{i}}(\lambda,2\pi R) = \det{}_{x^{0}}\left( \left( \frac{\partial}{\partial x^{0}} \right)^{2}+\lambda^{2} \right)
\,,
\end{equation}
where we have used the identity $\log\det M= \tr\log M$.
The quantity $\delta\rho_{q_{i}}(\lambda)=\rho_{+,q_{i}}(\lambda)-\rho_{-,q_{i}}(\lambda)$ is the difference between the densities of eigenvalues of the operators $\Delta_{+,q_{i}}$ and $\Delta_{-,q_{i}}$.
This can be worked out using the index theorem following \cite{Kaul}, yielding
\begin{equation}
\label{states density difference matter}
d\lambda^{2}\delta\rho_{q_{i}}(\lambda) =
-\frac{|a+m_{i}|}{2\pi\lambda^{2}\sqrt{\lambda^{2}-|a+m_{i}|^{2}}} \, \theta(\lambda^{2}-|a+m_{i}|^{2}) \, d\lambda^{2}
\,.
\end{equation}
As noted in chapter \ref{ch: pure}, the integration kernel $\mathcal{K}_{q_{i}}(\lambda,2\pi R)$ is the partition function of harmonic oscillator with frequency $\varpi_{i}=\lambda$ at inverse temperature $\beta=2\pi R$.
$\theta_{e}$ and $\psi_{i}$, which are non-vanishing components along the compactified direction, can be minimally coupled in the operators $\mathbb{D}_{\pm,q_{i}}$ given in (\ref{Delta matter}):
\begin{equation}
\frac{\partial}{\partial x_{0}} \to
\frac{\partial}{\partial x_{0}} + \gamma_{e} \, \frac{\theta_{e}+\psi_{i}}{2\pi R}
\,,
\end{equation}
introducing a chemical potential to the harmonic oscillator and shifting its frequencies to the complex values $\varpi_{i}=\lambda\pm i(\theta_{e}+\psi_{i})/2\pi R$.
Summing the contributions, we find that $\mathcal{K}_{q_{i}}^{2}=\mathcal{K}_{+,q_{i}}\mathcal{K}_{-,q_{i}}$ is determined by
\begin{equation}
\label{kernel matter}
\mathcal{K}_{\pm,q_{i}}(\lambda,\theta_{e},2\pi R)^{-1} = \frac{\exp(-\pi R\lambda\pm i(\theta_{e}+\psi_{i})/2)}{1-\exp(-2\pi R\lambda\pm i(\theta_{e}+\psi_{i}))}
\,.
\end{equation}

\paragraph{}

Substituting (\ref{kernel matter}) and (\ref{states density difference matter}) into (\ref{loop factor matter 2}) and setting $\lambda=\left|a+{m_{i}}\right|\cosh t$, we re-express the one-loop factor for quarks $\mathcal{R}_{q}$ as
\begin{equation}
\label{loop factor matter 3}
\begin{aligned}
\log\mathcal{R}_{q} & =
\sum_{i=1}^{N_{f}} \left(
-R |a+m_{i}|\arcosh\frac{\Lambda_{\rm UV}}{|a+m_{i}|}
+\frac{1}{2}\log(2\pi R)
\right.
\\
& \left.
-\frac{1}{2\pi} \int_{0}^{+\infty} \frac{dt}{\cosh t}
\log\left( 1-e^{-2\pi R |a+m_{i}|\cosh t+i(\theta_{e}+\psi_{i})} \right)
\right.
\\
& \left.
-\frac{1}{2\pi} \int_{0}^{+\infty} \frac{dt}{\cosh t}
\log\left( 1-e^{-2\pi R |a+m_{i}|\cosh t-i(\theta_{e}+\psi_{i})} \right)
\right)
\end{aligned}
\end{equation}
where we have introduced the ultraviolet cutoff $\Lambda_{\rm UV}$. 
We immediately recognise that the integral here is precisely the same as the integral in $\log\mathcal{D}_{q}(-i)$ (\ref{D factor q matter 2}) in the semiclassical expansion of the GMN equation.
The terms depending on $\Lambda_{\rm UV}$ should be cancelled by the corresponding counter-term for the coupling constant.
We can follow similar steps to evaluate $\mathcal{R}_{\tilde q}$ by changing $m_{i},\tilde m_{i}$ to $-m_{i},-\tilde m_{i}$, while for the $W$ boson contribution $\mathcal{R}_{W}$, we can recover the result by setting $m_{i}=\tilde m_{i}=0$ and replacing $|a|,\theta_{e}$ in the operators $\mathbb{D}^{q_{i}}_{\pm}$ by $2|a|,2\theta_{e}$.
A further factor given as a power of $2\pi R$ needs to be introduced into $\log\mathcal{R}$ to account for the removal of zero modes in the functional determinant, as well as matching with the three-dimensional limit computed in \cite{DKMTV} (see chapter \ref{ch: pure} for more details).
The resulting $\mathcal{R}_{W}$ and $\mathcal{R}_{\tilde q}$ match with (\ref{D factor W matter 2}) and (\ref{D factor qt matter 2}) up to $\Lambda_{\rm UV}$-dependent terms, which combine with the $\Lambda_{\rm UV}$-dependent term in $\mathcal{R}_{q}$ (\ref{loop factor matter 3}) to give the renormalised $\tau_{\rm eff}$ in the monopole mass.

\paragraph{}

Putting all the pieces together and summing over electric charges $\gamma_{e}$ and flavour charges $\vec s$, we express the large-distance behaviour of the four-fermion correlation function as
\begin{equation}
\label{correlator 4 fermions matter 3}
\begin{aligned}
\mathcal{G}_{4}(y_{1},y_{2},y_{3},y_{4}) & =
\frac{2^{10}\pi}{R |a|^{1/2}} \,
\mathcal{D}(-i) \left( \frac{2\pi R}{g_{\rm eff}^{2}} \right)^{-1/2} \exp\left( -S_{\rm mon} \right)
\sum_{\gamma=(\gamma_{e},1,\vec s\,)\in\Gamma} \Omega(\gamma,a) \exp\left( -S_{\varphi}^{(\gamma_{e},\vec s\,)} \right)
\\
& \int d^{3}X \epsilon^{\alpha'\beta'}\epsilon^{\gamma'\delta'}
S_{\rm F}(y_{1}-X)_{\alpha\alpha'}S_{\rm F}(y_{2}-X)_{\beta\beta'}S_{\rm F}(y_{3}-X)_{\gamma\gamma'}S_{\rm F}(y_{4}-X)_{\delta\delta'}
\,,
\end{aligned}
\end{equation}
where we have substituted (\ref{propagator fermion 3D matter}), (\ref{measures}), and (\ref{loop factor matter 3}) into (\ref{correlator 4 fermions matter 2}), and the actions $S_{\rm mon}$ and $S_{\varphi}^{(\gamma_{e},\vec s\,)}$ are given in (\ref{action monopole matter}) and (\ref{action angular matter}).
For consistency, we should also use the same renormalised $g^{2}_{\rm eff}(a)$ wherever the coupling appears.
The four-fermion correlator computed here corresponds to the following four-fermion interaction vertex in the low-energy effective action:
\begin{equation}
\begin{aligned}
\label{action 4 fermions matter semiclassical}
S_{\rm 4F} =
\frac{2^{7}\pi}{R |a|^{1/2}}
& \left( \frac{2\pi R}{g_{\rm eff}^{2}} \right)^{7/2}
\mathcal{D}(-i) \exp\left( -S_{\rm mon} \right)
\\
& \sum_{\gamma=(\gamma_{e},1,\vec s\,)\in\Gamma} \Omega(\gamma,a) \exp\left( -S_{\varphi}^{(\gamma_{e},\vec s\,)} \right)
\int d^{3}x \left( \psi\cdot\bar\psi \right) \left( \lambda\cdot\bar\lambda \right)
\,.
\end{aligned}
\end{equation}
This exactly matches the prediction (\ref{action 4 fermions matter}) obtained from the GMN equation (\ref{RH}).

%%%%%%%%%%%%%%%%%%%%%%%%%%%%%%%%%%%%%%%%%%%%%%%%%%%%%%%%%%%%%

\section{Interpolating to three dimensions}

\paragraph{}

In this section, we demonstrate how some of the three-dimensional physical quantities can be recovered from our earlier results taking the limit $R\to 0$ and reproduce some of the results in \cite{DTV, BHO}.
First, we focus on finding the explicit expressions for the one-loop determinants (\ref{D factor W matter 2}, \ref{D factor q matter 2}, \ref{D factor qt matter 2}).
Then, we use these results to find the four-fermion action.

\paragraph{}

The three-dimensional limit of $\mathcal{D}_{W}(-i)$ has already been calculated in the pure theory (\ref{D factor 3D pure}).
Here, we need to find the one-loop determinants $\mathcal{D}_{q}(-i)$ and $\mathcal{D}_{\tilde q}(-i)$ (\ref{D factor q matter}, \ref{D factor qt matter}), corresponding to quarks.
We should treat massive and massless hypermultiplets separately: for each massless hypermultiplet (when $m_{i}=0$, $\psi_{i}=0$), the appropriate factors in (\ref{D factor q matter 2}) and (\ref{D factor qt matter 2}) have the same form as (\ref{D factor W matter 2}); on the other hand, for massive hypermultiplets, setting $R|a|\to 0$ would lead to divergences in the integrals in (\ref{D factor q matter 2}, \ref{D factor qt matter 2}), and it is necessary to Poisson-resum (\ref{Poisson resummation}) the expressions to obtain finite values.
We decompose the one-loop factor as
\begin{equation}
\begin{aligned}
& \mathcal{D}(-i) = \mathcal{D}_{W}(-i) \, \prod_{i=1}^{N_{f}} \, \mathcal{D}_{H_{i}}(-i)
\,,
\\
& \mathcal{D}_{H_{i}}(-i) = \mathcal{D}_{q_{i}}(-i) \, \mathcal{D}_{\tilde q_{i}}(-i)
\end{aligned}
\end{equation}
where $W$ denotes the vector multiplet, $H_{i}$ denotes $i$-th hypermultiplet, $q_{i}$ and $\tilde q_{i}$ mean restricting (\ref{D factor q matter 2}) and (\ref{D factor qt matter 2}) to the $i$-th term.

\paragraph{}

First, consider the one-loop factors for $W$ bosons and quarks corresponding to the massless flavours.
For each massless flavour, labelled by $i$, its factor can be obtained by scaling the coefficients in the result for $W$ bosons (\ref{D factor W matter 2}):
\begin{align}
\label{D factor W 3D matter}
\mathcal{D}_{W}(-i) & = (4\pi RM_{W})^{2}
\,,
\\
\label{D factor hyper 3D massless}
\mathcal{D}_{q_{i}}(-i) & = \mathcal{D}_{\tilde q_{i}}(-i) = (2\pi RM_{W})^{-1/2}
\,,
\quad
\mathcal{D}_{H_{i}}(-i) = (2\pi RM_{W})^{-1}
\,.
\end{align}

\paragraph{}

Second, let us Taylor-expand the exponents in $\log\mathcal{D}_{q}(-i)$ and $\log\mathcal{D}_{\tilde q}(-i)$ corresponding to the massive flavours, ignoring all massless ones, and Poisson-resum each term of the resulting series.
After this procedure, we obtain
\begin{equation}
\begin{aligned}
\log\mathcal{D}_{q}(-i) =
-\frac{1}{2} & \sum_{i=1}^{N_{f}} \sum_{n\in\mathbb{Z}} \left( \Arsinh\left( \frac{|a+m_{i}|}{\frac{\theta_{e}+\psi_{i}}{2\pi R}+\frac{n}{R}} \right) - \kappa_{n}R|a+m_{i}| \right)
\\
+ & \sum_{i=1}^{N_{f}}R|a+m_{i}| \left( \log\frac{|\Lambda_{\rm UV}|}{|a+m_{i}|}+1 \right)
\,,
\end{aligned}
\end{equation}
\begin{equation}
\begin{aligned}
\log\mathcal{D}_{\tilde q}(-i) =
-\frac{1}{2} & \sum_{i=1}^{N_{f}} \sum_{n\in\mathbb{Z}} \left( \Arsinh\left( \frac{|a-m_{i}|}{\frac{\theta_{e}-\psi_{i}}{2\pi R}+\frac{n}{R}} \right) - \kappa_{n}R|a-m_{i}| \right)
\\
+ & \sum_{i=1}^{N_{f}}R|a-m_{i}| \left( \log\frac{|\Lambda_{\rm UV}|}{|a-m_{i}|}+1 \right)
\end{aligned}
\end{equation}
where $\kappa_{n}$ are regularisation constants; this expression is applicable only for $\theta_{e}\pm\psi_{i}\ne 0$, when the term with $n=0$ does not diverge.
We now restrict ourselves to the $\Arsinh$ terms since the other terms, being proportional to $R|a\pm m_{i}|$, vanish in three dimensions.
Rewriting the inverse hyperbolic functions in terms of logarithms~\footnote{
We are using the following identity: $x=\sinh(\pm\log(\pm x+\sqrt{x^{2}+1}))$; note that the expression for $\mathcal{D}_{q}(-i)\mathcal{D}_{\tilde q}(-i)$ is dimensionless with respect to $R$.
} and exponentiating, we express the product of the one-loop quark factors $\mathcal{R}_{H}=\mathcal{D}_{q}(-i)\mathcal{D}_{\tilde q}(-i)$ as
\begin{equation}
\label{D factor hyper 3D massive}
\mathcal{R}_{H} =
\prod_{i=1}^{N_{f}} \prod_{n\in\mathbb{Z}} \left(
\frac{\sqrt{\left| a-m_{i} \right|^{2}+\left( \frac{\theta_{e}-\psi_{i}}{2\pi R}+\frac{n}{R} \right)^{2}}-\left| a-m_{i} \right|}{\sqrt{\left| a+m_{i} \right|^{2}+\left( \frac{\theta_{e}+\psi_{i}}{2\pi R}+\frac{n}{R} \right)^{2}}+\left| a+m_{i} \right|}
\right)^{1/2}
\end{equation}
where $n/R$ should be regarded as Kaluza--Klein momentum over the compactified $S^{1}$.
Let us now take the limit $R\to 0$: in order to do this, one should keep $\theta_{e}/2\pi R$ and $\psi_{i}/2\pi R$ fixed, then all $n\ne 0$ terms in (\ref{D factor hyper 3D massive}) yield $1$.
In three dimensions, there is a symmetry under which $(\re a,\im a,\theta_{e}/2\pi R)$ and $(\re m_{i},\im m_{i},\tilde m_{i})$ transform as vectors (we have defined $\psi_{i}=2\pi R\tilde m_{i}$).
In particular, this allows us to exchange $|a\pm m_{i}|$ and $(\theta_{e}\pm\psi_{i})/2\pi R$ in (\ref{D factor hyper 3D massive}) and then rotate the VEV into the new vacuum for which $|a|=0$; after these manipulations, (\ref{D factor hyper 3D massive}) can be rewritten as
\begin{equation}
\label{D factor hyper 3D massive 2}
\mathcal{R}_{H} =
\prod_{i=1}^{N_{f}} \left(
\frac{\sqrt{\left( \tilde m_{i}-\frac{M_{W}}{2} \right)^{2}+|m_{i}|^{2}}+\tilde m_{i}-\frac{M_{W}}{2}}{\sqrt{\left( \tilde m_{i}+\frac{M_{W}}{2} \right)^{2}+|m_{i}|^{2}}+\tilde m_{i}+\frac{M_{W}}{2}}
\right)^{1/2}
\,.
\end{equation}
This formula correctly reproduces the one-loop factor for quarks given by eq.\ (34) in \cite{DTV} (this factor also appears in a calculation of three-dimensional superpotentials in \cite{BHO}).
In these notations, sending one flavour mass $m_{i}$ to infinity removes the corresponding factor, i.e., reduces the number of massive flavours, $N_{f}$, by 1, as expected.

\paragraph{}

Summarising the results above, the one-loop factor in three dimensions is given by the product of three components: $\mathcal{D}_{W}(-i)$ (\ref{D factor W 3D matter}) for the $W$ bosons, $\mathcal{D}_{H_{i}}(-i)$ (\ref{D factor hyper 3D massless}) for each massless flavour labelled by $i$, and $\mathcal{R}_{H}$ (\ref{D factor hyper 3D massive 2}) for the remaining massive flavours.

\paragraph{}

To determine which corrections (\ref{symplectic form dyon matter 2}) survive in the limit $R\to 0$, we can use dimensional analysis: the leading order of the metric in three dimensions must depend only on the effective coupling $e_{\rm eff}=g_{\rm eff}/\sqrt{2\pi R}$, which is held fixed, but not on the infinitely small value of $R$ itself since setting $R$ to zero would cancel the correction.
In chapter \ref{ch: pure}, we have shown that for one-instanton in three dimensions, the four-fermion correction is non-zero; this must also be the case in theories with fundamental hypermultiplets when all flavours are massive since $\mathcal{R}_{H}$ is of order one, and all corrections for BPS states with magnetic charge $1$ are of the same order with respect to $a$ and $R$.
However, the situation is different if there is a massless flavour.
The overall power of $R$ in the metric depends on the perturbative coefficient $\left( \mathcal{D}(-i) \right)^{k |\gamma_{m}'|}$.
As we know, for the $W$ bosons, $\mathcal{D}_{W}(-i)\sim R^{2}$, for massive hypermultiplets, $\mathcal{R}_{H}\sim R^{0}$, and for massless hypermultiplets, $\mathcal{D}_{H_{i}}(-i)\sim R^{-1}$.
Note that adding massive flavours does not change the order with respect to $R$; this is in agreement with the fact that massive flavours do not affect the number of zero modes.
From the case of only massive flavours, it follows that the perturbative factor must behave as $\left( \mathcal{D}(-i) \right)^{k |\gamma_{m}'|}\sim R^{2}$ in order to yield a non-zero correction.
Denoting the number of massless flavours $N_{f}^{(0)}$, we can conclude that this requirement leads to
\begin{equation}
(2-N_{f}^{(0)}) \, k |\gamma_{m}'| = 2
\,.
\end{equation}
This equality can be satisfied only in three cases:
\begin{align}
|\gamma_{m}'| = 1
\,,
\quad
k = 1
\,,
\quad
N_{f}^{(0)} = 0
\,,
\\
|\gamma_{m}'| = 1
\,,
\quad
k = 2
\,,
\quad
N_{f}^{(0)} = 1
\,,
\\
|\gamma_{m}'| = 2
\,,
\quad
k = 1
\,,
\quad
N_{f}^{(0)} = 1
\,.
\end{align}
The first case corresponds to one-instanton corrections (\ref{metric instanton matter}), the other two correspond to two-instanton corrections (\ref{metric two-instanton matter}), where the last condition can hold only for $N_{f}=3$ and $N_{f}=4$; no higher-order corrections exist in three dimensions.

%%%%%%%%%%%%%%%%%%%%%%%%%%%%%%

\subsection{One-instanton corrections}

\paragraph{}

Let us Poisson-resum the metric component $g_{a\bar{a}}^{\rm inst}$ (\ref{metric instanton matter}) over the electric charges $\gamma_{e}'$: in the present case, to get the exact formula, even and odd electric charges need to be resummed separately as the associated flavour charges are different.
The sum of the two resulting series yields the one-instanton metric:
\begin{equation}
\label{metric instanton resummed matter}
g_{a\bar a}^{\rm inst} =
\frac{4\pi}{g_{\rm eff}^{4}}
\sum_{p\in\{0,1\}} \sum_{\vec s\,^{(p)}\in S^{(p)}} \sum_{n\in\mathbb{Z}}
\frac{|a|^{2}\mathcal{D}(-i)}{|M(n)|^{3}}
\exp\left( -\frac{16\pi^{2}R}{g_{\rm eff}^{2}}|M(n)|+i\Psi(n)+2\pi iR\sum_{i=1}^{N_{f}}s_{i}^{(p)}F_{i}(n) \right)
\end{equation}
where we have defined
\begin{align}
|M(n)| & = \sqrt{|a|^{2}+\left( \frac{\theta_{e}+2\pi n}{2\pi R} \right)^{2}}
\,,
\\
\Psi(n) & = \theta_{m}-\frac{\Theta_{\rm eff}}{\pi}(\theta_{e}+2\pi n)
\,,
\\
F_{i}(n) & = \tilde m_{i}-\frac{\theta_{e}+2\pi n}{2\pi R} \, \frac{|m_{i}|}{|a|}
\,,
\end{align}
$S^{(p)}$ denotes the set of flavour charges for dyons with magnetic charge $1$ where $p=0$ and $p=1$ correspond to even and odd electric charges, respectively.
The dependence on $n$ appearing in (\ref{metric instanton resummed matter}) corresponds to the Euclidean action of the twisted monopole found in \cite{Lee Yi}, which can be generated by applying large gauge transformation on the monopole compactified on $\mathbb{R}^{3}\times S^{1}$.

\paragraph{}

After taking the limit $2\pi R\to 0$, the only $M(n)$ that does not diverge has $n=0$, therefore, only the $n=0$ terms survive in the summation. Furthermore, as $\theta_{e}/2\pi R$ and $\tilde m_{i}$ are also kept fixed, $2\pi RF_{i}(0)$ becomes negligible.
We keep the combination $1/e_{\rm eff}^{2}=2\pi R/g^{2}_{\rm eff}$ fixed, and the mass of the $W$ boson is $M_{W}=2M(0)$.
It has been established that the one-instanton metric is non-zero when all flavours are massive, hence, the perturbative factor is
\begin{equation}
\mathcal{D}(-i) = (4\pi RM_{W})^{2} \, \mathcal{R}_{H}
\,.
\end{equation}
Collecting all factors together and applying the three-dimensional $SU(2)_{N}$ symmetry, we find that the Poisson-resummed metric (\ref{metric instanton resummed matter}) leads to the following four-fermion vertex in the effective Lagrangian in three dimensions:
\begin{equation}
\label{action 4 fermions 3D matter}
S_{\rm 4F} =
\frac{2^{11}\pi^{3}M_{W}\mathcal{R}_{H}\Omega(1)}{e_{\rm eff}^{8}}
\exp\left( -\frac{4\pi}{e_{\rm eff}^{2}}M_{W}+i\theta_{m} \right)
\int d^{3}x \left( \psi\cdot\bar\psi \right) \left( \lambda\cdot\bar\lambda \right)
\end{equation}
where $\Omega(1)$ is the multiplicity of dyons with charge $(\gamma_{e},1)$ over all flavour charges.
For $N_{f}<4$, $\Omega(1)$ appears in the wall-crossing formulae (\ref{WCF2 1f}, \ref{WCF2 2f}, \ref{WCF2 3f}); for $N_{f}=4$, $\Omega(1)=8$ as all dyons with magnetic charge $1$ transform under $\bf 8_{s}$ or $\bf 8_{c}$ representations of the flavour group; for any $N_{f}$, the multiplicity can be expressed as
\begin{equation}
\label{multiplicity one matter}
\Omega(1) = 2^{N_{f}-1}
\,,
\end{equation}
which is the number of dimensions of the irreducible representation of the $SO(2N_{f})$ Clifford Algebra.
This, after rescaling the gauge coupling and electric charges, matches with the semiclassical result for the four-fermion vertex \cite{DTV, DKMTV}.

%%%%%%%%%%%%%%%%%%%%%%%%%%%%%%

\subsection{Two-instanton corrections}

\paragraph{}

To calculate the two-instanton corrections, which are present only if exactly one flavour is massless, we need to resum the metric with $k\gamma_{m}'=2$ (\ref{metric two-instanton matter}) in the same manner.
From the very beginning, we restrict our attention to the terms surviving in the limit $R\to 0$: at the leading order, flavour charges are irrelevant because the dominant correction to the monopole mass is given by the infinite tower of electric charges with $\gamma_{e}'\in\mathbb{Z}$; however, the number of states for each pair of $\gamma_{e}'$ and $\gamma_{m}'$ affects the overall factor.

\paragraph{}

We need to consider corrections coming from dyons with magnetic charge $\gamma_{m}'=1$ and winding number $k=2$ and from dyons with magnetic charge $\gamma_{m}'=2$ and winding number $k=1$ in (\ref{metric two-instanton matter}).
The latter case applies only to the theories with $N_{f}=3$ and $N_{f}=4$.
All such corrections are proportional to $(\mathcal{D}(-i))^{2}$.
The perturbative factor $\mathcal{D}(-i)$ in theories with one massless flavour, which we denote as $H_{0}$, is
\begin{equation}
\mathcal{D}(-i) = \mathcal{D}_{W}(-i) \, \mathcal{D}_{H_{0}}(-i) \, \mathcal{R}_{H} = 8\pi RM_{W} \, \mathcal{R}_{H} 
\,.
\end{equation}
This factor behaves as $\sim R$ rather than as $\sim R^{2}$ (as was the case for massive flavours), and the overall coefficient behaves as $(\mathcal{D}(-i))^{2}\sim R^{2}$.

\paragraph{}

We can Poisson-resum the metric repeating the steps in our one-instanton computation: to get the $\gamma_{m}'=1$ contributions, all central charges in the one-instanton formulae need to be multiplied by $2$; to get the $\gamma_{m}'=2$ contributions, all magnetic charges (or, equivalently, $1/g_{\rm eff}^{2}$) need to be multiplied by $2$; in addition, both expressions are proportional to the multiplicity of dyons with magnetic charge $\gamma_{m}'$.
After these modifications in (\ref{metric instanton resummed matter}), we obtain the two-instanton metric in three dimensions:
\begin{equation}
g_{a\bar a}^{\rm inst(2)} =
\frac{4\pi}{g_{\rm eff}^{4}}
\frac{(\mathcal{D}(-i))^{2}\left( \Omega(1)+8\Omega(2) \right)}{|a|}
\exp\left( -2S_{\rm mon}+2i\theta_{m} \right)
\end{equation}
where $\Omega(2)$ is the average between degeneracies of dyons with charges $(2p,2)$ and $(2p+1,2)$, $\Omega(1)$ has been defined and calculated above (\ref{multiplicity one matter}).
In the case of $N_{f}=4$, BPS states with charges $(2p,2)$ have multiplicity $-2$ (they transform as $W$ bosons), BPS states with charges $(2p+1,2)$ have multiplicity $8$ (they transform as ${\bf 8_{v}}$); for $N_{f}<4$, these multiplicities can be read off from the wall-crossing formulae (\ref{WCF2 1f}, \ref{WCF2 2f}, \ref{WCF2 3f}); summing up, we get
\begin{equation}
\begin{aligned}
\Omega(2) & = 0
\,,
\quad
N_{f} < 3
\,,
\\
\Omega(2) & = \frac{1}{2}
\,,
\quad
N_{f} = 3
\,,
\\
\Omega(2) & = 3
\,,
\quad
N_{f} = 4
\,.
\end{aligned}
\end{equation}

\paragraph{}

Combining all corrections and expanding the one-loop factor, we see that the four-fermion vertex for two-instantons is
\begin{equation}
S_{\rm 4F}^{(k=2)} =
\frac{2^{12}\pi^{3}M_{W}\mathcal{R}_{H}^{2}\left( \Omega(1)+8\Omega(2) \right)}{e_{\rm eff}^{8}}
\exp\left( -2S_{\rm mon}+2i\theta_{m} \right)
\int d^{3}x \left( \psi\cdot\bar\psi \right) \left( \lambda\cdot\bar\lambda \right)
\,.
\end{equation}
This formula reproduces the semiclassical result in eq.\ (45, 46) of \cite{DTV} up to a numerical coefficient.
This discrepancy requires further investigation.

%%%%%%%%%%%%%%%%%%%%%%%%%%%%%%

\subsection{Brane configuration}

\paragraph{}

\begin{figure}[ht]
\centering
\includegraphics[width=0.5\textwidth]{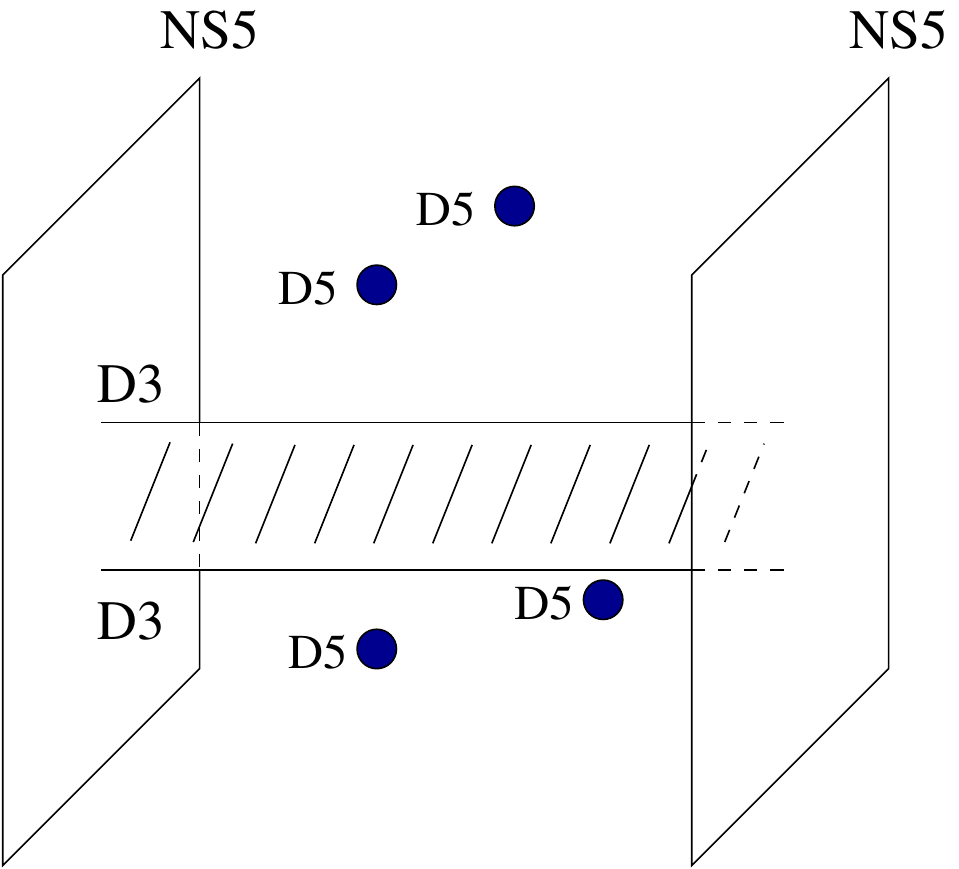}
\caption{
Hanany--Witten brane configuration for the $\mathcal{N}=2$ $SU(2)$ theory with flavours in three dimensions.
}
\label{fig: brane configuration}
\end{figure}

It is possible to understand the form of many of the previous field theory results in an elegant way in terms of Hanany--Witten brane configurations \cite{Hanany Witten} (figure~\ref{fig: brane configuration}).
In order to make the discussion more transparent, we will work in terms of a T-dual picture, in which instead of a four-dimensional theory compactified on a circle, we have a three-dimensional field theory localised in a compact transverse direction.
Consider IIB theory in the presence of two D3 branes with world volume coordinates $x^{0},x^{1},x^{2},x^{6}$ suspended between two NS5 branes with world volume coordinates $x^{0},x^{1},x^{2},x^{3},x^{4},x^{5}$ and located $L_{6}$ apart in the $x^{6}$ direction.
Additional $N_{f}$ D5 branes with world volume coordinates $x^{0},x^{1},x^{2},x^{7},x^{8},x^{9}$ provide the flavours.
The Coulomb branch of the gauge theory in the $x^{0},x^{1},x^{2}$ directions is realised when the D3 branes are split along $x^{3},x^{4},x^{5}$ with separation $\Delta\vec x$.
We take $x^{3}$ to be the compact direction in which we have T-dualised our original four-dimensional theory; it has dual radius $\tilde R=1/R$ where $R$ is the compactification radius of the original four-dimensional theory.

\paragraph{}

The open string stretching between the D3 brane at $\vec x$ and the additional D5 brane sitting at $\vec m_{i}$ where $\vec x$ and $\vec m_{i}$ are their respective positions in the $x^{3},x^{4},x^{5}$ directions, gives fundamental quark of mass $|\vec x-\vec m_{i}|$.
Again, due to the compactification in $x^{3}$, it is necessary to periodically identify
\begin{equation}
x^{3}-m_{i}^{3}\sim x^{3}-m_{i}^{3}+\frac{2\pi n}{R}
\end{equation}
and sum over the copies.
Combining with the mass of the $W$ boson, $|\Delta\vec x(n)|$, and using appropriate weights for different representations, they explain the form of the perturbatively corrected gauge coupling (\ref{coupling shift matter}), which can alternatively be obtained from explicit one-loop computation \cite{Seiberg Shenker, DKMTV, DTV}.

\paragraph{}

Instanton corrections to the Coulomb branch metric come from Euclidean D1 strings stretching between D3 and NS5 branes whose world volume is bounded between the intervals $\Delta\vec x$ in $x^{3},x^{4},x^{5}$ and $L_{6}$ in $x^{6}$ \cite{BHO, Hanany Witten, BHOOY}.
The D1 DBI action is given by the tension of the D1:
\begin{equation}
S_{\rm D1} \simeq -\frac{L_{6}\|\Delta\vec x(n)\|}{g_{s}}
\end{equation}
where $\|\Delta\vec x(n)\|$ is the norm of the vector $(\Delta x^{3}+2\pi n\tilde R=\Delta x^{3}+2\pi n/R,\Delta{x}_4,\Delta{x}_{5})$.
Due to the periodicity in $x^{3}$, we need to sum over all multiply wound D1 branes with winding number given by $n\in\mathbb{Z}$.
Using the expression
\begin{equation}
\frac{8\pi}{e^{2}_{\rm eff}} = \frac{L_{6}}{g_{s}}
\end{equation}
for the gauge coupling, we see that $S_{\rm D1}$ coincides with the real part of the twisted monopole action in (\ref{metric instanton resummed matter}).
To account for the phase $i\theta_m$ in (\ref{metric instanton resummed matter}), we recall that the D1 action also receives a boundary contribution since D1 branes are charged magnetically under the D3 gauge fields.
This was made explicit in \cite{BHO}, where the dual photon was identified with the $x^{6}$ component of the magnetic gauge potential $A_{D\,6}$.
Integrating over the boundary of D1 branes, we obtain the desired phase.

\chapter{Moduli space in higher-rank gauge theories}
\label{ch: gauge}

\paragraph{}

{\it This chapter is based on \cite{CDP2}.}

\paragraph{}

In chapter \ref{ch: supersymmetry}, we have demonstrated that $\mathcal{N}=2$ supersymmetric theories with gauge group of higher rank (i.e., those with Cartan subalgebra of rank higher than $1$) have walls of marginal stability extending into the weak-coupling region.
The spectrum of composite dyons changes as the vacuum expectation value crosses these walls (whereas the $W$ bosons and simple dyons exist everywhere in the weak-coupling region).
This feature allows us to study the moduli space and decay processes at these walls explicitly using semiclassical methods.
We focus on theories with gauge group $SU(n)$ where $n>2$, for which we have shown in chapter \ref{ch: walls} that the Kontsevich--Soibelman wall-crossing formula agrees with the weak-coupling BPS spectrum and its jumps across the walls; the general formula is given by (\ref{WCFn}), and the jumps are governed by (\ref{SU(n) pentagon}).
The hyper-K\"ahler moduli space metric in $\mathbb{R}^{3}\times S^{1}$ can be obtained by introducing the BPS spectrum from (\ref{WCFn}) into the Gaiotto--Moore--Netitzke solution (\ref{RH}); the resulting metric must remain continuous everywhere in the moduli space by construction.

\paragraph{}

As before, the dyons of the original four-dimensional theory give rise to multi-instantons in the compactified theory.
We expand the integral equations (\ref{RH}): in addition to the single-instanton corrections to the moduli space metric, we also extract the two-instanton mixing terms, ignoring higher-order corrections, and demonstrate smoothness of the resulting metric at the walls of marginal stability via a direct calculation.
More specifically, requiring that the two-instanton metric is continuous is equivalent to imposing a constraint on the degeneracies of composite dyons.
This constraint is satisfied by the pentagon formulae (\ref{SU(n) pentagon}).
Then, we compare the predicted one-instanton contribution with the result of an explicit field theory calculation.
The main challenge here is to evaluate the non-cancelling ratio of functional determinants \cite{DKMTV} arising from fluctuations around the instanton background, which is a complicated function of the compactification radius.
We express this contribution in closed form and find a precise match with the prediction from the expansion of the integral equations.

\paragraph{}

Finally, after performing the Poisson resummation, we continue our results to zero radius and make contact with an earlier semiclassical computation in the three-dimensional theory \cite{Fraser Tong}.
The latter results also provide a direct check of the smoothness of the metric in three dimensions.

%%%%%%%%%%%%%%%%%%%%%%%%%%%%%%%%%%%%%%%%%%%%%%%%%%%%%%%%%%%%%

\section{Semiclassical expansion of the moduli space metric}

\paragraph{}

Let us consider the $\mathcal{N}=2$ supersymmetric theory with any gauge group compactified on $\mathbb{R}^{3}\times S^{1}$, where $R$ is the radius of $S^{1}$.
As discussed in chapter \ref{ch: walls}, it was conjectured that the hyper-K\"ahler metric of the moduli space of compactified theory is determined by a set of integral equations.
In the case of gauge group of rank $r$ (for gauge group $SU(n)$, $r=n-1$), the Gaiotto--Moore--Netitzke solution for Darboux coordinates (\ref{RH}) is given by a set of $2r$ integral equations:
\begin{align}
\mathcal{X}_{e}^{I}(\zeta) & =
\mathcal{X}_{e}^{I\,\rm sf}(\zeta)
\exp\left( -\frac{1}{2\pi i} \sum_{\gamma'\in\Gamma(\vec a)} c_{e}^{I}(\gamma') \, \mathcal{I}_{\gamma'}(\zeta) \right)
\,,
\quad
c_{e}^{I}(\gamma') = -\Omega(\gamma',\vec a) \gamma_{m}^{I}{}'
\,,
\\
\mathcal{X}_{m\,I}(\zeta) & =
\mathcal{X}_{m\,I}^{\rm sf}(\zeta)
\exp\left( -\frac{1}{2\pi i} \sum_{\gamma'\in\Gamma(\vec a)} c_{m\,I}(\gamma') \, \mathcal{I}_{\gamma'}(\zeta)\right)
\,,
\quad
c_{m\,I}(\gamma') = \Omega(\gamma',\vec a) \gamma_{e\,I}'
\,,
\end{align}
where the integrals, which give rise to BPS corrections, are
\begin{equation}
\mathcal{I}_{\gamma'}(\zeta) =
\int_{l_{\gamma'}} \frac{d\zeta'}{\zeta'} \frac{\zeta'+\zeta}{\zeta'-\zeta}
\log\left( 1-\sigma(\gamma')\mathcal{X}_{\gamma'}(\zeta') \right)
\,.
\end{equation}
Using these equations, we shall perform semiclassical expansion of the metric in the $SU(n)$ theory at weak coupling: taking the limit $g_{\rm eff}\to 0$, we approximate the solution by performing iterations using $\mathcal{X}_{\gamma'}^{\rm sf}(\zeta')$ as the initial approximation for $\mathcal{X}_{\gamma'}(\zeta')$.
In addition to the one-instanton corrections to the moduli space metric, we shall also extract the two-instanton mixing terms and demonstrate smoothness of the moduli space metric at the walls of marginal stability.
These results will then be compared with calculations based on first principles.
We will approximate the corrected symplectic form as
\begin{equation}
\omega(\zeta) \approx \omega^{\rm sf}(\zeta)+\omega^{\rm P}(\zeta)+\omega^{\rm NP}(\zeta)
\end{equation}
where $\omega^{\rm P}(\zeta)$ denotes perturbative corrections, corresponding to $W$ bosons, $\omega^{\rm NP}(\zeta)$ denotes non-perturbative corrections, corresponding to dyons.

%%%%%%%%%%%%%%%%%%%%%%%%%%%%%%

\subsection{Perturbative corrections}

\paragraph{}

Analogously to the approximation used in chapter \ref{ch: pure}, our first step is to find the perturbative contributions to the Darboux coordinates: we decompose the perturbatively corrected coordinates as
\begin{equation}
\mathcal{X}_{\gamma}^{(0)}(\zeta) = \mathcal{X}_{\gamma}^{\rm sf}(\zeta)\mathcal{D}_{\gamma}(\zeta)
\end{equation}
where $\mathcal{D}_{\gamma}(\zeta)$ will be related to the one-loop determinants in the semiclassical calculation.
The electric components remain unchanged:
\begin{equation}
\mathcal{X}_{(\vec\gamma_{e},\vec 0)}^{(0)}(\zeta) =
\mathcal{X}_{(\vec\gamma_{e},\vec 0)}^{\rm sf}(\zeta) \mathcal{D}_{(\vec\gamma_{e},\vec 0)}(\zeta) =
\mathcal{X}_{(\vec\gamma_{e},\vec 0)}^{\rm sf}(\zeta)
\,,
\quad
\forall \, \vec\gamma_{e}
\,,
\end{equation}
whereas the magnetic components receive corrections:
\begin{equation}
\mathcal{X}_{(\vec 0,\vec\gamma_{m})}^{(0)}(\zeta) =
\mathcal{X}_{(\vec 0,\vec\gamma_{m})}^{\rm sf}(\zeta) \mathcal{D}_{(\vec 0,\vec\gamma_{m})}(\zeta)
\,,
\quad
\forall \, \vec\gamma_{m}
\,,
\end{equation}
\begin{equation}
\label{D factor}
\begin{aligned}
\log\mathcal{D}_{(\vec 0,\vec\gamma_{m})}(\zeta) =
\frac{1}{\pi i} \sum_{\vec\alpha_{A}\in\Phi_{+}} \vec\gamma_{m}\vec\alpha_{A}
\left( 
\int_{l_{(\vec\alpha_{A},\vec 0)}} \frac{d\zeta'}{\zeta'} \frac{\zeta'+\zeta}{\zeta'-\zeta}
\log\left( 1-\mathcal{X}_{(\vec\alpha_{A},\vec 0)}^{\rm sf}(\zeta') \right)
\right.
\\
\left.
-\int_{l_{(-\vec\alpha_{A},\vec 0)}} \frac{d\zeta'}{\zeta'} \frac{\zeta'+\zeta}{\zeta'-\zeta}
\log\left( 1-1/\mathcal{X}_{(\vec\alpha_{A},\vec 0)}^{\rm sf}(\zeta') \right)
\right)
\end{aligned}
\end{equation}
where $l_{(\pm\vec\alpha_{A},\vec 0)}$ means integrating from zero to infinity along the BPS ray $\vec\alpha_{A}\vec a/\zeta'\in\mathbb{R}_{\mp}$ in the $\zeta'$ plane.
The semiflat Darboux coordinate for electric charges is given by
\begin{equation}
\mathcal{X}_{(\vec\alpha,\vec 0)}^{\rm sf}(\zeta) = \exp\left( \pi R\frac{\vec\alpha\vec a}{\zeta}+i\vec\alpha\vec\theta_{e}+\pi R\vec\alpha\bar{\vec a}\zeta \right)
\,.
\end{equation}
Rotating the contours of integration via introducing $y=-\zeta'/\exp(i\phi_{W_{A}})$ where $\phi_{W_{A}}=\arg(\vec\alpha_{A}\vec a)$ in the first term and $1/y=-\zeta'/\exp(-i\phi_{W_{A}})$ in the second term, we rewrite (\ref{D factor}) as
\begin{equation}
\label{D factor 2}
\begin{aligned}
\log\mathcal{D}_{(\vec 0,\vec\gamma_{m})}(\zeta) =
\frac{1}{\pi i} \sum_{\vec\alpha_{A}\in\Phi_{+}} \vec\gamma_{m}\vec\alpha_{A}
\int_{0}^{+\infty} \frac{dy}{y}
\left( 
\frac{y-\zeta \, e^{-i\phi_{W_{A}}}}{y+\zeta \, e^{-i\phi_{W_{A}}}}
\log\left( 1-e^{-\pi R |\vec\alpha_{A}\vec a| (y+1/y)+i\vec\alpha_{A}\vec\theta_{e}} \right)
\right.
\\
\left.
-\frac{y-\zeta^{-1}e^{i\phi_{W_{A}}}}{y+\zeta^{-1}e^{i\phi_{W_{A}}}}
\log\left( 1-e^{-\pi R |\vec\alpha_{A}\vec a| (y+1/y)-i\vec\alpha_{A}\vec\theta_{e}} \right)
\right)
\,.
\end{aligned}
\end{equation}
The expression is real if and only if $|\zeta|=1$.
Expanding the corrected magnetic Darboux coordinates, we can extract the perturbative corrections:
\begin{equation}
\label{symplectic form P}
\omega^{\rm P}(\zeta) = -\frac{1}{4\pi^{2}R} \sum_{I=1}^{r}
d\log\mathcal{X}_{e}^{I\,\rm sf}(\zeta)
\wedge
d\log\mathcal{D}_{\vec E_{I}}(\zeta)
\end{equation}
where $\vec E_{I}$ is defined as a unit vector along the $I$-th direction.
It is straightforward to modify the $SU(2)$ calculations to find the explicit formula for the perturbative corrections:
\begin{equation}
\begin{aligned}
\omega^{\rm P}(\zeta) = &
-\frac{1}{4\pi^{2}R} \sum_{I=1}^{r} \sum_{\vec\alpha_{A}\in\Phi_{+}} \,
d\left(\pi R\zeta^{-1} a^{I}+i\theta_{e}^{I}+\pi R\zeta \bar a^{I} \right)
\\
& \wedge
\left(
2\pi i A_{I\,\vec\alpha_{A}}^{\rm P}(\vec a,\bar{\vec a})+\pi i V_{IJ\,\vec\alpha_{A}}^{\rm P}(\vec a,\bar{\vec a})
\left( \zeta^{-1} d a^{J}-\zeta d\bar a^{J} \right)
\right)
\end{aligned}
\end{equation}
where the potentials $A_{I}^{\rm P}=\sum_{\vec\alpha_{A}\in\Phi_{+}}A_{I\,\vec\alpha_{A}}^{\rm P}$ and $V_{IJ}^{\rm P}=\sum_{\vec\alpha_{A}\in\Phi_{+}}V_{IJ\,\vec\alpha_{A}}^{\rm P}$ are given as
\begin{align}
A_{I\,\vec\alpha_{A}}^{\rm P} & =
\frac{R}{\pi} \, (\vec\alpha_{A}\vec E_{I}) \,
\sum_{k\ne 0} |\vec\alpha_{A}\vec a| e^{ik\vec\alpha_{A}\vec\theta_{e}} K_{1}(2\pi R |k \vec\alpha_{A}\vec a|)
\left( \frac{d(\vec\alpha_{A}\vec a)}{\vec\alpha_{A}\vec a}-\frac{d(\vec\alpha_{A}\bar{\vec a})}{\vec\alpha_{A}\bar{\vec a}} \right)
\,,
\\
V_{IJ\,\vec\alpha_{A}}^{\rm P} & =
-\frac{2R}{\pi} \,
(\vec\alpha_{A}\vec E_{J}) (\vec\alpha_{A}\vec E_{I}) \,
\sum_{k\ne 0} e^{ik\vec\alpha_{A}\vec\theta_{e}} K_0(2\pi R |k \vec\alpha_{A}\vec a|)
\,.
\end{align}
The resulting shift of the effective complex coupling receives contributions from all massive bosons, counted by subscript $A$ (for $SU(n)$, $1\le A\le n(n-1)/2$).

\paragraph{}

The corrections can be rewritten in another form using the Gibbons--Hawking ansatz for generic gauge group \cite{GMN}:
\begin{equation}
\begin{aligned}
\label{Gibbons Hawking}
& g = \left( V^{-1}(\vec x) \right)^{IJ} \left( \frac{d\theta_{m\,I}}{2\pi}+A_{I}(\vec x) \right) \left( \frac{d\theta_{m\,J}}{2\pi}+A_{J}(\vec x) \right) + V_{IJ}(\vec x)d\vec x^{I}\vec x^{J}
\,,
\\
& a^{I} = x_{1}^{I}+ix_{2}^{I}
\,,
\quad
\theta_{e}^{I} = 2\pi Rx_{3}^{I}
\,,
\quad
\vec x^{I} = (x_{1}^{I},x_{2}^{I},x_{3}^{I})
\end{aligned}
\end{equation}
where each component of these potentials is defined as in (\ref{Gibbons Hawking potentials pure}).

%%%%%%%%%%%%%%%%%%%%%%%%%%%%%%

\subsection{Non-perturbative corrections and smoothness of the metric}

\paragraph{}

The non-perturbative corrections to the metric are given by the following iterative expansion of the integral equation:
\begin{equation}
\begin{aligned}
\label{RH correction}
\delta\log\mathcal{X}_{\gamma}^{(n)}(\zeta) & =
-\frac{1}{2\pi i} \sum_{\gamma'\in\Gamma(\vec a)}
\Omega(\gamma',\vec a) \langle\gamma,\gamma' \rangle
\int_{l_{\gamma'}} \frac{d\zeta'}{\zeta'} \frac{\zeta'+\zeta}{\zeta'-\zeta}
\log\left( 1-\sigma(\gamma) \, \mathcal{X}_{\gamma'}^{(n-1)}(\zeta') \right)
\,,
\\
\mathcal{X}_{\gamma}^{(n)}(\zeta) & = \mathcal{X}_{\gamma}^{\rm sf}
\exp\left( \delta\log\mathcal{X}_{\gamma}^{(n)}(\zeta) \right)
\end{aligned}
\end{equation}
with $n=-1$ meaning the semiflat values ($\mathcal{X}^{(0)}_{\gamma}(\zeta)$ has already been derived above).
To be consistent, one need to show that for each iteration, the associated correction to the moduli space metric is much smaller than for all previous iterations.
Indeed, if $\gamma'$ is a dyon, the integration in (\ref{RH correction}) creates an extra exponentially suppressed term (this can be seen by expanding the logarithm in a Taylor series and taking the limit $g_{\rm eff}\to 0$ and will be demonstrated explicitly for $n=1$ and $n=2$); if $\gamma'$ is a $W$ boson, the correction can be obtained by substituting $\mathcal{X}_{(\vec\alpha_{A},\vec 0)}^{\rm sf}$ in (\ref{D factor}) with $\mathcal{X}_{(\vec\alpha_{A},\vec 0)}^{(n-1)}$, where the latter Darboux coordinate is equal to its semiflat value corrected by exponentially suppressed terms coming from the dyons, leading to an extra small correction to (\ref{symplectic form P}).
This iterative process will be used below to find the contributions from one and two dyons, i.e., we will consider $n\in\{0,1,2\}$.
For our purposes, we can ignore the mixing terms between $W$ bosons and dyons since they are smaller than the one-instanton corrections (because (\ref{symplectic form P}) is smaller than the semiflat metric) and are manifestly continuous in the two-instanton approximation (because all terms involving both $W$ bosons and composite dyons produce corrections smaller than the two-instanton corrections).
In other words, having found $\mathcal{X}_{\gamma}^{(0)}$, which is corrected by the $W$ bosons, instead of the full spectrum $\Gamma(\vec a)$, we will consider only the dyons, denoted collectively as $\tilde\Gamma(\vec a)$.
Then, the superscript $n$ in $\mathcal{X}^{(n)}_{\gamma}(\zeta)$ should be understood as keeping up to $n$-instanton terms in the series expansion.
Of course, the smoothness property of $\mathcal{X}_{\gamma}(\zeta)$ is built in by construction \cite{GMN}, our expansion merely makes this explicit and suitable for the semiclassical instanton checks.

\paragraph{}

Using (\ref{RH correction}), we can write out the explicit expression for $\mathcal{X}_{\gamma'}^{(1)}(\zeta')$:
\begin{equation}
\mathcal{X}_{\gamma'}^{(1)}(\zeta') = \mathcal{X}_{\gamma'}^{(0)}(\zeta')
\exp\left(
\frac{1}{2\pi i} \sum_{\gamma''\in\tilde\Gamma(\vec a)}
\Omega(\gamma'',\vec a) \langle\gamma',\gamma''\rangle
\int_{l_{\gamma''}} \frac{d\zeta''}{\zeta''} \frac{\zeta''+\zeta'}{\zeta''-\zeta'}
\sum_{l=1}^{+\infty} \frac{1}{l} \left( \sigma(\gamma'')\mathcal{X}_{\gamma''}^{(0)}(\zeta'') \right)^{l} \right)
\,.
\end{equation}
In our two-instanton calculation, it is sufficient to set $l=1$.
Furthermore, since $|\delta\log\mathcal{X}^{(1)}_{\gamma'}(\zeta')|$ is small, we make the following approximation:
\begin{equation}
\label{Darboux 1}
\mathcal{X}^{(1)}_{\gamma'}(\zeta') \approx
\mathcal{X}^{(0)}_{\gamma'}(\zeta')
\left( 1+\delta\log\mathcal{X}^{(1)}_{\gamma'}(\zeta') \right)
\,,
\end{equation}
where we are only keeping up to two-instanton terms in the expansion.
Higher order terms in the $\exp(\delta\log\mathcal{X}^{(1)}(\zeta'))$ expansion will contribute as $n>2$ instanton terms.

\paragraph{}

The non-perturbative corrections to the symplectic form include terms corresponding to one dyon and two dyons (distinguished by an upper index):
\begin{equation}
\label{2 dyons correction}
\omega^{\rm NP}(\zeta) \approx \omega^{\rm NP(1)}(\zeta)+\omega^{\rm NP(2)}(\zeta)+\omega^{\rm NP(\tilde 2)}(\zeta)
\end{equation}
where the first two terms and the last term are
\begin{equation}
\begin{aligned}
\omega^{\rm NP(1)}(\zeta) + \omega^{\rm NP(2)}(\zeta) = -\frac{1}{4\pi^{2}R} \sum_{I=1}^{r}
\left(
d\log\mathcal{X}_{e}^{I\,(0)}(\zeta)
\wedge
d\delta\log\mathcal{X}_{m\,I}^{(2)}(\zeta)
\right.
\\
\left.
+
d\delta\log\mathcal{X}_{e}^{I\,(2)}(\zeta)
\wedge
d\log\mathcal{X}_{m\,I}^{(0)}(\zeta)
\right)
\,,
\end{aligned}
\end{equation}
\begin{equation}
\omega^{\rm NP(\tilde 2)}(\zeta) = -\frac{1}{4\pi^{2}R} \sum_{I=1}^{r}
d\delta\log\mathcal{X}_{e}^{I\,(1)}(\zeta)
\wedge
d\delta\log\mathcal{X}_{m\,I}^{(1)}(\zeta)
\,.
\end{equation}
All terms in $\omega^{\rm NP(\tilde 2)}(\zeta)$ are continuous to the leading order because dyons charged under simple roots exist everywhere in the moduli space, and the corresponding integrals do not have singularities.
Therefore, to check that the symplectic form is continuous, $\omega^{\rm NP(\tilde 2)}(\zeta)$ does not have to be calculated.

\paragraph{}

Let us start by calculating the first two terms in (\ref{2 dyons correction}):
\begin{equation}
\label{discontinuous correction}
\begin{aligned}
\omega^{\rm NP(1)}(\zeta)+\omega^{\rm NP(2)}(\zeta)
=
-\frac{1}{4\pi^{2}R} \sum_{\gamma'\in\tilde\Gamma(\vec a)}
\frac{d\mathcal{X}_{\gamma'}^{(0)}(\zeta)}{\mathcal{X}_{\gamma'}^{(0)}(\zeta)} \,
\wedge \,
\frac{\Omega(\gamma',\vec a)}{2\pi i}
\int_{l_{\gamma'}} \frac{d\zeta'}{\zeta'} \frac{\zeta'+\zeta}{\zeta'-\zeta}
\\
\sum_{k=1}^{+\infty} \left( \sigma(\gamma') \, \mathcal{X}_{\gamma'}^{(1)}(\zeta') \right)^{k}
\frac{d\mathcal{X}_{\gamma'}^{(1)}(\zeta')}{\mathcal{X}_{\gamma'}^{(1)}(\zeta')}
\,.
\end{aligned}
\end{equation}
This expression implicitly contains $(1+\delta\log\mathcal{X}^{(1)}_{\gamma'}(\zeta'))$ in $\mathcal{X}_{\gamma'}^{(1)}(\zeta')$ (\ref{Darboux 1}): substituting 1, we obtain $\omega^{\rm NP(1)}(\zeta)$; substituting the exponentially suppressed expression $\delta\log\mathcal{X}^{(1)}_{\gamma'}(\zeta')$, we obtain $\omega^{\rm NP(2)}(\zeta)$.
Generally speaking, the spectrum $\tilde\Gamma(\vec{a})$ here consists of both simple and composite dyons; $k$ is the winding number of the dyon world line over the compactified $S^{1}$.
In order to check that the sum of one- and two-instanton terms is continuous, we need to consider only the singly wound states, i.e., the terms with $k=1$.
To the leading order, $\omega^{\rm NP(1)}(\zeta)$ is a series of terms proportional to the instanton suppression factor $\exp\left(-2\pi kR |Z_{\gamma'}|+ik\theta_{\gamma'}\right)$:
\begin{equation}
\label{symplectic form NP1}
\begin{aligned}
\omega^{\rm NP(1)}(\zeta)
=
-\frac{1}{4\pi^2 R} \, \frac{1}{2\pi i}
\sum_{\gamma'\in\tilde\Gamma(\vec a)}
\Omega(\gamma',\vec a) \,
\frac{d\mathcal{X}_{\gamma'}^{(0)}(\zeta)}{\mathcal{X}_{\gamma'}^{(0)}(\zeta)} \,
\wedge \,
\int_{l_{\gamma'}} \frac{d\zeta'}{\zeta'} \frac{\zeta'+\zeta}{\zeta'-\zeta}
\\
\sum_{k=1}^{+\infty} \left( \sigma(\gamma') \, \mathcal{X}_{\gamma'}^{(0)}(\zeta') \right)^{k}
d\log\mathcal{X}_{\gamma'}^{(0)}(\zeta')
\,.
\end{aligned}
\end{equation}
The two-instanton correction $\omega^{\rm NP(2)}(\zeta)$ in (\ref{discontinuous correction}) contains integrating along two different BPS rays:
\begin{equation}
\label{symplectic form NP2}
\begin{aligned}
\omega^{\rm NP(2)}(\zeta)
= &
-\frac{1}{4\pi^{2}R} \left( \frac{1}{2\pi i} \right)^{2}
\sum_{\{\gamma',\gamma''\}\subset\tilde\Gamma(\vec a)}
\Omega(\gamma',\vec a) \, \Omega(\gamma'',\vec a) \, \sigma(\gamma') \, \sigma(\gamma'') \, \langle\gamma',\gamma''\rangle
\\
& \frac{d\mathcal{X}_{\gamma'}^{(0)}(\zeta)}{\mathcal{X}_{\gamma'}^{(0)}(\zeta)} \,
\wedge
\int_{l_{\gamma'}} \frac{d\zeta'}{\zeta'} \frac{\zeta'+\zeta}{\zeta'-\zeta} \,
\int_{l_{\gamma''}} \frac{d\zeta''}{\zeta''} \frac{\zeta''+\zeta'}{\zeta''-\zeta'} \,
\left(
\mathcal{X}_{\gamma''}^{(0)}(\zeta'') \, d\mathcal{X}_{\gamma'}^{(0)}(\zeta') +
\mathcal{X}_{\gamma'}^{(0)}(\zeta') \, d\mathcal{X}_{\gamma''}^{(0)}(\zeta'')
\right)
\,.
\end{aligned}
\end{equation}
For later convenience, we decompose (\ref{symplectic form NP1}) and (\ref{symplectic form NP2}) as
\begin{equation}
\omega^{\rm NP(1)}(\zeta) = \sum_{\gamma'\in\tilde\Gamma(\vec a)} \omega_{\gamma'}^{\rm NP(1)}(\zeta)
\,,
\quad
\omega^{\rm NP(2)}(\zeta) = \sum_{\{\gamma',\gamma''\}\subset\tilde\Gamma(\vec a)} \omega_{\gamma',\gamma''}^{\rm NP(2)}(\zeta)
\,.
\end{equation}

\paragraph{}

For the continuous terms in (\ref{2 dyons correction}), the expression can be expanded as
\begin{equation}
\label{symplectic form NP2t}
\begin{aligned}
\omega^{\rm NP(\tilde 2)}(\zeta)
= &
\frac{1}{4\pi^{2}R} \left( \frac{1}{2\pi i} \right)^{2}
\sum_{\{\gamma',\gamma''\}\subset\tilde\Gamma(\vec a)}
\Omega(\gamma',\vec a) \, \Omega(\gamma'',\vec a) \, \sigma(\gamma') \, \sigma(\gamma'') \,
\sum_{I=1}^{r}
\gamma_{m}^{I}{}' \, \gamma_{e\,I}''
\\
& \int_{l_{\gamma'}} \frac{d\zeta'}{\zeta'} \frac{\zeta'+\zeta}{\zeta'-\zeta}
\frac{\mathcal{X}^{(0)}_{\gamma'}(\zeta')}{1-\mathcal{X}^{(0)}_{\gamma'}(\zeta')}
\frac{d\mathcal{X}_{\gamma'}^{(0)}(\zeta')}{\mathcal{X}_{\gamma'}^{(0)}(\zeta')}
\wedge
\int_{l_{\gamma''}}\frac{d\zeta''}{\zeta''}\frac{\zeta''+\zeta}{\zeta''-\zeta}
\frac{\mathcal{X}^{(0)}_{\gamma''}(\zeta'')}{1-\mathcal{X}^{(0)}_{\gamma''}(\zeta'')}
\frac{d\mathcal{X}_{\gamma''}^{(0)}(\zeta'')}{\mathcal{X}_{\gamma''}^{(0)}(\zeta'')}
\,.
\end{aligned}
\end{equation}

\paragraph{}

We need to find out how (\ref{symplectic form NP2}) changes at the wall, when $l_{\gamma''}$ goes through $l_{\gamma'}$ clockwise.
First of all, at the wall, $Z_{\gamma'}/Z_{\gamma''}\in\mathbb{R}_{+}$, and there is a singularity at $\zeta''=\zeta'$, so that (\ref{symplectic form NP2}) diverges.
The only components that jump are $\omega_{\gamma',\gamma''}^{\rm NP(2)}$ and $\omega_{\gamma'',\gamma'}^{\rm NP(2)}$. Let us restrict our attention to $\omega_{\gamma',\gamma''}^{\rm NP(2)}$.
When we cross the wall, the contour $l_{\gamma''}$ passes through the singularity.
The jump of the second (internal) integral in (\ref{symplectic form NP2}) is given by the residue at $\zeta''=\zeta'$:
\begin{equation}
\label{wall residue}
\begin{aligned}
\frac{1}{2\pi i} &
\left( \lim_{\arg\frac{Z_{\gamma''}}{Z_{\gamma'}}\to 0-} - \lim_{\arg\frac{Z_{\gamma''}}{Z_{\gamma'}}\to 0+} \right)
\left(
\int_{l_{\gamma''}} \frac{d\zeta''}{\zeta''} \frac{\zeta''+\zeta'}{\zeta''-\zeta'} \,
\left(
\mathcal{X}_{\gamma''}^{(0)}(\zeta'') \, d\mathcal{X}_{\gamma'}^{(0)}(\zeta') +
\mathcal{X}_{\gamma'}^{(0)}(\zeta') \, d\mathcal{X}_{\gamma''}^{(0)}(\zeta'')
\right)
\right)
=
\\
& \res_{\zeta''=\zeta'}
\left(
\frac{\zeta''+\zeta'}{\zeta''}
\left(
\mathcal{X}_{\gamma''}^{(0)}(\zeta'') \, d\mathcal{X}_{\gamma'}^{(0)}(\zeta') +
\mathcal{X}_{\gamma'}^{(0)}(\zeta') \, d\mathcal{X}_{\gamma''}^{(0)}(\zeta'')
\right)
\right)
=
2 \, d\mathcal{X}_{\gamma'+\gamma''}^{(0)}(\zeta')
\,.
\end{aligned}
\end{equation}
The jump of $\omega_{\gamma'',\gamma'}^{\rm NP(2)}$ is the same as the jump of $\omega_{\gamma',\gamma''}^{\rm NP(2)}$ (symplectic product and contour orientation both give an extra minus factor).

\paragraph{}

In particular, suppose that we cross the wall where a dyon with charge $\gamma_{1}+\gamma_{2}$ changes its multiplicity by $\Delta\Omega(\gamma_{1}+\gamma_{2},\vec a)$.
To ensure smoothness of the metric, one needs to make sure that
\begin{equation}
\label{continuity condition}
\left(
\lim_{\arg\frac{Z_{\gamma_{2}}}{Z_{\gamma_{1}}}\to 0+} -
\lim_{\arg\frac{Z_{\gamma_{2}}}{Z_{\gamma_{1}}}\to 0-}
\right)
\left(
\omega_{\gamma_{1}+\gamma_{2}}^{\rm NP(1)}(\zeta)+
\omega_{\gamma_{1},\gamma_{2}}^{\rm NP(2)}(\zeta)+
\omega_{\gamma_{2},\gamma_{1}}^{\rm NP(2)}(\zeta)
\right)
= 0
\,.
\end{equation}
This condition imposes a constraint on multiplicities on both sides of the wall.
After finding this constraint, we will see that it is indeed satisfied by the pentagon wall-crossing formulae.

\paragraph{}

Let us calculate the jumps of $\omega_{\gamma_{1},\gamma_{2}}^{\rm NP(2)}(\zeta)$ and $\omega_{\gamma_{2},\gamma_{1}}^{\rm NP(2)}(\zeta)$ when the VEV crosses the wall.
In order to employ our results in (\ref{symplectic form NP2}) and (\ref{wall residue}), we need to set $\gamma'=\gamma_{1}$ and $\gamma''=\gamma_{2}$.
Using the fact that at the wall, $l_{\gamma_{1}}$, $l_{\gamma_{2}}$, and $l_{\gamma_{1}+\gamma_{2}}$ coincide, we see that the jump of $\omega^{\rm NP(2)}(\zeta)$ can be written as
\begin{equation}
\begin{aligned}
\left( \lim_{\arg\frac{Z_{\gamma_{2}}}{Z_{\gamma_{1}}}\to 0-} - \lim_{\arg\frac{Z_{\gamma_{2}}}{Z_{\gamma_{1}}}\to 0+} \right)
\left(
\omega_{\gamma_{1},\gamma_{2}}^{\rm NP(2)}(\zeta) +
\omega_{\gamma_{2},\gamma_{1}}^{\rm NP(2)}(\zeta)
\right)
=
- \, \Omega(\gamma_{1},\vec a) \, \Omega(\gamma_{2},\vec a) \, \sigma(\gamma_{1}) \, \sigma(\gamma_{2}) \, 2 \, \langle\gamma_{1},\gamma_{2}\rangle
\\
\frac{1}{4\pi^{2}R} \, \frac{1}{2\pi i} \,
\frac{d\mathcal{X}_{\gamma_{1}+\gamma_{2}}^{(0)}(\zeta)}{\mathcal{X}_{\gamma_{1}+\gamma_{2}}^{(0)}(\zeta)} \,
\wedge
\int_{l_{\gamma_{1}+\gamma_{2}}} \frac{d\zeta'}{\zeta'} \frac{\zeta'+\zeta}{\zeta'-\zeta} \,
d\mathcal{X}_{\gamma_{1}+\gamma_{2}}^{(0)}(\zeta')
\,.
\end{aligned}
\end{equation}
The increment of $\omega^{\rm NP(1)}(\zeta)$ across the wall can be easily seen from (\ref{symplectic form NP1}) setting $\gamma'=\gamma_{1}+\gamma_{2}$:
\begin{equation}
\begin{aligned}
\left( \lim_{\arg\frac{Z_{\gamma_{2}}}{Z_{\gamma_{1}}}\to 0-} - \lim_{\arg\frac{Z_{\gamma_{2}}}{Z_{\gamma_{1}}}\to 0+} \right)
\omega_{\gamma_{1}+\gamma_{2}}^{\rm NP(1)}(\zeta)
=
- \, \Delta\Omega(\gamma_{1}+\gamma_{2},\vec a) \, \sigma(\gamma_{1}+\gamma_{2})
\\
\frac{1}{4\pi^{2}R} \, \frac{1}{2\pi i} \,
\frac{d\mathcal{X}_{\gamma_{1}+\gamma_{2}}^{(0)}(\zeta)}{\mathcal{X}_{\gamma_{1}+\gamma_{2}}^{(0)}(\zeta)} \,
\wedge
\int_{l_{\gamma_{1}+\gamma_{2}}} \frac{d\zeta'}{\zeta'} \frac{\zeta'+\zeta}{\zeta'-\zeta} \,
d\mathcal{X}_{\gamma_{1}+\gamma_{2}}^{(0)}(\zeta')
\,.
\end{aligned}
\end{equation}
Using the relation between two quadratic refinements,
\begin{equation}
\sigma(\gamma_{1})\,\sigma(\gamma_{2}) = (-1)^{2\langle\gamma_{1},\gamma_{2}\rangle} \sigma(\gamma_{1}+\gamma_{2})
\,,
\end{equation}
we can see that the continuity condition (\ref{continuity condition}) is equivalent to
\begin{equation}
\Delta\Omega(\gamma_{1}+\gamma_{2},\vec a) =
2 \, \langle\gamma_{1},\gamma_{2}\rangle \,
(-1)^{2\langle\gamma_{1},\gamma_{2}\rangle-1} \,
\Omega(\gamma_{1},\vec a) \,
\Omega(\gamma_{2},\vec a)
\,.
\end{equation}
It is satisfied by the pentagon formulae (\ref{SU(3) pentagon}) and (\ref{SU(n) pentagon}):
\begin{equation}
\Delta\Omega(\gamma_{1}+\gamma_{2},\vec a) = \Omega(\gamma_{1},\vec a) = \Omega(\gamma_{2},\vec a) = 1
\,,
\quad
\langle\gamma_{1},\gamma_{2}\rangle = 1/2
\,.
\end{equation}

\paragraph{}

This allows us to conclude that the moduli space metric remains continuous to the two-instanton order across the walls where composite dyons decay.
This analysis can be repeated to ensure that higher-instanton mixing terms are smooth across the walls by expanding the higher $\mathcal{X}^{(n)}_{\gamma}(\zeta)$ terms.

%%%%%%%%%%%%%%%%%%%%%%%%%%%%%%

\subsection{Saddle-point approximation of the metric}

\paragraph{}

Knowing the general expressions for one- and two-instanton corrections (\ref{symplectic form NP1}, \ref{symplectic form NP2}, \ref{symplectic form NP2t}), we can extract the moduli space metric using the saddle-point approximation.
To approximate (\ref{symplectic form NP2}), this method can only be used far from the walls, where the integrands do not have poles near the contour of integration.
The approximation that we will be using is (\ref{saddle point}).
For the terms of first type (\ref{symplectic form NP1}), the peak is at $\zeta'=-Z_{\gamma'}/|Z_{\gamma'}|=-e^{i\phi_{\gamma'}}$ (where $\phi_{\gamma'}$ is the complex argument of $Z_{\gamma'}$).
Reusing our one-instanton results from chapter \ref{ch: pure}, we have
\begin{equation}
\label{symplectic form NP1 saddle}
\begin{aligned}
\omega^{\rm NP(1)}(\zeta)
= &
\frac{i}{8\pi^{2}}
\sum_{\gamma'\in\tilde\Gamma(\vec a)}
\sum_{k=1}^{+\infty}
\left( \mathcal{D}_{\gamma'}(-e^{i\phi_{\gamma'}}) \right)^{k} \,
\frac{1}{\sqrt{kR|Z_{\gamma'}|}} \exp\left( -2\pi kR |Z_{\gamma'}|+ik\theta_{\gamma'} \right)
\\
& \frac{d\mathcal{X}_{\gamma'}^{\rm sf}(\zeta)}{\mathcal{X}_{\gamma'}^{\rm sf}(\zeta)}
\wedge
\left(
|Z_{\gamma'}| \left( \frac{dZ_{\gamma'}}{Z_{\gamma'}}-\frac{d\bar{Z}_{\gamma'}}{\bar{Z}_{\gamma'}} \right)
-\left( \frac{dZ_{\gamma'}}{\zeta}-\zeta d\bar{Z}_{\gamma'} \right)
\right)
\,,
\end{aligned}
\end{equation}
where the global definition of $\vec\theta_{m}$ leads to the shift
\begin{equation}
\vec\theta_{m}\to\vec\theta_{m}+\re\hat\tau_{\rm eff} \, \vec\theta_{e}
\end{equation}
in $\theta_{\gamma}$ in order to define it consistently at infinity.
Now, let us approximate the two-instanton terms, (\ref{symplectic form NP2}) and (\ref{symplectic form NP2t}).
Using the same saddle-point approximation (\ref{saddle point}) in both integrals here (the maxima of the integrands are at $\zeta'=-Z_{\gamma'}/|Z_{\gamma'}|=-e^{i\phi_{\gamma'}}$ and $\zeta''=-Z_{\gamma''}/|Z_{\gamma''}|=-e^{i\phi_{\gamma''}}$), we see that both two-instanton terms behave as $\exp\left( -2\pi R(|Z_{\gamma'}|+|Z_{\gamma''}|) \right)$, correctly reproducing the two-instanton action.
At weak coupling, when masses of dyons are large, (\ref{symplectic form NP2}) and (\ref{symplectic form NP2t}) give higher order corrections with respect to (\ref{symplectic form NP1 saddle}).
Computing the third component of the symplectic form,
\begin{equation}
\omega_{3} = \frac{\omega(i)+\omega(-i)}{2}
\,,
\end{equation}
we express the contribution for two dyons in terms of their central charges:
\begin{equation}
\label{symplectic form 3 NP2}
\begin{aligned}
\omega_{3}^{\rm NP(2)}
= &
-\frac{1}{4\pi^{2}R} \left( \frac{1}{2\pi i} \right)^{2}
\sum_{\{\gamma',\gamma''\}\subset\tilde\Gamma(\vec a)}
\mathcal{S}_{\gamma',\gamma''} \,
\frac{1}{R\sqrt{|Z_{\gamma'}Z_{\gamma''}|}} \,
\sigma(\gamma') \, \sigma(\gamma'') \,
\frac{\vec\gamma_{m}' \, \vec\gamma_{e}'' - \vec\gamma_{e}' \, \vec\gamma_{m}''}{2} \,
\frac{e^{i\phi_{\gamma'}}+e^{i\phi_{\gamma''}}}{e^{i\phi_{\gamma'}}-e^{i\phi_{\gamma''}}}
\\
&
\left(
\frac{e^{i\phi_{\gamma'}}-i}{e^{i\phi_{\gamma'}}+i}
\left( i\pi R\left( -dZ_{\gamma'}+d\bar Z_{\gamma'} \right)+id\theta_{\gamma'} \right)
+
\frac{e^{i\phi_{\gamma'}}+i}{e^{i\phi_{\gamma'}}-i}
\left( i\pi R\left( dZ_{\gamma'}-d\bar Z_{\gamma'} \right)+id\theta_{\gamma'} \right)
\right)
\wedge
\\
&
\left(
-\pi R\left( e^{-i\phi_{\gamma'}}dZ_{\gamma'}+e^{i\phi_{\gamma'}}d\bar Z_{\gamma'} \right)
-\pi R\left( e^{-i\phi_{\gamma''}}dZ_{\gamma''}+e^{i\phi_{\gamma''}}d\bar Z_{\gamma''} \right) +
id\theta_{\gamma'+\gamma''}
\right)
\,,
\end{aligned}
\end{equation}
\begin{equation}
\label{symplectic form 3 NP2t}
\begin{aligned}
\omega_{3}^{\rm NP(\tilde 2)}
= &
\frac{1}{4\pi^{2}R} \left( \frac{1}{2\pi i} \right)^{2}
\sum_{\{\gamma',\gamma''\}\subset\tilde\Gamma(\vec a)}
\mathcal{S}_{\gamma',\gamma''} \,
\frac{1}{R\sqrt{|Z_{\gamma'}Z_{\gamma''}|}} \,
\sigma(\gamma') \, \sigma(\gamma'') \,
\frac{\vec\gamma_{m}' \, \vec\gamma_{e}''}{2}
\\
&
\left(
\frac{(e^{i\phi_{\gamma'}}-i)(e^{i\phi_{\gamma''}}-i)}{(e^{i\phi_{\gamma'}}+i)(e^{i\phi_{\gamma''}}+i)} +
\frac{(e^{i\phi_{\gamma'}}+i)(e^{i\phi_{\gamma''}}+i)}{(e^{i\phi_{\gamma'}}-i)(e^{i\phi_{\gamma''}}-i)}
\right)
\\
&
\left(
-\pi R\left( e^{-i\phi_{\gamma'}}dZ_{\gamma'}+e^{i\phi_{\gamma'}}d\bar Z_{\gamma'} \right) +
id\theta_{\gamma'}
\right)
\wedge
\left(
-\pi R\left( e^{-i\phi_{\gamma''}}dZ_{\gamma''}+e^{i\phi_{\gamma''}}d\bar Z_{\gamma''} \right) +
id\theta_{\gamma''}
\right)
\end{aligned}
\end{equation}
where the common factor for two dyons reproducing the action and non-zero modes determinants is
\begin{equation}
\mathcal{S}_{\gamma',\gamma''} =
\mathcal{D}_{\gamma'}(-e^{i\phi_{\gamma'}}) \, \mathcal{D}_{\gamma''}(-e^{i\phi_{\gamma''}})
\exp\left( -2\pi R(|Z_{\gamma'}|+|Z_{\gamma''}|)+i\theta_{\gamma'+\gamma''} \right)
\,.
\end{equation}
Note that (\ref{symplectic form 3 NP2}) is applicable only far from the walls of marginal stability: it diverges at the walls, where the saddle-point approximation cannot be used; (\ref{symplectic form 3 NP2t}) has no singularities at the walls.

\paragraph{}

Let us extract the dominant metric components, $g_{a^{I}\bar a^{J}}$, from these symplectic forms.
At weak coupling, all central charges can be approximated as
\begin{align}
Z_{\gamma} & = \vec\gamma_{e}\vec a+\vec\gamma_{m}\hat\tau_{\rm eff}\vec a
\,,
\\
\hat\tau_{\rm eff} & \simeq \frac{i}{\pi} \sum_{\vec\alpha_{A}\in\Phi_{+}}
\vec\alpha_{A}\otimes\vec\alpha_{A}
\log\left( \frac{\vec\alpha_{A}\vec a}{\Lambda} \right)^{2}
\,.
\end{align}
Furthermore, everywhere, except the exponents, we can approximate central charges for dyons as
\begin{equation}
Z_{\gamma}\simeq\vec\gamma_{m}(i\im\hat\tau_{\rm eff})\vec a
\,,
\quad
\bar Z_{\gamma}\simeq -\vec\gamma_{m}(i\im\hat\tau_{\rm eff})\bar{\vec a}
\,.
\end{equation}
For the symplectic product of central charges, we have
\begin{equation}
\begin{aligned}
dZ_{\gamma'}\wedge d\bar Z_{\gamma''} & \simeq
(\vec\gamma_{m}'\im\hat\tau_{\rm eff})_{I} \, (\vec\gamma_{m}''\im\hat\tau_{\rm eff})_{J} \,
da^{I}\wedge d\bar a^{J}
\,,
\\
\im\hat\tau_{\rm eff} & \simeq
\frac{2}{\pi} \sum_{\vec\alpha_{A}\in\Phi_{+}}
\vec\alpha_{A}\otimes\vec\alpha_{A} \,
\log\left| \frac{\vec\alpha_{A}\vec a}{\Lambda} \right|
\,.
\end{aligned}
\end{equation}
The resulting correction for single dyons is
\begin{equation}
\label{metric NP1}
\begin{aligned}
g_{a^{I}\bar a^{J}}^{\rm NP(1)}
= &
\frac{1}{4\pi}
\sum_{\gamma'\in\tilde\Gamma(\vec a)}
\sum_{k=1}^{+\infty}
\left( \mathcal{D}_{\gamma'}(-e^{i\phi_{\gamma'}}) \right)^{k}
\exp\left( -2k\pi R |Z_{\gamma'}|+ik\theta_{\gamma'} \right)
\sqrt{\frac{R}{k|Z_{\gamma'}|}}
\\
&
(\vec\gamma_{m}'\im\hat\tau_{\rm eff})_{I} \, (\vec\gamma_{m}'\im\hat\tau_{\rm eff})_{J}
\,.
\end{aligned}
\end{equation}
In (\ref{symplectic form 3 NP2t}), since the wedge-product is antisymmetric, we substitute $\vec\gamma_{m}'\vec\gamma_{e}''\to(\vec\gamma_{m}'\vec\gamma_{e}''-\vec\gamma_{e}'\vec\gamma_{m}'')/2$.
After some tedious but straightforward calculations, we obtain the dominant components of the moduli space metric coming from pairs of dyons:
\begin{equation}
\label{metric NP2}
\begin{aligned}
g_{a^{I}\bar a^{J}}^{\rm NP(2)}
= &
-\frac{1}{16\pi^{2}}
\sum_{\{\gamma',\gamma''\}\subset\tilde\Gamma(\vec a)}
\mathcal{S}_{\gamma',\gamma''} \,
\frac{1}{\sqrt{|Z_{\gamma'}Z_{\gamma''}|}} \,
\sigma(\gamma') \, \sigma(\gamma'') \,
i(\vec\gamma_{m}' \, \vec\gamma_{e}'' - \vec\gamma_{e}' \, \vec\gamma_{m}'') \,
\frac{e^{i\phi_{\gamma'}}+e^{i\phi_{\gamma''}}}{e^{i\phi_{\gamma'}}-e^{i\phi_{\gamma''}}}
\\
&
\left(
2 \, (\vec\gamma_{m}'\im\hat\tau)_{I} \, (\vec\gamma_{m}'\im\hat\tau)_{J} +
\frac{\exp(i\phi'')}{\cos\phi'} \, (\vec\gamma_{m}'\im\hat\tau)_{I} \, (\vec\gamma_{m}''\im\hat\tau)_{J} +
\right.
\\
& \qquad
\left.
\frac{\exp(-i\phi'')}{\cos\phi'} \, (\vec\gamma_{m}''\im\hat\tau)_{I} \, (\vec\gamma_{m}'\im\hat\tau)_{J}
\right)
\,,
\end{aligned}
\end{equation}
\begin{equation}
\label{metric NP2t}
\begin{aligned}
g_{a^{I}\bar a^{J}}^{\rm NP(\tilde 2)}
= &
\frac{1}{32\pi^{2}}
\sum_{\{\gamma',\gamma''\}\subset\tilde\Gamma(\vec a)}
\mathcal{S}_{\gamma',\gamma''} \,
\frac{1}{\sqrt{|Z_{\gamma'}Z_{\gamma''}|}} \,
\sigma(\gamma') \, \sigma(\gamma'') \,
i(\vec\gamma_{m}' \, \vec\gamma_{e}'' - \vec\gamma_{e}' \, \vec\gamma_{m}'') \,
\exp(-i\phi_{\gamma'}+i\phi_{\gamma''})
\\
&
\left(
\frac{(e^{i\phi_{\gamma'}}-i)(e^{i\phi_{\gamma''}}-i)}{(e^{i\phi_{\gamma'}}+i)(e^{i\phi_{\gamma''}}+i)} +
\frac{(e^{i\phi_{\gamma'}}+i)(e^{i\phi_{\gamma''}}+i)}{(e^{i\phi_{\gamma'}}-i)(e^{i\phi_{\gamma''}}-i)}
\right)
(\vec\gamma_{m}'\im\hat\tau)_{I} \, (\vec\gamma_{m}''\im\hat\tau)_{J}
\,.
\end{aligned}
\end{equation}
The reality condition for these expressions can be checked using the fact that these summations are symmetric under $\gamma'\to-\gamma'$, $\gamma''\to-\gamma''$.

\paragraph{}

Let us find the perturbative one-loop factor extracted from \cite{GMN}, i.e., $\mathcal{D}_{\gamma}(\zeta)$ in (\ref{D factor 2}), explicitly; then, we will explain how it can be reproduced from semiclassical analysis.
First, we notice that in the semiclassical limit, the phase $\phi_{\gamma}$ is given via
\begin{equation}
\exp(i\phi_{\gamma}) =
\frac{(\gamma_{e\,I}+\tau_{{\rm eff}\,IJ}\gamma_{m}^{J})a^{I}}{|(\gamma_{e\,I}+\tau_{{\rm eff}\,IJ}\gamma_{m}^{J})a^{I}|} \simeq
\frac{\tau_{{\rm eff}\,IJ}\gamma_{m}^{J}a^{I}}{|\tau_{{\rm eff}\,IJ}\gamma_{m}^{J}a^{I}|}
\,.
\end{equation}
This differs from the rank one case where $\exp(i\phi_{\gamma})\simeq ia/|a|$: in the $SU(n)$ case, even in the semiclassical limit, the phase $\phi_{\gamma}$ remains different for monopoles and dyons charged under different roots, and so, we need to carefully re-evaluate the one-loop factors.

\paragraph{}

For a given monopole $\gamma_{A}=(\vec 0,\vec\alpha_{A})$ charged under root $\vec\alpha_{A}$ (simple or composite), we can split the summation over different $W$ bosons into the term where the boson is charged under $\vec\alpha_{A}$ and all other terms where the boson is charged under $\vec\alpha_{B\ne A}$ roots.
We can then rewrite $\log\mathcal{D}_{(\vec\gamma_{e},\vec\gamma_{m})}(\zeta)$ (for any charges $(\vec\gamma_{e},\vec\gamma_{m})$) at the saddle point $\zeta=-e^{i\phi_{\gamma_{A}}}$ as
\begin{equation}
\label{D factor 3}
\log\mathcal{D}_{(\vec\gamma_{e},\vec\gamma_{m})} (-e^{i\phi_{\gamma_{A}}}) =
\log\mathcal{D}_{(\vec\gamma_{e},\vec\gamma_{m}),A} (-e^{i\phi_{\gamma_{A}}})+
\sum_{B\ne A} \log\mathcal{D}_{(\vec\gamma_{e},\vec\gamma_{m}),B} (-e^{i\phi_{\gamma_{A}}})
\,.
\end{equation}
Introducing $y=e^{t}$ in (\ref{D factor 2}), we re-express the first term (coming from the $W_{A}$ boson and its antiparticle):
\begin{equation}
\label{D factor simple}
\begin{aligned}
\log\mathcal{D}_{(\vec\gamma_{e},\vec\gamma_{m}),A} (-e^{i\phi_{\gamma_{A}}}) & =
\frac {2 \, \vec\alpha_{A}\vec\gamma_{m}}{\pi} \int_{0}^{+\infty} \frac{dt}{\cosh t}
\\
& \left( \log\left( 1-e^{-2\pi R |Z_{W_{A}}|\cosh t+i\theta_{W_{A}}} \right) + \log\left( 1-e^{-2\pi R |Z_{W_{A}}|\cosh t-i\theta_{W_{A}}} \right) \right)
\,.
\end{aligned}
\end{equation}
This term is analogous to the $SU(2)$ one-loop factor evaluated in chapter \ref{ch: pure}.
For generic gauge group, $\mathcal{D}_{(\vec\gamma_{e},\vec\gamma_{m})}(\zeta)$ also has contributions from other roots $\vec\alpha_{B\ne A}$.
To calculate them, we set $\vec\alpha_{B}$ in the summation in (\ref{D factor 2}) and substitute $y=e^{t}$ as above, then, at the saddle point $\zeta=-e^{i\phi_{\gamma_{A}}}$, we express these terms in terms of complex phases $\phi_{\gamma_{A}}$ and $\phi_{W_{B}}$:
\begin{equation}
\label{D factor mixed}
\begin{aligned}
\log\mathcal{D}_{(\vec\gamma_{e},\vec\gamma_{m}),B} (-e^{i\phi_{\gamma_{A}}}) & =
\frac{2 \, \vec\alpha_{A}\vec\gamma_{m}}{\pi} \int_{0}^{+\infty} dt \, \rho(t,\Delta\phi)
\\
& \left( \log\left( 1-e^{-2\pi R |Z_{W_{B}}|\cosh t+i\theta_{W_{B}}} \right)+\log\left( 1-e^{-2\pi R |Z_{W_{B}}|\cosh t-i\theta_{W_{B}}} \right) \right)
\end{aligned}
\end{equation}
where the integration kernel is given by
\begin{align}
\label{D factor kernel}
\rho(t,\Delta\phi) & =
\frac{\cosh t\cos\Delta\phi}{\cosh^{2}t-\sin^{2}\Delta \phi} =
\frac{\cos\Delta\phi}{2}\left( \frac{1}{\cosh t-\sin\Delta\phi}+\frac{1}{\cosh t+\sin\Delta\phi} \right)
\,,
\\
\Delta\phi & = \phi_{W_{B}}-\phi_{\gamma_{A}}+\frac{\pi}{2}=\phi_{W_{B}}-\phi_{W_{A}}
\end{align}
(the case that we are dealing with is $(\vec\gamma_{e},\vec\gamma_{m})=\gamma_{A}$). 
In three dimensions, $(\re\vec a,\im\vec a,\vec\theta_{e}/2\pi R)$ form a vector of enhanced $SO(3)$ triplets, and the one-loop factor should be invariant under such rotations.
In the next section, we will use this property to match this expression with the semiclassical result for non-zero mode fluctuations.

%%%%%%%%%%%%%%%%%%%%%%%%%%%%%%%%%%%%%%%%%%%%%%%%%%%%%%%%%%%%%

\section{Semiclassical derivation of one-loop determinants}

\paragraph{}

In chapter \ref{ch: pure}, it was shown that the one-loop factor from $W$ bosons with electric charges $\pm\vec\alpha_{A}$, i.e., $\mathcal{D}_{\gamma_{A},A}(-e^{i\phi_{\gamma_{A}}})$ in (\ref{D factor simple}), can be derived directly by considering the non-zero mode fluctuations around the associated $SU(2)$ monopole.
To see how additional contributions $\mathcal{D}_{\gamma_{A},B\ne A}(-e^{i\phi_{\gamma_{A}}})$ in (\ref{D factor mixed}, \ref{D factor kernel}) can also be obtained from semiclassical analysis, the key is to adapt the difference of the densities of states $\delta\rho_{A}(x^{2})$ in the pure $SU(2)$ theory to the $SU(n)$ case.
We can work this out by considering the index function
\begin{equation}
\mathcal{I}(\mu^{2}) = \sum_{B} \mathcal{I}_{B}(\mu^{2})
\end{equation}
(we imply summation over all possible indices $B$ corresponding to positive roots) counting the zero modes in the context of three-dimensional instanton computation for gauge groups of any rank \cite{Fraser Tong}.
For completeness, we first write down the index function for the zero mode fluctuations charged under the same root $\vec\alpha_{A}$:
\begin{equation}
\mathcal{I}_{A}(\mu^{2}) =
\frac{2 M_{W_{A}}}{(M_{W_{A}}^{2}+\mu^{2})^{1/2}}
\end{equation}
where $M_{W_{A}}=|Z_{W_{A}}|$ is the mass of the $W$ boson charged under $\vec\alpha_{A}$.
After compactifying the theory, each of the $r$ components of $\vec v\,^{i}=(\re\vec a,\im\vec a,\vec\theta_{e}/2\pi R)^{i}$ (consisting of three adjoint scalars, as in the case of gauge group $SU(2)$) belongs to a three-dimensional multiplet (with respect to superscript $i$).
We can then define the three-dimensional analogue of the four-dimensional phase angle corresponding to the VEV:
\begin{equation}
\lambda_{B}^{i} = \frac{\vec v\,^{i}\vec\alpha_{B}}{||\vec v\,^{l}\vec\alpha_{B}||_{l}}
\,.
\end{equation}
For the fluctuations charged under $\vec\alpha_{B\ne A}$, a simple manipulation gives the index function (see eq.\ (15) in \cite{Fraser Tong}):
\begin{equation}
\label{index function}
\mathcal{I}_{B}(\mu^{2}) =
\frac{2(\vec\alpha_{A}\cdot\vec\alpha_{B})M_{W_{B}}}{\left( M_{W_{B}}^{2}+\mu^{2} \right)^{1/2}}
\frac{(\lambda_{A}^{i}\lambda_{B}^{i}) \, \mu^{2}}{\left( \mu^{2}+{M_{W_{B}}^{2}}(1-(\lambda_{A}^{i}\lambda_{B}^{i})^{2}) \right)}
\end{equation}
where $M_{W_{B}}=|Z_{W_{B}}|$ is the mass of the $W$ boson charged under $\vec\alpha_{B}$.
We can now use the identity for the index function from \cite{Kaul} to derive the difference in the density of states in this case:
\begin{equation}
\mathcal{I}_{B}(\mu^{2})-\mathcal{I}_{B}(0) = \int^{+\infty}_{0}dx^{2}\frac{\mu^{2}}{x^{2}+\mu^{2}}\delta \rho_{B}(x^{2})
\,.
\end{equation}
Using our earlier results for gauge group $SU(2)$, we can derive the required $\delta\rho_{B}(x^{2})$:
\begin{equation}
\begin{aligned}
\delta\rho_{B}(x^{2}) = & -\frac {2(\vec\alpha_{A}\cdot\vec\alpha_{B})M_{W_{B}}}{\pi}\frac{\theta(x^{2}-M_{W_{B}}^{2})}{x^{2}(x^{2}-M_{W_{B}}^{2})^{1/2}}
\frac{x^{2}(\lambda_{A}^{i}\lambda_{B}^{i})}{x^{2}-M_{W_{B}}^{2}(\lambda_{A}^{i}\lambda_{B}^{i})^{2}}
\\
& +2\delta(x^{2}-M_{W_{B}}^{2}(1-(\lambda_{A}^{i}\lambda_{B}^{i})^{2}))
\end{aligned}
\end{equation}
where $\theta(y)$ is a step function.
We can now set $x=M_{W_{B}}\cosh t$ and rearrange $dx^{2}\delta\rho_{B}(x^{2})$ as
\begin{equation}
\int^{+\infty}_{0} dx^{2}\delta\rho_{B}(x^{2}) = -\frac{4(\vec\alpha_{A}\cdot\vec\alpha_{B})}{\pi} \int^{+\infty}_{0} dt \, \frac{\cosh t \, (\lambda_{A}\lambda_{B})}{\cosh^{2}t-(1-(\lambda_{A} \lambda_{B})^{2})}
\,.
\end{equation}
By using the $SO(3)$ symmetry to rotate into the vacuum $\vec\theta_{e}=0$, the difference in the densities of states $\delta \rho_{B}(x^{2})$ obtained here for $\mathbb{R}^{3}$ can be identified with the corresponding quantities for $\mathbb{R}^{3}\times S^{1}$. 
From the definition of $\lambda_{A}^{i}$, it follows that
\begin{equation}
\lambda_{A}^{i}\lambda_{B}^{i} = \cos\Delta\phi
\,,
\end{equation}
where $\Delta\phi$ was introduced in (\ref{D factor kernel}), and we see that $dx^{2}\delta\rho_{B}(x^{2})$ can be identified with $dt\rho(t,\Delta\phi)$ given in (\ref{D factor mixed}, \ref{D factor kernel}) up to an overall numerical factor, we also match the scalar product by setting $\vec\gamma_{m}=\vec\alpha_{B}$ in (\ref{D factor mixed}).
At this point, we can repeat our analysis in chapter \ref{ch: pure} where enumeration of non-zero mode fluctuations in the monopole background in $\mathbb{R}^{3}\times S^{1}$ was mapped to the partition function of harmonic oscillators with inverse temperature $2\pi R$ and background chemical potential $\theta_{e}/2\pi R$.
This yields the additional logarithmic integrands appearing in (\ref{D factor mixed}).
The overall factor can be fixed by requiring that for $B=A$, the formula reproduces the $SU(2)$ one-loop factor.
This completes our semiclassical derivation of the additional one-loop factor $\mathcal{D}_{\gamma_{A},B}(-e^{i\phi_{\gamma_{A}}})$.

\paragraph{}

We have calculated the ratio of one-loop determinants in $\mathbb{R}^{3}\times S^{1}$ and matched it with the GMN prediction extracted in the previous section (\ref{D factor 3}, \ref{D factor simple}, \ref{D factor mixed}, \ref{D factor kernel}).
All other steps required to find the one-instanton action and the overall coefficient for the moduli space metric are essentially equivalent to the $SU(2)$ case.
Summing up, we conclude that the one-instanton metric calculated semiclassically coincides with the prediction (\ref{metric NP1}) obtained in the previous section.

%%%%%%%%%%%%%%%%%%%%%%%%%%%%%%%%%%%%%%%%%%%%%%%%%%%%%%%%%%%%%

\section{Interpolating to three dimensions}

\paragraph{}

In \cite{Fraser Tong}, it was shown how the corresponding three-dimensional metric remains smooth as the VEV crosses the wall of marginal stability; this can be shown by Poisson-resumming the metric in $\mathbb{R}^{3}\times S^{1}$. 
In addition, will we also show that in this limit, our one-instanton correction coincides with the result obtained in \cite{Fraser Tong}.

\paragraph{}

We consider the strict three-dimensional limit
\begin{equation}
\label{3D limit}
2\pi R\to 0
\,,
\quad
\left( \re\vec a,\ \im\vec a,\ \frac{\vec\theta_{e}}{2\pi R} \right) = \const
\,.
\end{equation}
The semiclassical one-loop factor is then given by \cite{DKMTV}
\begin{equation}
\mathcal{R}^{\rm (3D)} = \lim_{\kappa\to 0} \left( \kappa^{2}\exp\left( \int_{\kappa}^{+\infty} \frac{d\nu}{\nu} \mathcal{I}(\nu) \right) \right)^{1/2}
\,.
\end{equation}
Substituting the index function $\mathcal{I}_{B}(\mu^{2})$ (\ref{index function}) into this expression and exchanging the order of $x^{2}$ and $\nu$ integrations, we obtain
\begin{equation}
\label{loop factor 3D}
\begin{aligned}
\log {\mathcal{R}}^{\rm (3D)}
& = \lim_{\kappa \to 0} \left( \log\kappa+\frac{1}{2} \sum_{B} \left( \int_{0}^{+\infty} dx^{2} \delta\rho_{B}(x^{2})\left( \log(\nu+x^{2}) \right)_{\kappa}^{+\infty}+\mathcal{I}_{B}(0)\left( \log\nu \right)_{\kappa}^{+\infty} \right) \right)
\\
& = -\frac{1}{2} \sum_{B} \int_{0}^{+\infty} dx^{2}\delta\rho_{B}(x^{2})\log(x^{2}) + ({\rm cutoffs})
\,.
\end{aligned}
\end{equation}
The same result can be obtained by considering the one-loop factor $\mathcal{D}_{\gamma_{A},B}(-e^{i\phi_{\gamma_{A}}})$ given in (\ref{D factor mixed}): in the three-dimensional limit (\ref{3D limit}), the logarithmic integrands in $\mathcal{D}_{\gamma_{A},B}(-e^{i\phi_{\gamma_{A}}})$ become
\begin{equation}
\label{D factor 3D}
\begin{aligned}
& \log\left( 1-e^{-2\pi R |Z_{W_{B}}|\cosh t+i\theta_{W_{B}}} \right)+\log\left( 1-e^{-2\pi R |Z_{W_{B}}|\cosh t-i\theta_{W_{B}}} \right)
\\
& \to \log\left( |Z_{W_{B}}|^{2}\cosh^{2}t+\left( \frac{\theta_{W_{B}}}{2\pi R} \right)^{2} \right)+2\log(2\pi R)
\,,
\end{aligned}
\end{equation}
then, after substituting $x=M_{W_{B}}\cosh t$, rotating into the vacuum where $\vec\theta_{e}/2\pi R=\vec 0$, and combining with the earlier identification of the density of states, we can see that in the limit $R\to 0$, $\mathcal{D}_{\gamma_{A},B}(-e^{i\phi_{\gamma_{A}}})$ corresponds to the ratio of determinants (\ref{loop factor 3D}). 

\paragraph{}

It is fairly straightforward to Poisson-resum the two-instanton terms and demonstrate that the metric remains smooth in three dimensions (to show this, it is convenient to perform the resummation before integrating over auxiliary parameters).
In \cite{Fraser Tong}, it was noted that the one-loop factor in three dimensions has singularities at the wall of marginal stability; this singularity is cancelled by the factor arising from integration over the soft modes (the modes that become zero modes at the wall of marginal stability).
In the context of our semiclassical expansion, this corresponds to the multiplicity $\Omega(\gamma,\vec a)$ of composite dyons jumping to zero as the VEV approaches the wall, so that the metric remains continuous.

\paragraph{}

To extrapolate the one-instanton metric (\ref{metric NP1}) to three dimensions, for each positive root $\vec\alpha_{A}$, we can Poisson-resum the terms corresponding to dyons with magnetic charge $\vec\alpha_{A}$ (the terms corresponding to $-\vec\alpha_{A}$ are their complex conjugates).
Again, we split the relevant one-loop factor, $\mathcal{D}_{(\vec 0,\vec\alpha_{A})}(-ie^{i\phi_{W_{A}}})$, into the $A$ term and $B\ne A$ terms (\ref{D factor 3}).
The $A$ term in this limit was considered in chapter \ref{ch: pure} and is known to give
\begin{equation}
\mathcal{D}_{(\vec 0,\vec\alpha_{A}),A}(-ie^{i\phi_{W_{A}}}) = (4\pi RM_{W_{A}})^{2}
\,,
\end{equation}
the $B\ne A$ terms were calculated above.
After Poisson-resumming and taking the limit $R\to 0$, we find the following result:
\begin{equation}
\label{metric instanton resummed}
g_{a^{I}\bar a^{J},A} =
\frac{16\pi}{e_{\rm eff}^{4}}
M_{W_{A}} \left( \prod_{B\ne A}\mathcal{D}_{(\vec 0,\vec\alpha_{A}),B}(-ie^{i\phi_{W_{A}}}) \right)
\exp\left( -\frac{4\pi}{e_{\rm eff}^{2}}M_{W_{A}}+i\vec\alpha_{A}\vec\theta_{m} \right)
\end{equation}
where $1/e_{\rm eff}^{2}=2\pi R/g_{\rm eff}^{2}$ is the effective gauge coupling in three dimensions.
Having matched the one-loop factor derived semiclassically (\ref{loop factor 3D}) and the corresponding expression in the GMN expansion (\ref{D factor mixed}, \ref{D factor 3D}), we conclude that the one-instanton metric (\ref{metric instanton resummed}) reproduces the first-principles result in \cite{Fraser Tong}.

\chapter{The BPS spectrum at the root of the Higgs branch}
\label{ch: root}

\paragraph{}

{\it This chapter is based on \cite{DP}.}

\paragraph{}

We study the BPS spectrum of the $\mathcal{N}=2$ SQCD with gauge group $SU(n)$ and $n\le N_{f}<2n$ fundamental flavours at the root of the Higgs branch in four dimensions \cite{Hanany Hori, Hanany Tong, Hanany Tong 2}.
The central charge of this theory was shown \cite{Dorey 1998, DHT} to be the same as in the $\mathcal{N}=(2,2)$ supersymmetric $\mathbb{CP}^{2n-N_{f}-1}$ sigma model in two dimensions \cite{CV, Witten phases}; the BPS spectra of these two theories were then conjectured to be the same \cite{Dorey 1998, DHT}.
In this chapter, we confirm the conjecture using the Kontsevich--Soibelman wall-crossing formula in four dimensions.
The finite set of BPS states of the two-dimensional theory in the strong-coupling regime is known \cite{BSY}; assuming that the strong-coupling spectrum of the four-dimensional theory is the same, we find the walls of marginal stability and, employing the wall-crossing formula, extrapolate the spectrum, recovering the semiclassical spectrum derived in \cite{Dorey 1998, DHT}, thus confirming that our assumption is correct.

\paragraph{}

For a given magnetic charge, there is a (``primary'') wall separating the strong-coupling region from the rest of the moduli space.
Outside this wall, the spectrum expands and includes an infinite (``primary'') tower of dyons, quarks, and $W$ bosons.
In addition, we show that if a particular condition on the complex masses is satisfied, there is one extra (``secondary'') tower of bound states consisting of a dyon and one or more quarks from different towers of charges.
According to the wall-crossing formula, an extra tower cannot be created at a single wall (considered in \cite{BSY}), rather, for every bound state, there is a unique (``secondary'') wall where it is created.
We also show that each secondary wall separates the primary wall from the weak-coupling region.

\paragraph{}

A particular configuration of $\mathbb{Z}_{n}$-symmetric masses, when all $n$ masses form a regular polygon in the complex plane, can be analysed more explicitly: we find that there exists one secondary tower of bound states with one quark for odd $n$ and no bound states for even $n$.

\paragraph{}

Let us introduce our conventions in the four-dimensional theory: $N_{f}=n+\tilde n$ is the total number of flavours, $\vec q_{e}$ and $\vec q_{m}$ are the vectors of electric and magnetic charges with $n$ components (counted by $I$), $\vec S$ is the vector of flavour charges with $n+\tilde n$ components (counted by $i$).
The central charge is given by
\begin{equation}
\label{central charge general}
Z_{(\vec q_{e},\vec q_{m},\vec M)} = \vec a\vec q_{e}+\vec a_{D}\vec q_{m}+\vec S\vec M = \sum_{I=0}^{n-1} \left( a^{I}q_{e\,I}+a_{D\,I}q_{m}^{I} \right)+\sum_{i=0}^{N_{f}-1}S_{i}M_{i}
\end{equation}
where $\vec a$ is the vacuum expectation value, $\vec a_{D}$ is its magnetic dual, $\vec M$ is the vector of flavour masses.
We divide $\vec S$ and $\vec M$ into two pieces: $\vec s$ and $\vec m$ contain the first $n$ components corresponding to the massless quarks at the root of the Higgs branch, $\vec{\tilde s}$ and $\vec{\tilde m}$ contain the remaining $\tilde n$ components; we distinguish the remaining $\tilde n$ flavour components by putting a tilde above their masses, charges, and indices.
The root of the Higgs branch is determined by setting $\vec a=-\vec m$; analogously, we define $\vec m_{D}=-\vec a_{D}$.
Therefore, the central charge (\ref{central charge general}) reduces to
\begin{equation}
\label{central charge root}
\begin{aligned}
& Z_{(\vec\gamma_{e},\vec\gamma_{m},\vec{\tilde s})} = \vec m\vec\gamma_{e}+\vec m_{D}\vec\gamma_{m}+\vec{\tilde s} \, \vec{\tilde m} = \sum_{I=0}^{n-1} \left( m^{I}\gamma_{e\,I}+m_{D\,I}\gamma_{m}^{I}\right)+\sum_{\tilde i=0}^{\tilde n-1}s_{\tilde i}m_{\tilde i}
\,,
\\
& \vec\gamma_{e} = -\vec q_{e}+\vec s
\,,
\quad
\vec\gamma_{m} = -\vec q_{m}
\end{aligned}
\end{equation}
(we will be using $m_{I}=m^{I}$).
Now, for each BPS state, the complete set of (electric, magnetic, and flavour) charges is $\gamma=(\vec\gamma_{e},\vec\gamma_{m},\vec{\tilde s}\,)$; if $\vec{\tilde s}=\vec 0$, we will omit it.

%%%%%%%%%%%%%%%%%%%%%%%%%%%%%%%%%%%%%%%%%%%%%%%%%%%%%%%%%%%%%

\section{The wall at strong coupling}

\paragraph{}

Our starting point is the strong-coupling spectrum of the theory.
We consider only the BPS states corresponding to kinks interpolating between two neighbouring vacua in the two-dimensional theory.
Without loss of generality, we can set the magnetic charge to be equal to $(-1,1,0,0,\dots)$.
It is known that the electric charges are determined up to a fixed shift \cite{DHT}; for our purposes, it is convenient to normalise the strong-coupling spectrum as \cite{BSY}
\begin{equation}
\label{spectrum strong root}
\begin{aligned}
\pm\gamma_{1} & = \pm ((1,0,0,0,0,\dots),(-1,1,0,0,0,\dots))
\,,
\\
\pm\gamma_{2} & = \pm ((0,1,0,0,0,\dots),(-1,1,0,0,0,\dots))
\,,
\\
\pm\gamma_{3} & = \pm ((0,0,1,0,0,\dots),(-1,1,0,0,0,\dots))
\,,
\\
& \dots
\end{aligned}
\end{equation}
These are the states that become massless at Argyres--Douglas points \cite{Argyres Douglas} located in the strong-coupling region.

\paragraph{}

\begin{figure}[ht]
\centering
\includegraphics[width=65mm]{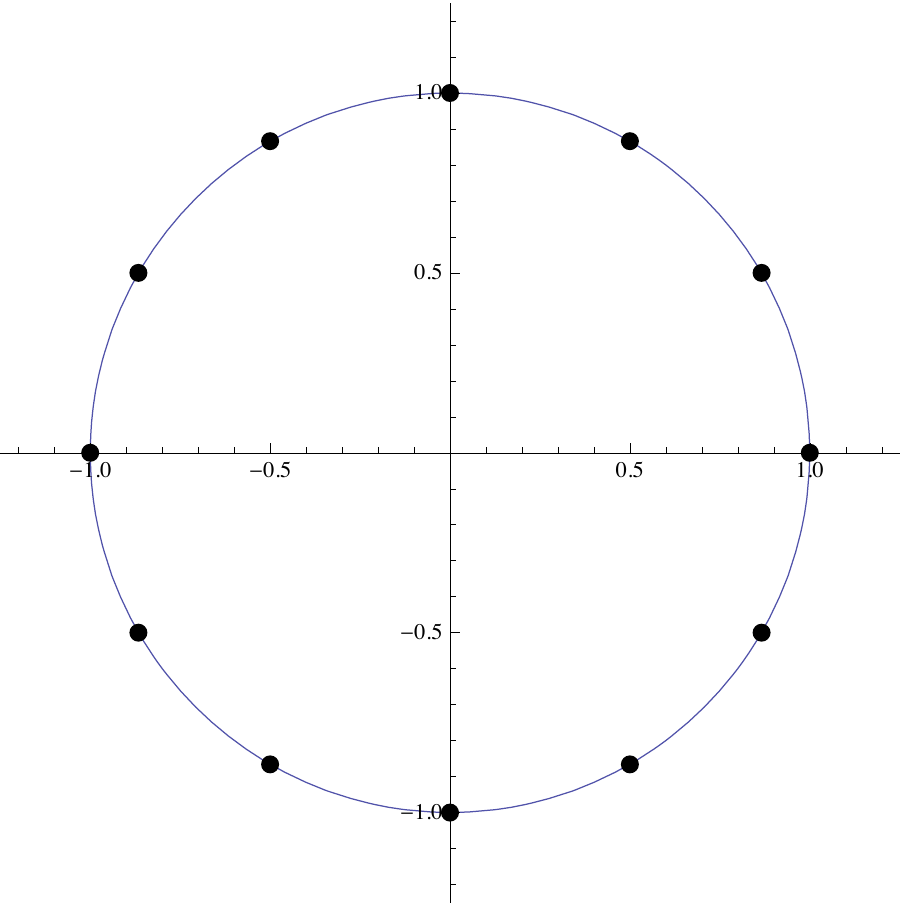}
\caption{
$\mathbb{Z}_{12}$-symmetric masses with $m_{0}=1$.
}
\label{fig: masses root}
\end{figure}

\begin{figure}[ht]
\centering
\includegraphics[width=65mm]{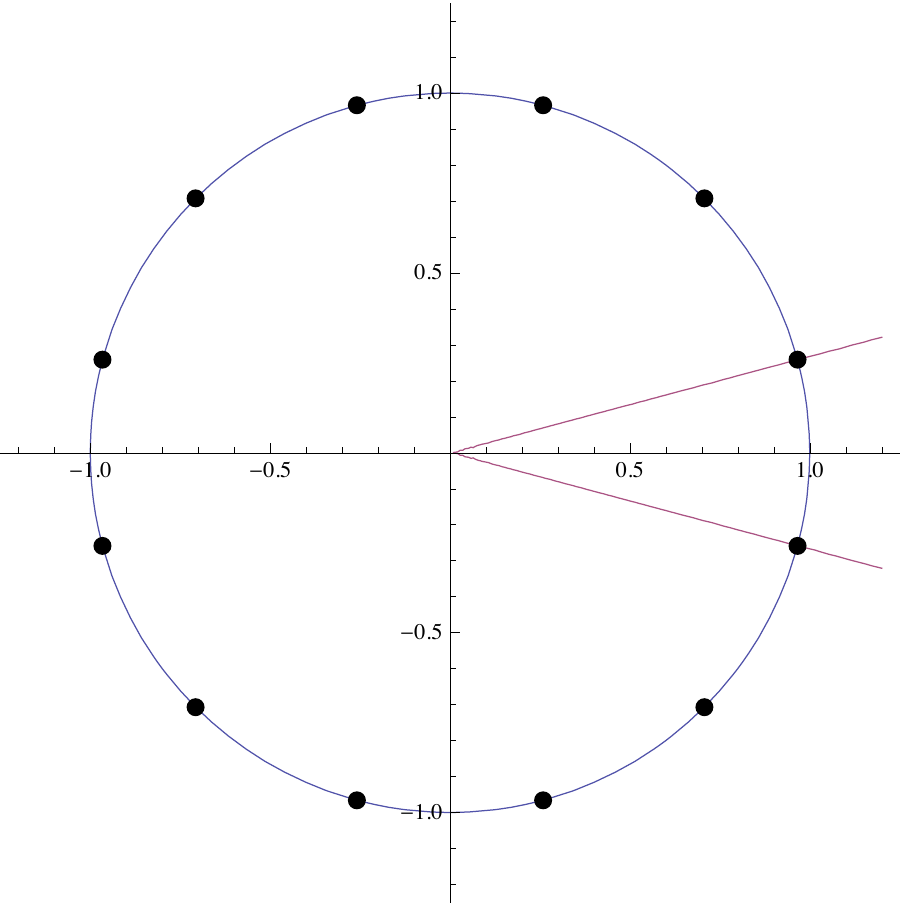}
\caption{
Argyres--Douglas points for $\mathbb{Z}_{12}$-symmetric masses in the $m_{0}$ plane.
}
\label{fig: massless states root}
\end{figure}

Consider the case of $\mathbb{Z}_{n}$-symmetric masses.
We set
\begin{equation}
m_{I} = m_{0} \, \exp\frac{2\pi iI}{n}
\,,
\end{equation}
where $m_{0}$ is not fixed (see figure \ref{fig: masses root} for an example).
Then, all central charges and walls of marginal stability depend only on $m_{0}$, and the masses automatically obey
\begin{equation}
\sum_{I=0}^{n-1} m_{I} = 0
\,.
\end{equation}
The magnetic duals of $\mathbb{Z}_{n}$-symmetric masses are given by \cite{Veneziano Yankielowicz, DDDS, CV, Hori Vafa}
\begin{equation}
m_{D\,I} = e^{2\pi iI/n}
\left(
n\sqrt[n]{m_{0}^{n}+\Lambda^{n}} +
\sum_{k=0}^{n-1} m_{0}e^{2\pi ik/n} \log\frac{\sqrt[n]{m_{0}^{n}+\Lambda^{n}}-m_{0}e^{2\pi ik/n}}{\Lambda}
\right)
\end{equation}
where we have transformed the masses as $m_{i}\to -m_{i}$.
Then, the Argyres--Douglas points, where the strong-coupling states (\ref{spectrum strong root}) become massless, are located at \cite{BSY}
\begin{equation}
m_{0} = \Lambda \, \exp\frac{\pi i(2j+1)}{n}
\,,
\quad
j \in \mathbb{Z}
\end{equation}
(figure \ref{fig: massless states root}).
As explained in \cite{Olmez Shifman}, for $\mathbb{Z}_{n}$-symmetric masses, it is sufficient to consider $m_{0}$ belonging to the sector between two neighbouring Argyres--Douglas points, $\Lambda e^{\pi i/n}$ and $\Lambda e^{-\pi i/n}$, where $\gamma_{1}$ and $\gamma_{2}$ from (\ref{spectrum strong root}) are massless.

\paragraph{}

\begin{figure}[ht]
\centering
\includegraphics[width=65mm]{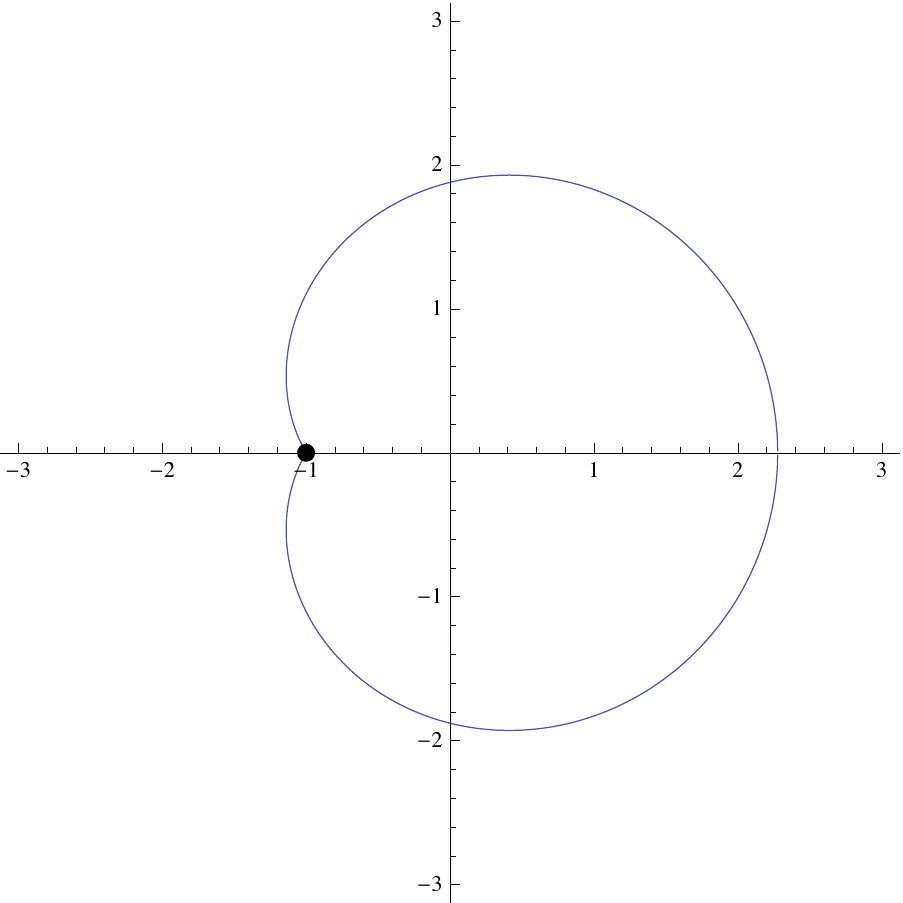}
\qquad
\includegraphics[width=65mm]{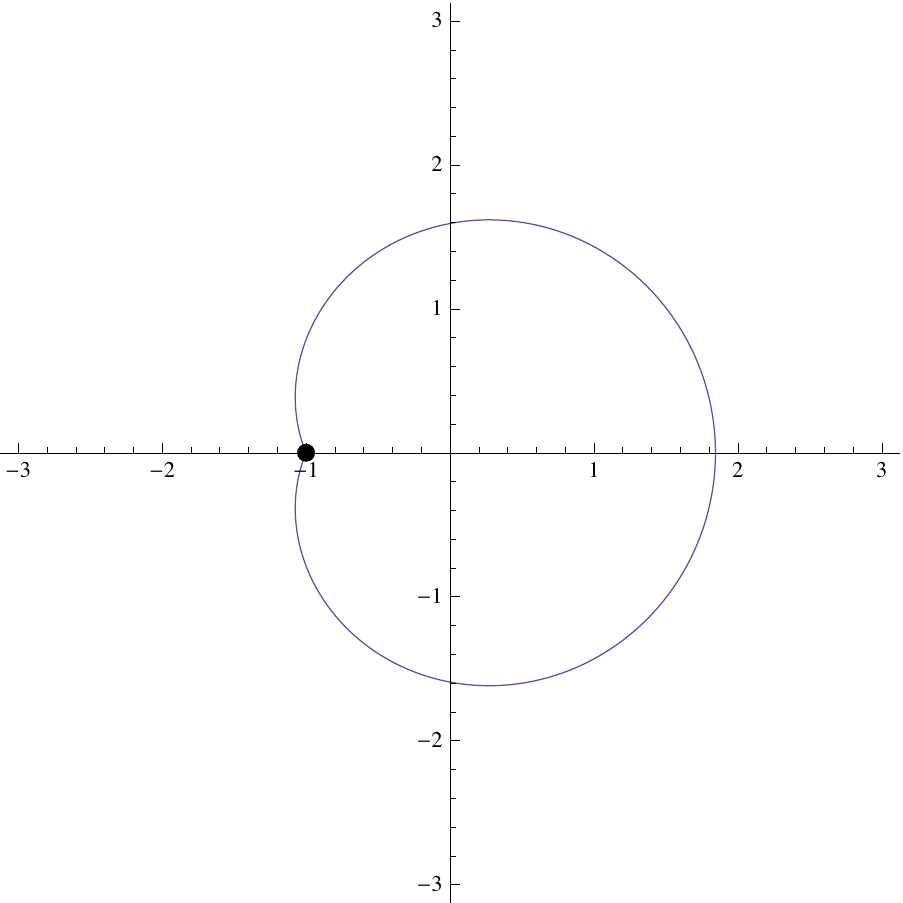}
\caption{
The primary walls of marginal stability for $\mathbb{Z}_{2}$- and $\mathbb{Z}_{3}$-symmetric masses \cite{Shifman Vainshtein Zwicky, BSY}.
}
\label{fig: walls primary root}
\end{figure}

Let us find out how the spectrum changes when $\vec M$ crosses the primary wall of marginal stability, where the central charges of the first two dyons in (\ref{spectrum strong root}), $\gamma_{1}$ and $\gamma_{2}$, become aligned:
\begin{equation}
\frac{Z_{\gamma_{1}}}{Z_{\gamma_{2}}} \in \mathbb{R}_{+}
\end{equation}
(figure \ref{fig: walls primary root}).
Using the wall-crossing formula, we can compute the spectrum on the external side of the wall.
The symplectic product is
\begin{equation}
\langle\gamma_{1},\gamma_{2}\rangle = -2
\,,
\end{equation}
hence, the relevant wall-crossing formula is a modification of the formula relating the strong- and weak-coupling spectra of the pure $SU(2)$ theory in four dimensions \cite{KS}:
\begin{equation}
\tilde{\mathcal{K}}_{-\gamma_{2}} \tilde{\mathcal{K}}_{\gamma_{1}}
=
\tilde{\mathcal{K}}_{\gamma_{1}} \tilde{\mathcal{K}}_{2\gamma_{1}-\gamma_{2}} \tilde{\mathcal{K}}_{3\gamma_{1}-2\gamma_{2}} \tilde{\mathcal{K}}_{4\gamma_{1}-3\gamma_{2}}
\dots \tilde{\mathcal{K}}_{\gamma_{1}-\gamma_{2}}^{-2} \dots
\tilde{\mathcal{K}}_{3\gamma_{1}-4\gamma_{2}} \tilde{\mathcal{K}}_{2\gamma_{1}-3\gamma_{2}} \tilde{\mathcal{K}}_{\gamma_{1}-2\gamma_{2}} \tilde{\mathcal{K}}_{-\gamma_{2}}
\,.
\end{equation}
In our notations, the part of the wall-crossing formula that changes across the wall takes the following form:
\begin{equation}
\label{WCF root abstract}
\begin{aligned}
\tilde{\mathcal{K}}_{-((0,1),(-1,1))} \tilde{\mathcal{K}}_{((1,0),(-1,1))}
& =
\tilde{\mathcal{K}}_{((1,0),(-1,1))} \tilde{\mathcal{K}}_{((2,-1),(-1,1))} \tilde{\mathcal{K}}_{((3,-2),(-1,1))} \tilde{\mathcal{K}}_{((4,-3),(-1,1))}
\\
\dots \tilde{\mathcal{K}}_{((-1,1),(0,0))}^{-2} & \dots
\tilde{\mathcal{K}}_{-((-3,4),(-1,1))} \tilde{\mathcal{K}}_{-((-2,3),(-1,1))} \tilde{\mathcal{K}}_{-((-1,2),(-1,1))} \tilde{\mathcal{K}}_{-((0,1),(-1,1))}
\end{aligned}
\end{equation}
where we consider only the first two electric and magnetic charges because all other charges are equal to zero.
This relation shows that the spectrum outside the wall consists of a tower of dyons and a finite number of quarks and $W$ bosons:
\begin{equation}
\label{WCF root}
\begin{aligned}
& \pm ((-\nu+1,\nu, 0,0,\dots),(-1,1,0,0,\dots))
\,,
\\
& \pm ((-1,1,0,0,\dots),(0,0,0,0,\dots))
\,.
\end{aligned}
\end{equation}

%%%%%%%%%%%%%%%%%%%%%%%%%%%%%%%%%%%%%%%%%%%%%%%%%%%%%%%%%%%%%

\section{Bound states}

\paragraph{}

The complete BPS spectrum is not limited to the primary tower of states found above: depending on the values of masses, there can also be secondary towers of bound states formed by a dyon and $p$ quarks \cite{DHT}.
Creation (or, conversely, destruction) of these extra states is described by the pentagon formula:
\begin{equation}
\label{pentagon root}
\tilde{\mathcal{K}}_{\gamma_{1}}\tilde{\mathcal{K}}_{\gamma_{2}} =
\tilde{\mathcal{K}}_{\gamma_{2}}\tilde{\mathcal{K}}_{\gamma_{1}+\gamma_{2}}\tilde{\mathcal{K}}_{\gamma_{1}}
\,,
\quad
\forall \ \langle\gamma_{1},\gamma_{2}\rangle = \pm 1
\end{equation}
where the new state $\gamma_{1}+\gamma_{2}$ is created from $\gamma_{1}$ and $\gamma_{2}$ where one of the initial states is a quark, and the other one is either a dyon or a bound state consisting of a dyon and $p-1$ quarks.
This process occurs when
\begin{equation}
\label{secondary wall root}
\frac{Z_{\gamma_{1}}}{Z_{\gamma_{2}}} \in \mathbb{R}_{+}
\,.
\end{equation}
We will find the secondary walls and prove that they are located outside the primary wall and have to be crossed as the VEV moves from strong to weak coupling.
The restriction on the wedge-product of the two interacting states in (\ref{pentagon root}) allows us to determine which states can combine to form a bound state if the corresponding secondary wall exists.

\paragraph{}

Starting with the states constructed in the previous section, when $\tilde n=0$, we can see that there can be two possible types of creation processes, both leading to the same set of new states:
\begin{equation}
\label{first process root}
\begin{aligned}
1:
\quad
((-\nu+1,\nu,0,0,0,\dots),(-1,1,0,0,0,\dots)) + ((-1,0,1,0,0,\dots),(0,0,0,0,0,\dots))
\\
\leftrightarrow
((-\nu,\nu,1,0,0,\dots),(-1,1,0,0,0,\dots))
\,,
\end{aligned}
\end{equation}
\begin{equation}
\label{second process root}
\begin{aligned}
2:
\quad
((-\nu,\nu+1,0,0,0,\dots),(-1,1,0,0,0,\dots)) + ((0,-1,1,0,0,\dots),(0,0,0,0,0,\dots))
\\
\leftrightarrow
((-\nu,\nu,1,0,0,\dots),(-1,1,0,0,0,\dots))
\,.
\end{aligned}
\end{equation}
These are the bound states formed by a dyon and one quark~\footnote{
These extra towers were considered in \cite{BSY}; they are the same as the states obtained in \cite{DHT} up to shifting $\gamma_{e\,0}$ by $-1$ here.
}.
Explicitly, the walls of marginal stability for these processes (\ref{secondary wall root}) are determined by
\begin{align}
\label{first wall root}
1:
\quad
\frac{-m_{0}+m_{2}}{(-\nu+1)m_{0}+\nu m_{1}-m_{D\,0}+m_{D\,1}} & \in \mathbb{R}_{+}
\,,
\\
2:
\quad
\label{second wall root}
\frac{-m_{1}+m_{2}}{-\nu m_{0}+(\nu+1)m_{1}-m_{D\,0}+m_{D\,1}} & \in \mathbb{R}_{+}
\,.
\end{align}

\paragraph{}

For general $N_{f}$, there are additional bound states involving the remaining $\tilde n$ flavours:
\begin{equation}
\begin{aligned}
((-\nu+1,\nu,0,0,0,\dots),(-1,1,0,0,0,\dots)) + ((-1,0,0,0,0,\dots),(0,0,0,0,0,\dots),(1,0,0,\dots))
\\
\leftrightarrow
((-\nu,\nu,0,0,0,\dots),(-1,1,0,0,0,\dots),(1,0,0,\dots))
\,,
\end{aligned}
\end{equation}
\begin{equation}
\begin{aligned}
((-\nu,\nu+1,0,0,0,\dots),(-1,1,0,0,0,\dots)) + ((0,-1,0,0,0,0,\dots),(0,0,0,0,0,\dots),(1,0,0,\dots))
\\
\leftrightarrow
((-\nu,\nu,0,0,0,\dots),(-1,1,0,0,0,\dots),(1,0,0,\dots))
\,.
\end{aligned}
\end{equation}
They are completely analogous to the ones above: the walls of marginal stability for these processes can be obtained by changing $m_{2}$ to $\tilde m_{0}$ in the previous formulae.

\paragraph{}

As has been discussed above, there can also be bound states formed by a dyon and $p$ quarks:
\begin{equation}
\label{bound state general root}
\begin{aligned}
& ((-\nu+1+p,\nu,j_{3},j_{4},j_{5},\dots),(-1,1,0,0,0,\dots),(\tilde j_{\tilde 1},\tilde j_{\tilde 2},\tilde j_{\tilde 3},\dots))
\,,
\\
& j_{i}(j_{i}-1) = \tilde j_{\tilde i}(\tilde j_{\tilde i}-1) = 0
\,,
\quad
p + \sum_{i=2}^{n-1} j_{i} + \sum_{\tilde i=0}^{\tilde n-1} \tilde j_{\tilde i} = 0
\,.
\end{aligned}
\end{equation}
They exist if, starting with the strong-coupling spectrum and moving into the weak-coupling region, $|p|$ different secondary walls of marginal stability (\ref{secondary wall root}) are crossed.

\paragraph{}

We need to find out if the processes (\ref{first process root}) and (\ref{second process root}), which we rewrite as
\begin{equation}
\label{first process root 2}
\begin{aligned}
1:
\quad
d_{1} + q_{1}
\leftrightarrow
b
\,,
\end{aligned}
\end{equation}
\begin{equation}
\label{second process root 2}
\begin{aligned}
2:
\quad
d_{2} + q_{2}
\leftrightarrow
b
\,,
\end{aligned}
\end{equation}
actually take place when the masses move from strong to weak coupling: to do this, we should check whether the secondary walls (\ref{secondary wall root}) are crossed, i.e., if the following conditions are satisfied somewhere outside the primary wall of marginal stability:
\begin{equation}
\label{first wall root 2}
1:
\quad
\arg Z_{d_{1}} = \arg Z_{q_{1}}
\,,
\end{equation}
\begin{equation}
\label{second wall root 2}
2:
\quad
\arg Z_{d_{2}} = \arg Z_{q_{2}}
\,.
\end{equation}
Note that $Z_{q_{j}}$ ($j=1$ or $j=2$) is independent of the region in the moduli space, and $\arg Z_{d_{j}}$ changes continuously between the primary wall and the weak-coupling region, therefore, (\ref{first wall root 2}) (with $j=1$) and (\ref{second wall root 2}) (with $j=2$) are satisfied somewhere if in the complex plane, $Z_{q_{j}}$ lies between the values of $Z_{d_{j}}$ at the primary wall and in the weak-coupling limit.
To check if this is the case, it is convenient to start at the Argyres--Douglas point where $d_{1}$ with $\nu=0$, i.e., $\gamma_{1}$ in (\ref{spectrum strong root}), becomes massless: near this point, the central charge of a dyon can be approximated as
\begin{equation}
\label{central charge dyon initial root}
(Z_{d_{1}})_{s} \simeq \nu(-m_{0}+m_{1})
\,.
\end{equation}
Then, we continuously move the masses into the semiclassical region, where
\begin{equation}
\label{central charge dyon final root}
(Z_{d_{1}})_{w} \simeq i(-m_{0}+m_{1})
\,.
\end{equation}
From (\ref{central charge dyon initial root}) and (\ref{central charge dyon final root}), we have
\begin{equation}
\label{central charge dyon change root}
\lim_{g_{\rm eff}\to 0} \arg\frac{(Z_{d_{1}})_{w}}{(Z_{d_{1}})_{s}} = \frac{\pi}{2}
\,,
\quad
\arg\frac{(Z_{d_{1}})_{w}}{(Z_{d_{1}})_{s}} < \frac{\pi}{2}
\end{equation}
where the inequality is strict for any $g_{\rm eff}$ as the central charge receives corrections from its electric components at weak coupling.
All these statements also hold if $d_{1}$ is substituted by $d_{2}$.
Comparing $Z_{q_{j}}$ with (\ref{central charge dyon initial root}) and (\ref{central charge dyon final root}) and using (\ref{central charge dyon change root}), we conclude that the walls (\ref{first wall root}, \ref{second wall root}) exist if
\begin{equation}
\label{first bound state constraint root}
\begin{aligned}
1:
\quad
& \nu > 0:
\quad
\arg\frac{m_{k}-m_{0}}{m_{1}-m_{0}} \in \left( 0,\frac{\pi}{2} \right)
\,,
\\
& \nu < 0:
\quad
\arg\frac{m_{k}-m_{0}}{m_{1}-m_{0}} \in \left( \frac{\pi}{2},\pi \right)
\,,
\end{aligned}
\end{equation}
\begin{equation}
\label{second bound state constraint root}
\begin{aligned}
2:
\quad
& \nu > 0:
\quad
\arg\frac{m_{k}-m_{1}}{m_{1}-m_{0}} \in \left( 0,\frac{\pi}{2} \right)
\,,
\\
& \nu < 0:
\quad
\arg\frac{m_{k}-m_{1}}{m_{1}-m_{0}} \in \left( \frac{\pi}{2},\pi \right)
\,.
\end{aligned}
\end{equation}
Since the bound states with $\nu\ne 0$ do not exist at strong coupling, they appear at weak coupling if exactly one of the two walls (\ref{first wall root 2}, \ref{second wall root 2}) is crossed.
For both $\nu>0$ and $\nu<0$, this means that one of the following conditions must be satisfied:
\begin{equation}
\label{bound state constraint root}
\begin{aligned}
& 0 < \re\frac{m_{k}-m_{0}}{m_{1}-m_{0}} < 1
\,,
\\
& 0 < \re\frac{m_{k}-m_{1}}{m_{1}-m_{0}} < 1
\end{aligned}
\end{equation}
(again, the inequalities are strict because of (\ref{central charge dyon change root})).
In fact, the conditions are equivalent to each other~\footnote{
This can be seen by reflecting the complex plane so that $m_{0}$ and $m_{1}$ exchange their positions.
}, and the first equation in (\ref{bound state constraint root}) is precisely the semiclassical constraint derived in \cite{DHT} from first principles.
Finally, we can conclude that the secondary tower of bound states exists if and only if the complex masses obey the equivalent inequalities in (\ref{bound state constraint root}).
Analogously, if (\ref{bound state constraint root}) holds for $p$ different indices $k$ and $\tilde k$, there are towers of bound states with $p$ quarks (\ref{bound state general root}) having $j_{k}=1$ and $\tilde j_{\tilde k}=1$ for these indices and $j_{i}=0$ and $\tilde j_{\tilde i}=0$ for all other $i$ and $\tilde i$, in accordance with \cite{DHT}.

\paragraph{}

\begin{figure}[ht]
\centering
\includegraphics[width=65mm]{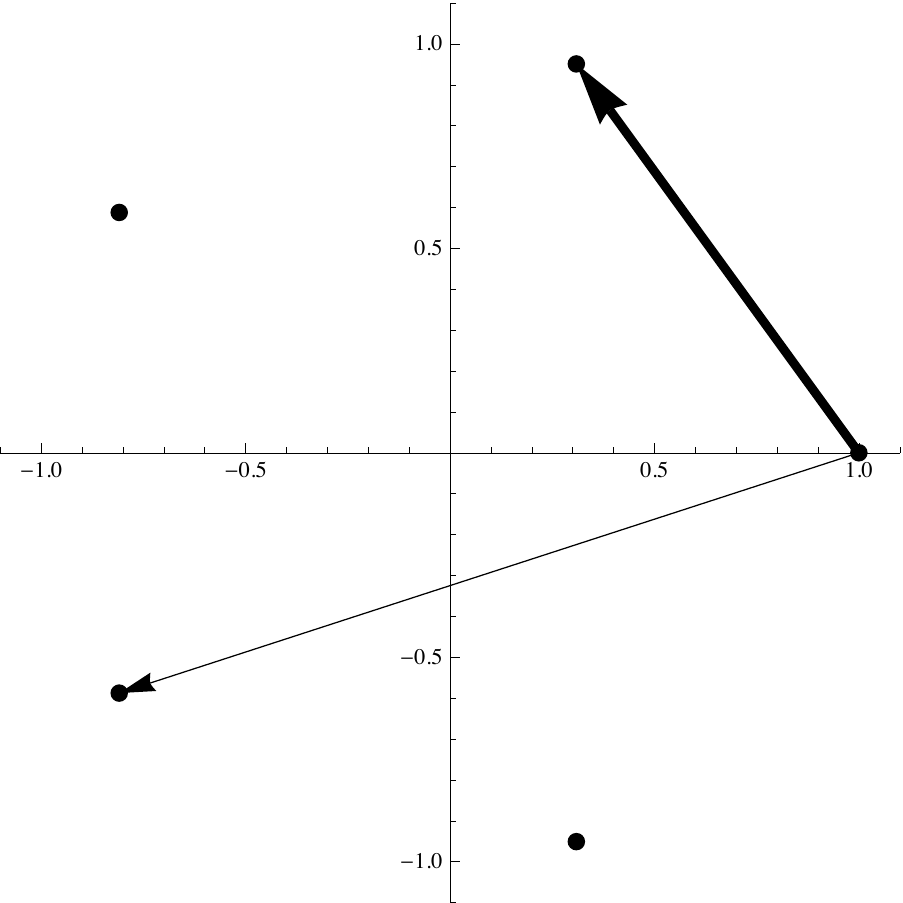}
\caption{
A dyon (thick vector connecting $m_{0}$ and $m_{1}$, representing its central charge near the massless point at strong coupling for $\nu>0$) and the quark that can form bound states with it (thin vector connecting $m_{0}$ and $m_{3}$, equal to its central charge) in the case of $\mathbb{Z}_{5}$-symmetric masses.
}
\label{fig: states root}
\end{figure}

\begin{figure}[ht]
\centering
\includegraphics[height=65mm]{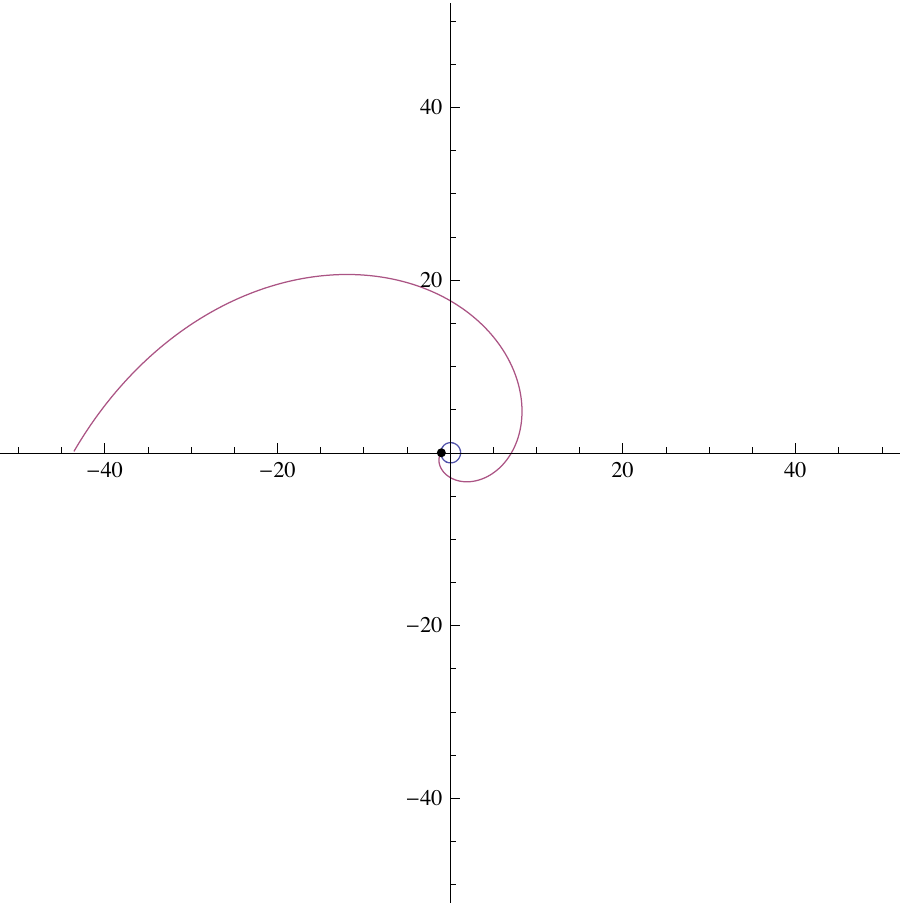}
\qquad
\includegraphics[height=65mm]{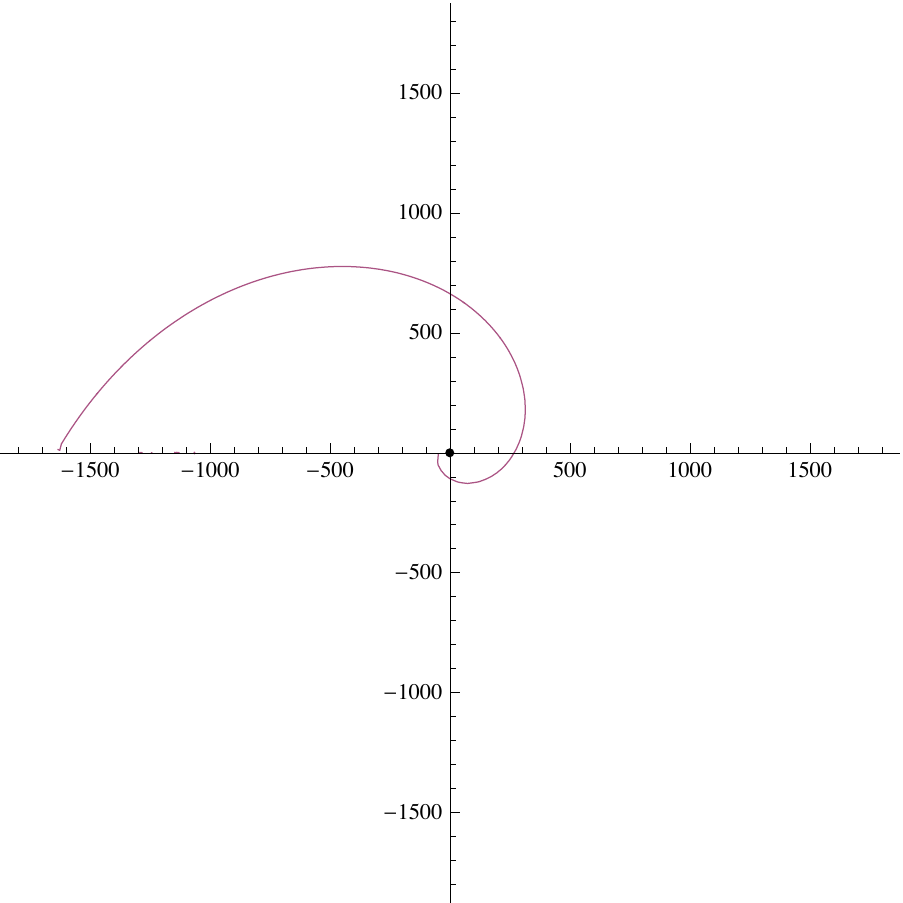}
\\[10pt]
\includegraphics[height=65mm]{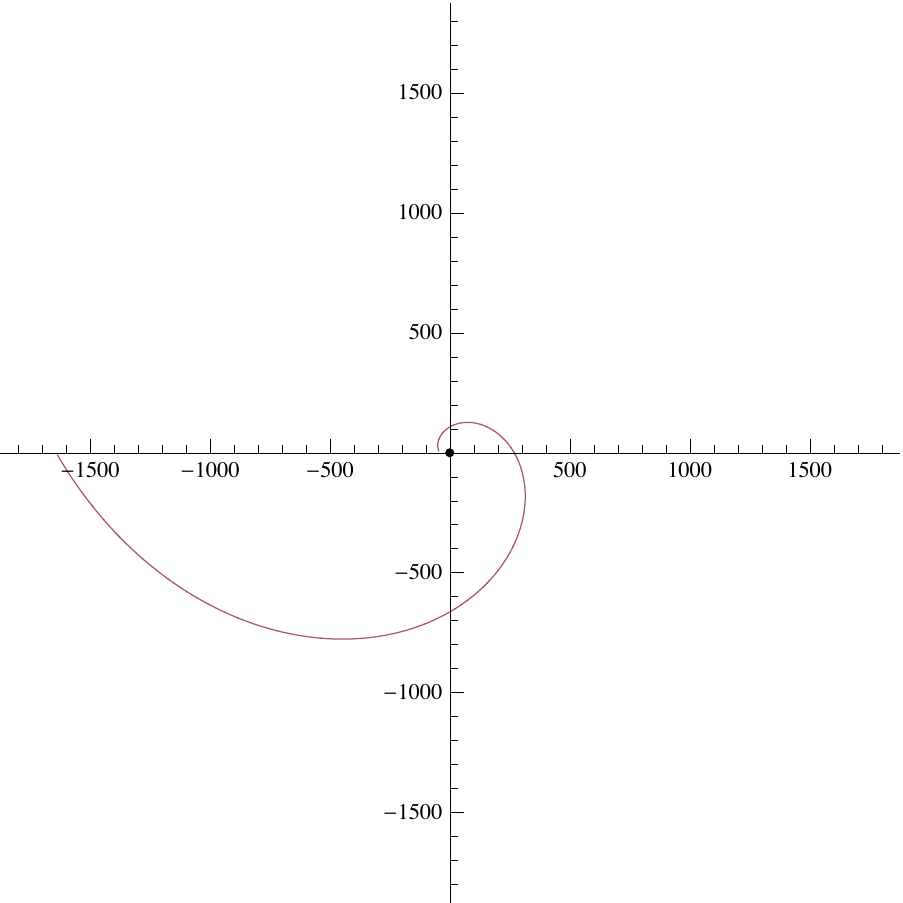}
\qquad
\includegraphics[height=65mm]{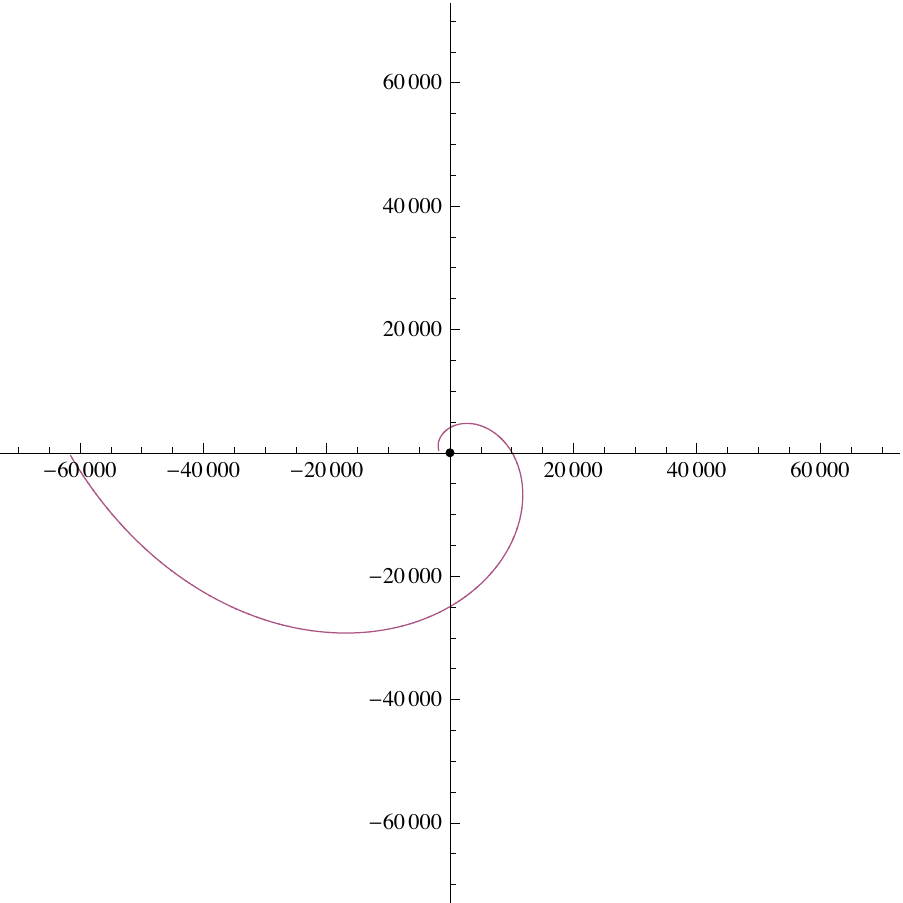}
\caption{
Secondary walls of marginal stability for $\mathbb{Z}_{3}$-symmetric masses in the $m_{0}^{3}/|m_{0}|^{2}$ plane (left column: $|\nu|=1$, right column: $|\nu|=2$; first row: bound state with the quark $(-1,0,1)$, second row: bound state with the quark $(0,-1,1)$).
The plots corresponds to the sector in figure \ref{fig: massless states root}.
}
\label{fig: walls secondary root}
\end{figure}

Applying this result to $\mathbb{Z}_{n}$-symmetric masses, it is easy to determine which bound states exist in the weak-coupling limit.
The constraint (\ref{bound state constraint root}) reduces to
\begin{equation}
0 < \re\frac{e^{2\pi ki/n}-1}{e^{2\pi i/n}-1} = \re\frac{e^{2\pi(k-1/2)i/n}-e^{-\pi i/n}}{2i\sin(\pi/n)} < 1
\quad \iff \quad
-1 < \frac{\sin\frac{2\pi k-\pi}{n}}{\sin\frac{\pi}{n}} < 1
\,.
\end{equation}
Here, $(2\pi k-\pi)/n$ is a multiple of $\pi/n$, therefore, the inequality holds only for $(2\pi k-\pi)/n=\pi$, that is, for $k=(n+1)/2$.
For $\mathbb{Z}_{2l+1}$-symmetric masses with $l\in\mathbb{N}$, this means that only the bound states formed by one quark with $\gamma_{e\,(l+1)}=1$ are present (figure \ref{fig: states root}); for $\mathbb{Z}_{2l}$-symmetric masses, there are no bound states.
We can go back to equations (\ref{first bound state constraint root}, \ref{second bound state constraint root}) to find out which secondary walls of marginal stability exist in the case of $\mathbb{Z}_{2l+1}$-symmetric masses: (\ref{first process root}) is realised for $\nu>0$, (\ref{second process root}) is realised for $\nu<0$, and the corresponding walls are determined by (\ref{first wall root}) with $\nu>0$ and (\ref{second wall root}) with $\nu<0$ (plotted for $\mathbb{Z}_{3}$ in figure \ref{fig: walls secondary root}).

\paragraph{}

In \cite{BSY}, it was noted that the weak-coupling spectrum of the two-dimensional theory can also contain extra towers of charges which are given as $((-\nu,\nu,1,0,0,\dots),(-1,1,0,0,0,\dots))$.
The simplest theoretically possible decay process for this tower is \cite{BSY}
\begin{equation}
\begin{aligned}
((-\nu,\nu,1,0,0,\dots),(-1,1,0,0,0,\dots))
\overset{?}{\leftrightarrow}
((0,0,1,0,0,\dots),(-1,1,0,0,0,\dots))
\\
+\nu((-1,1,0,0,0,\dots),(0,0,0,0,0,\dots))
\,.
\end{aligned}
\end{equation}
However, we can argue that in the four-dimensional theory, such process would be inconsistent with the wall-crossing formula.
To see this, suppose that the spectrum on right-hand side is correct, and we cross the wall in the other direction.
The symplectic product of the two charges on the right-hand side is $2$; then, we can use (\ref{WCF root abstract}) to find the infinite tower of states on the left-hand side: we can see that (\ref{WCF root abstract}) implies that magnetic charges of these states are $(-\nu,\nu, 0,0,0,\dots)$ for all $\nu\in\mathbb{Z}$, and not $(-1,1, 0,0,0,\dots)$, as they should be.

\paragraph{}

Finally, assuming that the BPS spectrum of the four-dimensional theory considered above is identical to the BPS spectrum of the two-dimensional $\mathcal{N}=(2,2)$ $\mathbb{CP}^{2n-N_{f}-1}$ sigma model with $\vec\gamma_{e}$ and $\vec\gamma_{m}$ being the vectors of global and topological charges, as conjectured in \cite{Dorey 1998, DHT}, we can also claim that we have found the BPS spectrum of the two-dimensional theory (which does not agree with the result obtained in \cite{BSY}).


\begin{thebibliography}{200}


\addcontentsline{toc}{chapter}{Bibliography}


\bibitem{SW}
  N.~Seiberg and E.~Witten,
  ```Electric-magnetic duality, monopole condensation, and confinement in N=2 supersymmetric Yang--Mills theory,''
  Nucl.\ Phys.\  B {\bf 426}, 19 (1994)
  [Erratum-ibid.\  B {\bf 430}, 485 (1994)]
  [\href{http://arxiv.org/pdf/hep-th/9407087}{arXiv:hep-th/9407087}].

\bibitem{SW2}
  N.~Seiberg and E.~Witten,
  ``Monopoles, duality and chiral symmetry breaking in N=2 supersymmetric QCD,''
  Nucl.\ Phys.\  B {\bf 431}, 484 (1994)
  [\href{http://arxiv.org/pdf/hep-th/9408099}{arXiv:hep-th/9408099}].

\bibitem{SW3}
  N.~Seiberg and E.~Witten,
  ``Gauge dynamics and compactification to three dimensions,''
  \href{http://arxiv.org/pdf/hep-th/9607163}{arXiv:hep-th/9607163}.

\bibitem{KS}
  M.~Kontsevich and Y.~Soibelman,
  ``Stability structures, motivic Donaldson--Thomas invariants and cluster transformations,''
  \href{http://arxiv.org/pdf/0811.2435}{arXiv:0811.2435 [math.AG]}.

\bibitem{KS2}
  M.~Kontsevich and Y.~Soibelman,
  ``Motivic Donaldson--Thomas invariants: summary of results,''
  \href{http://arxiv.org/pdf/0910.4315}{arXiv:0910.4315 [math.AG]}.

\bibitem{GMN}
  D.~Gaiotto, G.~W.~Moore and A.~Neitzke,
  ``Four-dimensional wall-crossing via three-dimensional field theory,''
  Commun.\ Math.\ Phys.\  {\bf 299}, 163 (2010)
  [\href{http://arxiv.org/pdf/0807.4723}{arXiv:0807.4723 [hep-th]}].

\bibitem{GMN2}
  D.~Gaiotto, G.~W.~Moore and A.~Neitzke,
  ``Wall-crossing, Hitchin systems, and the WKB approximation,''
  \href{http://arxiv.org/pdf/0907.3987}{arXiv:0907.3987 [hep-th]}.

\bibitem{Seiberg}
  N.~Seiberg,
  ``Supersymmetry and non-perturbative beta functions,''
  Phys.\ Lett.\  B {\bf 206}, 75 (1988).

\bibitem{Witten dyon charge}
  E.~Witten,
  ``Dyons of charge e theta/2 pi,''
  Phys.\ Lett.\  B {\bf 86}, 283 (1979).

\bibitem{Witten Olive}
  E.~Witten and D.~I.~Olive,
  ``Supersymmetry algebras that include topological charges,''
  Phys.\ Lett.\  B {\bf 78}, 97 (1978).

\bibitem{Bilal}
  A.~Bilal,
  ``Duality in N=2 SUSY SU(2) Yang--Mills theory: a pedagogical introduction to the work of Seiberg and Witten,''
  \href{http://arxiv.org/pdf/hep-th/9601007}{arXiv:hep-th/9601007}.

\bibitem{Bilal Ferrari}
  F.~Ferrari and A.~Bilal,
  ``The strong-coupling spectrum of the Seiberg--Witten theory,''
  Nucl.\ Phys.\  B {\bf 469}, 387 (1996)
  [\href{http://arxiv.org/pdf/hep-th/9602082}{arXiv:hep-th/9602082}].

\bibitem{Bilal Ferrari 2}
  A.~Bilal and F.~Ferrari,
  ``Curves of marginal stability and weak and strong-coupling BPS spectra in N=2 supersymmetric QCD,''
  Nucl.\ Phys.\  B {\bf 480}, 589 (1996)
  [\href{http://arxiv.org/pdf/hep-th/9605101}{arXiv:hep-th/9605101}].

\bibitem{Bilal Ferrari 3}
  A.~Bilal and F.~Ferrari,
  ``The BPS spectra and superconformal points in massive N=2 supersymmetric QCD,''
  Nucl.\ Phys.\  B {\bf 516}, 175 (1998)
  [\href{http://arxiv.org/pdf/hep-th/9706145}{arXiv:hep-th/9706145}].
  
\bibitem{Henningson}
  M.~Henningson,
  ``Discontinuous BPS spectra in N=2 gauge theory,''
  Nucl.\ Phys.\  B {\bf 461}, 101 (1996)
  [\href{http://arxiv.org/pdf/hep-th/9510138}{arXiv:hep-th/9510138}].

\bibitem{CFIV}
  S.~Cecotti, P.~Fendley, K.~A.~Intriligator and C.~Vafa,
  ``A new supersymmetric index,''
  Nucl.\ Phys.\  B {\bf 386}, 405 (1992)
  [\href{http://arxiv.org/pdf/hep-th/9204102}{arXiv:hep-th/9204102}].

\bibitem{Klemm Lerche Yankielowicz Theisen}
  A.~Klemm, W.~Lerche, S.~Yankielowicz and S.~Theisen,
  ``Simple singularities and N=2 supersymmetric Yang--Mills theory,''
  Phys.\ Lett.\  B {\bf 344}, 169 (1995)
  [\href{http://arxiv.org/pdf/hep-th/9411048}{arXiv:hep-th/9411048}].

\bibitem{Argyres Faraggi}
  P.~C.~Argyres and A.~E.~Faraggi,
  ``The vacuum structure and spectrum of N=2 supersymmetric SU(n) gauge theory,''
  Phys.\ Rev.\ Lett.\  {\bf 74}, 3931 (1995)
  [\href{http://arxiv.org/pdf/hep-th/9411057}{arXiv:hep-th/9411057}].

\bibitem{Danielsson Sundborg}
  U.~H.~Danielsson and B.~Sundborg,
  ``The Moduli space and monodromies of N=2 supersymmetric SO(2r+1) Yang--Mills theory,''
  Phys.\ Lett.\  B {\bf 358}, 273 (1995)
  [\href{http://arxiv.org/pdf/hep-th/9504102}{arXiv:hep-th/9504102}].

\bibitem{Brandhuber Landsteiner}
  A.~Brandhuber and K.~Landsteiner,
  ``On the monodromies of N=2 supersymmetric Yang--Mills theory with gauge group SO(2n),''
  Phys.\ Lett.\  B {\bf 358}, 73 (1995)
  [\href{http://arxiv.org/pdf/hep-th/9507008}{arXiv:hep-th/9507008}].

\bibitem{Argyres Shapere}
  P.~C.~Argyres and A.~D.~Shapere,
  ``The Vacuum structure of N=2 superQCD with classical gauge groups,''
  Nucl.\ Phys.\  B {\bf 461}, 437 (1996)
  [\href{http://arxiv.org/pdf/hep-th/9509175}{arXiv:hep-th/9509175}].
  
\bibitem{Danielsson Sundborg 2}
  U.~H.~Danielsson and B.~Sundborg,
  ``Exceptional equivalences in N=2 supersymmetric Yang--Mills theory,''
  Phys.\ Lett.\  B {\bf 370}, 83 (1996)
  [\href{http://arxiv.org/pdf/hep-th/9511180}{arXiv:hep-th/9511180}].

\bibitem{Abolhasani Alishahiha Ghezelbash}
  M.~R.~Abolhasani, M.~Alishahiha and A.~M.~Ghezelbash,
  ``The Moduli space and monodromies of the N=2 supersymmetric Yang--Mills theory with any Lie gauge groups,''
  Nucl.\ Phys.\  B {\bf 480}, 279 (1996)
  [\href{http://arxiv.org/pdf/hep-th/9606043}{arXiv:hep-th/9606043}].

\bibitem{Fraser Hollowood}
  C.~Fraser and T.~J.~Hollowood,
  ``On the weak coupling spectrum of N=2 supersymmetric SU(n) gauge theory,''
  Nucl.\ Phys.\  B {\bf 490}, 217 (1997)
  [\href{http://arxiv.org/pdf/hep-th/9610142}{arXiv:hep-th/9610142}].
  
\bibitem{Haag Lopuszanski Sohnius}
  R.~Haag, J.~T.~Lopuszanski, M.~Sohnius,
  ``All possible generators of supersymmetries of the S matrix,''
  Nucl.\ Phys.\  {\bf B88}, 257 (1975).

\bibitem{Grimm Sohnius Wess}
  R.~Grimm, M.~Sohnius, J.~Wess,
  ``Extended supersymmetry and gauge theories,''
  Nucl.\ Phys.\  {\bf B133}, 275 (1978).

\bibitem{Seiberg 2}
  N.~Seiberg,
  ``Naturalness versus supersymmetric nonrenormalization theorems,''
  Phys.\ Lett.\  {\bf B318}, 469-475 (1993)
  [\href{http://arxiv.org/pdf/hep-th/9309335}{hep-ph/9309335}].

\bibitem{Seiberg 3}
  N.~Seiberg,
  ``The power of holomorphy: exact results in 4D SUSY field theories,''
  [\href{http://arxiv.org/pdf/hep-th/9408013}{hep-th/9408013}].

\bibitem{Wess Bagger}
  J.~Wess, J.~Bagger,
  ``Supersymmetry and supergravity,''
  Princeton University Press, Princeton, USA (1992).

\bibitem{Alvarez-Gaume Hassan}
  L.~Alvarez-Gaume and S.~F.~Hassan,
  ``Introduction to S duality in N=2 supersymmetric gauge theories: a pedagogical review of the work of Seiberg and Witten,''
  Fortsch.\ Phys.\  {\bf 45}, 159 (1997)
  [\href{http://arxiv.org/pdf/hep-th/9701069}{arXiv:hep-th/9701069}].

\bibitem{Alvarez-Gaume Freedman}
  L.~Alvarez-Gaume and D.~Z.~Freedman,
  ``Geometrical structure and ultraviolet finiteness in the supersymmetric sigma model,''
  Commun.\ Math.\ Phys.\  {\bf 80}, 443 (1981).

\bibitem{CV}
  S.~Cecotti and C.~Vafa,
  ``On classification of N=2 supersymmetric theories,''
  Commun.\ Math.\ Phys.\  {\bf 158}, 569 (1993)
  [\href{http://arxiv.org/pdf/hep-th/9211097}{arXiv:hep-th/9211097}].

\bibitem{Denef}
  F.~Denef,
  ``Supergravity flows and D-brane stability,''
  JHEP {\bf 0008}, 050 (2000)
  [\href{http://arxiv.org/pdf/hep-th/0005049}{arXiv:hep-th/0005049}].

\bibitem{Denef 2}
  F.~Denef,
  ``On the correspondence between D-branes and stationary supergravity solutions of type II Calabi--Yau compactifications,''
  \href{http://arxiv.org/pdf/hep-th/0010222}{arXiv:hep-th/0010222}.

\bibitem{Denef 3}
  F.~Denef and G.~W.~Moore,
  ``Split states, entropy enigmas, holes and halos,''
  \href{http://arxiv.org/pdf/hep-th/0702146}{arXiv:hep-th/0702146}.

\bibitem{Montonen Olive}
  C.~Montonen and D.~I.~Olive,
  ``Magnetic monopoles as gauge particles?''
  Phys.\ Lett.\  B {\bf 72}, 117 (1977).

\bibitem{Hooft}
  G.~'t~Hooft,
  ``Magnetic monopoles in unified gauge theories,''
  Nucl.\ Phys.\  B {\bf 79}, 276 (1974).

\bibitem{Polyakov}
  A.~M.~Polyakov,
  ``Particle spectrum in the quantum field theory,''
  JETP Lett.\  {\bf 20}, 194 (1974).

\bibitem{Julia Zee}
  B.~Julia and A.~Zee,
  ``Poles with both magnetic and electric charges in nonabelian gauge theory,''
  Phys.\ Rev.\  D {\bf 11}, 2227 (1975).

\bibitem{Prasad Sommerfield}
  M.~K.~Prasad and C.~M.~Sommerfield,
  ``An exact classical solution for the 't Hooft monopole and the Julia--Zee dyon,''
  Phys.\ Rev.\ Lett.\  {\bf 35}, 760 (1975).

\bibitem{Callias}
  C.~Callias,
  ``Index theorems on open spaces,''
  Commun.\ Math.\ Phys.\  {\bf 62}, 213 (1978).

\bibitem{Kaul}
  R.~K.~Kaul,
  ``Monopole mass in supersymmetric gauge theories,''
  Phys.\ Lett.\  B {\bf 143}, 427 (1984).

\bibitem{Gauntlett}
  J.~P.~Gauntlett,
  ``Low-energy dynamics of N=2 supersymmetric monopoles,''
  Nucl.\ Phys.\  B {\bf 411}, 443 (1994)
  [\href{http://arxiv.org/pdf/hep-th/9305068}{arXiv:hep-th/9305068}].

\bibitem{Blum}
  J.~D.~Blum,
  ``Supersymmetric quantum mechanics of monopoles in N=4 Yang-Mills theory,''
  Phys.\ Lett.\  B {\bf 333}, 92 (1994)
  [\href{http://arxiv.org/pdf/hep-th/9401133}{arXiv:hep-th/9401133}].

\bibitem{Bernard}
  C.~W.~Bernard,
  ``Gauge zero modes, instanton determinants, and quantum-chromodynamic calculations,''
  Phys.\ Rev.\  D {\bf 19}, 3013 (1979).

\bibitem{Weinberg}
  E.~J.~Weinberg,
  ``Parameter counting for multi-monopole solutions,''
  Phys.\ Rev.\  {\bf D20}, 936-944 (1979).

\bibitem{Weinberg Yi}
  E.~J.~Weinberg and P.~Yi,
  ``Magnetic monopole dynamics, supersymmetry, and duality,''
  Phys.\ Rept.\  {\bf 438}, 65 (2007)
  [\href{http://arxiv.org/pdf/hep-th/0609055}{arXiv:hep-th/0609055}].

\bibitem{Ooguri Vafa}
  H.~Ooguri and C.~Vafa,
  ``Summing up D-instantons,''
  Phys.\ Rev.\ Lett.\  {\bf 77}, 3296 (1996)
  [\href{http://arxiv.org/pdf/hep-th/9608079}{arXiv:hep-th/9608079}].

\bibitem{DKMTV}
  N.~Dorey, V.~V.~Khoze, M.~P.~Mattis, D.~Tong and S.~Vandoren,
  ``Instantons, three-dimensional gauge theory, and the Atiyah--Hitchin manifold,''
  Nucl.\ Phys.\  B {\bf 502}, 59 (1997)
  [\href{http://arxiv.org/pdf/hep-th/9703228}{arXiv:hep-th/9703228}].

\bibitem{DKM}
  N.~Dorey, V.~V.~Khoze and M.~P.~Mattis,
  ``Multi-instanton calculus in N=2 supersymmetric gauge theory,''
  Phys.\ Rev.\  D {\bf 54}, 2921 (1996)
  [\href{http://arxiv.org/pdf/hep-th/9603136}{arXiv:hep-th/9603136}].

\bibitem{DKM2}
  N.~Dorey, V.~V.~Khoze and M.~P.~Mattis,
  ``Multi-instanton calculus in N=2 supersymmetric gauge theory. II: coupling to matter,''
  Phys.\ Rev.\  D {\bf 54}, 7832 (1996)
  [\href{http://arxiv.org/pdf/hep-th/9607202}{arXiv:hep-th/9607202}].

\bibitem{DHKM}
  N.~Dorey, T.~J.~Hollowood, V.~V.~Khoze and M.~P.~Mattis,
  ``The calculus of many instantons,''
  Phys.\ Rept.\  {\bf 371}, 231 (2002)
  [\href{http://arxiv.org/pdf/hep-th/0206063}{arXiv:hep-th/0206063}].

\bibitem{Dorey 1999}
  N.~Dorey,
  ``An elliptic superpotential for softly broken N=4 supersymmetric Yang--Mills theory,''
  JHEP {\bf 9907}, 021 (1999)
  [\href{http://arxiv.org/pdf/hep-th/9906011}{arXiv:hep-th/9906011}].

\bibitem{Dorey 2000}
  N.~Dorey,
  ``Instantons, compactification and S-duality in N=4 SUSY Yang--Mills theory. I,''
  JHEP {\bf 0104}, 008 (2001)
  [\href{http://arxiv.org/pdf/hep-th/0010115}{arXiv:hep-th/0010115}].

\bibitem{Dorey Parnachev}
  N.~Dorey and A.~Parnachev,
  ``Instantons, compactification and S-duality in N=4 SUSY Yang--Mills theory. II,''
  JHEP {\bf 0108}, 059 (2001)
  [\href{http://arxiv.org/pdf/hep-th/0011202}{arXiv:hep-th/0011202}].

\bibitem{DTV}
  N.~Dorey, D.~Tong and S.~Vandoren,
  ``Instanton effects in three-dimensional supersymmetric gauge theories with matter,''
  JHEP {\bf 9804}, 005 (1998)
  [\href{http://arxiv.org/pdf/hep-th/9803065}{arXiv:hep-th/9803065}].

\bibitem{DHK}
  N.~Dorey, T.~J.~Hollowood and V.~V.~Khoze,
  ``Notes on soliton bound-state problems in gauge theory and string theory,''
  \href{http://arxiv.org/pdf/hep-th/0105090}{arXiv:hep-th/0105090}.

\bibitem{BHO}
  J.~de~Boer, K.~Hori and Y.~Oz,
  ``Dynamics of N=2 supersymmetric gauge theories in three dimensions,''
  Nucl.\ Phys.\  B {\bf 500}, 163 (1997)
  [\href{http://arxiv.org/pdf/hep-th/9703100}{arXiv:hep-th/9703100}].

\bibitem{Harvey}
  J.~A.~Harvey,
  ``Magnetic monopoles, duality, and supersymmetry,''
  \href{http://arxiv.org/pdf/hep-th/9603086}{arXiv:hep-th/9603086}.

\bibitem{Hanany Witten}
  A.~Hanany and E.~Witten,
  ``Type IIB superstrings, BPS monopoles, and three-dimensional gauge dynamics,''
  Nucl.\ Phys.\  B {\bf 492}, 152 (1997)
  [\href{http://arxiv.org/pdf/hep-th/9611230}{arXiv:hep-th/9611230}].

\bibitem{BHOOY}
  J.~de Boer, K.~Hori, H.~Ooguri, Y.~Oz and Z.~Yin,
  ``Mirror symmetry in three-dimensional gauge theories, SL(2,Z) and D-brane moduli spaces,''
  Nucl.\ Phys.\  B {\bf 493}, 148 (1997)
  [\href{http://arxiv.org/pdf/hep-th/9612131}{arXiv:hep-th/9612131}].

\bibitem{DHIZ}
  O.~DeWolfe, T.~Hauer, A.~Iqbal and B.~Zwiebach,
  ``Constraints on the BPS spectrum of N=2, D=4 theories with A-D-E flavor symmetry,''
  Nucl.\ Phys.\  B {\bf 534}, 261 (1998)
  [\href{http://arxiv.org/pdf/hep-th/9805220}{arXiv:hep-th/9805220}].

\bibitem{Gauntlett Harvey}
  J.~P.~Gauntlett and J.~A.~Harvey,
  ``S duality and the dyon spectrum in N=2 superYang-Mills theory,''
  Nucl.\ Phys.\  B {\bf 463}, 287 (1996)
  [\href{http://arxiv.org/pdf/hep-th/9508156}{arXiv:hep-th/9508156}].

\bibitem{Gauntlett Kim Lee Yi}
  J.~P.~Gauntlett, C.~j.~Kim, K.~M.~Lee and P.~Yi,
  ``General low-energy dynamics of supersymmetric monopoles,''
  Phys.\ Rev.\  D {\bf 63}, 065020 (2001)
  [\href{http://arxiv.org/pdf/hep-th/0008031}{arXiv:hep-th/0008031}].

\bibitem{Lee Yi}
  K.~M.~Lee and P.~Yi,
  ``Monopoles and instantons on partially compactified D-branes,''
  Phys.\ Rev.\  D {\bf 56}, 3711 (1997)
  [\href{http://arxiv.org/pdf/hep-th/9702107}{arXiv:hep-th/9702107}].

\bibitem{Dimofte Gukov Soibelman}
  T.~Dimofte, S.~Gukov and Y.~Soibelman,
  ``Quantum wall crossing in N=2 gauge theories,''
  Lett.\ Math.\ Phys.\  {\bf 95}, 1 (2011)
  [\href{http://arxiv.org/pdf/0912.1346}{arXiv:0912.1346 [hep-th]}].

\bibitem{Alday Maldacena}
  L.~F.~Alday and J.~Maldacena,
  ``Null polygonal Wilson loops and minimal surfaces in anti-de-Sitter space,''
  JHEP {\bf 0911}, 082 (2009)
  [\href{http://arxiv.org/pdf/0904.0663}{arXiv:0904.0663 [hep-th]}].

\bibitem{Gaiotto}
  D.~Gaiotto,
  ``N=2 dualities,''
  [\href{http://arxiv.org/pdf/0904.2715}{arXiv:0904.2715 [hep-th]}].

\bibitem{GKPY}
  J.~P.~Gauntlett, N.~Kim, J.~Park and P.~Yi,
  ``Monopole dynamics and BPS dyons N=2 super Yang--Mills theories,''
  Phys.\ Rev.\  D {\bf 61}, 125012 (2000)
  [\href{http://arxiv.org/pdf/hep-th/9912082}{arXiv:hep-th/9912082}].

\bibitem{Taylor}
  B.~J.~Taylor,
  ``On the strong-coupling spectrum of pure SU(3) Seiberg--Witten theory,''
  JHEP {\bf 0108}, 031 (2001)
  [\href{http://arxiv.org/pdf/hep-th/0107016}{arXiv:hep-th/0107016}].

\bibitem{Taylor 2}
  B.~J.~Taylor,
  ``On the moduli space of SU(3) Seiberg--Witten theory with matter,''
  JHEP {\bf 0212}, 040 (2002)
  [\href{http://arxiv.org/pdf/hep-th/0211086}{arXiv:hep-th/0211086}].

\bibitem{FP}
  D.~Finnell and P.~Pouliot,
  ``Instanton calculations versus exact results in four-dimensional SUSY gauge theories,''
  Nucl.\ Phys.\  B {\bf 453}, 225 (1995)
  [\href{http://arxiv.org/pdf/hep-th/9503115}{arXiv:hep-th/9503115}].

\bibitem{Nekrasov}
  N.~A.~Nekrasov,
  ``Seiberg--Witten prepotential from instanton counting,''
  Adv.\ Theor.\ Math.\ Phys.\  {\bf 7}, 831 (2004)
  [\href{http://arxiv.org/pdf/hep-th/0206161}{arXiv:hep-th/0206161}].

\bibitem{Zamolodchikov}
  A.~B.~Zamolodchikov,
  ``Thermodynamic Bethe ansatz in relativistic models. Scaling three state potts and Lee--Yang models,''
  Nucl.\ Phys.\  B {\bf 342}, 695 (1990).

\bibitem{Alexandrov Roche}
  S.~Alexandrov and P.~Roche,
  ``TBA for non-perturbative moduli spaces,''
  JHEP {\bf 1006}, 066 (2010)
  [\href{http://arxiv.org/pdf/1003.3964}{arXiv:1003.3964 [hep-th]}].

\bibitem{HKLR}
  N.~J.~Hitchin, A.~Karlhede, U.~Lindstrom and M.~Rocek,
  ``Hyperkahler metrics and supersymmetry,''
  Commun.\ Math.\ Phys.\  {\bf 108}, 535 (1987).

\bibitem{Ivanov Rocek}
  I.~T.~Ivanov and M.~Rocek,
  ``Supersymmetric sigma models, twistors, and the Atiyah--Hitchin metric,''
  Commun.\ Math.\ Phys.\  {\bf 182}, 291 (1996)
  [\href{http://arxiv.org/pdf/hep-th/9512075}{arXiv:hep-th/9512075}].

\bibitem{Seiberg Shenker}
N.~Seiberg and S.~H.~Shenker,
  ``Hypermultiplet moduli space and string compactification to three dimensions,''
  Phys.\ Lett.\  B {\bf 388}, 521 (1996)
  [\href{http://arxiv.org/pdf/hep-th/9608086}{arXiv:hep-th/9608086}].

\bibitem{Fraser Tong}
  C.~Fraser and D.~Tong,
  ``Instantons, three dimensional gauge theories and monopole moduli  spaces,''
  Phys.\ Rev.\  D {\bf 58}, 085001 (1998)
  [\href{http://arxiv.org/pdf/hep-th/9710098}{arXiv:hep-th/9710098}].

\bibitem{Tomboulis Woo}
  E.~Tomboulis and G.~Woo,
  ``Semiclassical quantization for gauge theories,''
  Nucl.\ Phys.\  B {\bf 107}, 221 (1976).

\bibitem{Gradshteyn Ryzhik}
  I.~S.~Gradshteyn and I.~M.~Ryzhik,
  ``Table of integrals, series, and products,''
  Academic Press.

\bibitem{Atiyah Hitchin}
  M.~F.~Atiyah and N.~J.~Hitchin,
  ``The geometry and dynamics of magnetic monopoles. M.~B.~Porter lectures,''
  Princeton University Press, Princeton, USA (1988).

\bibitem{Witten phases}
  E.~Witten,
  ``Phases of N=2 theories in two dimensions,''
  Nucl.\ Phys.\  B {\bf 403}, 159 (1993)
  [\href{http://arxiv.org/pdf/hep-th/9301042}{arXiv:hep-th/9301042}].

\bibitem{Dorey 1998}
  N.~Dorey,
  ``The BPS spectra of two-dimensional supersymmetric gauge theories with twisted mass terms,''
  JHEP {\bf 9811}, 005 (1998)
  [\href{http://arxiv.org/pdf/hep-th/9806056}{arXiv:hep-th/9806056}].

\bibitem{DHT}
  N.~Dorey, T.~J.~Hollowood and D.~Tong,
  ``The BPS spectra of gauge theories in two and four dimensions,''
  JHEP {\bf 9905}, 006 (1999)
  [\href{http://arxiv.org/pdf/hep-th/9902134}{arXiv:hep-th/9902134]}.

\bibitem{Hanany Hori}
  A.~Hanany and K.~Hori,
  ``Branes and N=2 theories in two dimensions,''
  Nucl.\ Phys.\  B {\bf 513}, 119 (1998)
  [\href{http://arxiv.org/pdf/hep-th/9707192}{arXiv:hep-th/9707192}].

\bibitem{Hanany Tong}
  A.~Hanany and D.~Tong,
  ``Vortices, instantons and branes,''
  JHEP {\bf 0307}, 037 (2003)
  [\href{http://arxiv.org/pdf/hep-th/0306150}{arXiv:hep-th/0306150}].

\bibitem{Hanany Tong 2}
  A.~Hanany and D.~Tong,
  ``Vortex strings and four-dimensional gauge dynamics,''
  JHEP {\bf 0404}, 066 (2004)
  [\href{http://arxiv.org/pdf/hep-th/0403158}{arXiv:hep-th/0403158}].

\bibitem{BSY}
  P.~A.~Bolokhov, M.~Shifman and A.~Yung,
  ``BPS spectrum of supersymmetric CP(N$-$1) theory with Z$\rm _N$ twisted masses,''
  Phys.\ Rev.\  {\bf D84}, 085004 (2011)
  [\href{http://arxiv.org/pdf/1104.5241}{arXiv:1104.5241 [hep-th]}].

\bibitem{Veneziano Yankielowicz}
  G.~Veneziano and S.~Yankielowicz,
  ``An effective lagrangian for the pure N=1 supersymmetric Yang--Mills theory,''
  Phys.\ Lett.\  B {\bf 113}, 231 (1982).

\bibitem{Hori Vafa}
  K.~Hori and C.~Vafa,
  ``Mirror symmetry,''
  \href{http://arxiv.org/pdf/hep-th/0002222}{arXiv:hep-th/0002222}.
  
\bibitem{DDDS}
  A.~D'Adda, A.~C.~Davis, P.~Di~Vecchia and P.~Salomonson,
  ``An effective action for the supersymmetric CP(n$-$1) model,''
  Nucl.\ Phys.\  B {\bf 222}, 45 (1983).

\bibitem{Argyres Douglas}
  P.~C.~Argyres and M.~R.~Douglas,
  ``New phenomena in SU(3) supersymmetric gauge theory,''
  Nucl.\ Phys.\  B {\bf 448}, 93 (1995)
  [\href{http://arxiv.org/pdf/hep-th/9505062}{arXiv:hep-th/9505062}].

\bibitem{Olmez Shifman}
  S.~Olmez and M.~Shifman,
  ``Curves of marginal stability in two-dimensional CP(N$-$1) models with Z$\rm _N$-symmetric twisted masses,''
  J.\ Phys.\ A  {\bf 40}, 11151 (2007)
  [\href{http://arxiv.org/pdf/hep-th/0703149}{arXiv:hep-th/0703149}].

\bibitem{Shifman Vainshtein Zwicky}
  M.~Shifman, A.~Vainshtein and R.~Zwicky,
  ``Central charge anomalies in 2D sigma models with twisted mass,''
  J.\ Phys.\ A  {\bf 39}, 13005 (2006)
  [\href{http://arxiv.org/pdf/hep-th/0602004}{arXiv:hep-th/0602004}].

\paragraph{}

\bibitem{CDP}
  H.-Y.~Chen, N.~Dorey and K.~Petunin,
  ``Wall crossing and instantons in compactified gauge theory.''

\bibitem{CP}
  H.-Y.~Chen and K.~Petunin,
  ``Notes on wall crossing and instanton in compactified gauge theory with matter.''

\bibitem{CDP2}
  H.-Y.~Chen, N.~Dorey and K.~Petunin,
  ``Moduli space and wall-crossing formulae in higher-rank gauge theories.''

\bibitem{DP}
  N.~Dorey and K.~Petunin,
  ``On the BPS spectrum at the root of the Higgs branch.''

\end{thebibliography}
\end{document}